\documentclass[chap,letterpaper,10pt]{thesis}

\usepackage{mathtools}   
\usepackage{amssymb}
\usepackage{amsmath}
\usepackage{geometry}
\usepackage{amsfonts}
\usepackage{tikz}
\usetikzlibrary{arrows.meta,positioning,shapes.geometric}

\usepackage{graphicx}
\usepackage{subcaption}
\usepackage{xcolor}
\usepackage{listings}
\usepackage{float}

\usepackage{braket}

\usepackage{mdframed}
\usepackage{epigraph}
\usepackage[toc]{appendix}
\usepackage{url}       
\usepackage[T1]{fontenc}
\usepackage[side,ragged]{footmisc}

\lstdefinestyle{matlabnoframe}{
  language=Matlab,
  basicstyle=\ttfamily\footnotesize,
  columns=fullflexible,
  keepspaces=true,
  upquote=true,
  showstringspaces=false,
  breaklines=true,
  breakatwhitespace=true,
  numbers=left,
  numberstyle=\tiny,
  numbersep=8pt,
  xleftmargin=1.5em
}

\newmdenv[
  topline=true,
  bottomline=true,
  leftline=false,
  rightline=false,
  linewidth=0.5pt,
  skipabove=1.0\baselineskip,
  skipbelow=1.0\baselineskip,
  innertopmargin=6pt,
  innerbottommargin=6pt
]{apsrulebox}

\usepackage{hyperref}
\hypersetup{
  colorlinks=true,
  linkcolor=blue!50!black,
  citecolor=blue!50!black,
  urlcolor=blue!50!black,
  pdfauthor={Your Name},
  pdftitle={Your Thesis Title}
}

\definecolor{grey}{gray}{0.5}
\usepackage{fancyhdr} 
\usepackage[toc, acronym]{glossaries}

\newcommand{\ignore}[1]{}

\makeglossaries
\makeglossaries

\newglossaryentry{eregex}{name={eregex},
description={An extended regular expression, i.e. a regular expression with the addition of backreferences}}

\newacronym{SRE}{SRE}{Synchronized Regular Expression}

\makeatletter

\renewcommand{\@makechapterhead}[1]{%
  \vspace*{24pt}%
  {%
    \parindent\z@
    \raggedright
    \normalfont

    {\fontsize{10.5pt}{13pt}\selectfont
     \bfseries
     \MakeUppercase{\chaptername}\space\thechapter\par}

    \vskip 7pt

    {\fontsize{17pt}{20pt}\selectfont
     \bfseries
     #1\par}

    \nobreak
    \vskip 24pt
  }%
}

\renewcommand{\@makeschapterhead}[1]{%
  \vspace*{24pt}%
  {%
    \parindent\z@
    \raggedright
    \normalfont

    {\fontsize{17pt}{20pt}\selectfont
     \bfseries
     #1\par}

    \nobreak
    \vskip 24pt
  }%
}

\renewcommand{\section}{%
  \@startsection
    {section}%
    {1}%
    {\z@}%
    {-2.8ex plus -0.8ex minus -0.2ex}%
    {1.3ex plus 0.2ex}%
    {\normalfont\fontsize{13.5pt}{16pt}\selectfont\bfseries}%
}

\renewcommand{\subsection}{%
  \@startsection
    {subsection}%
    {2}%
    {\z@}%
    {-2.3ex plus -0.6ex minus -0.2ex}%
    {1.0ex plus 0.2ex}%
    {\normalfont\fontsize{11.5pt}{14pt}\selectfont\bfseries}%
}

\renewcommand{\subsubsection}{%
  \@startsection
    {subsubsection}%
    {3}%
    {\z@}%
    {-2.0ex plus -0.5ex minus -0.2ex}%
    {0.8ex plus 0.2ex}%
    {\normalfont\fontsize{10.5pt}{13pt}\selectfont\bfseries}%
}

\makeatother


\begin{document}








\thispagestyle{empty}

\clearpage
\newgeometry{
  left=0.80in,
  right=0.80in,
  top=0.78in,
  bottom=0.80in
}

\thispagestyle{empty}

\begingroup
\raggedright

{\fontsize{16pt}{19pt}\selectfont
Varun Immanuel\par}

\vspace*{1.55in}

{\fontsize{32pt}{38pt}\selectfont
 Quantum Processes\\
Under Epistemic Constraints\par}

\vspace{0.35in}

{\fontsize{15pt}{19pt}\selectfont\itshape
New Directions in the Foundations of Quantum Theory\par}

\vfill

\endgroup

\clearpage

\thispagestyle{empty}
\begingroup
\raggedright

\vspace*{\fill}

{\fontsize{10.5pt}{14pt}\selectfont

Copyright \textcopyright\ 2026

\vspace{1.25em}

Varun I. Premkumar Immanuel

\vspace{1.25em}

This work is licensed under the Creative Commons
Attribution 4.0 International License (CC BY 4.0).

\vspace{1.8em}

An arXiv version of a dissertation submitted to the
College of Arts and Sciences, Department of Physics,
University at Albany, State University of New York,
in partial fulfillment of the requirements for the degree
of Doctor of Philosophy. Originally published under the title ``On Quantum Processes and the Epistemic Constraints:
Some New Directions in the Foundations of Quantum Mechanics'' in the Scholar's Archive of the University at Albany, on August 4th, 2026.

\vspace{1.25em}

Summer 2026

}

\vspace*{1.25in}

\endgroup

\restoregeometry
\clearpage

\setstretch{1.00}
\setlength{\parskip}{0.05em}

\pagenumbering{roman}
\setcounter{page}{2}
\pagestyle{plain}

\tableofcontents   
\specialhead{Common Audience Abstract}
This doctoral dissertation on the foundations of quantum theory isolates and then formalizes a physically relevant concept that I have called “Epistemic Constraint.” Here, epistemic constraints are the definite, intersubjectively agreeable, ordinary-language conditions under which experiments are described.

The usual formulation of the quantum measurement problem, which I call the Schrodingerian measurement problem, has the structure of an anomaly: if we take quantum theory at face value, we expect no definite values, and yet we see definite values in experiments. The responses to this problem have been either to solve it or to dissolve it. These responses, which have taken the form of interpretation, modification, or reconstruction of quantum mechanics, seek either to derive (conceptually or mathematically) epistemic constraints from within quantum mechanics, as is the case with certain interpretations and modifications, or to posit the epistemic constraint, or parts of it, as a primitive assumption with the goal of deriving quantum mechanics, as is the case in some reconstruction programs.

In contrast to the Schrodingerian measurement problem, which had the structure of an anomaly, this dissertation develops the Bohrian Program, which (for lack of a better comparison) has a structure similar to the problem historically associated with Euclid’s fifth postulate. It seeks to keep epistemic constraints as primitive in an onto-epistemic sense. It then seeks new physical conclusions from the joint consideration of quantum mechanics and epistemic constraints, without seeking to derive one from the other. Among other results, this leads to a notion of the probability of instantiability of the Born Rule that specifies when to apply the Born Rule and when to apply a unitary transformation to a quantum state.

 
\specialhead{Technical Abstract}
The fact that we always observe definite outcomes in experiments, and that the definiteness of occurrences of events is inter-subjectively agreeable among all inter-communicating observers, can be called epistemic constraints. Epistemic constraints are not the same as any particular classical physics; they concern the ordinary language description of what was done in a lab and what was found (a common phrase attributed to Bohr). 

Quantum theory was discovered to replace the classical description of matter and energy, leaving it as an approximation. It appears that it did not replace the need for epistemic constraints. It seems it did not even render it ``approximate''. 

One can argue that epistemic constraints concerning ordinary-language descriptions of experiments have remained as intact today as they were for a physicist in, say, 1690 or in 1860. 

The motivating question is then, ``What can we conclude from the observation that classical mechanics is no longer fundamental, while epistemic constraints, especially the ordinary-language descriptions of experiments, remain indispensable?'' This question contrasts with the question raised by the familiar formulation of the quantum measurement problem, which demands an explanation of the definiteness of experimental outcomes within unitary quantum mechanics. We can call this latter question the Schrodingerian measurement problem to emphasize its historical origin with Schrodinger's analysis of the measurement process in 1935 (though the measurement problem was coined by Pascual Jordan in 1949). 

Epistemic constraints, therefore, appear to be another pan-theoretic restriction, alongside those imposed by the space-time transformation laws and the conservation laws. Pan-theoretic constraints are physically significant constraints that must be obeyed by any physically possible equation of motion.

This state of affairs suggests a rearrangement of the measurement problem in quantum mechanics, in which the assumptions and explanatory goals are reassessed. 

In this thesis, we develop the Bohrian program as a rearrangement of the Schrodingerian measurement problem. The Bohrian program seeks to keep quantum theory as it is, treats epistemic constraints as an independent assumption, and leaves the explanatory goals open-ended, seeking the physical conclusions that follow from the joint assumption of quantum theory and epistemic constraints. 

This treatment of the newly identified pan-theoretic constraint aligns with how, in the past, we sought new physical conclusions from the joint consideration of quantum theory and other well-known pan-theoretic constraints. Two familiar examples include the joint consideration of quantum theory and spacetime transformation laws, which resulted in relativistic quantum mechanics, and the joint consideration of quantum theory and conservation laws, which resulted in the discovery of selection rules.

The Schrodingerian measurement problem is a problem to be solved or dissolved. The Schrodingerian measurement problem can be thought of as having generated responses in the form of three agendas of research in the literature: interpret quantum theory, modify quantum theory, or reconstruct quantum theory.

In contrast, the Bohrian program is a program to be pursued. Its pursuit here can be viewed as a fourth agenda of research in the foundations of quantum mechanics.

Some new developments that follow are: a) (in Chapter 2) the notion of probability of instantiability of the Born Rule that specifies when to apply the Born Rule and when to apply unitary transformation to a quantum state; b) (in Chapter 7) a new definition of ensembles in Statistical Mechanics; c) (in Chapter 4) the consideration of macroscopic superpositions of agents; and finally, d) (in Chapter 6 and 7) the scenario in certain many-component quantum systems where the components undergo a probabilistic application of unitary evolution or a Born state update. Due to this process, there arises stochasticity in two respects: in the amount of entanglement between components, and in the evolution of the state, which can be assigned to a component at any given time.

As such, the present work is an attempt to connect quantum theory with the theory of meaning-containing information, as an extension of the ongoing attempts to connect quantum theory with Shannon's information theory, which is agnostic to meaning.

 
\specialhead{Acknowledgment}


I am very grateful to my committee for their feedback and encouragement. In particular, I would like to thank Professor Oleg Lunin for the interactions during my graduate school years that provided direction in my academic career. He was very kind to schedule Zoom meetings on several occasions when I needed career advice. 

I thank Professor Daniel Robbins for fruitful discussions on the mathematical aspects of my project before my November 2025 talk. His mathematical precision helped me to adopt the method of considering problems with simple, clear examples before generalizing.

I thank Professor Bradley Armour-Garb for allowing me to audit his Philosophy of Language class, which informed many of the ideas in this project. I also thank Professor Armour-Garb for his guidance and support during some academic hurdles in the past. 

I am grateful to Dr. Michael Cuffaro for taking the time to read my thesis with extraordinary care and precision, and for providing insightful comments and feedback. The conversations with him during the 2023 Graz Quantum Reconstruction Conference motivated me to partake in the exciting field of Quantum Foundations.

I am grateful to Professor Mathew Szydagis for the many big-picture questions about my project that further informed my perspective and clarified aspects that had previously been less clear. 

I am thankful to Professor Carlo Cafaro for his comments and feedback on my talk last November. 

I would like to thank the two important friends I have encountered over the years who have influenced my mind and heart. Special thanks to Raghav, my friend during my undergraduate years; the countless nights we spent on the bench overlooking the national highway, discussing physics, and watching the late-night traffic are still vivid in my memories. Special thanks to Felipe for the many nighttime walks, conversations, and warm support during times of distress when my father was ailing.

I am indebted to Professor Rongwei Yang, without whose supervision this dissertation would not have been possible. His openness to letting me pursue questions in science with careful attention and conceptual nuance has been a blessing. I will forever be grateful to Professor Yang for fostering in me the greatest virtue in scientific research: freedom of thought while maintaining rigor and epistemic discipline as much as possible. Professor Yang's frequent advice, ``Enjoy your research. That's more important than credentials,'' was very encouraging and inspiring.

I'm grateful to my middle school physics tuition teacher, Aruna, for inculcating in me an appreciation of the lucidity of Physics, and to my Quantum Mechanics Professor during my undergraduate years at Loyola College, Dr. Ravindhran, who showed me that solving maths problems is not as petrifying as it first seemed to me and showed\textbf{ }me how to do laboratory experiments critically.

I would like to thank the professors---Professor Herbert Fotso, Professor Ariel Caticha, and Professor Philip Goyal---under whom I apprenticed, gathering tools, skills, and ways of thinking from different fields.

I'm grateful to the UAlbany Physics Department for permitting me to investigate my own research questions. 

I'd like to thank my parents for their gift of life, love, and financial support and accommodation during the final year of my doctoral studies. I thank my father, Rev. Dr. Premkumar Immanuel Clement, for inculcating the love of Philosophy and history when I was a little boy, and my mother, Mrs. Sujatha Pauline, for teaching me soft skills. I'd finally like to thank my sister Mona and my brother-in-law Jai for their love and warmth, and my little but already imaginative niece Gia, for reminding me of my own wonder-filled, carefree, and curious childhood in the early 2000s, when the world appeared to me the way it appears to her now: constantly mysterious.

\listoffigures

\chapter*{A Note on the Meaning of Language}
\addcontentsline{toc}{chapter}{A Note on the Meaning of Language}

Every time I use the word `language', I assume a symbolic system used by a community of observers\footnote{A detailed definition of an observer would be provided in the thesis. For now, note that it need not be a human, but must be capable of language use. We will see that when considering quantum experiments, there is a unique and universal community of observers} to represent and communicate states of affairs. The symbolic system is defined as a collection of discrete symbols (such as $a, b, c, d, \ldots$), with a syntax that specifies which symbol combinations are well-formed, as opposed to a grammatically incorrect combination, and a semantics that specifies how well-formed sentences are interpreted by the community of observers. An aspect of ``interpretation'' relevant to us is the mapping of well-formed sentences to other well-formed sentences, which can include sentences that describe experimental states of affairs or interventions. An example would be the mapping between the sentence ``the spin experiment in the $Z$ direction has found the $\frac{\hbar}{2}$ eigenvalue,'' and the sentence ``the eigenstate of the spin was found to be $|S_z +>$,'' for example.

Allowed sentences in a language are not merely ``information-containing'', but are also meaningful, in the sense of being interpretable, and must rule out possible states of affairs.

This is sufficient for our purpose because, when we specify the measurement of the spin of an electron, for example, on the one hand, we have the problem statement describable as a sequence of sentences concerning the quantum states, the Hilbert space, and the relevant operators. On the other hand, we have the descriptive account of the experiment, which is also presented as a sequence of sentences, but in ordinary language, concerning what interventions were done. Meaning is then established between the two groups of sentences, according to the physical interpretation of the mathematical theory. 

The particular language of choice is, of course, immaterial, as long as the meaning is invariant across language choices. We can have different language choices on either side. On the mathematical side, it is immaterial whether we choose the path-integral approach, matrix mechanics, or the phase space formulation; and on the experimental description side, it is immaterial whether we use English or Sumerian, for example. The ``physical meaning'' of an experiment is an invariant across all of these possible representational choices on either side, in a rough analogy to the invariance of scalar quantities across all coordinate transformations. Some of the arguments and methods developed in this dissertation will suggest that this is more than a mere analogy and is physically consequential.
\chapter*{Chapter Organization}
\addcontentsline{toc}{chapter}{Chapter Organization}

\vspace{1cm}

\begin{center}
\begin{tikzpicture}[
  node distance=10mm and 18mm,
  every node/.style={font=\scriptsize},
  startstop/.style={
    rectangle,
    rounded corners,
    draw,
    align=center,
    text width=34mm,
    minimum height=8mm,
    inner sep=2pt
  },
  process/.style={
    rectangle,
    draw,
    align=center,
    text width=36mm,
    minimum height=8mm,
    inner sep=2pt
  },
  decision/.style={
    diamond,
    aspect=1.6,
    draw,
    align=center,
    text width=34mm,
    inner sep=1.5pt,
    minimum height=9mm
  },
  arrow/.style={-Latex, thick}
]

\node (start) [startstop]
  {Background and Motivation\\
   (Chapter 0)};

\node (p1) [process, below=of start]
  {The Role of Epistemic Constraints in the Practice of Quantum Theory\\
   (Chapter 1)};

\node (p2) [process, below=of p1]
  {Formalizing Language Use in the Practice of Quantum Theory\\
   (Chapter 2)};

\node (p3) [decision, below=of p2]
  {A Synthesis of Bohr and Heisenberg\\
   (Chapter 4)};

\node (p4) [process, below=of p3]
  {Does Quantum Theory Support the Atomistic Conception of Matter?\\
   (Chapter 6)};

\node (p5) [process, below=of p4]
  {Implications for Statistical Physics and the Origin of Mechanics\\
   (Chapters 6 and 7)};

\node (p6) [process, below=of p5]
  {Conclusion\\
   (Chapter 8)};

\node (l1) [process, left=of p3]
  {A Useful Classification of Known Quantum Processes
   \\
   (Chapter 3)};

\node (r1) [process, right=of p3]
  {What Kind of Complex Many-Component Systems
   Can Occur in Quantum Processes?\\
   (Chapter 5)};

\draw [arrow] (start) -- (p1);
\draw [arrow] (p1) -- (p2);
\draw [arrow] (p2) -- (p3);
\draw [arrow] (p3) -- (p4);
\draw [arrow] (p4) -- (p5);
\draw [arrow] (p5) -- (p6);

\draw [arrow] (p2.west) -| (l1.north);
\draw [arrow] (p2.east) -| (r1.north);

\draw [arrow] (l1.south) |- (p4.west);
\draw [arrow] (r1.south) |- (p4.east);

\end{tikzpicture}
\end{center}

\clearpage
\pagenumbering{arabic}
\setcounter{page}{1}
\pagestyle{fancy}

\setcounter{chapter}{-1}
\chapter{Background and Motivation}
\vspace{1cm}
\begin{flushright}
\itshape
\begin{minipage}{0.7\textwidth}
`` Whenever we proceed from the known into the unknown we may hope to understand, but we may have to learn at the same time a new meaning of the word `understanding' ''  \cite{Heisenberg1958}.  — Werner Heisenberg.
\end{minipage}
\end{flushright}
\vspace{1cm}

Newton's method of \textit{hypotheses non fingo} that was discussed in his general scholium  \cite{NewtonPrincipia} as a means to understand the physical world can be interpreted as saying, `Make  premises as needed only based on directly observable regularities, in terms of operationally well-demarcated physical quantities, and reason based on those premises, while avoiding the introduction of any further speculations beyond those observable regularities and the logical deductions from those regularities.'

\textit{Hypotheses non fingo} is not to be interpreted as `deny reality beyond direct observation'---that would edge towards solipsism or idealism.

For maximal empirical discipline, one might desire to utilize the method of \textit{hypotheses non fingo} for the measurement problem.

{The problem is that, when looking at atomic experiments, the phrase `directly observable regularities' in the above interpretation of the motto of \textit{hypotheses non fingo} has come to refer to the regularities in the measuring devices themselves, which is why it often evokes the accusation of `instrumentalism!'}

{If this reference to the measuring device itself for obtaining observable regularities is a practical limitation, then that accusation is valid, and it would be no different from the well deserved critique of a 17th-century scientist denying the existence of Saturn existing `out there', saying, ``I only observe colored patches in a telescope; therefore Saturn itself as some giant spherical object with giant rings around it, existing in outer space is an unwarranted fiction.'' 

If, on the other hand, the reference to measuring devices themselves in atomic experiments is an in-principle limit, then that `instrumentalist' accusation is perhaps misplaced.}

{In this work, based on the available evidence for how we acquire knowledge in atomic experiments, as opposed to some speculative future experimentalist's unforeseen capabilities, I make the case that this limit is of the in-principle kind, due to inherent constraints on how knowledge-acquisition processes can occur in nature, adding to the list of other such constraints, such as the inherent natural constraints on how computers or clocks can function. 

Any knowledge-acquisition system in the universe, whether one that could have evolved naturally in living systems, such as individual humans, or one that can be engineered, such as robotic experimentalists, appears to possess inherent constraints on what kind of knowledge it can acquire and how it can acquire it. This can be stated as a principle: 

{\textsc{Principle 0:} Any observer in the universe, whether naturally evolved or engineered by naturally evolved observers, is limited to describing nature in a language such that the experimental propositions concerning what was done and what was found operationally satisfy the laws of Boolean logic, and the definiteness of the answer to what was done and what was found is also, in principle, definite with respect to any other observer, even before intercommunication.}

Here, ``experimental propositions'' concern ordinary language declarative sentences such as `the detector clicked' and `the bubble formed'. A Boolean Algebra can be defined over such propositions. By that term, I do not mean statements corresponding to projectors acting on Hilbert space; those are not Boolean, as was already studied by von Neumann and Garrett Birkhoff \cite{Birkhoff1936QuantumLogic}.

{This principle has a form similar to laws that prohibit certain operations or the construction of certain machines. Just as energy conservation laws prohibit perpetual motion machines, this principle could be thought of as prohibiting the construction of an observer whose operation is governed by unitary evolution only, leading either to observers who access ``parallel worlds'' \cite{Wallace2012}, or to many observers such that the definiteness of occurrences of events does not hold for all observers before intercommunication \cite{FuchsMerminSchack2014}.}

Without intending to disparage these views, we can view the Everettian picture, along with the Qbist picture, as providing the useful conceptual vocabulary for what cannot be the case in knowledge-acquisition processes, similar to how the perpetual motion machine provided the conceptual vocabulary for what cannot be the case in thermodynamic processes, thereby illuminating impossibility principles in nature in either case.

{In simpler terms, Principle 0 therefore rules out, within the present framework, the natural evolution of observers, or the engineerability of observers, who will report a non-Boolean experience, such as reporting that they accessed ``parallel worlds,'' or reporting that what was a definite occurrence with respect to one observer was in a quantum-coherent pure state with respect to another observer. The idea now is to solidify this principle with more precisely defined terms, and to follow through the physical conclusions that follow from such a principle.}

Before I end this brief introduction, to avoid the risk of any perceived gesturing toward the wrong picture that Saturn does not exist ``out there'' beyond the telescope-screen patch, I will state the following partially useful paragraph, which will make only partial sense; but if the reader prefers to revisit this paragraph after their gracious gesture of reading through the entire thesis, it will make better sense:

In the case of Saturn, it does exist ``out there'' beyond the colored patch in the telescope pointing at Saturn, because we can always imagine a counterfactual, fictitious knowledge-acquisition system (i.e an observer) in the local vicinity of Saturn, performing an alignment position measurement of Saturn's position. But a photon in a Mach--Zehnder interferometer cannot be considered to exist ``out there'' beyond the clicks of the photodetector at the end of the interferometer, because we cannot consistently position a knowledge-acquisition system in the local vicinity of the photon, performing an alignment position measurement of the photon's position. Doing so would affect the interference fringes.

It is important to note that, here, the non-existence of the photon before detection is not in the same sense as the non-existence of, say, a Hippopotamus in my room (I don't have a Hippopotamus in my room); the photon's non-existence before detection is a different sense of ``non-existence''. 

This different sense of non-existence means that the photon before detection in the interferometer does exist in one sense: if an arm of the interferometer were probed with a photodetector, we would hear a click with some probability. That is, it exists as a ``potentiality'' \cite{Heisenberg1958}. 

This asymmetry between Saturn and the photon is not about size alone, because the same situation concerning observer placement in the local vicinity can, in principle, apply to a large cannonball or even to Saturn itself, depending on the experimental context. The placement of such a fictitious observer in the local vicinity is determined by the experimental context, not by any parameter regime. 

\section{Recognizing the Bohrian Program}

It is arguable that the so-called ``measurement problem''—which asks how definite events arise from deterministic unitary quantum evolution—is not, in fact, a problem in need of resolution when quantum mechanics is interpreted along the lines advocated by Bohr \cite{Bohr1928,Bohr1949,Bohr1958}. On this view, the problem emerges only once one adopts the von Neumann measurement model, in which the measuring apparatus itself is treated quantum mechanically by assigning a wavefunction to the pointer of a measuring device—something that Bohr never endorsed \cite{vonNeumann1932,Bohr1949}.

To be clear, Bohr did allow the quantum formalism to be applicable to any physical system, but unlike the implicit assumption behind the von-Neumann measurement scheme, for Bohr, the cut between the system and the apparatus is determined by the experimental context  \cite{Heisenberg1958}, and is not to be moved at will as a modeling choice (that stance is associated with Heisenberg).

Nevertheless, even in the absence of such von Neumann assumptions, conceptual tensions remained within Bohr’s own framework. Bohr and Heisenberg acknowledged and expressed conceptual unease \cite{Bohr1958} \cite{Heisenberg1958}.

Crucially, the source of this unease was not what later came to be called the measurement problem. The measurement problem as we know it crystallized with Schrodinger \cite{Trimmer1980} and von Neumann's work and was further developed by Everett \cite{vonNeumann1932}, but was coined as early as 1949 by Jordan \cite{Jordan1949}. 

Rather, for Bohr and Heisenberg, the surprise that quantum mechanics presented lay elsewhere: the apparent necessity of ``classical concepts'' \cite{bohr1938causality} for describing experimental contexts, and the observation that physical knowledge is possible only through such means. This seemed to indicate a fundamental role for classical concepts in physical theory, even though classical mechanics itself was no longer regarded as fundamental \cite{Bohr1928,Bohr1949}\cite{heisenberg1971physics}. 

Here, classical concepts and classical mechanics are to be distinguished. Classical mechanics is a dynamical theory for mechanical objects, whereas classical concepts is a term used by Bohr to stress the unavoidability of ordinary means of experimental description. To avoid the wrong interpretation that stressing the fundamental role of classical concepts is a retreat to the pre-quantum classical mechanical picture, henceforth, I replace classical concepts by the term ``\textsc{epistemic constraints}''. 

Epistemic constraints refer to the fact that all experimental propositions satisfy semantic bivalence (can have one of two possible truth values); and the fact that the experimental events that definitely occurred with respect to one observer also definitely occurred with respect to any other observer.

Bohr’s and Heisenberg’s bewilderment concerned, in particular, how such a primitive status of epistemic constraints was to be understood or accommodated within the methodological framework inherited from pre-quantum approaches to developing and practicing physical theories \cite{Bohr1958}\cite{heisenberg1971physics}.

Bohr wrote extensively and corresponded to address the doubts of skeptics of quantum theory, consistently maintaining that the mathematical formalism of quantum mechanics was complete and internally consistent \cite{Bohr1949,Plotnitsky2012RealityObserverComplementarity}. Nevertheless, he and Heisenberg appear to have remained reflective and unsettled about what the continued necessity of epistemic constraints ultimately signified. 

The aim of this work is to revisit this original unease, as articulated by Bohr and Heisenberg, and to investigate what follows if the semantic and epistemological aspects they identified are taken to play a fundamental role in natural processes themselves, rather than being regarded merely as features of our descriptions of nature \cite{Bohr1928,Bohr1958,heisenberg1971physics}. This stance is guided by the principle that it is not physically meaningful to make claims about what entities the universe fundamentally contains beyond the limits set by the preconditions for empirical knowledge \cite{Bohr1958}.

We will explore the consequences of jointly keeping standard quantum theory and the non-reductionist empirical regularities \footnote{These are not regularities in the purely information-theoretic sense; that is, they are not some information-theoretic principle. This is because expressing these regularities requires semantic considerations, whereas Shannon information theory is agnostic to semantics. Rather, these are intended as regularities that are expressible in terms of tools from the newly developing semantic information theory, namely, the theory of meaning-containing informative sentences \textbf{ \cite{shao2025theory}.}} of how epistemic systems (the ``observers''—which we formalize later) use quantum theory. These are regularities concerning how observers communicate in ordinary language about their operational interventions. 

This is the Bohrian Program.

We do not take classical mechanics as primitive. Classical mechanics remains an effective theory whose origin must be explained jointly in terms of quantum theory and the principles concerning epistemic constraints. 

Moreover, this is not a restraint on the universality of quantum theory itself; quantum theory can still apply to any physical system, even to the physical substrate constituting an observer, but there are some inevitable consequences of doing so. See Chapter 4.

The Bohrian program has a tradition in physics, and the methodology is not brand new. There is a recognizable tradition where pan-theoretic constraints, which constrain any physically possible dynamics, such as space-time transformation laws or conservation laws, were successfully considered in combination with quantum theory. The Bohrian program seeks to follow that path for epistemic constraints, which are yet another pan-theoretic constraints. See Fig. 1.

\begin{figure*}
  \centering
  \includegraphics[
    width=\textwidth,
    alt={A schematic illustrating the plan for the Bohrian program to jointly consider quantum theory and epistemic constraints. This methodology, which considers a pan-theoretic constraint that constrains any physically possible dynamics and a dynamical theory, is familiar from how relativistic quantum theory and the selection rules were conceived.}
  ]{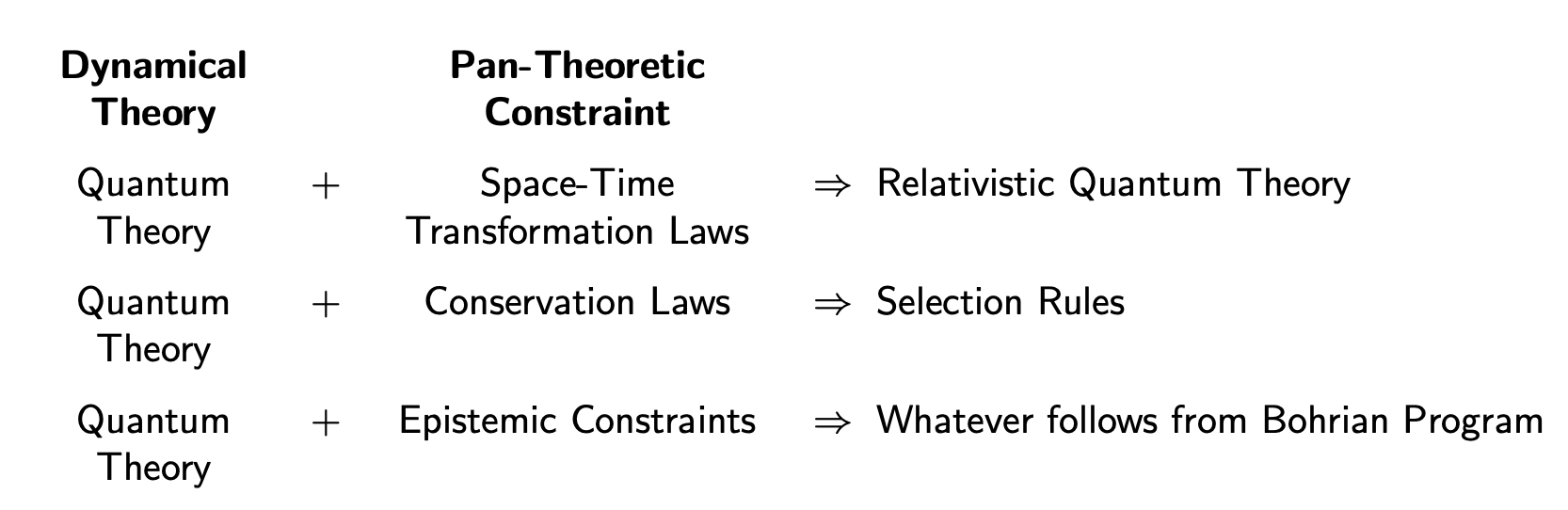}
  \caption{Bohrian Program follows a familiar tradition}
  \label{fig:born-rule-instantiability-scenario}
\end{figure*}

\section{A Panorama of the Conceptual Foundations of Quantum Measurements}

There are roughly two camps regarding how Bohr is received in the foundations of quantum mechanics. There is the Anti-Bohr camp, which either accuses Bohr of being vague and obscure \cite{bell1990againstmeasurement} and of having halted progress \cite{becker2018whatisreal}\cite{fuchs2019copenhagen}, or believes that Bohr gave up too soon and that we should strive for a realist picture of what is going on ``under the hood'' \cite{Bohm1952}\cite{deutsch1997fabric}. The second camp is what can be called Quietist Bohr, which believes that Bohr's lessons were epistemological in nature \cite{Osnaghi2022BohrEpistemologicalLesson} and that we have no new physics to search in the foundations of quantum theory. Qbism can also be, arguably, placed in this camp because it's an interpretation \cite{FuchsSchack2013} with the goal of reconstructing quantum mechanics, so at the operational level nothing changes. The stance developed here is closer to a third camp, which, to the best of our knowledge, is relatively underexplored: it posits that Bohr saw the right constraint for how we can know what we come to know about the world (the epistemic constraints), but that its constructive consequences for physics remain underdeveloped.

Let us soar a little bit higher to view the landscape more panoramically.

The classical mechanical worldview understood the totality of physical existence in terms of point particles, each with definite position and momentum, or field configurations with definite real-valued field variables specified at each point in a space--time coordinate chart \cite{NewtonPrincipia,LandauLifshitzFields}. In addition, there exist intrinsic properties such as mass and electric charge that are not reducible to position, momentum, or field variables \cite{LandauLifshitzFields,LandauLifshitzMechanics}. In the operationalist view (of Percy Bridgman) \cite{tal2015measurement}, a measurement was regarded as the passive act of comparing one of these properties of an object with a chosen, calibrated standard, referred to as the ``scale'' or the ``measuring device''.

In the classical mechanical worldview, the values read by the scale were assumed to be representative of properties that existed independently of the act of measurement. In other words, I risk no contradiction by assuming that a counterfactual observer, possessing abilities equivalent to my own, could access the values of these properties prior to my own experimental intervention on the same object, and that my subsequent intervention on that object would be such that I could not, even in principle, discern whether another observer had determined the object properties (including it's state) prior to my own measurement.

We take this criterion as the operational meaning of the phrase ``properties existing objectively'', or more colloquially ``properties existing out-there in the world''. Thus, when we say that tables, planets, and chairs exist objectively with definite classical mechanical properties, we invoke the following epistemic criterion: we may assume—without affecting our subsequent observations—that a collection of imaginary observers, equipped with capabilities like our own, could have measured the properties in question at an earlier time, or simultaneously.

Quantum mechanics makes demands on objects so that the set of all properties generally fails to satisfy the above criterion of objectivity \cite{Dirac1930}. For example, when a particle with spin is prepared in the state $\ket{+_z}$, an eigenstate of $\hat{S}_z$, we cannot, without encountering contradiction, assume that there exists a definite value for the observable $\hat{S}_x$; indeed, there exists no observer with respect to whom $\hat{S}_x$ possesses a definite value \cite{Dirac1930,vonNeumann1932}. Note that this ``non-objectiveness'' does not entail any subjectivity—it simply entails non-objectiveness in the sense we defined.

On the other hand, the results of completed quantum measurements—for example, when we have completed measuring the observable $\hat{S}_x$ on the state mentioned above— yield outcomes that satisfy the criterion for objectivity just defined. This gives rise to the well-known dichotomy: on the one hand, there are systems with properties that are objective, and on the other hand, there are systems with properties that are not objective, in the operational sense introduced above. The former kinds of systems are always required in order to describe the latter kind of systems \cite{vonNeumann1932,Bohr1928,Bohr1949}. Combined with the fact that classical mechanics is only an approximation, this leads, in some reading of this situation, to the familiar measurement problem in quantum mechanics \cite{Everett1957}.

The reasoning is as follows. Since classical mechanics is not fundamental, the objective aspect characteristic of classical mechanics must also be (supposedly) non-fundamental\footnote{Note that here classical mechanics and Bohr's epistemic constraints are fused together. We separate them in this thesis}. Therefore, the objective reality we observe must be explicable purely in terms of fundamentally non-objective quantum systems—or, more formally, must be derivable solely from the unitary Schrodinger dynamics of quantum states \cite{Everett1957,Wallace2012}. This line of reasoning motivates the ``unitary-only'' paradigms for addressing the measurement problem \cite{Everett1957,Wallace2012}.  This reasoning becomes less defensible once we observe that objectivity, as defined here, merely requires what Bohr called classical concepts (and we call ``epistemic constraints''), and is not synonymous with classical mechanics \cite{Bohr1928,Bohr1949}. Consequently, we argue, one can evade the motivation behind unitary-only approaches.

It should be noted that the measurement problem is not regarded as a ``problem'' by all researchers, although, the presence of conceptual unease is widely acknowledged, as evidenced by the now vibrant field of quantum foundations. Regardless of whether one takes the situation to constitute a problem, there is a broad recognition that quantum mechanics requires a dichotomy between objective existence of experimental outcomes, which we shall later call actuality, and the non-objective existence of the pre-measurement situation such as that of electrons in flight, which we shall later call potentiality. The implications of this stubborn dichotomy must be addressed \cite{Everett1957}.

There are three broad schools of thought in the foundations of quantum theory, corresponding to three classes of claims regarding what is more fundamental between systems with objective properties or systems with non-objective properties. 

The first class of approaches maintains that only systems with non-objective properties, the potentialities, are fundamental, and that the appearance of objective properties, the actualities, is only an apparent phenomenon. This view typically comes either from the supposed existence of inaccessible parallel worlds \cite{Everett1957,Wallace2012} or the supposed existence of relativity of facts \cite{Rovelli1996}\cite{FuchsSchack2013}; we will refer to this as \textsc{School of Thought~1}. The second class maintains the converse: that systems with non-objective properties, the potentialities, are an illusion, and that what fundamentally exists are entities with definite objective properties, the actualities, that are inaccessible and therefore function as hidden variables; we will refer to this as \textsc{School of Thought~2} \cite{Bohm1952}\cite{Caticha2019_EntropicDynamics}. The third class of approaches admits the fundamental coexistence of both systems with objective properties and systems with non-objective properties; we will refer to this as \textsc{School of Thought~3} \cite{Bohr1928,Bohr1949,GRW1986,BassiEtAl2013}.

Within \textsc{School of Thought~3}, there are approaches that introduce a mathematical criterion—typically by modifying the Schrodinger equation—to determine the conditions under which actualities emerges from potentialities. The objective-collapse models \cite{GRW1986,Pearle1989,GhirardiPearleRimini1990,BassiEtAl2013} belong to this sub-category within the \textsc{School of Thought~3}. 

Within the \textsc{School of Thought~3}, there are also approaches that posit an epistemic criterion for distinguishing between the actualities and potentialities, among which Niels Bohr’s  interpretation\textbf{ }is the paradigmatic example \cite{Bohr1928,Bohr1949}.

These three {school of thoughts} have, since the 1950s, catalyzed three\footnote{the number of schools of thoughts matching the number of agendas is coincidental} broad research agendas that aim to provide a definitive answer to the question: ``What theoretical response is called for by the dichotomy suggested by quantum theory—is it a problem to be resolved, a fundamental feature to be acknowledged and exploited in future physics, or a basis for some entirely different form of theoretical development?'' \cite{Everett1957,GRW1986,Hardy2001}

The first agenda consists of approaches that aim to provide an interpretation of quantum theory without modifying the mathematical formalism; we will refer to this as \textsc{Agenda~1} \cite{Everett1957,FuchsSchack2013,Rovelli1996,Griffiths1984}. The second consists of approaches that aim to modify the Schrodinger equation; we will refer to this as \textsc{Agenda~2} \cite{GRW1986,Pearle1989,GhirardiPearleRimini1990,BassiEtAl2013}. The third consists of the quantum reconstruction program, which aims neither to modify nor to interpret quantum theory, but instead to rederive its mathematical formalism from assumptions different from those used in its original formulation during the 1920s; we will refer to this as \textsc{Agenda~3} \cite{Hardy2001,Chiribella2011}. These three agendas represent distinct strategies for addressing, accommodating, or rethinking the conceptual basis of measurements in quantum theory  \cite{Hardy2001,Chiribella2011,BassiEtAl2013}.

In each of these agendas, one finds influences from more than one of the schools described in the previous paragraphs. For example, the quantum reconstruction program has been pursued on the basis of School of Thought~3 and School of Thought~2 \cite{Hardy2001,Chiribella2011} \cite{Caticha2019_EntropicDynamics}. The interpretation program has been developed from all three schools \cite{Everett1957,FuchsSchack2013,Rovelli1996} \cite{Caticha2019_EntropicDynamics}. The modification program, however, has relied exclusively on School of Thought~3 \cite{GRW1986,Pearle1989,GhirardiPearleRimini1990,BassiEtAl2013}.

The motivation for this thesis is to explore and develop a fourth research agenda, \textsc{Agenda~4}—which, to our knowledge, has not been systematically studied. Its central question is: ``How can we recognize and formalize the dichotomy as a fundamental feature, and how can we exploit it in future physics?'' In other words, we pursue the Bohrian program by seeking what conclusions follow from the joint assumption of usual textbook quantum theory \cite{Dirac1930} and Bohr's epistemic constraints taken as inherent features of nature. 

This agenda, by design, restricts itself to \textsc{School of Thought~3} noted earlier, and in particular assumes only an epistemic criterion for distinguishing between actualities and potentialities, in the spirit of Bohr \cite{Bohr1928,Bohr1949}. However, we do not confine ourselves to providing an interpretation of quantum theory; rather, we seek constructive physical consequences of the epistemic criterion in a direction that we believe was left as a cliffhanger in the foundations of quantum theory by Bohr \cite{Bohr1958}.

\chapter{Epistemic Constraints in the Practice of Quantum Theory}

The purpose of this small chapter is to highlight the experimental situations where, as a matter of fact, epistemic constraints are already exercised, even if we don't explicitly recognize them or call them as such.

These are the situations in which we invariably rely on definite experimental events occurring in localized regions at definite times, whose occurrence is intersubjectively agreed upon by all intercommunicating observers.

These situations arise when assigning and updating quantum states \cite{Bohr1958}, in determining what constitutes a measuring device, in determining which unitaries can be realized experimentally \cite{Drossel2017Connecting}, and in selecting a tensor-product factorization for a many-body system (including identical-particle systems \cite{lloyd2004observable}).  

Let's elucidate all this in a little bit more detail.

\section{The Five Situations where Epistemic Constraints are Unavoidable}

We refer to the dependence that the practice of quantum theory (hereafter abbreviated PQT) exhibits on situations that satisfy epistemic constraints by the umbrella term \emph{Epistemic-constraint Dependency} of quantum theory, which we henceforth abbreviate as \textsc{ECD}.

For instance, the update of a quantum state upon noticing a Geiger count click is an instance of an ECD; so is the counting of the time interval during which a quantum system evolved unitarily between preparation and detection; and the subsystem partitioning of a bipartite system based on the available experimental access.

We use the abbreviation QT for quantum theory per se, namely the Dirac-von Neumann postulates. PQT is the phenomenological description of how QT is practiced.

The ECD is often given a qualitative, plain-language description, and its formalization requires a linguistic framing.

It is instructive, before proceeding, to recall Nielsen and Chuang \cite{nielsen2010quantum} in their seminal text concerning what QT does and does not provide, and therefore why epistemic constraints are indispensable:

\begin{quote}
``Quantum mechanics does not tell us, for a given physical system, what the state space of that system is, nor does it tell us what the state vector of the system is. Figuring that out for a specific system is a difficult problem for which physicists have developed many intricate and beautiful rules.'' \cite{nielsen2010quantum}
\end{quote}

and further \cite{nielsen2010quantum}:

\begin{quote}
``Just as quantum mechanics does not tell us the state space or quantum state of a particular system, it does not tell us which unitary operators $\hat{U}$ describe real-world dynamics. Quantum mechanics merely assures us that the evolution of any closed system may be described in such a way, and that the time evolution of a closed quantum system is governed by the Schrodinger equation.'' \cite{nielsen2010quantum}
\end{quote}

This absence of clear prescriptions within QT for determining what the quantum state of an arbitrary system is, or whether such a state can be assigned at all\footnote{%
We do have the field of quantum-state tomography for determining the quantum state of quantum systems such as atoms, photons, and even molecules by performing measurements on an ensemble across multiple noncommuting bases and collecting statistics. However, quantum-state tomography does not tell us whether and how quantum states may be assigned to \emph{arbitrary} physical systems, such as a table.%
}, together with the lack of a clear prescription for experimentally realizable unitaries, constitutes two of the five tasks for which epistemic constraints become unavoidable, as far as all our current observations go. 

The third task for which QT alone does not appear to suffice, and which invariably invokes epistemic constraints, is in defining a ``measuring device'' and deciding the division between a measuring device and the measured system. There is no formal definition within QT for what physical system qualifies as a measuring device; which system $M$ induces the following transformation \cite{vonNeumann1932}

\begin{equation}
\ket{\psi} = \sum_n c_n \ket{n} \longrightarrow \ket{n},
\label{eq:collapse_projection}
\end{equation}

when the $\{\ket{n}\}$ basis is observed using system $M$ to study another system $S$ in state $\ket{\psi}$, is not specified by QT. 

Moreover, there is no formal criterion for when $M$ must be included in, or can be excluded from, the quantum-mechanical description\footnote{The diagonal form of the reduced density matrix in some basis is not such a criterion, since we can have such a situation also in cases where the subsystem is still part of a larger quantum system for which coherence is preserved. In such cases, it is incorrect to treat the subsystem as a measuring device, since the supposed `measurement' can be reversed.}. In the above example, $M$ was excluded.  but in other experiments, $M$ must be included, yielding, under the von Neumann scheme, an evolution

\begin{equation}
\ket{\psi} = \sum_n c_n \ket{n} \otimes \ket{M_0}
\longrightarrow
\sum_n c_n \ket{n}\ket{M_n},
\label{eq:von_neumann_measurement}
\end{equation}

where $\ket{M_0}$ is the ready state of the measuring device (the pointer), and ${\ket{M_n}}$ are orthonormal pointer states.

The fourth task for which QT alone does not suffice is in determining how a many component quantum system, potentially involving identical particles, is partitioned into subsystems \cite{lloyd2004observable}.  It has been shown in previous work \cite{lloyd2004observable} that sub-system partitioning is dictated by the set of operationally accessible measurements and interactions.

And finally, for counting time, and determining spatial intervals, we have epistemic constraints dependency as well  \cite{Bohr1958}. 

Our successful use of QT stems from intuition and long-established laboratory practice, which implicitly enforces the epistemic constraints. For instance, a widely used rule of thumb identified by Bell \cite{Bell2004SpeakableUnspeakable}, for deciding which physical systems in an experiment to include in the quantum-mechanical treatment, is the following:
\begin{quote}
``Put sufficiently much into the quantum system that the inclusion of more would not significantly alter practical predictions.'' \cite{Bell2004SpeakableUnspeakable}
\end{quote}
For example, in studying the hydrogen atom, one obtains incorrect predictions if the nucleus is excluded from the quantum-mechanical treatment. In contrast, in Schrodinger’s cat thought experiment, excluding the cat from the quantum-mechanical description leads to no contradiction with experience.

\section{Stances Toward Epistemic Constraints in other Interpretations of Quantum Mechanics}

In the existing interpretations of QT, ECD is handled in one of two main ways: (i) to explain the \textsc{ECD} from within unitary Schrodinger dynamics, often at the expense of interpretations that reject certain seemingly self-evident features of everyday experience, or (ii) to explain it by modifying or supplementing the Schrodinger equation. The first group includes the Many-Worlds Interpretation  \cite{Wallace2012}, QBism\footnote{QBism also treats experience as primitive, but in a different sense from the present framework. In QBism, phenomenal experience is definite for the individual agent, and thus one may say that it incorporates a first-person-level epistemic constraint. However, QBism does not impose the intersubjective definiteness constraint defended here: namely, that observers can reason about the definite outcomes other observers may have observed even before communication. The present framework therefore differs from QBism by assigning a non-perspectival structure to intersubjective definiteness.} \cite{FuchsSchack2013}, and Relational Quantum Mechanics \cite{Rovelli1996}, since each rejects, in different ways, either the inter-subjective agreeability of facts or the definiteness of events: the Many-Worlds Interpretation does so by introducing parallel branches of the universe, while QBism and Relational Quantum Mechanics do so by denying individual observer-independent event-definiteness. The second group contains objective-collapse theories, such as the GRW model \cite{GRW1986}, which modify the Schrodinger equation, and hidden-variable models, such as the de~Broglie–Bohm theory \cite{Bohm1952}, which retain the Schrodinger equation but introduce an additional guiding equation for particles.

In all of the approaches above, the primary goal is to explain the appearance of the \textsc{ECD} from within QT, or through suitable supplementation or modification of it; A related but distinct goal is to account for the emergence of classical mechanical behavior, either via decoherence or via the application of collapse models to many-particle systems with large number of degrees of freedom.

In the approach developed in this thesis however, the \textsc{ECD} is taken as a primitive starting point, alongside the primitive status of QT. The task is not to reduce the \textsc{ECD} to QT, but rather to mathematically integrate their interplay in search of new physical effects. This may be viewed as a continuation of where Bohr left off: the \textsc{ECD} is treated as a matter-of-fact feature of experimental practice that coexists with unitary Schrodinger dynamics, with the former preconditioning the latter. 

 Our experience of the world, as well as our most precise experiments, provides no indication of any failure of Schrodinger dynamics for closed systems. At the same time, there is no empirical evidence for a failure of epistemic constraints in physical occurrences: we lack even a single instance in which two or more experimentalists have disagreed on the definiteness of the occurrence of the same experimental outcome, although the numerical values corresponding to that outcome may differ among observers due to the spacetime transformation laws. A parsimonious accommodation of both of these experimental truisms—the robustness of quantum mechanics in its current form and the validity of epistemic constraints—may require recognizing both as independent primitives that cannot be reduced to one another and are jointly necessary for a complete description of the physical world. The joint consideration of QT and ECD is PQT. On the present proposal, a complete physical description involves the description of PQT.

\chapter{An Onto-Epistemic Consideration of Language and Communication in Physics}
\vspace{1cm}
\begin{flushright}
\itshape
\begin{minipage}{0.7\textwidth}
``A measurement can mean
nothing else than the unambiguous comparison of some property
of the object under investigation with a corresponding property of
another system'' \cite{bohr1938causality} —Niels Bohr, in The Causality Problem In Atomic
Physics.
\end{minipage}
\end{flushright}
\vspace{1cm}

Bohr's epistemic constraints concern language and communication, and we want to treat them as primitive. This is taken as a first step in formalizing PQT.

What does it even mean to make epistemic constraints, and thus language and communication, ontologically primitive? Obviously, language is not a ‘substance’; that would be a category error. Instead, we can define language with reference to functionally defined devices called ``observers''\footnote{an abstract device instantiated by a language-using automaton that can navigate an environment and perform actions on it: we shall systematize this soon}, analogous to how time is not a substance but is defined with reference to the functionally defined device that we call a ``clock''. 

A clock is a system with a sequence of reliably distinguishable events and a counting mechanism that counts the elapsed distinguishable events, such that the counting is monotonic. The clock needs to be stable, in the sense of not being significantly affected by environmental perturbations, and approximately uniform: equal increments of counts correspond to equal intervals of time parameter read from the clock. 

The idea now is this: \emph{just as a universe with time will have clocks as sensible entities, a universe with laws of nature will have observers as sensible entities}.

The construct of observers should allows us to make some of Bohr’s epistemological observations, ontologically relevant. We must systematize observers with this requirement in mind.

We avoid the term ``agents'', since it has recently come to assume a more generic definition \cite{Rovelli1996} involving the state of one system labeled `memory' and another system, labeled `system', such that the mutual information between the state of the system and the memory is used to characterize observation. Language is not part of this definition; Even an electron getting entangled with a photon could be considered an agent. 

But this generic definition appears to omit something crucial about observation. During the act of observation, we aren't merely copying information from the environment, we are also interpreting the meaning of the information acquired in a context sensitive manner. That's what is meant by ``acquiring knowledge about a system''. The establishment of a correlation between the measured and the measurer is necessary but not sufficient for knowledge acquisition. Meaning is the missing ingredient.

Shannon information theory is agnostic to meaning; it's easy to come up with many highly informative sentences that is meaningless (for example, ``cicnsdnc237e379138831'') . Therefore, semantic information theory \cite{shao2025theory}, which is a generalization of Shannon's information theory to describe meaningful representation and communication, is better suited to describe observations.

 Moreover, in any given situation involving propositional knowledge  acquisition about natural phenomena, language appears to be a precondition\footnote{it doesn't matter which particular language, though}. Therefore, we explicitly require language-using capabilities for the act of observation. 
 
 This leads us to the definition of `observers' as an abstract device capable of language use. The observer also needs embodied cognition, since it must be able to act on it's environment and describe it's own actions in some language.

An observer is a functionally defined unit constituted by a localized, embodied device capable of language. It must be able to provide an unambiguous plain-language description of an experiment by outputting a set of experimental propositions. The unit must possess internal clocks for co-ordination. It's information processing can be either classical or quantum, but the Input/Output channel must carry meaningful classical information.  An example would be a movable automaton capable of language use. 

Note that language in everyday communication is not unambiguous; however, for our purpose, we restrict attention to a stripped-down language that encodes experimental occurrences as factual yes/no responses. These outcomes are defined within a shared theoretical framework, which can be treated as operationally unique \footnote{i.e at the operational level, any viable theoretical framework for describing a class of physical phenomenon must yield the same operational predictions} for any given fundamental physical phenomena, even if expressible in multiple (empirically equivalent) mathematical formalisms.

 There is a symbolic system processed in an observer, where the symbolic system is constituted by two discrete sets of symbols (e.g., The set \{Alphabets a,b,c,d …along with Set $\mathbb{R}$\} on the one hand, and the set of  Hilbert space objects \{$|\psi\rangle_S $, $\mathcal{H}_S$ , $\hat{H}_S$, $\hat{U}$ …\}  on the other hand). The first set describes the maneuvers that were done by a movable automaton in a lab and the latter set describes the physical states prepared and state transformations effected by those laboratory maneuvers.

All this is not to anthropomorphize nature, but simply to recognize the only means of knowledge we know, and posit, due to a lack of evidence for any alternative form of knowledge acquisition, that we ourselves—individual humans, a network of communicating humans, or the language-using robots we built—are simply a few terrestrial instances of the more abstract devices capable of language use that we have called an observer.

The observers can evolve in living systems, or be engineered in non-biological media, and can function anywhere in the universe, counterfactually determining local conditions concerning the experimental contexts; just as the the Earth-Sun system can be thought of as one instance of a ``naturally occurring'' clock (if we imagine a counting mechanism to count revolutions), even though clocks are more abstract devices that can occur and function anywhere in the universe, counterfactually determining local temporal conditions of phenomena.

Positing general features for observers from our localized experience here on earth, as being applicable to any possible knowledge-acquiring system in the rest of the universe appears no different from positing that the laws of nature observed terrestrially here on earth have to apply to any region of the universe \cite{NewtonPrincipia}.

Note that this has some partial overlap with Karen Barad's onto-epistemology \cite{Barad2007} ---hence the title of this chapter--- and more broadly to the post-Cartesian \cite{WiltscheBerghofer2020} views. But unlike Barad,  we will soon see that an ontology is given for epistemology by introducing functionally defined units, which we call observers, similar to how one might give ontology to time by introducing functionally defined units of clocks.

With this understanding, we can state the two principles concerning observers (along with principle 0 we stated in Chapter 0):

 \begin{enumerate}
     \item \textsc{Principle 1:} The definiteness of the occurrence of an event specifiable by an experimental proposition $E$, defined as the truth value $T(E)\in \{0,1\}$, where $T(E)=1$ if the event occurred and $T(E)=0$ otherwise, is universally invariant across all observers. Here, an `event' is an individuated occurrence in a region of space that is alignment-position-measurable.

    \item \textsc{Principle 2:}  All natural phenomena, even in regions of the universe without actual observers, will be found to abide by Principles 1, with counterfactual, fictitious observers, when a later evolved observer studies the past and those regions of the universe without observers.

       \end{enumerate}

Principle 2 allows us to consider fictitious observers even in regions of the universe where no actual observers are present.  As illustrated in the prelude, the reasoning is the same as for the other functionally defined devices in physics, such as clocks or magnetometers. In regions with no actual clocks around, we can still meaningfully talk about concepts such as the spacetime metric because we can counterfactually imagine a fictitious clock present in that region and reason about what it would have read. Events in the distant past had meaningful ``intervals of time,” even with no actual clocks around. A similar reasoning applies to magnetometers. For magnetic field determination in regions with no magnetometers nearby, we imagine counterfactual fictitious magnetometers. E.g., a counterfactual fictitious magnetometer at a point in vacuum with a non-zero magnetic field.

Just as we can consider the functionally defined devices of magnetometers and clocks counterfactually, we can also consider a counterfactual in the case of the abstract device of an observer. But there are restrictions when Principle 2 is applied to quantum theory. For example, we can assign a fictitious observer who recorded in their fictitious lab notes, in unambiguous language, the imprinting on a rock on Mars 2 billion years ago, a high-energy particle from outer space that created a blackening on the rock, which we can still observe today as a paleoarcheological fact, as though that fictitious observer never existed. However, we cannot consider a counterfactual observer co-moving with a particle from a quasar that interfered with itself around a gravitational lens, as that would affect subsequent observation by altering the interference that could have been observed in the absence of such a co-moving observer. Counterfactual, fictitious observers are allowed in certain situations but not others—in Bohr’s terms, such a counterfactual observer can be considered in cases where its introduction would not alter the experimental context of the quantum systems involved  \cite{Bohr1949}.

Notice that the two principles are not specific to quantum theory, but are true for any physical phenomenon, which justifies their independent status. 

The general idea is that while it has been customary in physical description to regard states, forces (or interactions) and associated symmetries as the primitive building blocks for the complete description of nature, it appears that a complete set of primitives for physical description consists of states, forces (or interactions), associated symmetries and the operational interventions, which concern the different maneuvers that a language using automaton may perform in a lab, the symbols used to represent the maneuvers, and the semantics which concern what those maneuvers may mean (such as the question pertaining to ``what observable does this experimental arrangement determine?''). 

See Appendix B for a dialog discussing some common concerns.

\section{The Apparatus and Observers as Functional Units}

To initiate this program, it is useful to clarify the three distinct meanings of the word ``measurement'' that arise in the description of quantum experiments. These three meanings will, in turn, illuminate the definitions needed to identify what counts as a ``measuring device''. 

Accordingly, we distinguish three usages of the term ``measurement,'' which we first list before elaborating on each, with the needed depth: (i) measurement as alignment measurement, (ii) measurement as information copying, and (iii) measurement as quantum state determination—or state tomography.

The first, alignment measurement, concerns the numerical comparison of a particular kind of physical property, or category of variation, in one object with the same kind of property in another object chosen as the standard. It is called an alignment measurement because it involves establishing an alignment between the property of the object and the corresponding property of a standard. A familiar example is the alignment measurement of position. In elementary situations, to determine the position of an object relative to an observer, or to determine its size, one may take a rigid measuring stick and bring it into alignment with the intervening distance between the object and the observer along a straight line connecting the two, or with the relevant spatial extent of the object. This procedure presupposes that the object under measurement can be distinguished as persisting with an identity of its own for the duration of the process.

More generally, we often employ measurement techniques such as this: There is a numerical parameter $X$ with units $x$, whose functional influence on another numerical parameter $Y$ with units $y$ is known. This allows us to infer $X$ by performing an alignment measurement of $Y$, having already defined a standard response of $Y$ to a known value of $X$.  We may now note the following  reasonably general scheme for position measurement:

In most measurements of moving bodies, the word ``position'' refers to a quantity that is accessible only through the detection of light\footnote{Light is emphasized here because all objects above absolute zero temperature emit some thermal radiation in accordance with the Stefan--Boltzmann law. Unlike other methods, light from objects seems to provide a universal means of measuring the position of bodies, perhaps with the exception of dark matter.} reflected from, or thermally emitted by the body. 

We can, in principle, determine the distance, velocity, acceleration, and all functions thereof for moving bodies using only local clocks, spectrometers, and the universally available spectral signatures of chemical elements that is in principle available in the local vicinity of where the observation is being made  \cite{RybickiLightman}.

The only conceivable justification\footnote{What else could we mean by `position of an object' as existing `out there', operationally?} for assuming that objects such as trees or tables possess definite positions and momenta—existing independently of our explicit optical measurements of those observables—is that we may always imagine a hypothetical observer who could have performed the corresponding alignment measurements. Importantly, we would expect no inconsistency with future observations of those same objects, in principle, regardless of whether such measurements were actually carried out by the hypothetical observer.

The role of alignment measurements already appears in Einstein’s work on the electrodynamics of moving bodies, where he correctly emphasized that the ``position’’ and ``time’’ of events can be meaningfully defined only through comparisons with locally present scales and clocks. For example, the only operationally coherent way to speak about the length of a moving rod is to register the light emitted from its endpoints \emph{simultaneously} (in the chosen frame) and to compare this inferred spatial separation with that of a locally available measuring scale.

The next usage of the word ``measurement’’ that we must examine is
\textit{measurement as information copying}.
In the quantum foundations literature, it is common to encounter the claim
that “a measurement is just an interaction that correlates states.’’
Formally, in this usage, if we have a system $s$ that can take one of two
values, $0_s$ or $1_s$, and a memory register $m$ that can likewise be in one
of two states $0_m$ or $1_m$, with the memory initialized in state $0_m$, then
a measurement is represented simply by the copying operation
\begin{equation}
(s,m) \longrightarrow (s, m \oplus s),
\qquad m = 0_m \ \text{initially}.
\end{equation}
(Here $\oplus$ is the XOR gate\textbf{ })
This usage underlies the von Neumann measurement scheme, in which all
measurements are modeled as unitary interactions.
Consider a quantum system $\mathcal{S}$ with Hilbert space $\mathcal{H}_S$ and
an observable
\begin{equation}
\hat{A} = \sum_a a |a\rangle\langle a|,
\end{equation}
with eigenvalues $a$ and eigenstates $|a\rangle$.
A measurement apparatus ${M}$ contains a pointer degree of freedom
with canonically conjugate operators $(\hat{Q},\hat{P})$.

The von Neumann measurement interaction is taken to be
\begin{equation}
\hat{H}_{\text{int}} = g(t) \hat{A}\otimes \hat{P},
\end{equation}
where $g(t)$ is nonzero only during the measurement interval.

With the
coupling constant
\begin{equation}
\kappa = \int g(t) dt.
\end{equation}

the associated unitary evolution is
\begin{equation}
\hat{U}
= \exp\left( -\frac{i}{\hbar}\kappa \hat{A}\otimes \hat{P} \right).
\end{equation}

If the initial composite state is
\begin{equation}
|\Psi(0)\rangle
= \left( \sum_a c_a |a\rangle \right)\otimes |\phi_0\rangle,
\end{equation}
where $|\phi_0\rangle$ is a narrow wavepacket in $Q$–space, then using the
fact that the operator
\begin{equation}
\exp\left(-\frac{i}{\hbar} \kappa a \hat P \right)
\end{equation}

acts as a translation of the pointer wavepacket,
\begin{equation}
\exp\left(-\frac{i}{\hbar}\kappa a \hat P \right) |\phi_0\rangle
= |\phi_0(Q - \kappa a)\rangle,
\end{equation}

one obtains

\begin{equation}
|\Psi(t)\rangle
= \sum_a c_a |a\rangle \otimes |\phi_0(Q - \kappa a)\rangle.
\end{equation}

Thus each system eigen-state labeled by $a$ correlates with the pointer, due to the displacement of the pointer by an amount $\kappa a$.

The Hamiltonian and the resulting unitary are designed precisely to establish a
correlation between the system state and the pointer state: information about
the system is ``copied’’ to the pointer.

This model also leads to more elaborate considerations, such as Wigner’s friend thought experiment, in which measurement is treated simply as a sequence of interactions resulting in the von Neumann chain: the system, apparatus, friend, and finally Wigner get entangled.

Although this model—motivated by the view of measurement as information
copying—plays a central role in understanding ancilla-assisted quantum
measurements, it is important to note that this usage of the term
``measurement’’ does \emph{not,} in general, coincide with the earlier notion of
alignment measurement.

We can now see that these two senses of the word ``measurement’’ diverge when viewed through the lens of special relativity (the divergence becomes even sharper in general relativity, though we will not require that here). Consider an event $E$ with spacetime coordinates $(\vec r,t)$ in an inertial frame $F$. An observer in $F$ may perform a local alignment measurement of some physical property $P$ of the event—one whose value depends on the spacetime coordinates—and obtain the value $a$. This value is meaningful \emph{in frame $F$}, because it is defined through comparison with $F$'s own co-located clocks, scales, and other instruments.

Now suppose there is another observer in an inertial frame $F’$, moving at relativistic velocity with respect to $F$. With respect to $F’$, the same event $E$ will in general be assigned a different value, say $a’ \neq a$, for the same property $P$, due to the Lorentz transformation laws that relate the descriptions in the two frames.

Let the observer in $F$ send a classical-information message to the observer in $F’$ informing them of the value $a$. This is a physical process consisting of an interaction between an emitting device in $F$ and a receiving device in $F’$, thereby correlating their respective states and realizing measurement as ``information copying’'. For concreteness, we may imagine that the observer in $F$ emails the message, ``the value is $a$.''

Since classical communication does not Lorentz-transform the \emph{meaning} of transmitted symbols, the observer in $F’$ will therefore receive the message ``the value is $a$'',  even though their own alignment measurement would yield $a’ \neq a$. Thus, from the perspective of $F’$, although the classical message is received without alteration, the reported value $a$ is not the value that $F’$ would assign to the same property $P$ of the same event $E$ using their own local measuring instruments.

The third and final usage of the word “measurement’’ that we must examine is the one in which measurement refers to the task of determining an unknown quantum state, commonly called \emph{state tomography}. This task typically relies on both alignment measurements and information copying  (for example, in ancilla-assisted schemes), but it additionally requires repeated preparations of the same experimental conditions in order to generate a collection of outcomes from which to infer the statistics and thus the quantum state. In the context of quantum measurements, alignment measurements may be viewed as supplying the eigenvalue data, whereas state tomography supplies the quantum state itself.

Having distinguished the three meanings of the word “measurement,” we can now define what counts as a measuring device in PQT.

\textsc{Definition 1: What counts as a ``detector'' /``measuring device''?}

\vspace{0.15cm}

A``detector'' or ``quantum measuring device'', henceforth referred to as an epistemically constrained measurer (ECM), is any object that possesses a definite, alignment position measurement value relative to an observer and satisfies the following conditions:
\begin{enumerate}
\item It possesses distinct classical information channels, each labeled by an index ${l}$.
\item It will determine the position eigenvalue—understood as localization within a finite spatial region—of a quantum system prepared in a state $\ket{j}$, provided that the system has a nonzero probability of entering the region occupied by the object and a nonzero transition probability
\begin{equation}
P = \sum_{i}\sum_{l} \left| \bra{i^{l}} \hat{U} \bigl( \ket{d}\otimes\ket{j} \bigr) \right|^{2} \leq 1
\end{equation}
to activate the classical channels. Here, ${\ket{i^{l}}}$ denote output states corresponding to activation of channel $l$, $\ket{d}$ is the initial state of the relevant degrees of freedom of the object, and $\hat{U}$ is the unitary evolution generated by the total Hamiltonian of the quantum system, the relevant degrees of freedom of the object, and their interaction.
\end{enumerate}

We will illustrate this definition in section 2.3. For now, picture the toy example of an object that can change in two different ways: it can change its color, from among a set of N possibilities, and it can make sounds from among a set of M possibilities. This toy object has two classical channels.

Note that $ECM \subset ECD$.

 Whether the system is a stone, a table, a Geiger counter, or a cloud chamber, all appear to share the features characterized in this definition.

As for the `observer' that appears in definition 1, we need another formal definition. What follows is a functional definition of an abstract object that can be instantiated in physical systems. 

{Whether we append phenomenal experience to the construct of an observer or not, at this point, seems more of an option than a necessity. We assume that language use does not require phenomenal experience and, therefore, drop the requirement for phenomenal experience. What we do need for the construct of the observer is only the language-using ability, among others, as we list in definition 2. 

It is true that the only instance of an autonomous language-using system currently known is the living human brain, which is also conscious when using language. However, we cannot, based on that fact alone, infer that the construct of an observer has to be conscious as well. That would be like saying ``All known mathematicians are oxygen-breathing; therefore, mathematics requires oxygen''. This is clearly absurd reasoning.

That said, here's a question for the cognitive science or machine intelligence researcher: There can be phenomenal experience without self-directed language processing (animals do this all the time, and we do it too when picturing objects without attempting to articulate what it is); conversely, can there be self-directed language processing without phenomenal experience?

If future research shows that autonomous language use always goes hand in hand with phenomenal experience, we would have to add that feature to the following list. If that addition happens, it would not, however, mean that 'consciousness is everywhere', nor idealism, nor would it mean that 'consciousness causes collapse'; rather, it would mean that the functionally defined unit of an observer who uses language to qualify experimental arrangements will happen to have phenomenal experience. It would simply be a non-reductionist feature of autonomous language-using systems. That said, we do not need phenomenal experience in the following definition.


\textsc{Definition 2: observer/ Knowledge Acquiring System}

\vspace{0.25cm}

 Observers are movable automata that possess the following primitive features, which define whether a system is an observer or not:
\begin{enumerate}
    \item observers are physical systems with an input interface, an output interface, memory, clocks, and information- and language-processing units.
    \item Memory and information processing can be either quantum or classical.
    \item The input and output interfaces are always classical information channels. The classical information sent and received should have meaning in the language of choice.
    \item A clock is attached to the observer to time their actions and information processing.
    \item Light pulses could be sent and intercepted to make distance measurements of the environment, and observers can perform specific operational tasks in their environment by maneuvering their bodies.
    \item Each observer has a definite spatial location, in the sense of alignment position measurement, relative to other observers.

\end{enumerate}

Note that we have required embodied cognition for the observer construct: observers are essentially language-using automata, not brains in a Vat \cite{Putnam2000-PUTBIA}. 

That is, a complex information-processing system alone is not enough; it also needs to be integrated with a body capable of navigating the environment in different ways. For instance, on paper, the sperm whale's brain is rather complex and larger than the human brain. It appears plausible, on paper, that their brains are more capable than ours, even for abstract tasks, but because they did not evolve bodies like ours, which are well optimized for more tactile and embodied cognition than whale bodies are, this theoretical capability may not have translated into human-level cognition. We can manipulate objects with precision, etch symbols on surfaces, and transmit records across generations in more reliable ways. 

Any knowledge-acquisition system that may have evolved anywhere in the universe, it seems, will be found to abide by this coupling between information processing and tactile abilities. 

One may now ask, do simpler intelligent systems, such as sperm whales, dogs, cats, insects, plants, rocks, lava, and so on, qualify as observers? To address this question, first consider the nature of clocks.

Would we say a pendulum is a clock? Would we say a waterfall is a clock? Would we say a piece of rusting metal is a clock? It is reasonable to suppose that they do, but they are not `ideal clocks', because they do not have enough precision and do not have the ability to count arbitrarily long intervals of time. 

Nevertheless, despite their limitations, in the immediate vicinity of such clocks, we can still say `time passes', and we can talk about a spacetime metric, because we can always imagine placing a counterfactual clock of arbitrary precision close to that non-ideal clock. Just because a particular system is not an idealized clock does not mean we cannot assume a spacetime metric in its vicinity.

The point is that, however non-ideal a system may be at functioning as a clock, we can nevertheless talk about time in its vicinity with arbitrary precision, given the aforementioned counterfactual consideration of a high-precision fictitious clock positioned next to it.

We can expect a similar argument to be applicable to the construct of observers. 

Analogous to the `unit of time' for a clock, for observers, we can define a `unit of meaningful action' corresponding to the unambiguous meaningful actions that can be performed, communicated, and reproduced elsewhere. Of course, just as the unit of time is chosen arbitrarily, this is also chosen arbitrarily.

A choice that we can make is this: the completion of the simplest quantum-mechanical experiment, specifically a two-level system, such as a photon, starting in an initial state \(\ket{0}\), evolving for 1 unit of time under free evolution, and being probed at the end of that unit of time for a specific choice of observable.

The ability to complete such tasks and communicate them unambiguously in some language could be taken as the certificate of an `ideal' observer. More precisely, we can take definition 2 to characterize an `ideal' observer, since such a defined system can carry out the unit of action just prescribed and communicate it.

A sperm whale or an insect or a rock, analogous to a non-ideal clock, can conceivably detect photon polarization using its vertebrate eye or compound eyes or imprints on it's crystal structure, respectively, thereby completing a simple quantum experiment. But just like a non-ideal clock, it has no sophistication to demarcate the completion of a simple quantum experiment. 

Nevertheless, we can counterfactually position an idealized, fictitious observer with language abilities in the vicinity of those non-ideal observer systems, to probe the sperm whale or the insect, and demarcate the completion of the simple quantum-mechanical experiment.

All this does not entail that human beings are `ideal-observers', but rather that such a conceptual category exists in nature, and we may approximate it to some degree.

\section{Heisenberg's Potentialities and Actualities}

At this juncture, it is useful to formalize a conceptual invention concerning the ontology of quantum systems, due to Heisenberg   \cite{heisenberg1971physics}. Namely, the concepts of actuality and potentiality. Actuality and Potentiality can be thought of as distinct categories of physical existence, even though they are both composed of the familiar chemical elements or particles of the Standard Model. 

We can interpret Heinseberg's usage of the word `actuality' to be the same as the concept of objective existence that we have discussed earlier: a property of an object is objective if a real or fictitious observer, could in principle have accessed the numerical value of that property, by a passive alignment position measurement by capturing the light from the object, such that, another observer's subsequent intervention would not allow the latter—even in principle—to discern that a prior determination had already occurred. 

It's passive in the sense that the object's position can be measured from its thermal radiation or from reflections of ambient light, rather than by actively probing it with a probe beam.

Note that this is a stricter criterion for actuality than macro-realism alone. A region with a classical electromagnetic field in vacuum would qualify as macro-real; however, it would not qualify as an actuality in the stricter sense defined, because the electromagnetic field in vacuum cannot be alignment position measured in the way described. 

We have defined actuality in terms of both macro-realism and alignment-position measurability, as prescribed.

For illustration, consider the following two examples: i) a region of vacuum with a non-zero classical magnetic field in it {;}  ii) the view from the direction orthogonal to the direction of propagation of a beam of laser in vacuum that originated in a source and gets detected at a detector screen. 

In neither of these cases is there any sense of alignment position measuring the field, unless we place a scattering medium and alignment position measure the scattering medium. In which case, the scattering media qualify as an actuality, and not the field in vacuum itself. 

The examples aren't restricted to electromagnetic fields. We can also have the same scenario apply to the subsystem of a larger quantum system, composed of matter, with the larger system in a pure quantum state, but the subsystem in a reduced state that is a diagonal improper mixture of several pointer states. Suppose we have \(N\) number of particles in a closed quantum system, and a subsystem of \(M \ll N\) particles is considered. Let \(M\) still be a large number when compared to 1. 

Then the coarse-grained observables of the \(M\)-particle subsystem, like the coarse-grained observables \(E\) and \(B\) in classical electromagnetism, will be approximately non-contextual and approximately macro-real. However, when we look at the \(N\)-particle system as a whole, described by a pure quantum state, its observables will be contextual and, in general, not macro-real, as one could have expected.

Under this situation, when there is still the possibility to coherently reverse the N particle quantum state, to some initial quantum state, it is not possible to alignment position measure the M particle subsystem.

Heisenberg initially considered potentialities only as an ontology for coherent quantum systems. The macro-real systems that nevertheless do not satisfy the requirement for alignment position measurability can be reinterpreted as the classical limit of Heisenberg's potentialities. 

It is in explaining this classical limit that decoherence is such a successful framework, but it is not sufficient to explain the classical mechanics of actualities, as discussed in Chapters 5, 6, and 7. For the origin of classical mechanics, which is a distinct problem from explaining the classical limit of potentialities, we present an alternative model in Chapter 6.

For Heisenberg, actuality included detector clicks, spots on photographic screens, and the localized presence of objects such as rocks. 

He considered potentiality to be the ontology of quantum systems before they were detected \cite{Heisenberg1958}. A form of existence when entities don't have spatio-temporal localization, and the properties of the entities exist only as propensities. 

We have characterized Heisenberg's potentiality and, in the process, identified a new category of existence, ``the classical limit of potentiality'', which is not fundamental and can be deduced from decoherence. We have also characterized actualities, which are distinct from the classical limit of potentialities. Actuality and Potentiality are fundamental categories of existence and cannot be reduced to each other.

\begin{enumerate}
    \item \textsc{Potentiality:} Entities in one or more enclosed regions of space, possibly disconnected, that are characterized by incompatible properties that are not macro-real, not non-invasively measurable, and counterfactual-indefinite. E.g., a 1000-particle Bose-Einstein condensate gas in a box; an entangled pair of particles in two distinct enclosed boxes.\\

    \item \textsc{Classical Limit of Potentiality:} Entities in an enclosed region of space that are characterized by effectively compatible properties that are effectively counterfactually definite, macro-real, and non-invasively measurable. E.g., a classical electromagnetic field in vacuum.\\

    \item \textsc{Actualities:} Entities in an enclosed region of space that are characterized by effectively compatible properties that are counterfactually definite, macro-real, non-invasively measurable, {and alignment-position-measurable by passively capturing light from the region}. E.g., the spot on a photographic screen.
\end{enumerate}

 And we understand macro-realism in the sense of Leggett and Garg  \cite{LeggettGarg1985}.

 In each of the above definitions, we use ``in an enclosed region of space'' to be specific about the thing we are talking about. So that we can operationally act on it by targeting that region of space.

 We can now look back at the earlier definition of an epistemically constrained measurer (ECM) and view it as a context that abides by epistemic constraints, in the sense that it can be stated in terms of experimental propositions.
 
 When a quantum system is incident on an ECM, the ECM only helps materialize actuality with some probability. Alignment position measurement is then performed on the materialized actuality, for example, by checking where on the screen a spot appears. 
 
 Calling an ECM a ``measuring device'' bundles together two distinct concepts that may cause confusion: the actuality that gets materialized versus the alignment position measurement on that materialized actuality. 

Furthermore, note that the criterion to distinguish between actuality and potentiality is based on operational intervention performed by a real or a counterfactual observer, and it is not based on a restriction on the quantum states involved or a restriction on the size of the system.

\section{A Meta Rule: Probability of Instantiability of Born Rule}

The probability in the title of this section is not the usual Born probability for realizing an eigen-state when we measure an observable, but it concerns the probability specifying when to apply the Born rule (or when it naturally gets applied anyway, even without any specific observer present). 

\textsc{The Probability of Instantiability of Born Rule} specifies the probability that the Born Rule is applicable under the given experimental conditions.

Let \(E\) denote the experimental conditions specified in ordinary language by a sequence of experimental propositions.

The Born Rule concerns:
  \[
  \begin{aligned}
  &P(\text{Outcome } o_i \mid \text{Born Rule instantiable}, E)
  \end{aligned}
  \]

 The Probability of instantiability of the Born Rule concerns:
  \[
  \begin{aligned}
  &P(\text{Born Rule instantiable} \mid E)
  \end{aligned}
  \]

The relationship between the Born rule and the probability of instantiability of the Born rule would then be:

  \[
  \begin{aligned}
  P(\text{Outcome } o_i \mid E)
  &=
  P(\text{Outcome } o_i
    \mid \text{Born Rule instantiable}, E) \\
  &\quad \times
  P(\text{Born Rule instantiable} \mid E).
  \end{aligned}
  \]

The probability of instantiability of the Born rule helps decide the answer to \textit{``Do we apply the Born rule for the behavior of S inside the box or not?"} in the scenario in Fig 2.1. 

Alternatively, the probability of instantiability of the Born rule helps decide the answer to this Bayesian question: \textit{Does a bet made based on state $\ket{j}$, a basis choice $\{\ket{o_i}\}$, and the Born Rule get
resolved/settled inside the box or not?}

\vspace{0.5cm}

\begin{figure*}
  \centering
  \includegraphics[
    width=\textwidth,
    alt={A schematic scenario illustrating the probability of instantiability of the Born rule. A quantum system is placed inside a box containing specified experimental conditions, and the question is whether those conditions instantiate a situation in which the Born rule applies.}
  ]{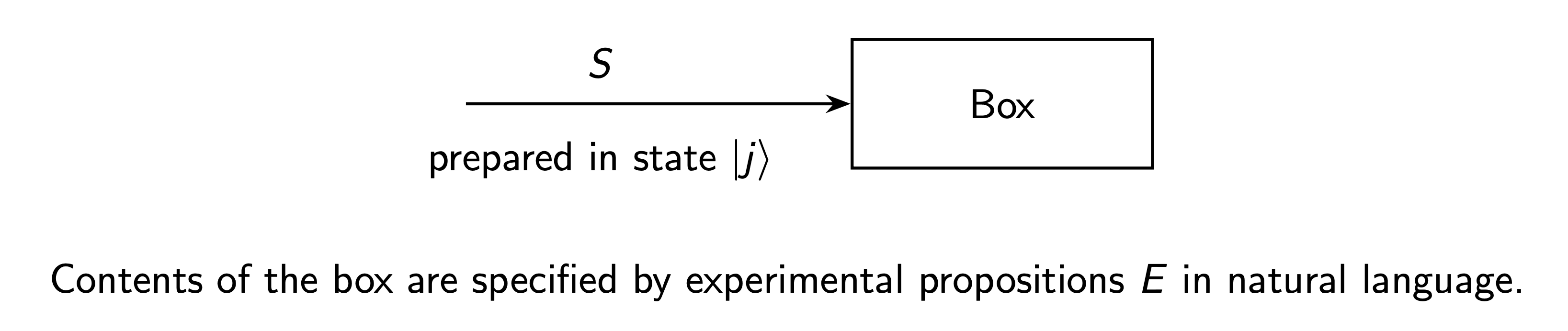}
  \caption{Relevant scenario for the concept of probability of instantiability of the Born rule}
  \label{fig:born-rule-instantiability-scenario}
\end{figure*}

Note that this probability for the instantiability of the Born rule can also have classical ignorance contribution relating to the classical efficiency of the detector (ECM) or more simply to the classical probability, from say a coin toss, that is used to determine whether we put a beam splitter inside the box or a photodetector inside the box (when S is a photon). 

We ignore that contribution to the probability of instantiability of Born Rule in the following constructions but it may be accommodated if desired.

The reader may have already anticipated that the probability of the applicability of the Born rule arguably has a Bernoulli form because of the binary possibilities

\[P(\text{Apply Born Rule}) = 1-Q\],
\[P(\text{Unitary Evolution}) = Q \]

\[P(\text{Apply Born Rule}) + P(\text{Unitary Evolution}) = 1\]

To arrive at the general form of the probability of instantiability of the Born Rule $1-Q$, let us look at an example, the Stern–Gerlach experiment, a familiar workhorse for highlighting features of quantum experiments that have no analog in classical mechanical experiments.

In the Stern–Gerlach experiment, the goal is to subject a beam of silver-atom vapor to an inhomogeneous magnetic field and to observe the subsequent probabilistic appearance of macroscopic traces on a detection screen. Instead of forming a continuous band, as one would expect for a classical magnetic moment traversing an inhomogeneous magnetic field, the impinging silver atoms produce two well-separated spots on the screen. These two characteristic spots directly reflect the quantization of the spin observable—and hence of the magnetic moment—of the silver atom.

The interaction of the silver atom with the magnetic field is governed by the Hamiltonian
\begin{equation}
\hat{H}_{\mathrm{int}} = \frac{2\mu_B}{\hbar}\hat{\vec{S}}\cdot\vec{B}(\mathbf{r}),
\end{equation}
where $\mu_B$ is the Bohr magneton and $\hat{\vec{S}}$ is the spin operator.\footnote{Up to an overall sign and $g$-factor, which play no role in the present discussion.}

For simplicity, let us assume that the oven is supplemented by additional preparation apparatus that ensures that the silver atoms emerge one at a time in a pure spin state $\ket{+_x}$, an eigen-state of $\hat{S}_x$.

If the inhomogeneous magnetic field is oriented along the $z$ direction, the state of an individual silver atom after passing through the interaction region will be

\begin{equation}
\frac{1}{\sqrt{2}}
\left(
\ket{+_z}\otimes\ket{\text{up beam}} + \ket{-_z}\otimes\ket{\text{down beam}} \right),
\end{equation}

here the states $\ket{\pm_z}$ are the eigen-states of the $\hat{S}_z$ observable, and the states $\ket{\text{up beam}}$ and $\ket{\text{down beam}}$ are the  states corresponding to the trajectories that carry the atom toward the upper and lower spots on the detection screen, respectively.

Following Bohr, the experimental description can be demarcated—in an epistemically determined way—into (i) the portion in which an alignment measurement is performed, which we have called an ECM, and (ii) the portion that fixes the relevant Hamiltonian and hence the unitary operator for the quantum-mechanical problem under study, which we will call the \emph{epistemically constrained unitary determination} (\textsc{ECU}). 

Just like the ECM, ECM is also describable as a sequence of experimental propositions and it specifies the Hamiltonian for the system. Moreover, ECU is an instance of the much broader ECD: $ECU \subset ECD$.

All this can be identified with the familiar `state-preparation' protocols, but the purpose of the ECU terminology is to emphasize the role of definite, intersubjectively agreeable experimental propositions, such as `I turned on the magnetic field by flipping the current switch on,' in the description of such protocols.

It is now a good time to define the notion of a quantum-mechanical problem. It is a definition that we will need for our present goals in this section, and also for our definition of what a ``quantum system'' means in Chapter 7, a definition that generalizes the usual understanding of a quantum system as a particle or a field. It is a definition based on meaningful ``acts'' involving what an observer knows, what intervention the observer performs, and what the observer finds as a consequence of that intervention. The ``unit of meaningful action'' performed by an observer, which we defined earlier as an analog of the unit of time interval counted by a clock, is one such standardized act. Of course, we can talk about these concepts even in places without actual observers or clocks present.

Consider a set, which we can call a \textsc{type}, characterized by a prepared quantum state of a specific kind of system (say an electron), evolved for a fixed interval of time under a repeatable Hamiltonian, time-dependent or time-independent, and the fixed observable interrogated at the end of the fixed interval of time. By a \textsc{quantum-mechanical problem}, we mean an instance of the experimental algorithm---carried out either by an actual observer or, counterfactually, by a fictitious observer---that implements the type, such that many such quantum problems for the same type can be repeated any number of times by any other observer(s), in any region of spacetime, such that the experimental outcomes from all those repetitions can be pooled together to instantiate the Schrodinger equation for the given type.

Every quantum problem culminates in an alignment measurement of the eigenvalue probabilistically realized for a measured observable \(\hat{O}\). We call this objective event---in principle accessible to any observer via alignment measurement---the \emph{materialization of the outcome} corresponding to the given quantum-mechanical problem. The outcome materializes in an ECM.

More will be discussed in Section 7.7 about quantum problems, the rationale behind them, their generalizations, and their role in defining a quantum system. For now, what we have defined is sufficient.

In the case of the Stern–Gerlach experiment, the \textsc{ECU} includes the surrounding arrangement of magnets and the nature of the intervening volume, namely, a vacuum region containing an inhomogeneous magnetic field $\vec{B}(\vec{r})$. The \textsc{ECM} includes the region with a silver-bromide screen.

The silver-bromide screen produces the materialization of an outcome through a microscopic amplification process. An incident silver atom initiates the formation of a small number of neutral silver atoms within a silver-bromide grain. This, in turn, triggers further clustering of neutral silver atoms, leading to the formation of ``image centers'',  that can subsequently be developed photographically.

Because the demarcation between \textsc{ECM} and \textsc{ECU} is taken to be epistemic, the boundary is found by an operational criterion formulated in terms of counterfactual actions that could be performed by real or fictitious observers. Specifically, it is the earliest point in the experiment at which one may consistently assume that some observer could already have acquired information about which outcome was realized, without affecting a later reading of the outcome for the same fixed quantum-mechanical problem. 

This counterfactual assumption cannot be pushed arbitrarily far back in the experimental process. When the silver atom is still in flight, making such an assumption would require introducing an additional Stern–Gerlach arrangement (or an equivalent intervention), thereby altering the phenomenon itself. Because the experimental arrangement defines the quantum phenomenon (in our vocabulary, a quantum-mechanical problem), after such a change, one would no longer be addressing the same quantum-mechanical problem.

In this sense, the demarcation is epistemic and not specifiable in a reductionist manner (for example, in terms of a threshold number of atoms required for a ``macroscopic change’'). Rather, it is fixed only through the counterfactual criterion above, formulated in terms of real or fictitious observers.

The use of counterfactual reasoning was emphasized by both Bohr and, later, Wheeler. This line of reasoning is also, we believe, part of what makes many physicists uncomfortable, since prior to quantum theory, physical reasoning appeared to be formulated primarily in terms of factual statements about actual events. Nevertheless, one may argue that counterfactual reasoning was already implicit in classical physics, particularly in the concept of fields. The reality of a field—for example, a magnetic field—is operationally defined in terms of what a magnetometer \emph{would} read if it were placed at a given point in otherwise empty space. It is precisely this counterfactual reading that is taken to represent the value of the magnetic field at that point. In this sense, field values are inferred from a network of counterfactual measurements.

In the present case, the outcome on the silver-bromide screen is taken to have materialized precisely at the stage when a fictitious observer could, in principle, have intervened and completed the algorithm defining the quantum-mechanical problem by performing an alignment measurement of the materialized record, and any subsequent observer's reading would not alter the quantum-mechanical problem that was completed.

 The silver-bromide screen constitutes an \textsc{ECM} by definition 1, since it has a definite position and spatial extent under alignment measurement. It also possesses distinct classical information channels. This includes the appearance of visible spots on the screen when a silver atom becomes incident and produces image centers in a silver-bromide grain.

These are \emph{classical} channels in the sense that the associated changes can be encoded into classical bits of information that, in principle, complete the algorithm constituting the quantum-mechanical problem under consideration. In the Stern–Gerlach experiment with a silver-bromide screen, for example, one may encode a spot appearing higher in the vertical direction as $0$ and a spot appearing lower as $1$.

For the silver bromide screen, the classical information channels in \emph{definition~1} may also refer to other kinds, as in the following list, which is not exhaustive. We may label each channel by an index $l$:
\begin{enumerate}
\item \emph{Material rearrangement}, $l=1$: classical information output in the form of definite, localized changes in the material, attributable to atomic or molecular rearrangements (e.g., droplet formation in a cloud chamber).
\item \emph{Electrical}, $l=2$: classical information output in the form of definite meter readings, attributable to electrical currents produced (e.g., an avalanche photo-diode).
\item \emph{Thermodynamic}, $l=3$: classical information output in the form of definite changes in thermodynamic state variables in a calorimeter (such as in a bolometer).
\item \emph{Radiation based}, $l=4$: classical information output in the form of radiation arising from light–matter interactions in definite localized regions (e.g., a phosphorescent screen).
\end{enumerate}

For the silver-bromide screen, the $l=1$ channel is directly relevant.

For more complex \textsc{ECM's}, there may be multiple channels described by a set $\{l\}$.

The probability appearing in \emph{definition~1},
\begin{equation}
P
= \sum_{i}\sum_{l}
\left|
\bra{i^{l}} \hat{U} \bigl( \ket{d}\otimes\ket{j} \bigr)
\right|^{2}
\leq 1
\end{equation}
becomes, in the present case where only $l=1$ is relevant,
\begin{equation}
P
= \sum_{i}
\left|
\bra{i^{l=1}} \hat{U} \bigl( \ket{d}\otimes\ket{j} \bigr)
\right|^{2}
\leq 1.
\end{equation}

Here $\hat{U}$ describes the interaction between the incoming silver atom and the silver-bromide molecules in the screen. The channel $l=1$ is activated when the interaction between the incoming atom and the coated screen leads to the formation of an image center in one of the grains. 

The mechanism by which this can happen is when the incoming silver atom (directly or indirectly) reduces a small number of $\mathrm{Ag}^{+}$ ions to neutral $\mathrm{Ag}^{0}$ atoms, which then cluster via metallic bonding to form an image center. Subsequent photographic development renders this as a localized ``spot’’ on the screen.

More generally, the post-unitary state may be written schematically as
\begin{equation}
\ket{\Psi_{\mathrm{out}}}=\sum_i c_i \ket{i},
\end{equation}
where the $\ket{i}$ are joint states of the detector degrees of freedom and the incoming quantum system. The set ${\ket{i^{l=1}}}$ is the subset of ${\ket{i}}$ that activates the channel $l=1$.

The components of $\ket{\Psi_{\mathrm{out}}}$ may include possible outcomes such as the following:
\begin{enumerate}
\item $\ket{i=1}$ with amplitude $c_1$ which could describe elastic reflection or scattering from the surface.
\item The $\ket{i=2}$ state with an amplitude $c_2$ which could describe the process where the incoming silver atom lands on the Silver Bromide molecules and doesn’t result in free electrons.
\item The $\ket{i=3}$ state with an amplitude $c_3$ could correspond to an inelastic scattering event that transfers kinetic energy of the incoming silver atom to electronic degrees of freedom in the silver bromide screen, which result in free electrons that reduce the $\mathrm{Ag}^{+}$ ions to $\mathrm{Ag}^{0}$ atoms.
\item $\ket{i=4}$ with an amplitude $c_4$ could correspond to the transfer of a $5s$ electron to an $\mathrm{Ag}^{+}$ ion, reducing it to $\mathrm{Ag}^{0}$.
\item $\ket{i=5}$ with an amplitude $c_5$ could describe an ionization followed by electron transport, which then reduces many other $\mathrm{Ag}^{+}$ ions to $\mathrm{Ag}^{0}$ atoms.
\item $\ket{i=6}$ with an amplitude $c_6$ could describe an inelastic scattering event which transfers momentum to lattice phonons, with the atom emerging at reduced momentum.
\end{enumerate}

Only those output states that lead to activation of the relevant channel—here $l=1$, via the formation of developable image centers—belong to the set ${\ket{i^{l=1}}}$. In the illustrative list above, the processes associated with $c_3$ and $c_5$ would qualify. In that case,
\begin{equation}
P = |c_3|^2 + |c_5|^2 \leq 1.
\end{equation}

The probability defined in \emph{definition~1}—apart from any contribution due to classical ignorance—is the probability that a physical system can function as a system capable of producing the materialization of an outcome for a given quantum-mechanical problem. In more familiar terms, it is the probability that a system functions as a measuring device (ECM) in the sense required to produce a Born state update.

This is not the same as the probability of obtaining a particular eigenvalue (or eigenstate) when an observable is measured. The latter is given by the Born rule; we refer to it simply as the \emph{Born probability}.

Although the probability in \emph{definition~1} is still computed using the squared norm amplitude, it is categorically different in its role, and its physical interpretation is different. 

It has a meta character in that its purpose is to determine whether—and with what probability—the Born rule is instantiable for the scenario in Fig. 2.1.

Only after ensuring that the probability of instantiability of the Born rule is nonzero, does the standard Born probability apply in the usual way for a well-defined quantum-mechanical problem. This is required in order to decide when to apply unitary evolution and when to apply the measurement update.

To make this explicit, we can now state the \emph{Probability of instantiability of the Born state update}, henceforth labeled by the symbol $P$, as a Postulate.

\textsc{Postulate 1: Probability of instantiability of Born state update:}

\begin{enumerate}
    \item  Removing classical ignorance, the probability of instantiability of the Born Rule has a Bernoulli form:
  \[
    P(\text{Born Rule instantiable situation present} \mid E) = 1-Q(E),
  \]
  \[
    P(\text{Unitary evolution continues} \mid E) = Q(E).
  \]
\[
  P(\text{Born Rule instantiable situation present} \mid E)
  +
  P(\text{Unitary evolution continues} \mid E)
  =
  1.
\]
\item The form of $1-Q(E)$  is
\begin{itemize}
\item For regions of space inside the box with potentiality,
  \[
    P\left(\text{Born Rule instantiable situation present}
    \mid \text{$E$ specifies potentiality}\right)
    =
    1-Q(E)
    =
    0.
  \]

  \item For regions of space inside the box with actuality,
  \[
  \small
  \begin{aligned}
  &P\left(\text{Born Rule instantiable situation present}
  \mid \text{$E$ specifies actuality}\right) \\
  &\qquad = 1-Q(E) \\
  &\qquad =
  \sum_i \sum_l
  \left|
  \left\langle i^l \right|
  \hat{U}
  \left(
  \left| d \right\rangle \otimes \left| j \right\rangle
  \right)
  \right|^2
  \leq 1 .
  \end{aligned}
  \]

   The label $l$ denotes distinct classical information channels discernible in an actuality. The state $\ket{d}$ is the initial state of the relevant degrees of freedom of the actuality with which the potentiality in state $\ket{j}$ interacts. The states $\left| i^{l} \right\rangle$ denote output states corresponding to the activation of channel $l$.
\end{itemize}
\item The relationship between the Born rule and the probability of instantiability of the Born rule can be found by applying the law of total probability. Let \(B\) denote the event that a Born Rule instantiable situation is present, and let \(o_i\) denote an eigenvalue outcome.

   Since
  \[
  \begin{aligned}
  P\left(o_i \mid E\right)
  &=
  P\left(o_i \mid B,E\right)P\left(B \mid E\right) \\
  &\quad +
  P\left(o_i \mid \neg B,E\right)P\left(\neg B \mid E\right),
  \end{aligned}
  \]
  and
  \[
    P\left(o_i \mid \neg B,E\right)=0,
  \]
  we have
  \[
    \boxed{
    P\left(o_i \mid E\right)
    =
    P\left(o_i \mid B,E\right)P\left(B \mid E\right)
    }.
  \]

 $P\left(o_i \mid B,E\right)$ is what we get from the Born rule for outcome i, in its usual formulation, which is agnostic about potentialities or actualities.
\end{enumerate}

\vspace{1cm}

Note that the probability of instantiability of the Born rule uses the Born rule (in the sense of interpreting the norm squared amplitudes as providing the probability), at the level of individual terms in the sum, which raises an apparent circularity concern. But there is no circularity, since our goal is not to derive the Born rule.

Moreover, the sum that appears in the postulate has a new physical interpretation, distinct from the usual interpretation of such sums. This new physical interpretation is central to the postulate.

Moreover, unlike the Born rule, the probability of its applicability hinges on the distinction between actuality and potentiality, a distinction that the Born rule does not make. That distinction relies on epistemic constraints and alignment measurability. 

The probability of instantiability of the Born rule illuminates a nested structure in the practical use of the Born rule. Namely, in any quantum mechanical problem involving a system \(S\) that we are investigating experimentally, we need pre-existing actualities with possibly more than one classical information channel. The degrees of freedom of this pre-existing actuality interact with the quantum system \(S\) to activate one or more of those channels, and this interaction is itself a quantum mechanical problem. This constitutes the outer layer in the nested structure. For this outer layer, the applicability of the Born rule is specified axiomatically: we apply the Born rule for the outer layer when a pre-existing actuality is involved. It is for the inner layer in that nested structure that the Probability of instantiability of the Born rule can be useful.

The physical content of \emph{postulate 1} is a certain \emph{non-specificity} property of \textsc{ECM}s.

The non-specificity is that, once a system—any system—qualifies as a \textsc{ECM} by \emph{definition~1}, it can, in principle, materialize outcomes for \emph{any} quantum system in \emph{any} quantum-mechanical problem, provided that there is a nonzero probability to find the system inside the box in Fig.2.1, and that the Probability of instantiability of the Born state update, $1-Q$, is nonzero.

For example, a radon-222 detector, which functions by counting $\alpha$-decay events per unit time inside a chamber via gas ionization and subsequent current production (classical channel $l=2$), can also be activated by particles other than radon-222 decay products, provided that they have nonzero probability of being found in the detector region and nonzero probability of ionizing the gas  to produce a detectable current.

More naturally occurring examples include the way a mica rock can preserve tracks of cosmic rays that arrived at its surface tens of thousands of years ago ($l=1$), or the way pleochroic halos form in geological crystals due to the embedding of radioactive nuclei during natural crystallization ($l=1$). One may also note the persistent activation of an $l=4$ channel throughout the history of such rocks via the scattering of photons from their surfaces. Such systems can serve as improvised quantum measuring devices (\textsc{ECM}s) even today for a wide range of quantum systems and quantum-mechanical problems.

\section{ Inter-Subjective Objectivity Postulate}

In addition to the postulate supplying the Probability of instantiability of\textbf{ }the\textbf{ }Born rule, the inter-subjective agreeability of occurrences of actualities (and, as a corollary, potentialities) among all observers has the form of a postulate, which will constrain theoretical constructions later. This postulate expresses a certain symmetry in the nature of actuality and potentiality.

\textsc{Postulate 2: Inter-subjective Objectivity of observers}
    
What is an actuality with respect to one observer is an actuality with respect to all observers. Similarly, what is a potentiality with respect to one observer is a potentiality with respect to all observers.

This means that in any situation where epistemic constraints become relevant, such as in subsystem partitioning and in the application of unitary versus intrinsically probabilistic pure state update, there is an invariance across all observers. 

In the case of intrinsically probabilistic pure state update, by the Born rule, versus unitary application, the invariance is in the fact as to whether pure state update happened or unitary evolution happened (e.g, if a pure state update happened with respect to one observer, it happened with respect to all observers—even though, there can additionally be a mixed state update to reflect classical ignorance); in the case of subsystem partitioning, the invariance is in the operationally accessible tensor product factorization, as we will see in section 6.6.

This inter-subjective objectivity postulate emphasizes that there are no privileged observers. The physical occurrence of a definite event, operationally accessed locally by one real, or counterfactual observer A, and expressible in plain language by A, is also an event expressible in plain language as a definite occurrence for any real or counterfactual observer B. That is, the definiteness of occurrences of events is invariant.

This postulate applies among literal observers, but also in situations when counterfactual observers could be invoked.

In an experiment, Alice can claim that the moon exists even when she does not look at it \cite{Pais1982}, since it is possible in principle to conceive of an observer Eva who could have witnessed the existence of the moon, by performing alignment position measurement of the moon's position and extent, unbeknownst to Alice, without contradicting subsequent observations and conclusions by Alice based on her initial claim. On the other hand, if Alice had assumed there to be a photon in a coherent superposition state in a Mach-Zehnder interferometer, then, unbeknownst to Alice, there cannot be an Eva who was able to decipher a definite photon existence in one arm of the interferometer by placing detectors in the arm and doing alignment position measurement on it. If such an Eva were to exist, then Alice's conclusion based on her initial assumption (namely that she would observe interference) would contradict future observations. 

This suggests the construction of a network of fictitious counterfactual observers who can intercommunicate, just as we may construct a grid of fictitious counterfactual clocks wherever temporal durations are physically meaningful, with the network's nodes in regions of actuality, so that such an assignment does not appreciably affect the future course of events. The nodes cannot be assumed to be in regions of potentiality, as that could contradict subsequent conclusions by affecting interference experiments that could have been performed.

\section{The Bohrian Program: A Concise Statement}

Having developed the necessary conceptual vocabulary, principles and postulates, we can now concisely restate the Bohrian Program as follows.

\begin{figure}[htbp]
\centering
\includegraphics[
  width=0.8\textwidth,
  alt={A schematic diagram concisely restating the Bohrian Program. The diagram represents the plan to combine epistemic constraints with quantum mechanics and ask what physical conclusions follow from their joint consideration.}
]{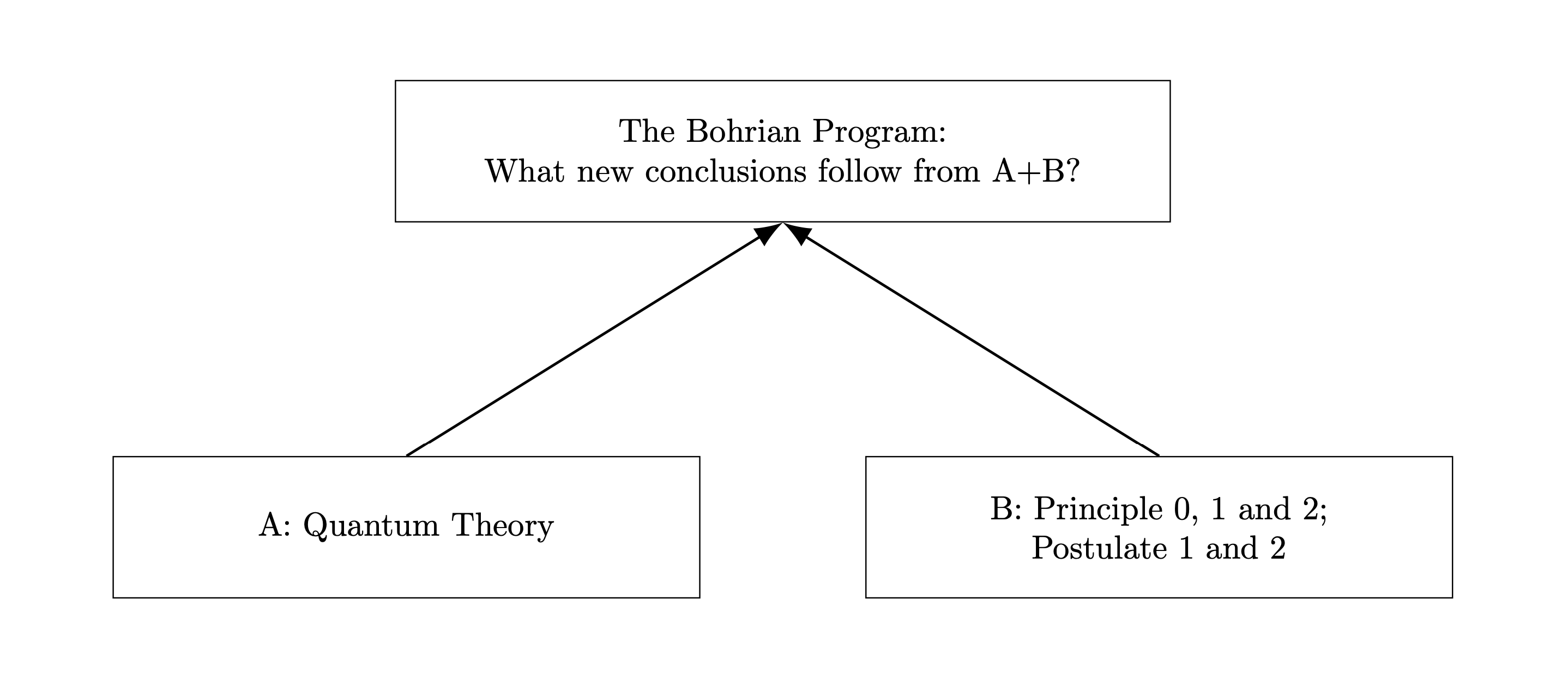} 
\caption{Concisely restating the Bohrian Program}
\label{fig:bohrian-program}
\end{figure}
This is what we aim to explore in the rest of the thesis. Unlike closed problems that have the form ``Given X, find Y'', the Bohrian Program is seemingly open-ended, taking the form ``Given X and Y, what new W follows?'' Therefore, the inquiry in the subsequent chapters may only be a part of a more expansive exploration that can be pursued in a later work. Nevertheless, in the conclusion, we specify a criterion for completing the Bohrian program based on what we can foresee.

We also need to pursue further inquiry into some background assumptions concerning what a quantum state means, how it is assigned, and other interpretative matters. This is pursued in Chapter 4. This interpretative clarification is necessary to unambiguously implement the Bohrian program for physical problems


 \chapter{Systematizing Quantum Dynamics into a Hierarchy}

 Having defined Heisenberg's notion of Actualities and Potentialities, we can obtain four broad categories of quantum processes, all of which are well known, in terms of the pattern in which actualities and potentialities are spatially and temporally arranged relative to each other.

But first, we must clarify what we mean by a `closed quantum system' and an `open quantum system'.

There are at least three different views on the relationship between open and closed quantum systems. Firstly, there are views that assume that all quantum systems are open quantum systems, and that what we call a closed quantum system is only an effective description. In these views\footnote{A related view due to Cuffaro et al. \cite{cuffaro2024open} posits that closed system dynamics need not be considered fundamental, not that they can never be.}, only the universe as a whole is a closed quantum system  \cite{Wallace2012}. 

Secondly, there are views that suppose that open quantum systems are not really fundamental, and that any open quantum dynamics can eventually be described by the closed dynamics of a larger system; this is based on the Stinespring dilation theorem  \cite{nielsen2010quantum}. 

Thirdly, in this work, we take the view that both open- and closed-system dynamics are separately fundamental. This is because, as the classification below shows, both open and closed quantum systems can be given non-overlapping definitions. 

We define, by a `` closed quantum system'', the unitarily evolving potentiality that is temporally bounded by two actualities, one at an initial time and another at a final time, counted by a locally positioned clock. 

In what follows, we denote actuality by the symbol `A ' and potentiality by the symbol` P ', and we use the em-dash symbol ` --- ' to denote temporal ordering, so that what comes to the left of ` --- ' came before what comes to the right of  ` --- '. The parenthesis ` (  ) ' is used to group units consisting of a single `A--P--A' sequence. The curly bracket `\{ \}'  is used to group the actualities and potentialities that occur simultaneously, in a give inertial frame. For instance,  `\{A, P, P \}---\{A, A, P \}' indicates an actuality and two distinguishable potentialities at the earlier time, say t=0 and two actualities and a potentiality at a later time, say t=1. The ordering of entries inside the curly bracket reflect the identity of the respective actuality or potentiality. This identity could be based on spatial location.\\

Recall that Potentiality can have a classical limit. In the notation we have introduced, however, the symbol P does not correspond to the classical limit of Potentialities. It corresponds neither to an improper mixture nor to a proper mixture. 

In what follows, a single P corresponds to the potentiality for which it is operationally meaningful to assign a pure state, based on the specification of the experimental arrangement. So, an `A--P--A' unit is to be interpreted as an experiment where a single system was prepared in a pure state, whereas a ` \{A, A \}---\{P, P \}---\{A, A \}'  sequence is to be interpreted as an experiment involving two systems each prepared in a separate pure state. The location of the A's and P's in the curly bracket distinguishes the prepared systems.

By definition, for closed quantum systems, we have the involvement of actualities and potentialities in a single A--P--A form. That is, when many identical preparations of a given A--P--A sequence are made, one can instantiate the Schrodinger equation in the frequentist probabilistic sense. A closed quantum system is what it takes to operationally instantiate the Schrodinger equation once. In this case, the Born rule simply provides the probabilities for the possible outcomes that can occur. 

An ``open system'' is one in which more than one A--P--A sequence can occur; namely, the Born rule and the Schrodinger equation together dictate such evolutions from the initial to the final state. Non-demolition, weakly coupled measurements \cite{bhattacharya2003continuous} are examples of these. For instance, for an evolution interspersed by one weak non-demolition measurement, we will have the A--P--A--P--A sequence. 

In terms of the classification that is about to follow, Class~1 is a closed quantum system; Classes~3 and~4 are open quantum systems; Class~2 can be either a closed or an open quantum system, depending on the environment's behavior. These classes are labeled in the following classification system.

In the diagrams below, the dark grey scale symbol
\raisebox{-0.2\height}{\includegraphics[height=3.2em]{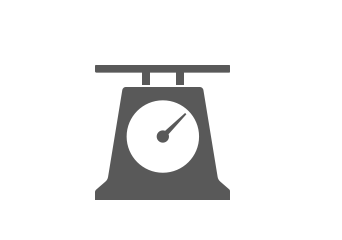}}
denotes clock-attached detectors (an ECM with a clock) or preparation devices (an ECM combined with a suitable ECU and a clock). We additionally assume that these systems are clocked, in the sense that a clock is attached to determine when the preparation was carried out, for how long unitaries were applied, and when the potentialities were intercepted to realize actualities. The dark Grey magnet symbol \raisebox{-0.2\height}{\includegraphics[height=2.2em]{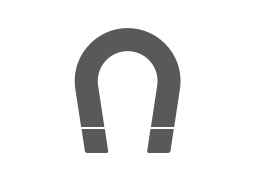}} is used to represent external unitary control (such as those provided by magnets for spin systems). The olive \raisebox{-0.2\height}{\includegraphics[height=3.2em]{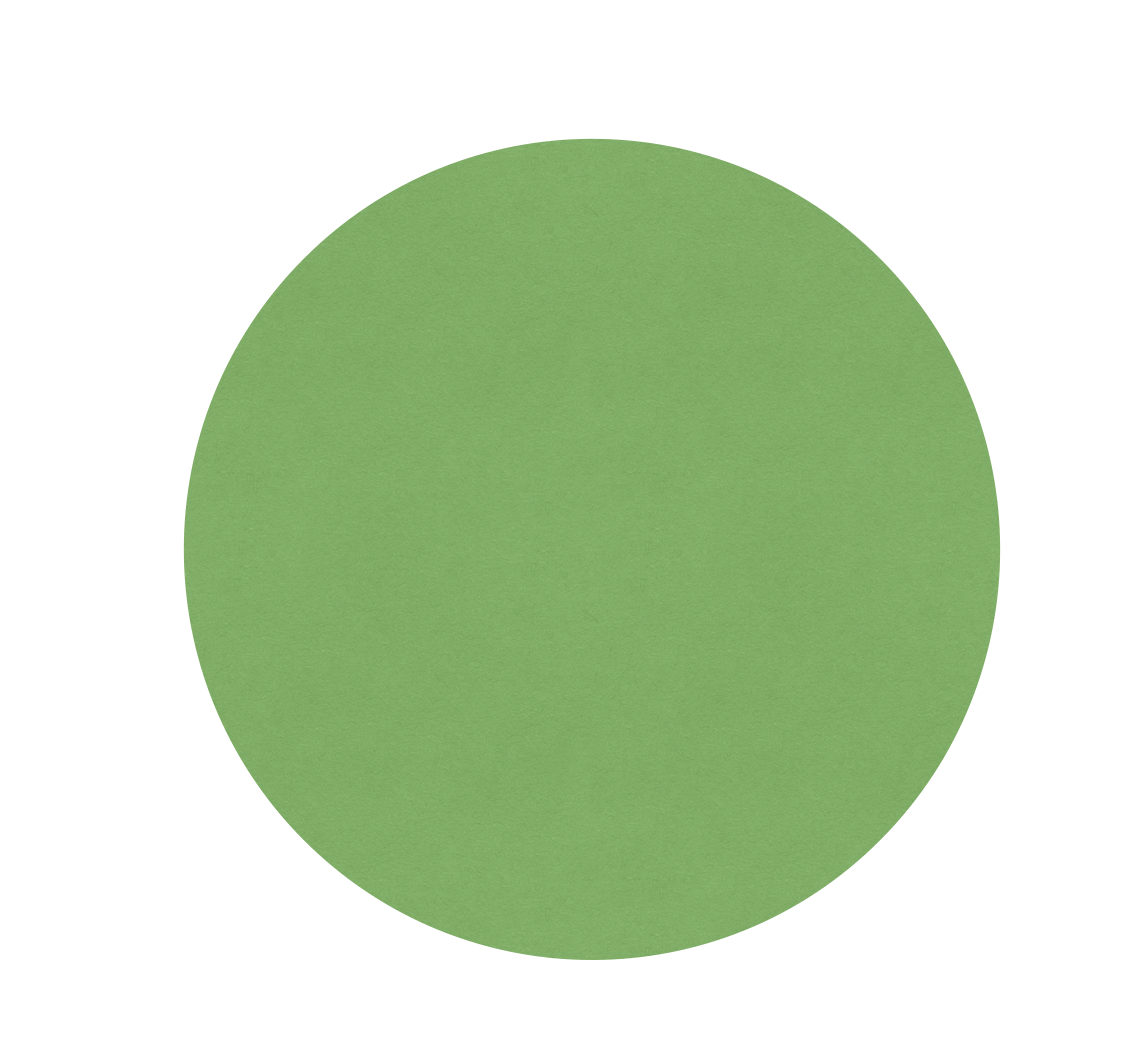}}, brown \raisebox{-0.2\height}{\includegraphics[height=3.2em]{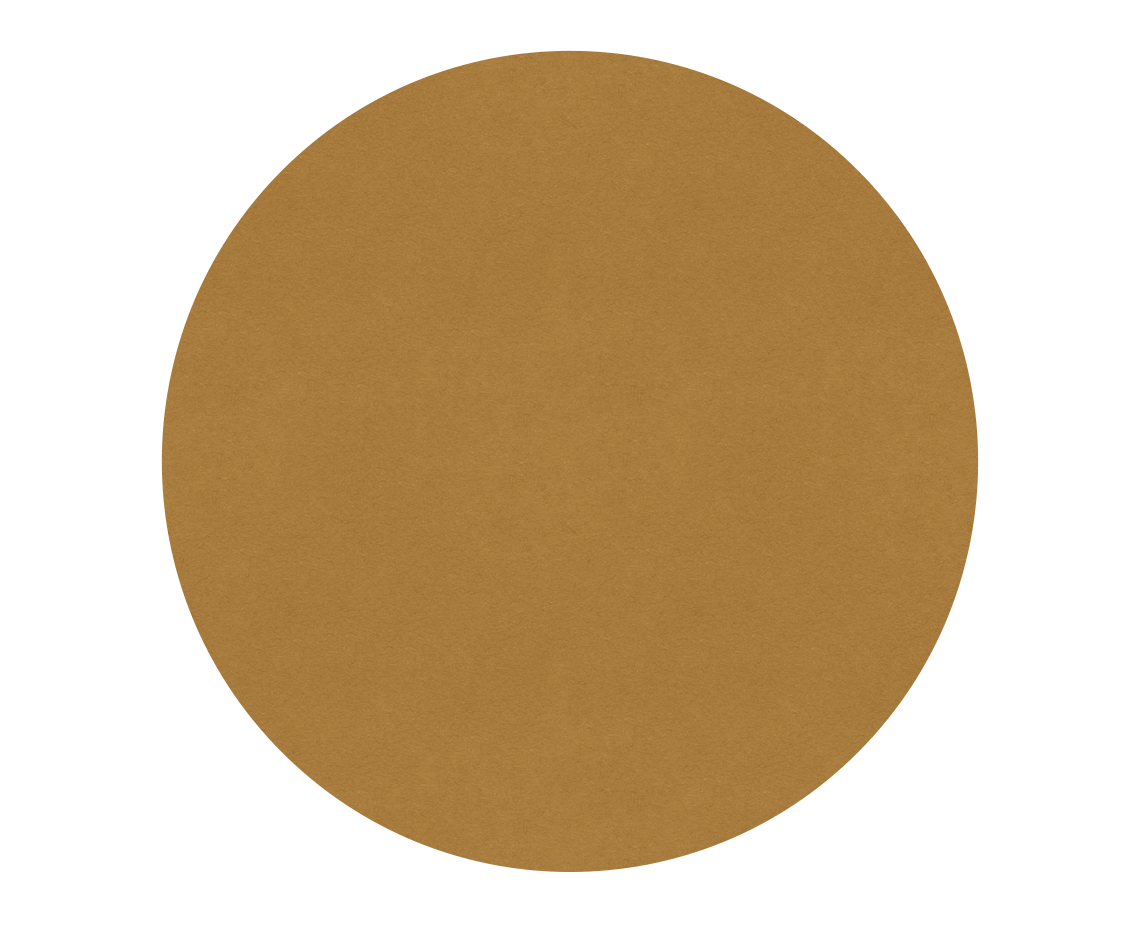}} and blue \raisebox{-0.2\height}{\includegraphics[height=3.2em]{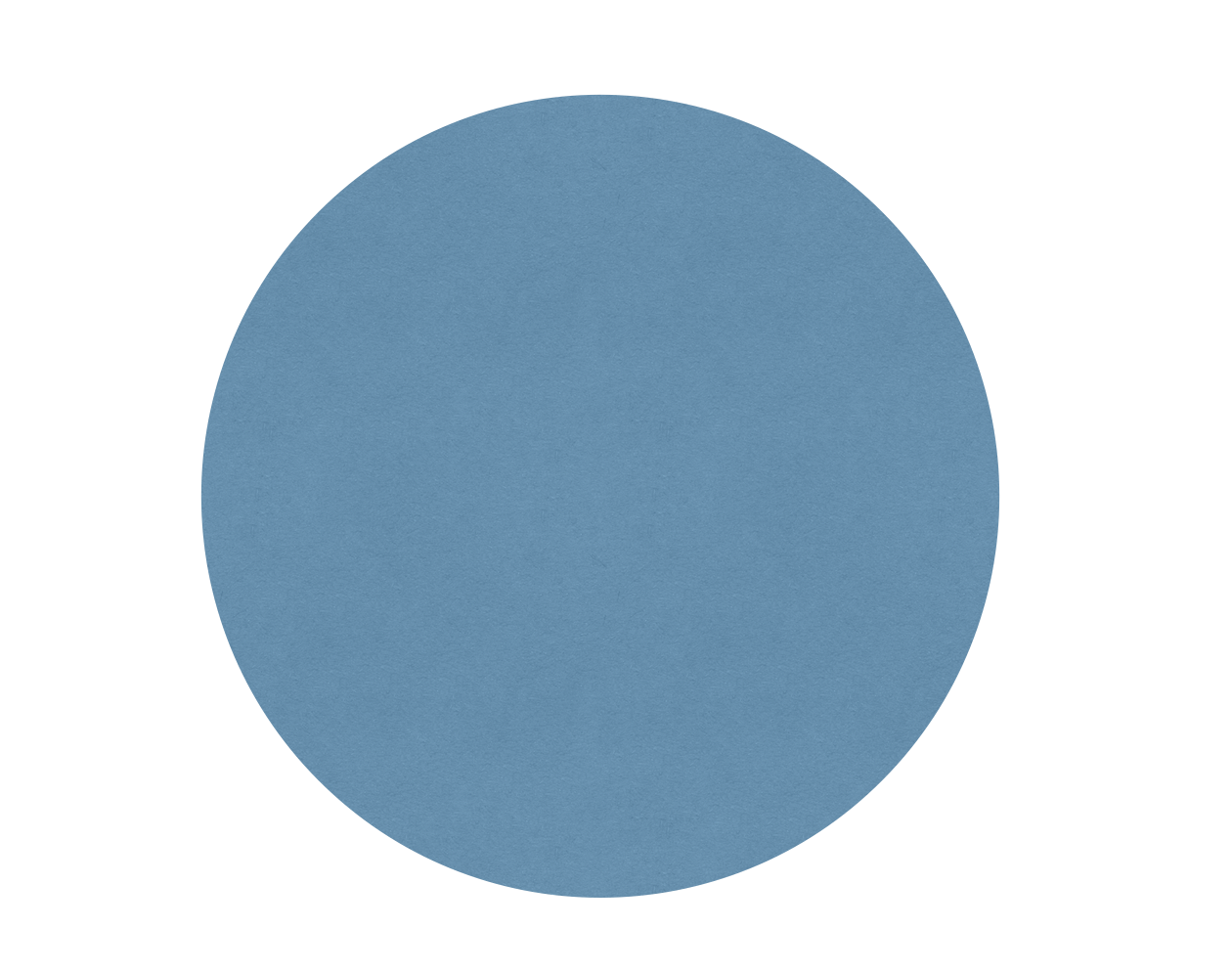}} circles denote distinguishable potentialities that we call `system', `environment/ancilla' and `additional system' respectively.

The classification is presented by using illustrative examples wherever necessary.

\section{ Dynamics Due to Schrodinger}
 \subsection{ Class 1: Unitary Deterministic Dynamics}

The Class 1 process is the dynamics of a potentiality between a state preparation and ending in a quantum measurement. It specifies the evolution of a potentiality described by a pure quantum state, in accordance with the Schrodinger equation.  If we rewrite the pure state in terms of a density matrix \(\rho = |\psi\rangle\langle\psi|\), to accommodate a lack of knowledge about the experimental preparation or update, the evolution is governed by the von Neumann equation. Therefore, more generally, closed quantum systems are characterized by unitary deterministic dynamics in accordance with the von Neumann equation
\begin{equation}
    \dot{\rho} = -\tfrac{i}{\hbar}[H,\rho].
\end{equation}

\begin{figure*}
  \centering
  \includegraphics[
    width=\textwidth,
    alt={A schematic diagram of a closed quantum system, represented by a green circle interlaced by a preparation and a measurement stage. The potentiality evolves unitarily according to the Schrodinger equation}
  ]{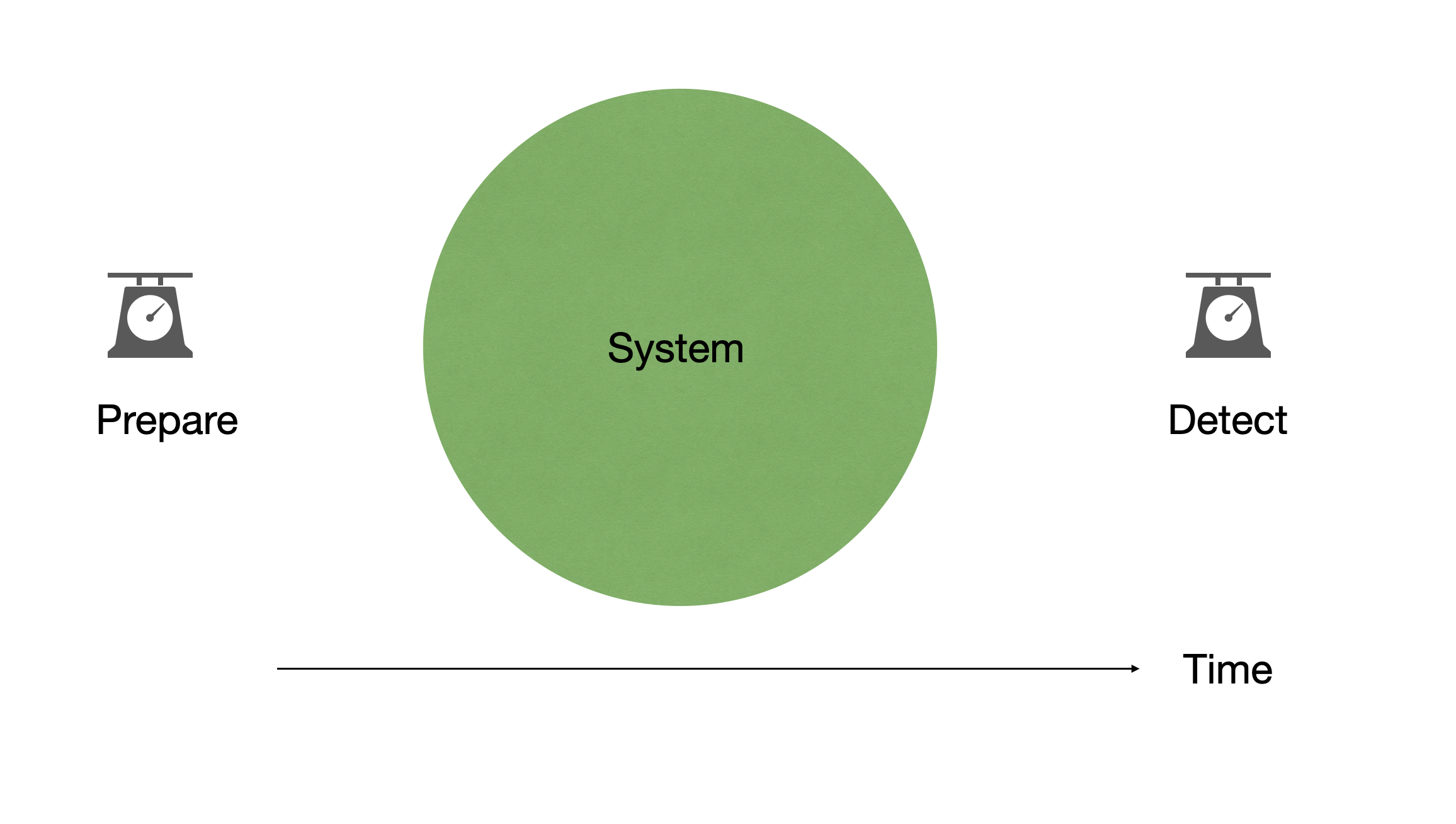}
  \caption{A closed quantum system: unitary deterministic dynamics}
  \label{fig:closed-quantum-system}
\end{figure*}

Under Eq.(3.1), the density matrix evolves unitarily:
\begin{equation}
    \rho(t) = \hat{U}(t)\,\rho(0)\,\hat{U}^\dagger(t).
\end{equation}
The cartoon in Fig.(3.1) illustrates the archetype of a closed quantum system. Notice that the definition of a closed quantum system always requires actualities at the beginning and at the end. Closed quantum systems cannot be defined as stand-alone potentialities.

Class 1, process, is in our notation, of the A--P--A type. One instance of the Class 1 process is one instance of the A--P--A type.

  \subsection{ Class 2: Non-Unitary Deterministic Dynamics}

Now suppose we have an environment $E$, which is another distinguishable potentiality represented by the Hilbert space $\mathcal{H}_E$, with orthonormal basis $\{\ket{e_\alpha}\}$, in addition to the system $S$, a potentiality represented by the Hilbert space $\mathcal{H}_S$. The state of $S$ is described by the reduced density matrix $\rho_S$, where
\begin{equation}
\rho_S \;=\; \mathrm{Tr}_E\!\left[\rho_{SE}\right]
\;=\; \sum_{\alpha} \langle e_\alpha |\, \rho_{SE}\, | e_\alpha \rangle.
\end{equation}

\begin{figure*}
  \centering
  \includegraphics[
    width=\textwidth,
    alt={A schematic diagram of a system that is ambiguous between closed and open descriptions. The diagram represents non-unitary deterministic dynamics, where the system couples with an environment, and the environment is traced out.}
  ]{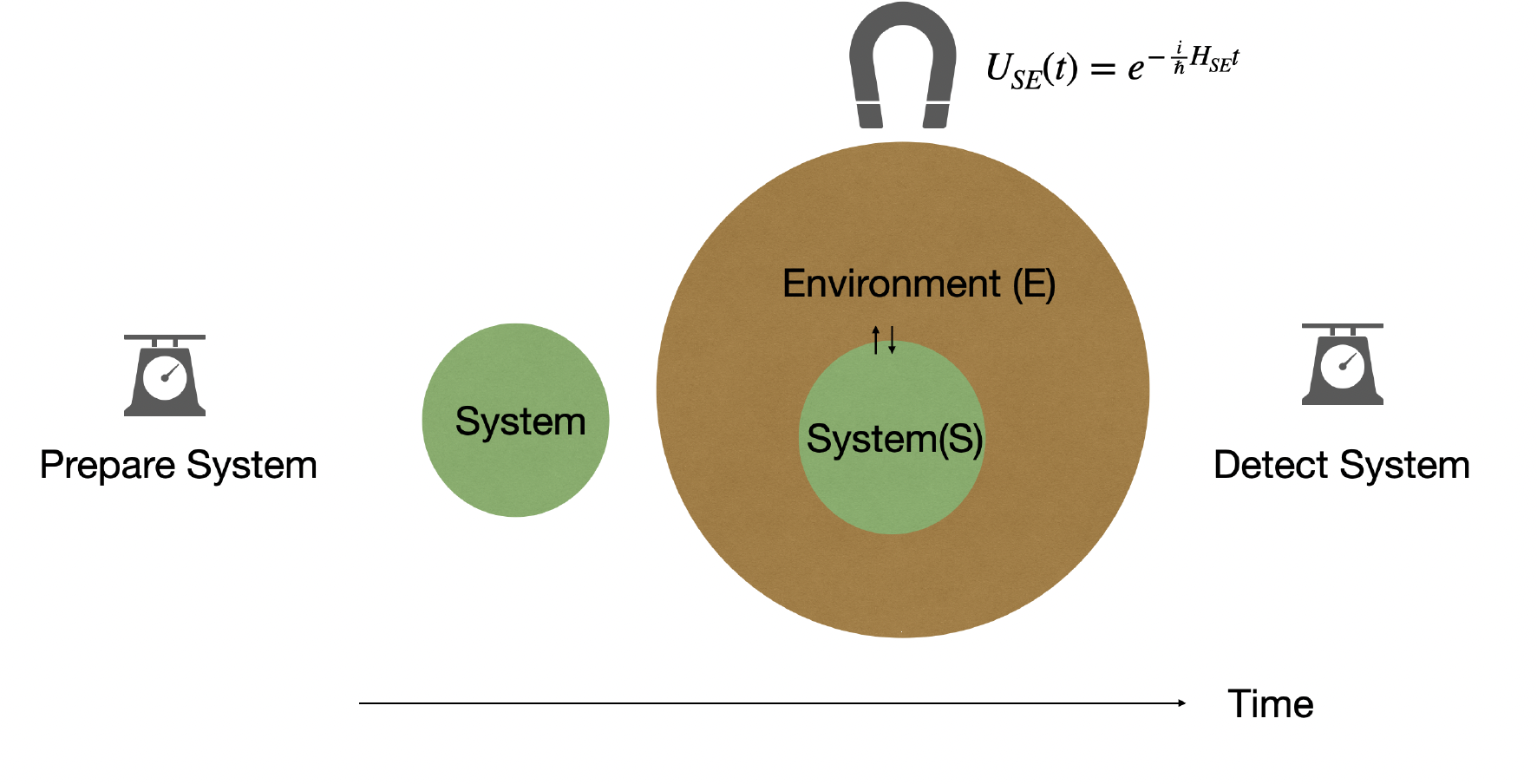}
  \caption{Closed/open ambiguous system: non-unitary deterministic dynamics}
  \label{fig:closed-open-ambiguous-system}
\end{figure*}

In the cartoon in Fig.(3.2), the olive circle denotes the potentiality that we consider to be our system, and the brown circle denotes another potentiality that we can distinguish operationally and that is considered the ``environment.'' The magnet represents the operational stage that couples the environment with the system.

When weak coupling to the environment is assumed, along with the Markov approximation and the rotating-wave approximation, the time evolution of the reduced density matrix is given by the GKSL equation \cite{opensystems} (Gorini--Kossakowski \\--Sudarshan--Lindblad, also known as the Lindblad master equation):
\begin{equation}
\frac{d\rho_S}{dt}
= -\,\mathrm{i}\,[H_S,\rho_S]
+ \sum_{i,j} c_{ij}\!\left(
F_i \rho_S F_j^\dagger
- \tfrac{1}{2}\{F_j^\dagger F_i,\rho_S\}
\right).
\end{equation}
Here, $H_S$ is the Hamiltonian of the system acting on $\mathcal{H}_S$, and $\{F_i\}$ are the Lindblad operators that form a basis in operator space and represent the different decoherence channels. The coefficients $c_{ij}$ are the elements of a matrix known as the Kossakowski matrix $C$, with $C$ positive semidefinite ($C \ge 0$), which makes sure that the time evolution is completely positive. When the matrix $C$ is diagonalized, the FGKSL equation takes a simpler form with a new set of Lindblad operators:
\begin{equation}
\frac{d\rho_S}{dt}
= -\,\mathrm{i}\,[H_S,\rho_S]
+ \sum_{k} \gamma_{k}\!\left(
L_k \rho_S L_k^\dagger
- \tfrac{1}{2}\{L_k^\dagger L_k,\rho_S\}
\right),
\end{equation}
in which case $\gamma_k$ has the physical interpretation of decay rates \cite{opensystems}.

The right-hand side can be written as a superoperator $\mathcal{L}[\rho_S]$, in which case the GKSL equation becomes
\begin{equation}
\frac{d\rho_S}{dt}
= \mathcal{L}[\rho_S].
\end{equation}
Thus, the evolution equation resembles that for pure states (for time independent superoperator):
\begin{equation}
\rho_S(t)= e^{\mathcal{L} t}\rho_S(0).
\end{equation}

To illustrate these ideas, let us consider a well known example involving a two-level system.

We can ignore the unitary part (which is not interesting for the present case), and so one can set
\[
H = 0.
\]
One can take the Lindblad operator to be
\begin{equation}
F = \sqrt{\frac{\gamma}{2}}\,\sigma_z.
\end{equation}
This represents decay of coherence. The Lindblad master equation for such situations is 
\begin{equation}
\dot{\rho}
= \frac{\gamma}{2}\!\left(\sigma_z \rho \sigma_z - \rho \right),
\end{equation}
 where $\gamma$ is the dephasing rate (a measure of how fast the off-diagonal terms go to zero)\textbf{.}

For the initial state,
\begin{equation}
\rho(0) =
\begin{pmatrix}
\rho_{00} & \rho_{01} \\
\rho_{10} & \rho_{11}
\end{pmatrix},
\end{equation}
the solution would look like
\begin{equation}
\rho(t) =
\begin{pmatrix}
\rho_{00} & e^{-\gamma t}\,\rho_{01} \\
e^{-\gamma t}\,\rho_{10} & \rho_{11}
\end{pmatrix}.
\end{equation}
which gives\textbf{ }the\textbf{ }time evolution of the population (diagonal elements) and coherence (off-diagonal elements).

The dynamics in Eq.(3.4) is deterministic, but in general non-unitary due to the second term. In the classification scheme introduced at the beginning of this chapter, the dynamics presented by Eq.(3.4) can be considered either a closed system or an open system, depending on whether the environment is monitored. Eq.(3.4) is agnostic with respect to monitoring of the environment. This is reflected in the two different ways of deriving Eq.(3.4).\\
One way to obtain the GKSL equation is to start from the closed quantum system description of a composite system and, after applying physically justified approximations such as weak coupling, the Born approximation, and the rotating-wave approximation, derive the GKSL equation for the dynamics of a subsystem of the composite. Another way to obtain the GKSL equation is to consider the stochastic Schrodinger equation (which will become relevant in the next section) describing continuously and weakly monitored quantum systems, and to perform ensemble averaging over the quantum trajectories,
\[
\mathbb{E}\!\left[\rho_S^{\mathrm{cond}}(t)\right]
=
\rho_S(t).
\]

Where the ensemble averaging $\mathbb{E}$ is over the different stochastic realizations of the quantum trajectory.

Locally, when we ignore the environment, the two situations are equivalent, but globally the distinction remains. In the unmonitored case, the evolution has the A--P--A sequence, whereas in the monitored case, the evolution will in general have a sequence that looks like $(A\!-\!P\!-\!A)\!-\!P\!-\!(A\!-\!P\!-\!A)\!-\!P\!-\!(A\!-\!P\!-\!A)\!-\!P \ldots$
As qualified earlier, in the A--P--A sequence the system is closed and in the $(A\!-\!P\!-\!A)\!-\!P\!-\!(A\!-\!P\!-\!A)\!-\!P\!-\!(A\!-\!P\!-\!A)\!-\!P \ldots$ sequence the system is open.

One way to implement the A--P--A form for the dynamics described in Eq.(3.5) is to design a controlled experiment where we take a sequence of distinguishable environment fragments and couple them unitarily and sequentially to the system, in such a way that the the approximations that are needed to derive the GKSL equation is physically realized, but then ensure that the fragments are not measured. Then such a scenario when globally considered, as opposed to focusing on the system, would describe a  A--P--A form, since it would constitute a single quantum problem describable by the  Schrodinger equation. 

Note that this illustrates that the  A--P--A form is not necessarily about mono-partite systems. It can also be about multi-partite systems, as long as the dynamics instantiates a well-defined quantum problem in the sense defined earlier: i.e. instantiates the Schrodinger equation for an operationally well specified  Hamiltonian and initial state, whether it be on a mono-partite or on a multi-partite system.

For later reference, we shall label the non-unitary deterministic dynamics discussed in this section as ``Class~2.'' Class 1 and Class 2 so far can be classified under ``dynamics due to  Schrodinger'', since the Born rule functions mostly as an updating rule and does not play a role in the Dynamics.
\section{ Dynamics Due to Schrodinger and Born}
  \subsection{ Class 3: Non-Unitary Stochastic Dynamics}

The next class of dynamics is explicitly of the sequence $(A\!-\!P\!-\!A)\!-\!P\!-\!(A\!-\!P\!-\!A)\!-\!P\!(A\!-\!P\!-\!A)\!-\!P \ldots$ and is thus an open system, in that its time evolution is shaped by both the Schrodinger equation and the Born rule. The dynamics is stochastic, due to the involvement of measurement back-action \cite{wiseman2009quantum}, and also non-unitary due to the noise term—the second term—in the general form of the equation of motion for such dynamics, which has the form (for a single measurement channel and for homodyne detection) \cite{wiseman2009quantum}
\begin{equation}
d\lvert \psi_t \rangle
=
f(t)\,\lvert \psi_t \rangle\, dt
+
g(t)\,\lvert \psi_t \rangle\, dW(t),
\end{equation}
where the first term is the unitary part, and the second term arises from the stochasticity induced by measurement back-action. In this case, $dW(t)$ is a Wiener process. 

To illustrate how such processes work, one can consider a simple toy model. First, consider the cartoon in Fig.(3.3), which serves as an illustration of the process. We have two distinguishable potentialities, one represented by the color olive and another by brown. The olive potentiality is the ``system,'' and the brown potentiality is the ancilla (controlled environment). At equal intervals of time, as counted by a clock, the system is coupled to the ancilla through a weak interaction (symbolized here by the magnet), and the ancilla is projectively measured to gather information about the system. The system then evolves unitarily for a short interval of time (the region with the isolated olive circle). This process continues into the next interval of time, conditioned on the state of the system obtained in the preceding interval.

\begin{figure*}
  \centering
  \includegraphics[
    width=\textwidth,
    alt={A schematic diagram of an open quantum system undergoing non-unitary stochastic dynamics. The diagram represents a quantum trajectory in which the system evolves under stochastic measurement or environmental interactions rather than purely unitary dynamics.}
  ]{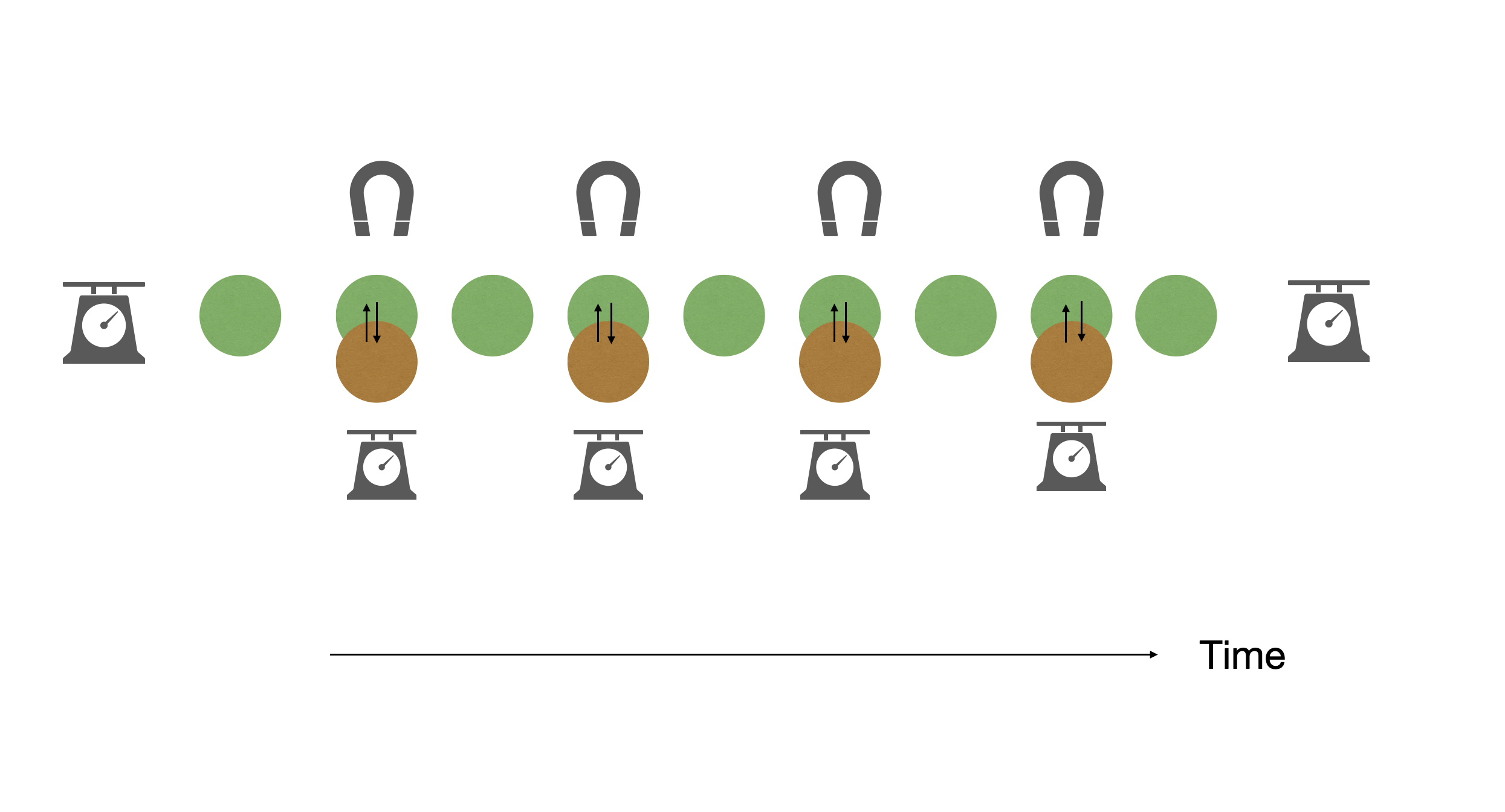}
  \caption{Open system: non-unitary stochastic dynamics}
  \label{fig:open-system-stochastic-dynamics}
\end{figure*}
For illustration, let us consider a two-level system, and let the initial state of the system be
\[
|\psi_0\rangle = |1\rangle.
\]
The initial values of the Bloch vector are therefore
\[
(x_0, y_0, z_0) = (0, 0, -1).
\]

We can conveniently choose the weak measurements to be defined by the following two Kraus operators:
\begin{equation}
K_{+} =
\begin{pmatrix}
\sqrt{(1+m)/2} & 0\\
0 & \sqrt{(1-m)/2}
\end{pmatrix},
\end{equation}

\begin{equation*}
K_{-} =
\begin{pmatrix}
\sqrt{(1-m)/2} & 0\\
0 & \sqrt{(1+m)/2}
\end{pmatrix}.
\end{equation*}

Notice that these operators satisfy the normalization condition
\[
K_{+}^\dagger K_{+} + K_{-}^\dagger K_{-} = I.
\]
In the Kraus operators, when $m=1$, we recover the limit of strong measurements, namely, projections, and when $m=0$, we obtain the limit when no measurement happens. Thus, $m$ quantifies the measurement strength. Between weak measurements, the system evolves unitarily:
\begin{equation}
U = \exp\!\left[-\,i\,\frac{\theta}{2}\,
\frac{\sigma_x + \sigma_y}{\sqrt{2}}\right].
\end{equation}
More precisely, at each odd time step, the evolution of the potentiality is described by the unitary operator in Eq.~(3.14). At each even time step, a weak measurement involving actualities occurs and is described by the Kraus operators in Eq.(3.13).

For state $|\psi_k\rangle$ at the beginning of an odd time step during the process, the probabilities of the two possible measurement outcomes $K_+$ or $K_-$ at the end of the next even time step when an actuality is realized, are given by the Born rule
\[
p_{\pm} = \langle \psi_k |\, U^{\dagger} K_{\pm}^{\dagger} K_{\pm} U \,| \psi_k \rangle,
\]
and the state updates according to
\[
|\psi_{k+1}\rangle =
\frac{K_{\pm} U |\psi_k\rangle}{\sqrt{p_{\pm}}}.
\]

In such a case, it can be shown that there will be a random walk executed by the Bloch vector. That is, the conditioned state and the associated expectation values of observables will undergo a stochastic evolution.
We label this process of the $(A\!-\!P\!-\!A)\!-\!P\!-\!(A\!-\!P\!-\!A)\!-\!P\!-\!(A\!-\!P\!-\!A)\!-\!P \ldots$ type as ``Class~3.''

  \subsection{Class 4: Non-Unitary Stochastic Dynamics with Competition between Entangling operations and Measurements}

 The next class of processes, depicted in\textbf{ }Fig.(3.4),  involves dynamics where, in addition to the stochastic dynamics of the state due to monitoring, the amount of measurable entanglement between subsystems also undergoes a stochastic evolution.

\begin{figure*}
  \centering
  \includegraphics[
    width=\textwidth,
    alt={A schematic diagram of entanglement cluster dynamics in a many-component quantum system. Subsystems are periodically subjected either to unitary evolution or to measurement-like processes, producing changing clusters of entangled components within an open-system description.}
  ]{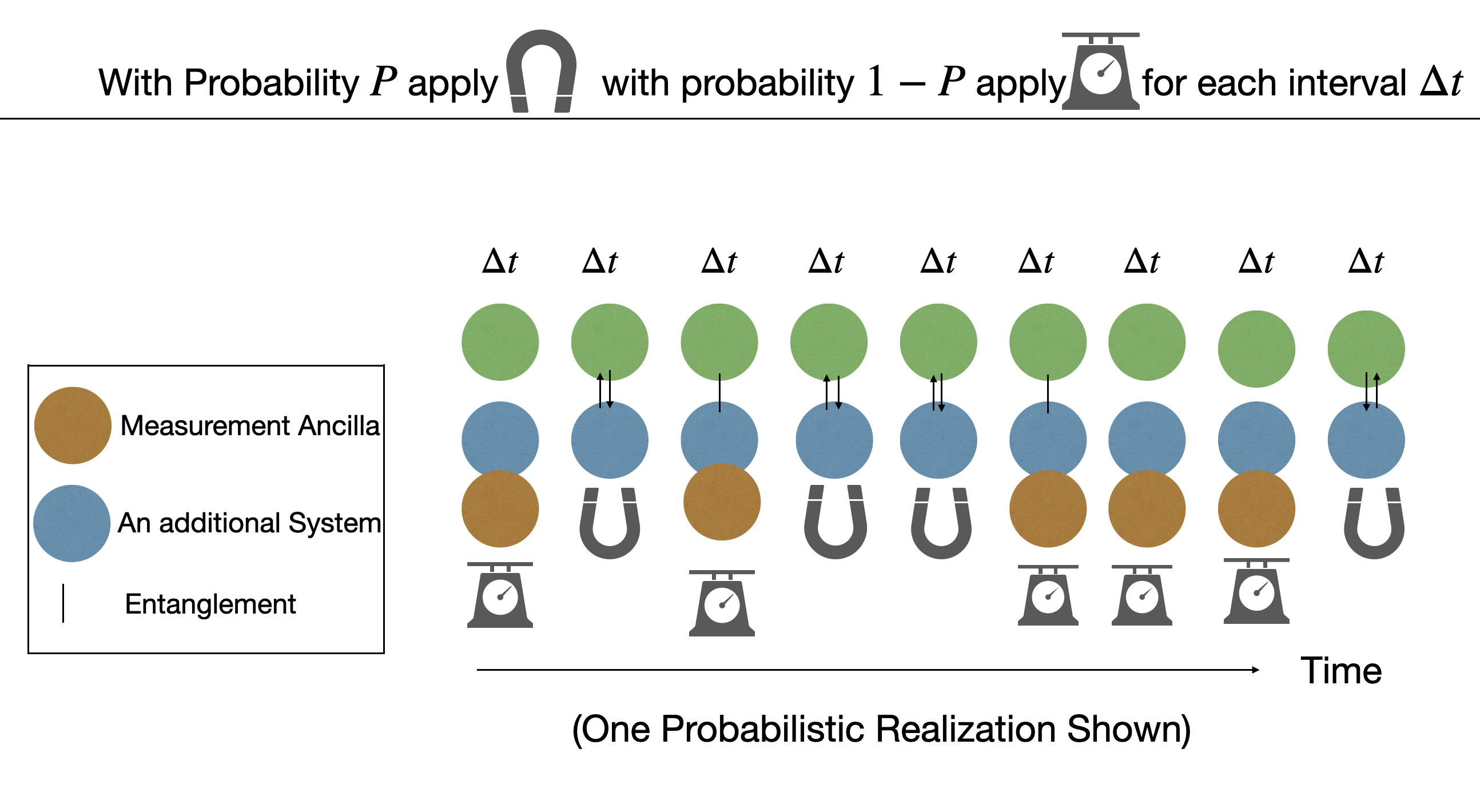}
  \caption{Open system: ``Entanglement Cluster Dynamics''}
  \label{fig:entanglement-cluster-dynamics}
\end{figure*}
 
 In such processes, there is competition between operations that generate entanglement between subsystems and operations that destroy entanglement between subsystems. This cannot be considered to be of the $(A\!-\!P\!-\!A)\!-\!P\!-\!(A\!-\!P\!-\!A)\!-\!P\!-\!(A\!-\!P\!-\!A)\!-\!P \ldots$ type, but more accurately of the type represented by $\{A, P, A \ldots\}\!-\!\{P, P, A \ldots\}\!-\!\{A, A, A \ldots\}\!-\!\{A, A, P \ldots\}\!-\!\{A, P, A \ldots\}\!-\!\{A, P, P \ldots\}\!-\!\{P, P, P \ldots\}\!-\!\{A, P, A \ldots\}\!-\!\{A, P, P \ldots\}\!-\!\{P, A, A \ldots\}\!-\!\{A, P, A \ldots\}\!-\!\{A, A, A \ldots\} \ldots$. Here, as before, the symbol `—' denotes temporal ordering, and the set $\{A, P, A \ldots\}$, denote that distinguishable actualities and potentialities can co-exist.  $\{A, P, A\}$ could mean system 1 is present as an actuality, system 2 as a potentiality, and system 3 as an actuality. 
  
Such processes were first described by Dorit Aharonov \cite{Aharonov2000NoisyQC} and was later studied in the context of measurement induced entanglement phase transitions \cite{Skinner2019}. The idea is the following. There exists an array of qubits in the random quantum circuit. In each small interval of time, for each lattice site there is a probability $1-P$ with which the qubit is measured, and a probability $P$ that an entangling unitary gate is applied that entangles the qubit with a fixed number of nearby qubits (perhaps the nearest neighbours). When the value of $P$ is high, the entanglement among the qubits tends to grow with time, and when the probability $P$ is low, the qubits tend not to be entangled with each other.

To illustrate this kind of dynamics, let us consider a toy model with just two operationally distinct systems represented in Fig.(3.4) by Olive and Blue circles. In Fig.(3.4) the brown circle still represents the ancilla, and the olive and blue circles represent system A and system B, which are mathematically described by the Hilbert space

$\mathcal{H} = \mathcal{H}_A \otimes \mathcal{H}_B$

Let the initial state of A and B be given by the product state
\[
|\psi_0\rangle = |00\rangle .
\]

Here is how the process proceeds. At each time step, one of two possibilities can happen probabilistically. An entangling unitary is applied with probability $P$. The entangling operation we choose is Hadamard followed by CNOT.
\[
U_{\mathrm{ent}} = \mathrm{CNOT}\,(H \otimes I),
\]
where the Hadamard is in matrix form
$H = \tfrac{1}{\sqrt{2}}\begin{pmatrix}1 & 1 \\ 1 & -1\end{pmatrix}$.
The CNOT gate acts with qubit A as control and qubit B as target; in this case the state updates as
\[
|\psi_t\rangle = U_{\mathrm{ent}}\,|\psi_{t-1}\rangle .
\]

A weak measurement is applied with probability $1-P$. One of the qubits is chosen at random, each with probability $(1-P)/2$, and a weak measurement in the $\hat{\sigma_z}$ basis is applied. The Kraus operators are
\begin{equation}
K_{\pm} = 
\begin{pmatrix}
\sqrt{\tfrac{1\pm m}{2}} & 0\\
0 & \sqrt{\tfrac{1\mp m}{2}}
\end{pmatrix},
\qquad 0 < m < 1,
\end{equation}

Now, the state evolves stochastically as in Class 3, but more interestingly we have an additional structure. The entanglement between A and B also undergoes a stochastic process.

We label the type of process in this section as ``Class 4''.

\chapter{A Synthesis of the Views of Bohr and Heisenberg}

In this chapter, we attempt a synthesis of Bohr and Heisenberg, each supplying his own contribution, which, along with the current author's view, together paints the picture that we cannot physically meaningfully say what entities the universe fundamentally contains beyond the borders set by the preconditions for empirical knowledge. And the borders set by the preconditions of empirical knowledge are themselves a physically relevant feature of the comprehensible universe.

Niels Bohr’s contribution to the conceptual foundations of quantum mechanics, often summarized under what is now called Bohr’s version of the Copenhagen interpretation, takes the form of epistemological\footnote{I have used ``epistemology'' or ``epistemic'' several times so far without really defining it. That is because it was assumed that the readers in the foundations of quantum mechanics have a sense of what it means. But I now want to define it, to emphasize that epistemology has an object of study too, as opposed to viewing it as the physically irrelevant background in which we study physical objects. Epistemology \cite{sep-epistemology} is the inquiry into how we know what we know—what conditions make knowledge possible and what justifies our claims about the world. In physics, such questions cannot be ignored, because the content and meaning of our theoretical descriptions depend on the ways in which knowledge of physical reality can be obtained. I also want to define ontology, to emphasize that any verifiable definition of ontology drags epistemic elements along with it \cite{Barad2007}: Ontology  \cite{sep-ontological-commitment} concerns the inquiry into what fundamentally exists—the basic, irreducible elements or categories that constitute reality. It asks what must be taken as primitive in describing the physical world.} lessons concerning the nature of meaning and objectivity in scientific practice. These lessons serve as a necessary supplement to the orthodox machinery of quantum mechanics, in the sense that they articulate the preconditions for acquiring quantum–experimental knowledge; preconditions that the machinery of quantum theory itself cannot self–supply. 

We can give a realist reinterpretation of Bohr’s interpretation of quantum mechanics, expanding what it means to be a `realist’ account\footnote{See Appendix B, conversations with Professor Bradley Armour Garb, for more on this matter.}. 

The usual sense of a ``realist’' \cite{sep-realism-about-metaphysics} stance supposes that objects (whether small or large) have states or properties regardless of measurements. This stance often brings with it, as an associated assumption, though not as a necessity, a tacit metaphysical commitment inherited all the way from Leucippus and his student Democritus, namely the assumption that what is fundamental lies only at the smallest physical scales and that all observable physical phenomena are “no more than”  \cite{sep-mereology}\cite{sep-scientific-reduction} the interactions of the universal, independently existing, smallest-scale fundamental constituents. The realist implementation of Bohr’s stance on quantum mechanics demands a more layered, functionally defined identification of “fundamental.” This shall be investigated further in Chapter 6. It is realist in the sense of first subsuming what we call an “observer” as a functionally defined conceptual primitive, just as clocks are for time, and thence using the following expanded conceptual base to define the objective—in the sense of inter-subjective objectivity among all observers— existence of systems with properties and states.

To clarify this expanded realist picture, it is useful to make explicit the ontological commitments.

In the earlier chapters, we initiated the construction of this realist picture by reintroducing the ontological commitments associated with Heisenberg's notion of actuality and potentiality. 

At the same time, it is true that, for Bohr, there was no `quantum world,' and he is reported to have said to Aage Peterson \cite{Petersen1963Philosophy}\cite{bub2025thereisnoquantumworld}: ``There is no quantum world. There is only an abstract quantum physical description. It is wrong to think that the task of physics is to find out how nature is. Physics concerns what we can say about nature.'' But here we adopt an ontic-potentiality realism \cite{kastner2018takingheisenbergspotentia}, as opposed to the seemingly anti-realist stance echoed in the above quote attributed to Bohr. Here, we attribute realism to potentialities not in the substance-dualist sense, but in terms of what operational interventions real or counterfactual observers can perform and the observed responses to those interventions. A beam of X-rays in vacuum, which qualifies as the classical limit of potentialities, is `real' because it can cause a change in my hand if I obstruct the beam, and the moon of Saturn, Titan, is real because a real or counterfactual fictitious observer can alignment-position-measure the location and extent of Titan.

The arising synthetic view then involves subsuming observers as primitive, functionally defined entities, and the notion of actualities and potentialities as irreducible categories of existence that can nevertheless be operationally defined.

The introduction of potentialities and actualities also provides an opportunity for reinterpretation of the usual primitives of particles or fields. Of course, it is clear that actuality and potentiality are not a mere terminological replacement of particles and fields. Rather, they are ontologically defined with respect to a different set of criteria. Whether we can fully replace the particle and field picture with a description in terms of actualities and potentialities remains not fully settled in this work.

In the next section, we make explicit the ontological commitments of the arising synthetic view as twelve assumptions

\section{Twelve Assumptions on the Foundations of Quantum Mechanics}

Additional explanation is given wherever necessary.

\textsc{Assumption 1: Concerning the Meaning of the wavefunction}:

\vspace{0.25cm}

The wavefunction $\psi$ is the probability amplitude, that describe the potentialities in closed quantum systems, as described by the Class 1 process—of the A-P-A type—and it's meaning has semantic character, in the sense of reflecting the experimental description of the ECM in the beginning A, and of the ECM in the concluding A, and finally in the description of the ECU determining the Hamiltonian to which the P is subject. Here it suffices that the experimental description be given by a counterfactual observer. The wavefunction can also be seen to have the ontic character in the strict sense that the demarcation between A and P is objective for all observers by postulate 2. 

A Bayesian meaning can also be attached to the wavefunction if we imagine the observers as gamblers, using the experimental preparation manifest in the beginning A and the concluding A. Here, the Bayesian reasoning concerns rational belief updating among observers. For instance in the state \[\frac{1}{\sqrt{2}}(\ket{\uparrow}+\ket{\downarrow}),\]  rational decision making based on the outcomes demands equal betting weights due to Born rule giving equal probability amplitude for both possibilities in the $\{\uparrow, \downarrow\}$ basis. 

There are no relative pure state assignments for the same system, due to postulate 2. Only the density matrix description can vary among observers. 

Moreover, the probabilities given by the Born rule, unlike classical probabilities, cannot be out-competed by any observer. That is, in the above example, since $P(\uparrow)=\frac{1}{2}$ and $P(\downarrow)=\frac{1}{2}$, there cannot exist any observer with respect to whom either $P(\uparrow)> P(\downarrow)$ or $P(\uparrow)< P(\downarrow)$. The probabilities obtained from wavefunctions via the Born rule represent intrinsic randomness.

 As for ``where the wavefunction exists ’’, the wavefunction exists in the information processing units of the observers, while describing the potentialities, contingent on the structure of the actualities (the experimental arrangements).

Bohr’s view of the wave function fits neither the $\psi$-epistemic nor the $\psi$-ontic categories as defined by Harrigan and Spekkens  \cite{HarriganSpekkens2010}. 

For Bohr, the wave function is \emph{epistemic} only in a semantic or representational sense: it is a symbolic expression required to specify and encode experimental arrangements and events for the purpose of making intrinsically probabilistic predictions about subsequent experimental occurrences. It is not epistemic in the modern sense of representing ignorance about underlying hidden variables—what may be called \emph{ignorance epistemicity}—but rather epistemic only in reflecting the limits of the means by which experimental knowledge of quantum systems is possible at all, which we may call \emph{semantic epistemicity}.

At the same time, despite the semantic epistemicity of the wave function, Bohr nevertheless regarded the wave function as \emph{ontic}. This is not because it describes a physical field in configuration space, as in $\psi$-ontic collapse models such as GRW, but because it provides an unambiguous, inter-subjective objective, communicable specification of experimental arrangements together with the statistical regularities of observed phenomena. In this sense, Bohr’s stance is simultaneously epistemic and objective, but in ways that do not align with the contemporary $\psi$-epistemic / $\psi$-ontic taxonomy.

Quantum Mechanics is complete and is a universal theory instantiable \footnote{Can be realized for any physical system, even relatively massive ones like rocks or cats} for any system regardless of its mass; however, unlike in classical mechanics (where any system is always associated with a classical state), not every system comes “in-built” with a quantum state, and quantum state assignment is contingent on the prevailing inter-relationship between the counterfactual or literal observer, the actualities and the potentialities. It must be noted, however, that by “Quantum Mechanics” (or alternatively QT, when referring to general quantum theories such as the relativistic quantum theory) we mean both of Von-Neumann mathematical processes on the wavefunction (Schrodinger dynamics and Born update), not just the Schrodinger dynamics; however, being mathematical processes, they occur as information processing in the computational units of  counterfactual or literal  observers, while reflecting the prevailing inter-relationship between observers, the actualities (i.e., experimental arrangements) and the potentialities.

\textsc{Assumption 2: Concerning Wavefunction Assignment, Unitary Determination, Subsystem Partitioning, and Space-Time Coordination}:

\vspace{0.25cm}

Epistemic constraints, which manifest as actuality, with respect to a real or a counterfactual observer is unavoidable in the updating and determination of the wavefunction corresponding to a potentiality, in determining sub-system partitioning \cite{lloyd2004observable}, in determining what observable is being measured, in the determination of the unitaries involved, and in counting time and measuring distances.

\textsc{Assumption 3: Complementarity}:

\vspace{0.25cm}

Complimentarity is an epistemic principle (here, the epistemic is in the semantic and representational sense, not in the ignorance sense).

More specifically, complementary principle expresses an unavoidable constraint in the operational manipulation of  experimental arrangements: The ECM and ECU corresponding to mutually incompatible observables $\hat{A}$ and $\hat{B}$ cannot coexist.

Let the experimental arrangements for an observable be specified by a set of experimental propositions. Let this be the specification S. The experimental arrangements for an incompatible observable is specified by the experimental propositions S'. There can be no arrangement that is specifiable as $S \wedge S'$ in the same experiment.

For instance, the experimental arrangement of the magnets for the measurement of $\hat{S_x}$ and $\hat{S_y}$ cannot co-occur, because if we aligned the magnets for the measurement of $\hat{S_x}$ and $\hat{S_y}$ in the same setup, we would no longer be measuring $\hat{S_x}$ and $\hat{S_y}$, but instead be measuring a different observable, namely $\hat{S_x} + \hat{S_y}$ . Note $[\hat{S_x}, \hat{S_y}] \neq 0$. 

This is unlike the classical mechanical case, where angular momentum components can simultaneously be determined in the same experimental arrangement because in $L_x= y p_z - z p_y$ and $L_y= z p_x - x p_z$, all the individual momenta and coordinate components exist in a counterfactual definite sense with respect to some observer.

The next assumption assumes the validity of Quantum theory as it stands \cite{Dirac1930}, as a complete theory of nature, with no modification.

\textsc{Assumption 4: The Dirac Von-Neumann Postulates apply to the assigned wavefunctions, Hamiltonian, and subsystems}:

The wavefunctions, unitaries, and subsystem partitioning that get objectively determined in terms of actualities, even in the absence of literal observers, as long as counterfactual fictitious observers can be assumed, are subject to the usual Dirac-Von Neumann postulates of quantum mechanics, and dictate the behavior of potentialities.

It must be noted, however, that the Born Rule and the Schrodinger equation are both relevant in describing general classes of dynamics. The Born Rule is both a normative rule for observers to reason and to intervene operationally, and a descriptive rule for the pattern in which outcomes materialize in actualities.

The next assumption organizes how causal relations are analyzed.

\textsc{Assumption 5: Nature of Events and Causality principles}:

Only actualities constitute events in spacetime, and are thus subject to the causal constraints imposed by relativity. The counting of intervals of time in clocks happens by the indexing of a sequence of distinguishable actualities. While potentialities can also be used to count time intervals, those intervals are always demarcated by actualities.

Note that even a quantum clock (even with quantum control) does not evade this constraint. In an atomic clock, we still always require interspacing actualities (in the form of classical control pulses that are sent to the atom) to infer the `clicks' counted by the quantum clock. One might argue that we can use another quantum clock to control a quantum clock; however, we will still need another interspacing of actualities to count the latter control quantum clock. 

Assumption 5 concerns why we have no trouble with relativistic causality when we interrogate the subsystem of an entangled pair, with the other subsystem in a 
region outside the lightcone of the interrogation region in space-time. The latter subsystems' state would have changed, but they remain potentialities and thus do not constitute events.

The next assumption concerns the constraint that any observer must satisfy. 

\textsc{Assumption 6: Operating constraints on Experimenters}:


An observer can either have quantum mechanical information processing or classical information processing, but the encoding and decoding of memories, the hardware for the processing of those memories, and the counting of time intervals will always require actualities. This assumption is being made because that appears to be the nature of the knowledge-acquisition system we have observed so far.


{\textsc{Assumption 7: Potentialities `inside' volumes bounded by actualities}:

Potentialities can exist inside (a localized volume in space) whose boundaries qualify as actualities. 


This assumption expresses a relationship between potentialities and actualities. Consider electricity that can pass through an Aluminum wire. The Aluminum wire as a whole has a definite location, and we can alignment position-measure it from the light emitted by the wire. At the same time, we can have a current that can flow in a superposition of two directions, even when the wire itself still occupies a localized position with respect to an observer. In this sense, potentialities can occur inside volumes bounded by actualities.

\textsc{Assumption 8: Construction of Miniature abstract Models for Time Independent Hamiltonians describing Potentialities}:

The time-independent Hamiltonian describing interactions between Potentialities can be described well by constructing miniature abstract models. These are conceptual miniatures of models informed by actualities (such as the solar system model for the atom of Rutherford). This is the case for non-identical constituents. For identical constituents, such miniature models can be used as well; however, we must additionally apply the symmetrization postulate to account for particle symmetry.

We can construct classically informed abstract `miniature models’ as existing `inside' potentialities, as long as we understand them not to correspond to events in the sense of assumption 5. The Rutherford and later the Bohr atom models initiated this trend, and they then subsequently informed many such miniature models in atomic and particle physics.

After all, it can be argued that the common presumption that quantum systems have `internal structure' that defines spatial orientation, in the same way we might say a coconut has internal structure, is a habit of thought inherited from the mechanistic philosophy. The internal structure in quantum systems is inherently abstract, and this realization led Heisenberg to Matrix mechanics \cite{heisenberg1971physics}. Matrix mechanics is not an instrumental evasion. That's the best any physically possible observer can do.

Let's say there is an extraterrestrial intelligence that had no knowledge of classical particle mechanics but was well-versed in advanced mathematics, including Hilbert spaces and matrices. Suppose they have access to spectrometers, single-photon detectors, single-electron detectors, and so on. Importantly, they never had an atomistic conception of matter (no Democritus or Pierre Gassandi like figure was born in their world). They do understand probability theory and statistics. They could have figured out quantum mechanics in Heisenberg's matrix mechanics form. For the Hydrogen atom, for example, they may also have figured out the Coulombic Hamiltonian by seeking the Hamiltonian operator that satisfies the relevant symmetry constraints and reproduces the Rydberg formula as an eigenvalue. They may have achieved this solely through strictly operational interventions, with no microscopic picture in mind. So for them, an `atom' would not mean `the tiny particles at the basis of everything' but instead, the `repeatable spectroscopic patterns that occur when any material object is probed at certain energy scales'. An `atom' would simply be the label (and so any other word could be used in its place) for a recognizable pattern consisting of a Hamiltonian Eigen spectrum $\{E_n\}$, that could be observed via spectrometers, and the observed transition rules and the transition amplitudes between the eigen values. Two distinct patterns of that kind may be classified as two distinct atoms. In other words, `atoms' would be for them a classification scheme/taxonomy for abstract objects consisting of the spectral lines, the allowed transitions and the transition amplitudes between the allowed spectral transitions.

Before we continue with this thread of thought, let us note an adjacent question that may arise: how might such extraterrestrial creatures do statistical mechanics? Given that statistical mechanics is a well-attested theory in nature, and thus universally relevant, they surely would have figured out statistical mechanics even without the Democritus-style atomistic conception of matter. This we discuss in the Chapter 7.

Now back to the thread. 
This argument may also be extended to more complex quantum experiments, such as preparing a C60 molecule in a coherent superposition of spatial paths, where `C60' again refers to a very complex spectrum with many lines and transition amplitudes. And when such a molecule is in a superposition of paths L and R with equal amplitude, those extraterrestrial scientists would understand the Born probabilities obtained from the spatial wavefunction as follows: with 50 percent amplitude if the spectrometers were put in L, the molecular signature characteristic of C60 would be found there and with 50 percent amplitude if the spectrometers were put in R, the molecular signature characteristic of C60 would be found there. Again, the aliens may need no `micro picture' in our mechanical philosophy-informed, traditional understanding of `realist' stance here.

This Alien scenario\footnote{See Appendix B, which involves conversations with Professor Armour-Garb, for counterarguments to our thought experiment and our responses.} reveals just how much of our conception of matter is historically and culturally influenced (notably that of Democritus, and the later Atomists). If we are to stress strictly on only the direct  observations which is arguably shared with any intelligence anywhere in the universe all of which can be assumed to be subject to principles 0, 1 and 2, many of our metaphysical suppositions are trimmed away and we may be left with an understanding of the universe without the cultural baggage that may be specific to earth. This further discussed in the next Chapter 6 and 7.

Notice that the distinction between potentiality and actuality, however, will likely apply to them as much as it does to us: Since the definition of those two categories is based on empirical access criteria, not on any additional unobserved metaphysics. In the above example of the C60 in a superposition, `potentiality' refers to the possibility to observe the spectral features characteristic of C60 either in L or in R, with no prior space-time existence, before we put a detector in L or R.

 Assumption 8 is needed only to accommodate working in the Schrodinger picture, which is now more common. In the Heisenberg picture, we can ignore any reference to unobservable `miniature models’ and rely solely on directly observable experimental results, such as spectral lines and their transitions. 

In the Heisenberg (matrix mechanics) approach, only observed phenomena, such as spectra, transitions, and intensities, are considered. The Schrodinger picture, by contrast, might give the false impression of offering a microscopic realist perspective. The same realist picture of a `microstructure’ can be reinterpreted in the Heisenberg style understanding without any ``under the hood’’ micro-picture at all.

It is not correct to say that the Schrodinger realist view gives a ``fuller explanation'' of what is happening under the hood, as though the situation were like that of a car whose internal machinery is oblivious to us. The analogy is incorrect.  The internal mechanics are abstract, not just ``miniature cranks moving around'' as such an analogy might suggest. Even without the Schrodinger account, the Heisenberg method could have independently discovered or predicted the fine structure. That is a sufficiently full explanation.

At the same time, it seems possible to do experiments where we can trap an atom in an atomic trap (or look at atoms on the surface of a thin film) and see in real time the electronic distribution matching approximately the intuitive visual picture of a central nucleus with a cloud of electrons around it, as can be seen with AFM (atomic force microscopes) or the STMs (scanning tunneling microscope). While what we are seeing is simply the detector response, we do, in fact, see the spatial structure corresponding to the intuitive picture provided by the abstract miniature model. It is not that there is a pre-existing electron cloud as an actuality; rather, the detector signal's pattern (such as the tunneling current in the tip of an STM) resembles the spatial arrangement suggested by the abstract miniature model.

A more accurate way to account for such instances is to employ Heisenberg's own potentiality language (which he employed for quantum systems before measurements) also for mereology, concerning parts and wholes: A notion of \emph{potential parts}, in the sense that in some experiments they can be literally seen as parts existing inside the whole, and in other experiments, that model is to be taken as an abstract representation with no actual spatio-temporally existing parts.

This potential parts idea is a broader version of Goyal’s potential parts idea \cite{goyal2026potentialparts} for identical particles. Identical particles, as is well known, are not re-identifiable in scattering experiments, and are thus not to be pictured as temporally persistent objects with individuality. In Goyal’s view \cite{goyal2026potentialparts}, the identical particles in a system with many identical particles are not to be seen as parts as we usually understand them in the mereology implicit in the atomistic conception of matter, but rather as potential parts, in the sense that their individuation can become manifest effectively in some experiments where the identical particles can be made effective distinguishable, but in other experimental context their individuation is only an abstract model, that is supplemented by the symmetrization postulate, to account for the lack of individuation. The distinction between Goyal's potential parts idea and ours is that his is the potential parts concerning the individuality of persisting entities, whereas ours is the potential parts concerning spatial arrangement of parts making a whole, and the two don't overlap. Because for non-identical particles only, Goyal's notion doesn't apply, but ours does, and for identical particles, both Goyal's notion and ours apply.

In the next assumption, we highlight what we saw in Chapter 3: even the definition of a closed quantum system involves actualities, and the potentialities cannot exist in silos.

\textsc{Assumption 9: Potentialities are always interlaced by actualities}:

The simplest quantum mechanical system has the form of A-P-A (where A corresponds to actuality, P to potentiality: see Chapter 3 for more context).

Having described potentialities, actualities, and observers, we now need to re-interpret the familiar concepts of particles and fields in the context of the conceptual constructions in this thesis. We can re-express the senses in which particles and fields are meaningful, in terms of potentialities and actualities.

\textsc{Assumption10: Particles and Fields can exist as entities in space and time only in the following two senses}:

\begin{enumerate}
    \item {As patterns exhibited by actualities, instantiated as spacetime events:} Particles and Fields are discrete localized patterns and de-localized, approximately continuous disturbances respectively, instantiated as spacetime events in various actualities: water waves, a small planet in a central potential around a star, or a grain of sand, a spot on a screen, etc. 
    \item {As abstract concepts instantiated as spacetime events in processing and memory units of observers:} This relates to Assumption 8. Particles and Fields are useful abstract concepts—namely the miniature models—that can be thought of as existing as events in space-time, only in the sense that they are conceptualized, stored, and processed by the physical memory and processing units of the observers, which, by Assumption 6, require actualities and are therefore localized events in space-time. Such localized representations then model potentialities, contingent on the Epistemically constrained situations of the experimental arrangement (depending on whether the experimental arrangement is arranged for the particle or the wave description, as per complementarity). 
\end{enumerate}

Note the non-overlap in the meaning of particles and waves versus potentiality and actuality:
What we conventionally call `particles' in quantum mechanics, such as wave-packets localized in some experimental basis, count as potentialities; what we call waves, such as coherently interfering wave-packets in interference experiments, also count as potentialities. 
 What counts as actuality are the alignment positions measurable objects in regions of space, with reference to observers. Examples include the spots on a screen or the clicks of a detector: those are not `wavefunctions' themselves. The demarcation between actuality and potentiality is defined by the operational access criterion provided by alignment measurements and the possibility to consider literal or counterfactual observers who could perform the alignment measurements.

It should be noted that the potentialities and actualities are not to be viewed as suggesting a form of substance dualism, as they both are supervened by the familiar elements of the periodic table and the primitive constituents of the standard model.

Next, the\textbf{ }Schrodinger equation has no size limit. It can, in principle, be found to apply to large systems with an arbitrary number of constituent atoms and an arbitrarily large mass. However, the Schrodinger equation does not provide a protocol for experimentally preparing such a state. 

\textsc{Assumption 11:  The concepts of actualities and potentialities can apply for any scale of objects}:

Neither is an actuality only about the `big objects' nor is potentialities about `small objects'. Individual atoms can be actualities in experiments where the atom is continually monitored by fluorescence, for example. Another example is the monitoring of atoms in the surface of a solid using a tunneling microscope. Furthermore, potentialities are not only about small objects. In principle, with ingenuity we can even make something like a chair a potentiality. Quantum theory does not forbid this. But quantum theory per-se does not appear to specify the experimental protocol to achieve such a preparation. See Section 4.4 for consideration of observers themselves in such preparation, in view of the framework in this thesis.

The reason big objects are more likely to be actualities is that they have a higher probability of thermal photon emission, which correlates with a higher tendency to be actualities in the sense we have defined. However, there appears to be no threshold in the likelihood and amount of thermal emission beyond which the system becomes an actuality; Instead, the decision as to whether an object is an actuality or a potentiality can only be made contextually.

This is not very different from Bohr's emphasis that the so called `cut' is determined by the experimental context, and by nature, it is neither arbitrarily moveable, nor is there a threshold on a parameter that yields the `cut'. 

There is an unspoken conceptual unease that demands for either a ``precise'' cut in terms of a threshold on a parameter, or demands no cut at all. However, this unease appears to be a habit of thought inherent to the atomistic conception of matter, where we picture complex objects as simply `made of' simpler constituents. However, as we argue later in Chapter 6, the atomistic conception of matter, and more broadly the mechanical philosophy is less justified in view of the practice of quantum theory. It appears to be a historically inherited framework, not an empirically constrained metaphysics and mereology.

The view that the ``experimental context decides the cut'' of Bohr, can be made more operationally precise by the potentialities vs actualities distinction, and the criterion involves counterfactual consideration of observers who track the system by performing alignment position measurements by capturing passively the radiation from the system: if that counterfactual consideration is consistent with later observations of the same system by actual or counterfactual observers, such that it attests to macro-realism of position of the same system, then the system is an actuality, otherwise it is a potentiality. The possibility for such a counterfactual consideration increases with the likelihood of position-revealing thermal emission and scattered emissions from the ambient light in the surroundings.
The ability to consider a counterfactual observer near a system is analogous to the ability to consider a counterfactual clock near an object undergoing a process, or a magnetometer around a magnet.

``If the likelihood of path-revealing photon emission by a system is sufficient to predict the likelihood of assuming a counterfactual observer, and thus calling the system an actuality, why do we need the counterfactual observer at all? Isn't that an unnecessary artifact?'',  one might contend.  
However, we believe there is one reason to justify the very ability for the consistent assumption of the counterfactual observer as the criterion for actualities: because, just phrasing in terms of the likelihood of path-revealing photon emission risks giving counterfactual definiteness to pure states (for example, a locally thermalized but globally pure many-body system). This would violate quantum theory, whereas positing a criterion in terms of the non-reductionist ability to assign counterfactual observers does not violate it.


Finally, in the list of assumptions, we need to specify what we really mean when we are talking about classical state variables, such as position, momentum, and so on. It appears that all classical state variables are determinable by optical means (with dark matter a possible exception). Therefore, we can define classical state variables in terms of light, either in the visible or the invisible region of the electromagnetic spectrum. This light can be used for the alignment position measurement. After all, `position', `momentum' are in practice never like labels attached to bodies, even for classical mechanical bodies; they are always quantities inferred either by tracking the light from the bodies, or by tracking non-optical transmissions from the bodies (such as sound). However, light is a more general means for tracking bodies, since it can travel through vacuum.

\textsc{Assumption 12: Optical basis\footnote{The reason for the optical basis was discussed in a footnote in Chapter 1. I reproduce it here for ease of reference: light is emphasized here because all objects above absolute zero temperature emit some thermal radiation in accordance with the Stefan--Boltzmann law. Unlike other methods, light from objects seems to provide a universal means of measuring the position of bodies, perhaps with the exception of dark matter. What else could we mean, operationally, by the `position of an object' as existing `out there'?} for the meaning of classical States}:

    All Classical mechanical state variables are definable from the alignment measurement of the properties of light emanating from it, by a co-moving counterfactual fictitious observer.

\section{Are Potentialities in isolation ``Invisible'', even in Principle?}
For a light beam in vacuum, viewed from the transverse direction, we won’t in any way ``see''—even with spectrometers—that there is an electromagnetic field in that region  (unless there is a scattering medium present). Instead, we will just see a vacuum.

That statement also holds true for potentialities constituted by beams of matter, as discussed earlier.

Interestingly, this extends to scenarios involving larger objects—that we usually consider ``visible''—when they are prepared as a potentiality, in some quantum state, even if such a preparation is not yet within our technological reach. Imagine we manage to prepare a large object, such as a table, in a pure coherent state, and the wavepacket of the table propagates as a narrow-profile wavepacket in some direction. Assuming the experimental control is so advanced that even internal sources of decoherence are suppressed for a duration (otherwise we would no longer have a coherent state for the table). Under such conditions, something interesting happens.

The table would be ``invisible'' to any region of the electromagnetic spectrum, without a scattering medium or ancillary particle. This is the closest we can get to making an ``object invisible,'' at least for some time, because under these conditions it’s not that the object ``is in the dark.'' The object, in general, has no counterfactual definite existence. The coherent state can even be split into a coherent superposition in some orthogonal basis.

Imagine we have a system inside a volume V in a pure quantum state, whether a single particle or a collection of particles. The system inside V does not interact with what lies outside V, nor with the boundary of V. There is also no incoming electromagnetic field from inside V. We remain outside, passively directing our spectrometers, cameras, etc., at the volume V, without actively probing it.

In this case, as long as the system inside V remains in the prepared pure state, no signals will be detected by our cameras or spectrometers. Even if the system inside V is a narrow wave packet in position space, it would remain undetectable. So a narrow position-space wave packet is not synonymous with an actualized outcome, as presumed in the decoherence-based accounts for the emergence of classical mechanics. This is discussed more in Chapters 5 and 7. A narrow position wave packet is still a potentiality.

Note that here invisibility is not with respect to some observer, but to any observer.

Consider a chunk of wood inside volume V in a sealed container. Although uncoupled from the outside and not directly visible from the outside due to its uncoupled nature with the radiation fields accessible from the outside, it is still meaningful to treat the chunk of wood's existence as localized actuality with respect to an observer inside V. This assumption is consistent, because a counterfactual fictitious observer can be assumed to be present alongside the wood without raising any inconsistency. 

For a quantum system inside V, however, such an internal observer assumption is inconsistent with the system staying in a pure state. Thus, while the wood’s invisibility to outside observers is a matter of practical unobservability, the quantum system is, in principle, unobservable to any possible observer while in a pure state.

If an internal observer inside V were to see a definite alignment position measurement for the wood, then the definite occurrence/existence of the position of the wood as an actuality would hold objectively, by postulate 2, including for any external observer. But for a quantum system inside \(V\), there is no epistemically constrained situations available relative to any such counterfactual fictitious observer inside \(V\).

This is what we mean by in-principle invisibility: a quantum system in a pure state inside V cannot be observed as a localized object—such as by capturing light—relative to any observer, as long as the system remains in a pure state.

For Bohr, the cut between quantum and the epistemically constrained situations (note that we have replaced Bohr's ``classical concepts'' with epistemic constraints earlier) is not arbitrary (as in Heisenberg \cite{Heisenberg1958} or more recent QBist/neo-Copenhagen approaches \cite{FuchsSchack2013}). The cut is specified contextually and is not at an individual observer's discretion. Whether the wood is treated quantum mechanically or classically depends on the experimental context.

\section{The meaning of `Universality' in the Universal Validity of Quantum Theory}

In this section, I argue that quantum theory is universally applicable, despite the primitive role of epistemic constraints. We just need to reassess the meaning of ``universal''.

The inverse-square law of gravity is a `universal' law. The same is true for Coulomb's law of electrostatic attraction. When we say a principle or law is universal, we mean that it applies to every instance of a particular class of identifiable operations that could be performed by an observer, either actual or counterfactual. It does not typically mean, in an unqualified sense, that ``it applies to everything''.

Notice that we say `particular class of identifiable operations that could be performed by an observer, either actual or counterfactual', as opposed to simply `property'. This is to avoid implying that `properties' are free-floating labels on objects, not accessible via operational interventions.

In the case of gravity (in Newtonian physics), it's mass, and it corresponds to the identifiable class of operations that are identified with `mass', such as the response of a test object when it is accelerated or its behavior in a Cavendish torsion balance with another test mass. There are many more operational methods that correspond to the determination of what we call ``mass'', and they can be considered to form an equivalence class.

When we say the law of gravity is `universal', we mean that it is applicable in every instance where the class of operations that can be distinctly identified with the determination of ``mass'' is performed by an actual or counterfactual observer.

In the case of quantum mechanics, as we saw in assumption 2, quantum states are assigned a posteriori, given the epistemically constrained experimental conditions, and not a priori—objects do not come `inbuilt' with a quantum state in any sensible way. Such an `inbuilt' understanding of quantum states is an extrapolation of the metaphysics left behind by the understanding of classical mechanical states. Where we immediately presume that any ``classical system'' possesses a ``classical state''.

The Schrodinger equation is universal in the sense that it applies to every instance where the wavefunction assignment is physically meaningful. In everyday experiments, this concerns assigning a wavefunction to electrons in a metal and to the light particles in a Mach-Zehnder interferometer. But it is conceivable that in a future specialized experiment we have an operationally meaningful attribution of the wavefunction to the center of mass degree of freedom of a 1 kg iron ball or a 1-ton iron ball, and the Schrodinger equation would apply to it. In circumstances where the iron ball could be alignment position measured with respect to at least one observer, however, the experimental context does not justify assigning a wavefunction to the center of mass degree of freedom of the ball.

``Universal'' does not mean we can go stick a wavefunction to every chunk of matter we come across. The wavefunction assignment needs to be justified and supplied by the prevailing experimental context. 

That is, just as the applicability of the word ``universality'' to the respective laws for masses and charges (namely the law of inertia, the law of gravity, and Coulomb's law), hinges on the corresponding classes of operations, we need to identify the operation or classes of operations for wavefunction assignment, and then `universality' would mean that wavefunction assignment and the Schrodinger dynamics governing closed systems are applicable to any such instance of that class of operations, performed by an actual or a counterfactual observer.

The operational criterion we specified for differentiating actualities and potentialities provides such an operation. Potentialities are those that can be described by pure quantum states or mixtures of quantum states. 

What is meant by the universality of quantum mechanics is then the fact that in a future experiment, there is no restriction on the mass or on the number of particles for which the potentiality description can apply. When we say quantum theory is universal, we mean that the Schrodinger equation is applicable in every instance where we come across a realization of this operational criterion that certifies a potentiality.

\section{Observers in a Quantum State}

The view developed so far does not prohibit macroscopic quantum superpositions, but at the same time, it is not the same as Everett's view, in the sense that, unlike Everett, we posit that quantum objects do not come `built in' with a quantum state, which we think is an incorrect extrapolation of classical metaphysics (---where we think of classical objects as always attached to a classical state, such as ``an apple `having' a position and momentum''---) to quantum objects. 

It is also distinct from the relational and QBist views, since we have retained inter-subjective objectivity across all observers, even before communication.

Rather, a quantum state is meaningful, conditional on the experimental context, and if there is no meaningful experimental context for, say, a chair, then it is not incorrect to assign a wavefunction to the chair, but meaningless to assign a wavefunction. For when the context warrants wavefunction assignment, see the previous section, which considers the meaning of `universal' in the universal validity of quantum mechanics.

The following consideration is to study what happens when an observer is placed in a quantum-mechanical superposition. We must note, however, that quantum theory, nor any other means, gives us a prescription on how to experimentally prepare such states, only that they don't seem to forbid such possibilities. However, the purpose of this section is to pursue the reasoning that follows by treating Bohr's epistemic constraints as laws of nature (though as non-reductionist ones at higher levels of complexity, like the laws of evolution) and by considering the possibility of macroscopic objects, including a computer or a robot, being in a macroscopic quantum superposition.

The recent experiments across atoms  \cite{atom}, optomechanical membranes  \cite{oscillator}, and levitated nanoparticles \cite{nano} have successfully prepared quantum states in the vibrational degrees of freedom, and even in the center-of-mass motional degrees of freedom of entire mesoscopic objects (often along one spatial direction). 

For nanoparticles, the whole object is treated as a single quantum harmonic oscillator and prepared into a narrow Gaussian wavepacket \cite{nano} , the ground state of its external motional Hamiltonian. This is the ground state of the center-of-mass sector's Hamiltonian for levitated particles or macromolecules —but in a harmonic trap potential, so the ground state is just a Gaussian— not just the vibrational mode as for cantilevers. 

The motional degrees of freedom differ from the internal phononic degrees of freedom. Such a system is, metaphorically, like a ``hot little balloon floating in a vast, silent vacuum'': internally it may still be thermally excited, with its internal phonon bath not cooled, often near room temperature, while externally it is placed in ultrahigh vacuum with thermal emission suppressed so that almost nothing in the environment is allowed to learn about its position. Internal thermal excitation is not the same as radiating position information. 

What matters for preparing macroscopic quantum states in positional degrees of freedom is not being `hot inside' (i.e the phononic modes are in thermal states) but being oblivious to being observed by any observer: That is, decoherence occurs when internal excitations produce outwardly emitted photons that encode positional information, or more generally, when any decoherence channel becomes correlated with position or momentum. In present experiments, the emission rate is so slow (or more generally the rate of coupling to any external bath is so slow) and the quantum evolution so short that decoherence does not occur in time — the only reason we get away with preserving quantum behavior for such a large object in motional degree of freedom is simply that the decoherence timescale is long and the experiment finishes before the spatial information becomes available. If radiation eventually `leaks' out with positional information, that ground state would decohere. So we merely outrun the leak of positional information. 

Yet nothing observed so far violates Bohr’s three central epistemic stances: 1) all evidence must be in the form of definite events in space and time, 2) all experimental results must be expressed in classical terms (i.e epistemic constraints), and 3) Intersubjective communicability of experimental results using a language.

Every observation still arrives to us as a single definite outcome; we only infer superposition states through interference that ends in a communicable, epistemically constrained record. If we push the experiment to observers' bodies—whether conscious or not— placed into such a superposition, two strictly Bohr-compatible possibilities remain: 

The first possibility is to impose super-selection rules that prevent observers themselves from being prepared in a quantum state. But this appears artificial. 
The remaining possibility, that stays consistent with quantum theory and the postulates of epistemic constraints, is then this:
The observer can be prepared in quantum states, but would fundamentally lose the ability to have an `experience' of the situation. During the potentiality phase, they would not exist functionally as an observer at all, only as quantum potentiality, and `experience' would reappear for the observer only at the moment of epistemically constrained definiteness at the end of the experiment. When being a potentiality, they would cease to exist as a definite, functioning, epistemic entity, in space and time, in the intervening time interval. 

This means configurations involving a functionally active observer in a spatial macroscopic superposition are fundamentally forbidden by nature due to natural laws concerning observer systems, even if macroscopic superposition remains possible for any scale. This is not to be viewed as a prohibition on macroscopic superposition, but instead as a prohibition on the nature of epistemic agency. This preserves Bohr’s epistemic constraints requirements and is compatible with the possibility that arbitrarily large objects can be prepared in a macroscopic quantum superposition.

No physical observer, it seems, can perform reasoning or generate epistemically constrained records while remaining in a quantum superposition; agency seems to exist only in epistemically constrained contexts, which are actualities. During the purely unitary, coherent phase of a quantum evolution, there cannot be an observer “inside” the superposition — only quantum potentiality evolving lawfully. The observer does not remain experientially present across multiple branches (as in Everett) or perspectives (as in relational quantum mechanics); they simply do not exist as an epistemic entity, and only reappear when a definite epistemically constrained outcome materializes. 

In Wigner's friend thought experiment, Wigner stands outside a sealed lab. A friend of Wigner is assumed to be present in a sealed lab, performing a quantum experiment that yields two possible outcomes, each with probability given by the Born rule. In the experiment, one then asks, from Wigner's perspective, whether the friend is in a definite epistemically constrained state or in an entangled state with the quantum system the friend intended to measure. 

Since quantum mechanics stays neutral about what counts as a measurement, there has been a manifold of views on this matter; some claim that the outcome can be definite to the friend, but not to Wigner  \cite{Rovelli1996}, thereby establishing a ``relativity of facts''. Others claim that the friend cannot be put in a macroscopic superposition in the first place. Yet other views claim that the universe splits every time a quantum measurement is made \cite{Wallace2012}. Our framework does not require any of these conclusions.
Applying the previous developments to Wigner's Friend thought experiment presents us with two experimentally possible scenarios: 

\begin{enumerate}
    \item Either the friend sees a definite outcome, then, by postulate 2, the definite outcome has 
occurred for Wigner as well, even before intercommunication.
\item In a special, isolated experiment, if the friend-detector-atom system is prepared in an entangled state, then, due to the natural principles concerning observers, the friend will 
unavoidably fail to continue functioning as a language-using system. Fail to function as a language-using system not merely in the sense of `malfunctioning', but fundamentally lose the ability to function as a knowledge-acquiring system, and cease to have definite space-time existence or experience.
\end{enumerate}

That is, for the experimental arrangement described in an epistemically constrained manner by the friend, Wigner will also agree. However, Wigner can engineer another experiment in which the friend and the detector are described quantum mechanically, in an entangled state, and described by larger epistemically constrained situations, but in this case the friend would have no say—in the sense that he won't have any possible experienced description of `what he observed'.  The friend can later, when they qualify as actuality, resume functioning, but there will 
have been an epoch when they lacked “experience” and definite space-time existence.

\chapter{The Nature of Emergence in Quantum Processes}

Classical physics, unlike epistemic constraints, concerns the dynamical laws of motion —or, more generally, change— applicable to different classes of phenomena whose quantities satisfy macro-realism. By this criterion, the population dynamics of a group of Zebrafish would be classified as `classical physics' too. However, it is conventional to mean by ``physics'' those formalisms concerning observables relevant to as wide a class of phenomena as is experienced. Such physical quantities include the mechanical quantities of position and momentum of point particles and material media, described by the classical mechanics of Newton; the electrodynamical quantities of magnetic and electric fields, described by the electrodynamics of Maxwell; the thermodynamic quantities such as energy, entropy, and volume, described by equations of state and the laws of thermodynamics; and finally, the classical informational quantities associated with the representation, processing, and communication of classical bits of information. It is known from experience that all the above four classes of phenomena are relevant aspects of matter and energy inside any volume. They are relevant for a slice of marble, just as much as they are relevant for a volume of interstellar gas or a volume containing biological entities.

Exhaustively, we have the following four distinct kinds of ``classical'' (in the sense that quantum counterparts of them have been sought in the literature):

\begin{enumerate}
 \item Classical in the sense of Classical Physics.
    \item Classical in the sense of Classical Propositional Logic
    \item Classical in the sense of Classical Probability Distribution
    \item Classical in the sense of Classical Computation and information.
\end{enumerate}

A central question that the program of Quantum-Classical transition asks is how classical physics emerges from Quantum Mechanics. This includes how classical mechanics, classical thermodynamics, and classical electrodynamics emerge from their quantum counterparts. Decoherence, along with the Correspondence Principle, is seen as a common thread across all four kinds of Quantum-Classical transitions \cite{Zurek2003Decoherence}\cite{popescu2006entanglement}.

Now, it must be noted that even in these derivations, the sole assumptions that are being made, when we look closely at the derivations, are not simply the Schrodinger equation and the Born rule. They are part of the assumptions being made, not the only assumptions. Epistemic constraints are also implicitly assumed in how we partition the subsystems and in how the unitaries are determined, and in determining what counts as a measurement in quantum theory. It is misplaced to seek the quantum description of epistemic constraints.

The position we take here is that epistemic constraints correspond to the second kind of `classical' we listed above, namely classical propositional logic. It has no quantum counterpart. It's a misplaced search \footnote{There is quantum logic, but in our view, quantum logic is really not a quantum version of experimental propositions that we use to describe experiments, but rather concerns the structure of statements we can make about quantum states in a Hilbert space.}

Not `everything' has a quantum counterpart, and it may be a category error to demand one. More concretely, it does not make sense to seek the quantum version of aspects of language, namely Bohr's epistemic constraints, for the same reason that it appears misplaced to seek a quantum version of the principle of relativity or the quantum version of the law of conservation of energy (while the laws of thermodynamics can be explored at the quantum scale, the statement of the law of conservation of energy still stands in quantum systems). Bohr's epistemic constraints, Einstein's relativity, and the law of conservation of energy are not restricted to any particular physical dynamics but provide the scaffold within which any existing, and arguably even future, physical dynamics must fit.

This was our motivation to pursue new directions from the joint consideration of quantum theory and epistemic constraints. Seeking the quantum-classical transition for the rest of the three kinds of classical in the list---classical physics, classical probability distribution, and classical computation— from the joint consideration of quantum theory and the epistemic constraints, is the first set of research problems in the implementation of the Bohrian program.

These three kinds of transitions, we argue, depend on the interplay of two distinct kinds of emergence mechanisms, which we will explain in more detail in this chapter: one is the limit and emergence of classical dynamical behavior for homogeneous systems of potentialities; another is the emergence of classical dynamics for heterogeneous systems of actualities and potentialities. There appear to be no `homogeneous systems of actualities', since, as we saw in assumption 7, potentialities can occur inside volumes demarcated by actualities, as in the flow of current inside an aluminium wire.

These two kinds of emergence mechanisms in turn rely on the following two micro-level processes:

\begin{enumerate}
    \item Quantum coherence to classical probability distribution via decoherence.
    \item Competition between unitaries and measurements arising from the probabilistic application of a unitary or a measurement operation to a quantum system within any given time window.
\end{enumerate}

1) is relevant for the classical limit of potentialities, which are the effective classical probability distribution, locally, over semi-classical pure states. This is what we have called the classical dynamical behavior for homogeneous systems of potentialities, and is particularly responsible for the classical limit of electromagnetic radiation. For the emergence of classical computation from quantum computation and for the origin of classical mechanics, both 1) and 2), we argue, play a role, not just 1). 2) concerns the heterogeneous systems of actualities and potentialities

\section{The Archaic and Complex Systems View}

Any general physical theory typically, under limiting cases, reduces to a more specialized empirically tested theory that was previously used to describe the same phenomenon with narrower empirical success. This is the general correspondence principle. General relativity reduces to special relativity in the absence of gravity; the Lorentz transformation reduces to the Galilean transformation for low velocities compared to the speed of light; and gravity in general relativity reduces to Newtonian gravity in the limit of weak gravity and for slow-moving bodies. Similarly, quantum mechanics can be seen, naively, as reducing to classical mechanics in the `classical limit' (in quotation marks because it needs closer conceptual and mathematical scrutiny).

We can label the older accounts \cite{Shankar1994} on the classical limit of quantum mechanics, in accordance with the general correspondence principle, as the Archaic view, no different from the classical limit of any other general theory.

Notwithstanding such archaic views, unlike the other general theories in nature, quantum mechanics is peculiar in two respects: 

Firstly, it's peculiar in that its governing equation, the Schrodinger equation, requires the irreducible, epistemic constraints for its meaningful operational application. This is unlike any other general theory. While epistemic considerations play a role in the application of other general theories too, the relevant state variables of the general theories can be assumed to objectively exist without reference to any epistemology. For example, the metric tensor can be assumed to objectively exist as a property of spacetime without reference to an experimental arrangement. In quantum mechanics, the determination of the state variable, the wavefunction, requires epistemic constraints that specify the experimental arrangements and thus the wavefunction.

Secondly, unlike the limits of other general theories, there are no fixed parameter regimes at which quantum behavior universally becomes unnoticeable. Although mass and action are often invoked as the relevant parameters for such a limiting consideration in quantum mechanics, especially in elementary treatments \cite{Shankar1994}, it remains in principle possible to realize macroscopic quantum superposition states of systems of arbitrarily large mass and action relative to the Planck constant.

As a result, the relationship between classical physics and quantum mechanics is more nuanced than what the general correspondence principle suggests. This leads us to the modern view, where classical physics is seen as emerging from the complex nature of real environmental conditions. 

This more recent view, starting with Zeh \cite{Zeh2000Meaning} and Zurek \cite{Zurek2003Decoherence} in the 1970s and 1980s, does not assume that large objects cannot be quantum-mechanical; rather, it shows why realistic environments nevertheless impose certain constraints on the realizability of macroscopic quantum states in large objects.

The homogeneous complex systems of potentialities, of which the Zurek-Zeh decoherence framework is the paradigmatic example, and the heterogeneous complex systems of potentialities and actualities, for which we introduce examples in this work, both belong to the more modern complex system view concerning the classical limit.

In what follows, we use the word ``complex system'' in a somewhat loose sense relative to what it means in complex-system science. Here, by complex system, we mean a system with many operationally accessible subsystems. In a homogeneous complex system of potentialities, all the accessible parts are potentialities. In heterogeneous complex systems of potentialities and actualities, some parts are potentialities, while others are actualities, and thus alignment measurable by optical means.

Notwithstanding its archaic character, the understanding of the classical limit of quantum theory is still relevant for the modern complex systems view, since it identifies which classes of quantum states exhibit behavior characteristic of classical mechanics. That is, they provide a characterization of the so-called semi-classical states. It is these characterized semi-classical states which are then explained as being ``preferred'' \cite{Zurek2003Decoherence} in nature by the modern complex systems view. 

So I review the characterization of the semi-classical states in the next section. 


\section{Classical Physics as a formal limit of Quantum Mechanics: The Archaic View}
  
The characterization of the classical limit is more convenient in terms of dynamics derived from the Schrodinger equation than in terms of the Schrodinger equation per se.
The classical limit of quantum mechanics refers to the conditions under which the description of a system effectively becomes a classical mechanical description in two different senses: in the sense of having a state, that under the given experimental conditions,  makes the quantum Hamilton-Jacobi description nearly equivalent to the classical Hamilton-Jacobi description and makes the Ehrenfest equation's description nearly equivalent to the description given by the Newtonian form for expectation values of position and momenta.  The latter ensures spatial localization of wave packets and linear approximation for potentials, and the former (along with the continuity equation) ensures that the dynamics is consistent with the Hamiltonian dynamics of classical waves in the geometrical optics limit.

Since the Ehrenfest equation is likely familiar to my audience, I won't elaborate on it here. Instead, I will briefly revisit the Hamilton-Jacobi equation in quantum mechanics.

The following discussion is based on \cite{bohm1951quantum}\cite{Shankar1994}\cite{Demme2017ClassicalLimitED}\cite{Bohm1952}.

The time-dependent Schrodinger equation, as is well known, has this form:

\begin{equation}
i\hbar \frac{\partial \psi}{\partial t}
=
\left[
-\frac{\hbar^2}{2m}\nabla^2 + V
\right]\psi
\end{equation}

Now the wavefunction can be written in the polar form as

\begin{equation}
\psi(\mathbf{x},t)=A(\mathbf{x},t)e^{\frac{i}{\hbar}S(\mathbf{x},t)}
\end{equation}

The time derivative then becomes\textbf{ }

\begin{equation}
\frac{\partial \psi}{\partial t}
=
\left(
\frac{\partial A}{\partial t}
+
\frac{i}{\hbar}A\frac{\partial S}{\partial t}
\right)e^{iS/\hbar}
\end{equation}

and the Laplacian,

\begin{equation}
\nabla^2\psi
=
\left[
\nabla^2 A
+
\frac{2i}{\hbar}\nabla A\cdot \nabla S
+
\frac{i}{\hbar}A\nabla^2 S
-
\frac{1}{\hbar^2}A(\nabla S)^2
\right]e^{iS/\hbar}
\end{equation}

Isolating the imaginary parts of the equation that results after we substitute Eq.(5.2), Eq.(5.3), and Eq.(5.4) into Eq.(5.1) yields the probability continuity equation, which already has the form we expect for the continuity equation in classical dynamics.

The real part yields,

\begin{equation}
-\frac{\partial S}{\partial t}
=
\frac{(\nabla S)^2}{2m}+V
-\frac{\hbar^2}{2m}\frac{\nabla^2 A}{A}
\end{equation}

or

\begin{equation}
\frac{\partial S}{\partial t}
+
\frac{(\nabla S)^2}{2m}
+
V
-
\frac{\hbar^2}{2m}\frac{\nabla^2 A}{A}
=0
\end{equation}

The last term on the left-hand side is the quantum potential.

\begin{equation}
Q=-\frac{\hbar^2}{2m}\frac{\nabla^2 A}{A}
\end{equation}

When \(A\) varies slowly, the quantum potential term will be negligible, and so we get the classical Hamilton-Jacobi equation.

\begin{equation}
\frac{\partial S}{\partial t}
+
\frac{(\nabla S)^2}{2m}
+
V(\mathbf{x},t)
=0
\end{equation}

We can identify the momenta with the gradient of \(S\):

\begin{equation}
\mathbf{p}=\nabla S
\end{equation}

And the Hamiltonian of the classical system is

\begin{equation}
H(\mathbf{x},\mathbf{p},t)=\frac{\mathbf{p}^2}{2m}+V(\mathbf{x},t)
\end{equation}

So the Hamilton-Jacobi equation can be written as

\begin{equation}
\frac{\partial S}{\partial t}
+
H\bigl(\mathbf{x},\nabla S,t\bigr)
=0
\end{equation}

The limit where this description applies concerns the geometrical optics limit of wave theory. It does not yet concern localized systems.

A good example for the quantum mechanical situation where this classical Hamilton-Jacobi description applies is that of a free electron prepared in a plane-wave state.  In quantum mechanics, the plane wave solution describing an electron, for example, can neither be given the classical point particle physical interpretation nor a spatially `smeared out' physical interpretation. Its physical interpretation would still be in terms of potentialities, albeit the classical limit of potentialities.

There is another classical limit in quantum mechanics apart from the derivation we have seen above. The derivation we have seen above concerns situations when the wavefunction has slowly varying amplitude; that is, when \(\frac{\nabla^2 A}{A}\) is small. This can be called the WKB (Wentzel---Kramers---Brillouin)  limit. The classical point-particle intuition can only be used when another classical limit associated with Ehrenfest also applies. The other classical limit occurs when the wavepacket is sufficiently narrow, so that the potential \(V(x)\) can be approximately considered to be linear across the packet. This latter classical limit is the limit in which the Ehrenfest equation of motion obtains the same form as Newton's equations. In this limit, we can use the classical mechanical intuition, but in reality, a narrow wavepacket system is still a potentiality in a classical limit, and not an actuality.

In the case of an \(N\)-particle system subject to the Schrodinger equation, described by the \(N\)---particle wavefunction, under the limit where the classical Hamilton-Jacobi equation applies, and the Ehrenfest limit also applies to the individual particles, we have a description that can be approximated by treating the \(N\)-particle system via the classical mechanics of Newton.

 That is applicable when the particles are individually in localized wave packets. Only in such cases can the \(N\)-particle system be modeled approximately as \(N\) point particles following Newtonian trajectories. Strictly speaking, decoherence arguments are also needed here to justify the factorization of the states, but this is not usually presented as a `mechanism' in the Archaic View; rather, one assumes, in somewhat ad hoc manner, that `big objects' come in localized, slowly spreading wavepackets.

For classical electrodynamics, the WKB limit is applied to quantum equations of motion for the wavefunctional, and one gets the classical Hamilton-Jacobi equation for classical fields, and the many characteristic curves correspond to a family of Maxwell's equation solutions. However, to get the behavior we expect from classical electrodynamics, we additionally need a localization in the field phase space. Such localization is satisfied by certain classes of states, such as the coherent states. Coherent states are those whose amplitudes and phases remain concentrated near a classically expected trajectory in the phase space parametrized by the field's amplitudes and phases  \cite{gerry_knight_2005_introductory_quantum_optics}.

There is a parallel to notice here: Earlier we saw that the WKB limit gives the Hamilton-Jacobi equation, but that is not sufficient to get the Newtonian point-particle approximation: we also need the Ehrenfest limit. Similarly, the WKB limit applied to the wavefunctional of quantum field states gives the classical field-theoretic Hamilton-Jacobi equation and a family of Maxwell equation solutions along characteristic curves, but that is not sufficient to get the classical electrodynamics approximation: we also need states such as coherent states.

Let us pause and realize that we identified potentialities and actualities as distinct ontological categories. And recall that even classical electromagnetic fields and unobserved decohered systems qualify as potentialities by that criterion. The classical limit calculation discussed so far concerns the regime of potentialities that can be modeled by classical physics---Maxwell's equations and Newton's point-particle dynamics. The relation between the Archaic View and the more modern decoherence framework, as we previously noted, is that the latter explains why the limits derived in the Archaic View tend to apply in many realistic environments.

\section{Emergence of Classical Physics as a Complex Systems Problem: The Modern View }

In contrast to the Archaic View is the newer understanding that recognizes that in any practical situation, objects are typically entrenched in an environment  \cite{Zurek2003Decoherence} and are never in an isolated vacuum. This insight was taken more seriously at a time that coincided with the emergence of the field of complex-system science in the 1970s and 1980s. In this work, we extend that understanding by recognizing that epistemic constraints cannot be avoided, and they lead to the definition of potentialities and actualities, and thence to the possibility of considering complex systems involving combinations of potentialities and actualities.

In this modern view, we are led to the conviction that there is no mass-based or complexity-based cut-off above which the dynamics described by the Schrodinger equation fail. 

Despite this conviction for the lack of any cut-off in the applicability of the Schrodinger equation, the role of epistemic constraints is still unavoidable. This relates to our consideration in the previous chapter of what `universal' means in the statement `quantum mechanics is universal'.

Let's broaden our perspective, and then refocus our analysis on why we need to expand this complex-systems approach to understanding the classical limit beyond what the decoherence framework alone offers.

Decoherence theory as it stands is primarily based on the Class 3 dynamics (Non-Unitary Deterministic Dynamics).

The criterion for having derived classical physics in decoherence is provided by the classical mixture of pointer states. 

Pointer states are `classical-like' quantum states, in the sense of satisfying the two kinds of classical limits we discussed in the previous section and being `robust' \cite{Zurek2003Decoherence} under environmental decoherence. Namely, for the system-environment unitary coupling that produces the copying of information about the system state,

\begin{equation}
U \big( |c_i\rangle \otimes |E_0\rangle \big) \;=\; |c_i\rangle \otimes |E_i\rangle
\end{equation}

where \(\ket{c_i}\) are the pointer states. If the system started in a superposition of such pointer states,

\begin{equation}
|\Psi(0)\rangle \;=\; \sum_i c_i \, |c_i\rangle \otimes |E_0\rangle
\end{equation}

the evolved state would be

\begin{equation}
|\Psi(t)\rangle \;=\; \sum_i c_i \, |c_i\rangle \otimes |E_i\rangle
\end{equation}

and because the environmental states can be assumed to be orthogonal  \cite{Zurek2003Decoherence},

\begin{equation}
\langle E_i | E_j \rangle \;\approx\; \delta_{ij}
\end{equation}

the reduced state, which we compute as

\begin{equation}
\rho_S \;=\; \mathrm{Tr}_E \big( |\Psi(t)\rangle \langle \Psi(t)| \big),
\end{equation}

\begin{equation}
\rho_S \;=\; \sum_{i,j} c_i c_j^* \, \langle E_j | E_i \rangle \, |c_i\rangle \langle c_j|,
\end{equation}

will become approximately diagonal  \cite{Zurek2003Decoherence}:

\begin{equation}
\rho_S \;\approx\; \sum_i |c_i|^2 \, |c_i\rangle \langle c_i| .
\end{equation}

In this sense, what decoherence delivers is the emergence of a classical probability distribution and the removal of quantum coherence when we restrict ourselves to the subsystem. It can be argued, given the criterion for potentialities, that what decoherence accomplishes is providing a mechanism for the classical limit of potentialities; the limit when potentialities can be modeled by classical physics, like Newton's laws. It can be considered to be about homogeneous ``complex systems'' of potentialities, because it concerns many parts that are interacting with each other, each part being a potentiality. It is homogeneously about potentialities because any possible role of actualities is removed by marginalizing over the records to give the dynamics of a reduced density matrix.

 A related idea from thermodynamics, namely that the reduced state of a subsystem of a larger quantum system in a pure state starts exhibiting characteristics that can be identified with a thermal state under certain conditions \cite{popescu2009quantum}, underlies work on the thermalization of quantum systems. Thermal light in a vacuum is an example. But the methods can also apply to matter, as in the thermalized reduced state of a part of a molecular gas, when the gas system as a whole is in a pure state. As long as the whole gas system is in an operationally meaningful pure state, that part of the system that was shown to be thermalized cannot be passively alignment position measured by any observer. Thermalization of potentialities is studied in Chapter 7. 

Thus, the important contribution of the decoherence framework can be reinterpreted as providing the classical dynamical and thermal limit of potentialities; Potentialities have classical dynamical and thermal limits, unlike the original conception of potentialities by Heisenberg, where they were seen purely as a quantum-mechanical ontological concept.

Decoherence per se, arguably, does not fully explain the classical mechanics and irreversible thermodynamics of material objects, which are best categorized as heterogeneous complex systems of actualities and potentialities. Decoherence per se is also not sufficient to explain the classical information-processing limit of quantum computers. For the latter, Dorit Aharonov already showed that a more elaborate, novel kind of mechanism (Class 4 dynamics) is necessary\cite{Aharonov2000NoisyQC} to explain the classical-computer limit of a quantum computer. This mechanism involves the competing effects of entangling operations, which entangle qubits, and local measurements, which erode that entanglement. This process is identifiable with the Class 4 type non-unitary stochastic dynamics with competition between entangling operations and measurements, which we discussed in Chapter 3.

As for the classical mechanics of material objects, we need to impose a more operationally aligned criterion to certify a derivation as having or not having derived the classical mechanics of actualities (namely Newtonian Mechanics). The criterion is that the optically inferred motion of moving bodies, which is a time series of alignment-position measurements of the moving body using the light from the moving body, must trace a path that we expect a light-emitting Newtonian material object to trace. After all, as noted in assumption 12 (Chapter 4), when we use the word ``position'' of a material object, what we mean is the alignment measurement of the position by capturing the light from the body. This criterion is considered in more detail in Chapter 7.

Now we discuss the homogeneous complex systems of potentialities and the heterogeneous complex systems of potentialities and actualities in more detail.

\section{Homogeneous ``Complex Systems'' of Potentialities}

The homogenous ``Complex Systems'' of Potentialities will include systems with many operationally addressable subsystems, each of which is a potentiality, whose state can be specified by a reduced density matrix. The Zeh-Zurek Decoherence theory  \cite{Zeh2000Meaning}\cite{Zurek2003Decoherence} describes such systems.

\subsection{Zeh-Zurek's Decoherence Framework}
Zeh and Zurek's decoherence framework is, within our interpretation, a mechanism for explaining the classical limit of potentialities. The framework has many interrelated aspects to it, and we discuss them one by one \cite{wiseman2009quantum}\cite{Zurek2003Decoherence}.

\subsubsection{Decoherence}
 Decoherence is said to happen when the reduced density matrix of a sub-system of a larger quantum system starts resembling proper mixtures of quantum states locally. The question pertaining to the basis in which the reduced density matrices attain a diagonal form is the preferred basis problem and is addressed by the nature of many realistic interactions  \cite{Zurek2003Decoherence}.

For the system-environment unitary coupling that produces the copying of information about the system state,

\begin{equation}
U \big( |c_i\rangle \otimes |E_0\rangle \big) \;=\; |c_i\rangle \otimes |E_i\rangle
\end{equation}

where \(\ket{c_i}\) are the pointer states. If the system started in a superposition of such pointer states,

\begin{equation}
|\Psi(0)\rangle \;=\; \sum_i c_i \, |c_i\rangle \otimes |E_0\rangle
\end{equation}

as discussed earlier, the goal is to show that the system state will become approximately diagonal:

\begin{equation}
\rho_S \;\approx\; \sum_i |c_i|^2 \, |c_i\rangle \langle c_i| .
\end{equation}

So far, this is just decoherence. There are a few other concepts relating to what may be called aspects of complex systems of potentialities.

\subsubsection{Einselection}

\begin{quote}
    ``Decoherence and einselection are two complementary
views of the consequences of the same process of environmental monitoring. Decoherence is the destruction
of quantum coherence between preferred states associated with the observables monitored by the environment. Einselection is its consequence---the de facto exclusion of all but a small set, a classical domain
consisting of pointer states---from within a much larger
Hilbert space. Einselected states are distinguished by
their resilience---stability in spite of the monitoring environment.''---Zurek \cite{Zurek2003Decoherence}
\end{quote}

Einselection is the dynamical consequence of decoherence \cite{Zurek2003Decoherence} and helps address the preferred basis problem. The dynamical consequence is that typical environmental interactions will `select' a certain form for the reduced state over others. As additional environmental interactions occur, more environmental factors may become correlated with each branch in the global state, but in the reduced state, the branch indexed by \(i\) is maintained rather than coherently mixed. That is, interference between distinct components \(\ket{c_i}\) and \(\ket{c_j}\) is suppressed in the system's reduced density matrix. 

Moreover, the pointer basis for many typical interactions tends to be the semi-classical states (typically, localized position-space wave packets) \cite{Zurek2003Decoherence}.

The reduced state of the system evolves toward a stable, classical mixture over semi-classical states,

\begin{equation}
\rho_S
\to
\sum_i |c_i|^2 |c_i (t)\rangle\langle c_i(t)| ,
\end{equation}

The interference between distinct pointer components remains suppressed for all practical purposes over subsequent time evolution, even while the global state continues to evolve unitarily and accumulate further entanglement with more and more environmental fragments.

\subsubsection{Quantum Darwinism}

Quantum Darwinism \cite{Zurek_2009} concerns a complex quantum system, where many environmental fragments interact with one system of interest. Consider the many distinguishable environmental fragments labeled by \(k\):

\begin{equation}
E = \bigotimes_{k=1}^{N} E_k .
\end{equation}

The general system-environment state would be, in view of einselection seen in the previous section,

\begin{equation}
\lvert \Psi \rangle = \sum_i c_i \lvert c_i \rangle \otimes \bigotimes_{k=1}^{N} \lvert E_k(i) \rangle .
\end{equation}

Where the $\lvert E_k(i) \rangle $ is the $k^{th}$ fragment of the environment.

An useful object of study in quantum Darwinism is the quantum mutual information, which is defined as

\begin{equation}
I(S:F) = H(\rho_S) + H(\rho_F) - H(\rho_{SF}) .
\end{equation}

where

\begin{equation}
H(\rho) = -\mathrm{Tr}(\rho \log \rho).
\end{equation}

The central result in quantum Darwinism is that the mutual information between even a small fragment of the environment and the system approaches

\begin{equation}
I(S:F) \approx H(\rho_S)
\end{equation},

when we have many environmental fragments. That is, there'll be redundant encoding  \cite{Zurek_2009} of information about the system in the environment's fragments.

\subsubsection{Role of Epistemic Constraints in Decoherence}

The meaning of decoherence models depends on one's interpretive commitments. However, regardless of interpretation, decoherence does not, by itself, specify \emph{when} and \emph{under what operational conditions} one can observe an arbitrary object in a macroscopic superposition. Beyond generic counsel to reduce uncontrolled couplings (``isolate the system''), it does not furnish an operational algorithm for preparing macroscopic superpositions. Such coherence must be \emph{prepared}, often from experimental ingenuity and trial and error. Even in near-vacuum conditions, systems typically preserve coherence only in certain degrees of freedom  \cite{nano}, often internal ones, rather than exhibiting spontaneous macroscopic superpositions.

Decoherence-rate calculations can yield quantitative estimates when one specifies an interaction Hamiltonian and an environmental model \cite{Zurek2003Decoherence}. As in any application of quantum theory, the burden of those specifications lies outside the formal machinery: the theory answers questions \emph{conditional} on a choice of well-specified interaction Hamiltonian and environmental model and a choice of what is taken to constitute the system and its environment---choices that, as far as we know, have only been made using epistemic constraints.

Furthermore, decoherence calculations require epistemic constraints when we specify experimental parameters such as  \cite{nielsen2010quantum} particle flux, background radiation characteristics, and environmental temperature. The experimental algorithm, whether for closed-system experiments or open-system experiments, is specified in ``classical terms'', which brings one back to Bohr's point about the indispensability of epistemic constraints in any possible quantum experiment.

Finally, decoherence requires, as modeling input, a system-environment decomposition; these choices are not unique. One may treat the ``system'' as the whole object or as a decomposition into a collection of parts. Which partition is physically relevant is not fixed by decoherence alone; it is supplied by noticing the experimentally accessible degrees of freedom. Once those specifications are in place, decoherence shows that the selected degrees of freedom evolve toward approximate pointer states.

  Decoherence answers the following question: given a specified system-environment partitioning, an interaction Hamiltonian, and initial prepared states, all fixed through epistemic constraints, what happens to phase coherence in the prepared quantum state?

The classification of decoherence's accomplishment by interpreting it as providing the classical limit of potentialities is an attempt to accommodate the empirical success of decoherence alongside a few explanatory gaps (in explaining the origin of classical mechanics and thermodynamic irreversibility) that are only partially addressed by decoherence. We shall discuss these in Chapter 7.

\section{Heterogeneous ``complex systems'' consisting of Potentialities and Actualities.}
The heterogeneous ``complex systems'' of potentialities and actualities will include systems with many operationally addressable subsystems, each of which can be either a potentiality or an actuality. If the subsystem is a potentiality, an operationally meaningful density matrix describing the subsystem can be identified; and if the subsystem is an actuality, alignment position  measurement of that subsystem is possible. There appears to be no general theory for such systems yet, but examples of such systems have been studied in the context of noisy quantum computers \cite{Aharonov2000NoisyQC}\cite{Skinner2019}. The mathematical theory of percolation seems better suited for such systems \cite{Skinner2019}, but in this thesis we explore the issue more heuristically with tools that many physicists are more commonly aware of, namely quantum trajectory methods. 

Since both actualities and potentialities are involved, it seems that there are two equivalent ways to study the system: either by looking at how the quantum states that can be locally assigned to the potentiality subsystems, conditioned on the records instantiated in the actuality, evolve with time. This is the route we try to utilize. The other route appears to be to study how the records instantiated in the actualities evolve with time. This dual picture is also present in the quantum trajectory description, where there are two complementary methods to study the dynamics: one can look at how the quantum state evolves stochastically, or one can look at how the records evolve stochastically \cite{wiseman2009quantum}.

In the absence of a Zeh-Zurek-style general theory, we enumerate two instances where such ``complex systems'' occur: one is in the context of noisy quantum computers  \cite{Aharonov2000NoisyQC}, providing an explanation for the emergence of classical computation  \cite{Aharonov2000NoisyQC}; and a few others, as we shall see in more detail in Chapter 6 and 7, are in the context of explaining the origin of classical mechanics and in providing a new statistical-mechanics construction.

\subsection{Dorit Aharonov's Model for Explaining the Transition from Quantum Computation to Classical Computation}
Dorit Aharonov considers a model \cite{Aharonov2000NoisyQC} for explaining the quantum-classical transition in quantum computers. To be precise, the quantum-classical transition in her model is the transition from quantum computation to classical computation, that is, how a quantum computer effectively becomes a classical computer, not the quantum-mechanics-to-classical-mechanics transition. The model she considers consists of an array of qubits in the quantum computer, with the needed quantum gates. Probabilistically, with probability \(1-\eta\), each of the qubits can undergo a measurement, and with probability \(\eta\), it can undergo a unitary transformation. It was shown by Aharonov that when the probability of measurement is higher, large-scale entanglement among the qubits cannot survive, and entanglement tends to `cluster' in localized patches consisting of a few qubits. When the measurement probability is lower, large-scale entanglement can form with larger clusters, with many qubits across larger distances in space involved in the larger entangled state. It is in this limit that the quantum computer can function reliably. In the high-probability-for-measurement limit, the quantum computer effectively becomes a classical computer.

This model can be seen as an instance of the $\{A, P, A \ldots\}\!-\!\{P, P, A \ldots\}\!-\!\{A, A, A \ldots\}\ldots$ type process (Class 4).


\subsection{The Origin of Classical Mechanics}

The $\{A, P, A \ldots\}\!-\!\{P, P, A \ldots\}\!-\!\{A, A, A \ldots\}\ldots$ type process also seems to underlie the origin of classical mechanics. This is studied in more detail in Chapter 7. It arises as a consequence of applying Postulate 1 and Postulate 2, along with Assumptions 1 and 2 and 7, to many-body quantum systems that qualify as an actuality.

When one considers a region \(R\) containing an actuality, then by Postulate 1, any potentiality having a nonzero probability of being found in the region \(R\) will probabilistically undergo either a unitary or a measurement in any finite interval of time. This leads to a Wiener- or Poisson-like process, depending on the probability of instantiability of the Born rule. When all the relevant potentialities are considered, we have an assembly of such processes.

The light emitted from a potentiality in a small region r, within R, will have information about the stochastic processes undergone by the potentiality in region \(r\). When the aggregate of all of such light from the potentialities in many small sub-regions of \(R\) is considered, the inferred alignment position measurement of the actuality in region \(R\) will follow a space-time trajectory that a light-emitting Newtonian particle would follow. This model is developed in Chapters 6 and 7.



\chapter{Is the Atomistic Conception of Matter Supported by Quantum Theory?}
\vspace{1cm}
\begin{flushright}
\itshape
\begin{minipage}{0.7\textwidth}
``The word `reality' is also a word, a word which we must learn to use correctly'' \cite{newton1997truth}. — Niels Bohr, as quoted in \cite{newton1997truth}.
\end{minipage}
\end{flushright}
\vspace{1cm}

In the title of this chapter, by the atomistic conception of matter, we do not mean the conception of atoms, which today has a specific operational meaning. We can think of the `atom' of a particular chemical species, say potassium, as the characteristic spectral lines, the allowed transitions, and the corresponding intensities for transformations manifest under a spectrometer. This is not what is meant here by the ``atomistic conception of matter''.

Instead, what we mean by that phrase is the implicit metaphysical assumption that all matter is composed of temporally persisting, spatio-temporally existing elementary constituent entities that exist independently of a `context'\footnote{Here, context can be understood as referring to the realization of the meaning of different operations that can be performed by an actual or counterfactual knowledge-acquisition system in an environment. Context is not necessarily about `macroscopic entities'; even the movement of a single-celled organism in an environment can be thought of as occurring in a context. Context is about the operational realization of meaning.} and belong to a set of universally recognizable types of elementary constituents, and that all of reality's behavior could be reduced to the aggregate behavior of those entities, given the initial conditions at the beginning of the universe.
The atomistic conception of matter implicitly assumes, to varying degrees, concepts such as individuality, ``possession of a state''\footnote{The idea that entities exist in some state without reference to a context; namely that the state is an a-priori (time-dependent) characteristic of the entity}, separability, location, persistence, and causality by contact for the elementary constituents of matter.

However, in view of the role of epistemic constraints, which we take to be primitive, the atomistic conception of matter in the form stated appears to be no longer supported by experience.

To illuminate the reason why that is so, a key question with which to begin the investigation is this:

\section{Should the Fundamental Be Only about the `Small'?}

What does `fundamental' mean? There are three senses to that question. First, it can be interpreted as the elementary starting points for a reasoning process; second, it can mean the elementary starting operational access points, such as primitive matter particles and allowed experimental interventions, available to an actual or counterfactual observer, from which they can synthesize everything else in a laboratory; third, it can concern what ultimately remains after an experiment aimed at breaking apart material things has been completed.

The following questions help sharpen our analysis further:

\begin{enumerate}
    \item Is the notion of the fundamental only about the very small that cannot be broken down further, in some operational sense, as Democritus envisioned?
    \item Or is it about space and time, which are relevant for any observable situation?
    \item Or is it also about certain complex phenomena that are naturally supervenient on atoms and molecules, in some operational sense, but are not logically supervenient on them \cite{chalmers1996conscious}? In other words, the phenomenon, when dissected physically with probes, reveals the familiar atoms and elementary particles in the form of appropriate experimental signatures, and yet the complex phenomenon is not explicable, even in principle, from the known micro-laws governing micro-behavior. An example of such complex phenomenon could be the inability to explain the definiteness of experience from the unitary evolution of atomic states.
    
    Here there is also the curious feature that, for the very meaning of the phrase ``micro-laws governing micro-behavior,'' we need, in an unavoidable way, the context provided by the complex phenomenon. This suggests a form of reverse supervenience.
    
    An example is the familiar quantum mechanical situation where, to even state the statement ``Schrodinger equation governing an atom'' meaningfully, we unavoidably need the Epistemically constrained situations of the experimental arrangement. There is no demonstrable exception to this reverse supervenience\footnote{Here we introduce the term `reverse supervenience' to describe cases in which changes in a lower level description are observed only in conjunction with changes in a higher-level description}.
    
    \item Or is it about the logical primitives of a conceptual system, where the logical primitives happen also to represent a real physical system, such as the murmuration of starlings or Conway's Game of Life, where the primitives are the individual starling and a Game of Life cell, respectively?
\end{enumerate}

In this thesis, the view that appears to be concordant with the arguments and constructions of the preceding chapters appears to be one that supposes all these four kinds as ``fundamental'' and that each is individually needed for a complete description of any given natural phenomenon.

Note that there are really two metaphysical commitments when it comes to the atomistic conception of matter: one is that the world is intelligible through analysis into simpler parts, and the second is stronger---that the world is made of a few universal fundamental primitives and that everything can be understood from the behavior of those primitives. The former can include studying how a fish moves through water by studying its fins and the fluid properties of water; the latter demands that the fish and its movements be understood from the smallest, universal, elementary primitives that apply to all physical systems, such as atoms and electrons. We retain the first kind of reduction, which could be called ``analysis,'' \cite{Mayr1988WhatMakesBiologyUnique} but we reject the second, and replace it with a more qualified stance: any physical system, when probed suitably, has the same set of fundamental primitives, but their existence is always contingent on higher-level entities such as epistemic constraints. This contingency cannot be avoided, even in principle, and thus reflects the way the world is, rather than a limitation of our experimental access. 

Analysis can also invoke those fundamental smallest-scale constituents, such as when we explain electricity in terms of the behavior of electrons, but the explanans will always include
\begin{enumerate}
    \item the higher-level epistemically constrained situations (ECM and ECU), such as in dictating the lattice structure and thus the Hamiltonian, and
    \item the conceptual miniaturization of potentialities—instantiated in the processing units of observers but not as spacetime events elsewhere until detection—such as that which allows the consideration of a hydrogen atom as a miniature planetary system or an electron gas as a bunch of bouncing billiard balls (Assumption 8).
\end{enumerate}

This choice of what we take to be fundamental now benefits from a study of emergence.

\section{The Complex Richness of the World and Where Bohr's Epistemic Constraints Fit}

In this section, we survey the complexity of the world and situate epistemic constraints in the surveyed landscape.

Consider a heap of sand. All properties of a heap of sand are ultimately determined by the properties of the individual grains and by the interactions among them. Some of these properties can be meaningfully attributed both to individual grains and to the heap as a whole. These may be called \emph{nonemergent} properties. Mass is a paradigmatic example: individual grains have mass, and the mass of the heap is simply the sum of the masses of its constituent grains. Nonemergent properties of a whole are typically \emph{compositional}, in the sense that they are built directly from the same kind of property instantiated by the constituents.

By contrast, some properties of a heap of sand are not meaningfully attributable to individual grains but are meaningful and useful only at the level of the aggregate. An example is the property of being a heap, or ``heapness''. The \emph{blinkers} and \emph{gliders} in Conway's Game of Life provide another illustration: these are effective patterns that arise from the local update rules governing individual cells in the cellular automaton. The collective behavior of ant colonies, arising from simple pheromone-based interaction rules among individual ants, provides yet another example. Classical statistical physics provides further examples, such as temperature and pressure: these properties are well-defined and explanatorily useful for a gas as a whole, but not for individual gas molecules.

In classical statistical physics, such properties are coarse-grained \cite{Chalmers2006StrongWeakEmergence} manifestations of more fine-grained underlying dynamics, and they can therefore be explained in terms of microlevel properties. In the case of temperature and pressure, one reduces them to a microlevel description in terms of a large collection of interacting molecules, for example, their momenta and interaction dynamics. Although temperature and pressure are not possessed by individual molecules, they are assumed to be completely determined by, and in principle derivable from, the underlying microphysical facts. Properties of this kind are therefore called \emph{weakly emergent} \cite{Chalmers2006StrongWeakEmergence}: they are emergent because they apply only at the system level consisting of many interacting parts, yet \emph{weak} because they are reducible in principle to the behavior and interactions of the constituents. Here, `system level' refers to the composite object.

The notion of weak emergence is contrasted with strong emergence  \cite{Chalmers2006StrongWeakEmergence}, which involves system-level properties and laws that are not only meaningless at the level of the constituents, but are also not reducible, even in principle, to the properties, interactions, and laws of those constituents. Weak emergence on the other hand is often characterized as an apparent effect that arises from limitations of human epistemic access, which restrict us to a coarse-grained description of reality while leaving finer-grained processes at more fundamental levels unobserved.

One can imagine a gradation within weak emergence. We can have situations where complete microphysical knowledge would, in principle, fix the behavior of the synchronic or diachronic emergent properties deterministically; or situations where complete microphysical knowledge would support only probabilistic predictions of emergent behavior. 

Even more comprehensively, weakly emergent phenomena can be assessed in terms of multiple characteristics. All these characteristics provide different senses in which we can study the part-whole relations. They may be characterized dynamically, in terms of whether the system exhibits ordered, critical, or chaotic behavior  \cite{Kauffman2000}; thermodynamically, in terms of whether the system is in equilibrium, metastable, near a thermodynamic critical point, or far from equilibrium; evolutionarily, in terms of how variations in constituent parts (when the parts are replicators capable of variation, reproduction and inheritance) respond to environmental pressures  \cite{Kauffman2000}; in terms of \emph{organizational complexity}, for example, the number of steps required to assemble  \cite{Sharma2023AssemblyTheory} the system from simpler constituents under a specified construction scheme; and computationally, for example, in terms of whether predicting the macroscopic phenomenon from the microscopic description is a decidable problem or, in some special cases, a formally undecidable problem. This is when a macroscopic weakly emergent phenomenon may be computationally irreducible relative to its microscopic description. That is, the emergent phenomenon cannot be predicted by a computational process given the knowledge about the microscopic description \cite{Hossenfelder2019}. Most physically relevant weakly emergent systems remain decidable \cite{Hossenfelder2019}.

In the context of quantum many-body systems, we define \emph{weak emergence} as the appearance of novel system-level properties and laws---namely, collective observables, such as phononic observables  \cite{Hossenfelder2019}, and effective coarse-grained laws  \cite{Hossenfelder2019}---that are nevertheless supposed to be fully reducible to the underlying microscopic dynamics and laws. These dynamics consist either of purely unitary evolution of interacting microscopic constituents, or of unitary evolution supplemented by effective stochastic state updates, as in decoherence and open-system descriptions. In either case, at a given time, a pure quantum state need not be assignable to individual constituents, but a reduced density matrix can always be assigned to each constituent subsystem.

By contrast, strong emergence posits emergent behavior that is novel in a more fundamental sense and not an artifact of coarse-graining a fine-grained description. The higher-level strongly emergent properties and laws are fundamental in that they are not reducible, even in principle  \cite{Chalmers2006StrongWeakEmergence}, to a description stated purely in terms of the properties ascribed to elementary constituents. Moreover, even a complete description of those constituents in terms of their properties will not anticipate the higher-level properties and behavior under any coarse-graining. To quote Chalmers,
\begin{quote}
    ``Strong emergence requires that high-level truths are not conceptually or
    metaphysically necessitated by low-level truths'' \cite{Chalmers2006StrongWeakEmergence}.
\end{quote}

At the same time, such properties are not taken to have an independent existence. Rather, they occur only in conjunction with the underlying constituents and their physical organization. This constant conjunction is often discussed under the heading of \emph{natural supervenience} \cite{chalmers1996conscious}. This is to be contrasted with the combination of logical and natural supervenience often associated with weak emergence  \cite{Chalmers2006StrongWeakEmergence}: weakly emergent properties are taken to be a logical consequence of the microscopic description, logical supervenience, while also occurring only in systems with an appropriate microstructure (a constant conjunction), natural supervenience. Here logical supervenience means the emergent properties and their laws can be predicted in principle by a computer fed with the information about the microscopic parts and their laws.

Note that natural supervenience does not, by itself, entail a reductionist bottom-up explanation in terms of elementary constituents. That is, natural supervenience does not entail logical supervenience. The impulse to demand such an explanation is a longstanding methodological habit that is neither demanded by logic nor plausibly a necessary character of nature. It is useful, but not without limitations, and it is precisely this limitation that motivates the notion of strong emergence in the foundations of complexity theory \cite{Chalmers2006StrongWeakEmergence}\cite{Hossenfelder2019}.

Strong emergent properties and laws are expected to constitute novel ingredients for the vocabulary of physics, or science in general:
\begin{quote}
    ``Strong emergence has much more radical consequences than weak emergence. If there are
    phenomena that are strongly emergent with respect to the domain of physics, then our
    conception of nature needs to be expanded to accommodate them.\ldots\ In any case like this,
    fundamental physical laws need to be supplemented with further fundamental laws to ground
    the connection between low-level properties and high-level properties'' \cite{Chalmers2006StrongWeakEmergence}.
\end{quote}

Chalmers' discussion of strong emergence is developed primarily in the context of consciousness, whereas here we employ some of his terminology and methodological distinctions for an unrelated problem. Consciousness is not required for Bohr's notion of ``epistemic constraints'': epistemic constraints can plausibly be instantiated by a classical computer with robotic actuators and language capabilities, even if the computer is not conscious. 

In the case of quantum theory, the low-level laws and description is given by unitary Schrodinger dynamics, and the high-level features concern the role and character of epistemic constraints in the operational access to quantum phenomena. Chalmers gestures to this distinct kind of strong emergence, and related directions can also be found in prior work by Ellis  \cite{Ellis2012TopDown} and Drossel  \cite{Drossel2017Connecting} (but unlike Ellis and Drossel, we retain the `universal' validity of quantum theory independent of the size or mass of the system):
\begin{quote}
    ``A key issue is that no one knows just what is the criterion for a measurement
    taking place. Yet it is clear that for the collapse interpretation to work, measurements must
    involve certain highly specific causal events, most likely at a high level. If so, then we can see
    the measurement postulate as itself a sort of configurational law, involving downward
    causation'' \cite{Chalmers2006StrongWeakEmergence}.
\end{quote}

Epistemic constraints, and the associated functional notion of agents who can intercommunicate using a shared language, are treated as \emph{strongly emergent} in the present work---not in the sense that, beyond some threshold number of atoms, something ``nonphysical''\footnote{in the sense of not being composed of atoms and elementary particles when probed}\footnote{notice that in the course of this work we have been developing a refined meaning of what ``physical'' means: we are adopting an onto-epistemic understanding of ``physical'', as opposed to the Descartes's `res-extensa' sense of ``physical'' which has dominated our view leading to the prevalent picture of ``stuff out there''. See Appendix B, coversations with Professor Armour Garb, for more on this topic} emerges mysteriously, but in a specific and operational sense: epistemic constraints are indispensable for providing the very context in which quantum particles and fields can be specified and experimentally accessed at all. Since there appears to be no alternative route by which quantum systems can be empirically known, we lose nothing, methodologically, by treating the inevitability and fundamentality of epistemic constraints---a higher-level feature---as a feature of nature itself, rather than merely a contingent limitation of our descriptive practices.

This differs from the situation in, for example, ant colonies. A couple of ants and an ant colony as a whole have a similar ontological status: both exist in space and time, both possess individuality, and both exhibit persistence of identity over time. Moreover, the higher-level entity---the colony---is not required in order to observe, identify, or define a pair of ants.

This added, narrower qualification of ``strong emergence,'' as applied to epistemic constraints, remains consistent with the broader definition stated earlier. In particular, on Bohr's interpretation, higher-level truths concerning epistemic constraints and pointer readings under alignment measurements are not reducible to truths about quantum states evolving unitarily according to the Schrodinger equation. The added qualification on the strong emergent nature of epistemic constraints emphasizes a further point: even the \emph{formulation} of the purportedly lower-level truths---those stated in terms of quantum states and unitary evolution---presupposes epistemically constrained contexts, because the experimental arrangements that fix what state is assigned and what unitary is implemented must be specified as a precondition.

Now, let us look at the character of explanation in both weak and strong emergence.

In weak emergence, explanations are invariably bottom-up  \cite{Chalmers2006StrongWeakEmergence}\cite{Ellis2012TopDown}. They begin with a microscopic description and explain the macroscopic phenomenon in terms of properties and behavior at the microscopic level.

Strong emergence, by contrast, is expected to provide an explanatory model in which strongly emergent properties operate in two ways simultaneously  \cite{Ellis2020ArxivCausalClosure}. First, they affect the lower-level structure via ``top-down causation''---a term we will henceforth replace with \emph{top-down realization/determination}, a term due to Ellis, to avoid the misleading suggestion that there is a direct causal influence between distinct ontological levels. The top-down effect of strongly emergent properties can be characterized, to quote Chalmers:
\begin{quote}
    ``The best way of thinking of this sort of possibility is as involving a sort of downward
    causation (downward realization). Downward causation means that higher-level phenomena are not only irreducible
    but also exert a causal efficacy of some sort. Such causation requires the formulation of basic
    principles which state that when certain high-level configurations occur, certain consequences
    will follow. (These are what McLaughlin (1993) calls configurational laws.) These
    consequences will themselves either be cast in low-level terms, or will be cast in high-level
    terms that put strong constraints on low-level facts'' \cite{Chalmers2006StrongWeakEmergence}. 
\end{quote}
This provides a route for incorporating strongly emergent features as useful and potentially predictive ingredients in science, rather than leaving them as a philosophical position about certain classes of emergence. In fact, the attempt to incorporate the role of Bohr's epistemic constraints that we have pursued in this thesis takes this direction.

To achieve this incorporation, especially given the success of weak emergence, one may expect strongly emergent properties to indirectly support bottom-up explanations for other higher-level emergent phenomena, after first shaping the microscopic constituents via top-down realization. An example of this abstract-sounding method could be this: Bohr's epistemic constraints determine the quantum-state assignment and subsystem partitioning in a many-body system in a top-down manner, and that determination in turn can determine bottom-up emergent behavior in the many-body system, due to what the Schrodinger equation and the Born rule dictate for the many-body system. Thus, as Chalmers anticipates,
\begin{quote}
    ``Cases of strong emergence will likely also be cases of weak emergence
    \ldots\ But cases of weak emergence need not be cases of strong emergence'' \cite{Chalmers2006StrongWeakEmergence}. 
\end{quote}

Therefore, one can likely still employ quantitative assessment criteria familiar from weak emergence, including dynamical criteria, whether the system is ordered, critical, or chaotic; thermodynamic criteria, whether the system is in equilibrium, metastable, close to a critical point, or out of equilibrium; evolutionary criteria, in terms of how variations in constituent parts respond to environmental pressures; \emph{organizational complexity}, the number of steps required to assemble the system from simpler constituents under a specified construction scheme \cite{Sharma2023AssemblyTheory}; and computational criteria, whether predicting the macroscopic phenomenon from the microscopic description is decidable or, in some special cases, formally undecidable.

So far, the analysis of strong emergence has remained abstract. This is understandable: although strong emergence is logically coherent, examples using the idea in physical theories in a productive way remain limited. In the last section of this chapter, we present an example that follows from the theoretical framework developed so far. It is framed as a `strong-weak hybrid emergent model', in the sense that epistemic constraints and the quantum state description together dictate a dynamics via top-down realization and bottom-up emergence.

\section{Quantum Theory and the Atomistic Conception of Matter}

Although there have been significant experimental advances in probing quantum events that happen at the highest energy scales, and in the elucidation of a fixed set of universally recognizable fundamental `particles', the way any realistic experiment is performed already suggests that we are still far from having indisputable evidence to suppose that what is fundamental occurs only at the smallest scales. The most important of these is the previously noted observation that, in any quantum mechanical phenomenon, a complete description always appears to require the experimental context, which allows itself to be described in plain language. This is not simply because we humans `lack the ability' to directly see or access the elementary particles, but because of a fundamental and natural epistemic limitation as to how they may ever be accessed and what it means for them to be accessed (Principles 0, 1 and 2).

More concretely, the atomistic conception of matter can be thought of as consisting of a mereology that posits, firstly, temporally persisting elementary entities, whether particles, fields, or strings, which are the parts that exist independently of the context that can be supplied by a whole—and therefore that the parts have ontological priority. Secondly, it posits that wholes are no more than the parts and the interactions between the parts. Thirdly, the genesis of the whole happens by a rule-bound, context-independent, step-by-step deterministic or probabilistic construction from the parts interacting with each other. The smallest constituents in the atomistic conception of matter need not be atoms, as was originally thought, but here we still use the term `atomistic conception of matter' to refer to the metaphysical commitment corresponding to the reductionist bottom-up account of nature that assumes that any macroscopic physical phenomenon could in principle be explicable in terms of the behavior of the aggregate of the smallest elementary constituents, whether those are elementary particles or fields or strings. This is a modern-day extension of the original Democritean ideal.

However, given the nature of epistemic constraints, it appears that experience suggests the following status for the atomistic conception of matter:

\begin{figure}[htbp]
\centering
\includegraphics[
  width=1.0\textwidth,
  alt={A schematic diagram revisiting the atomistic conception of matter. The diagram distinguishes aspects of atomism supported by evidence from historically inherited assumptions, including assumptions about elementary constituents as temporally persisting, spatiotemporally existing, context-independent entities with individuality, state-possession, separability, location, persistence, and causality by contact.}
]{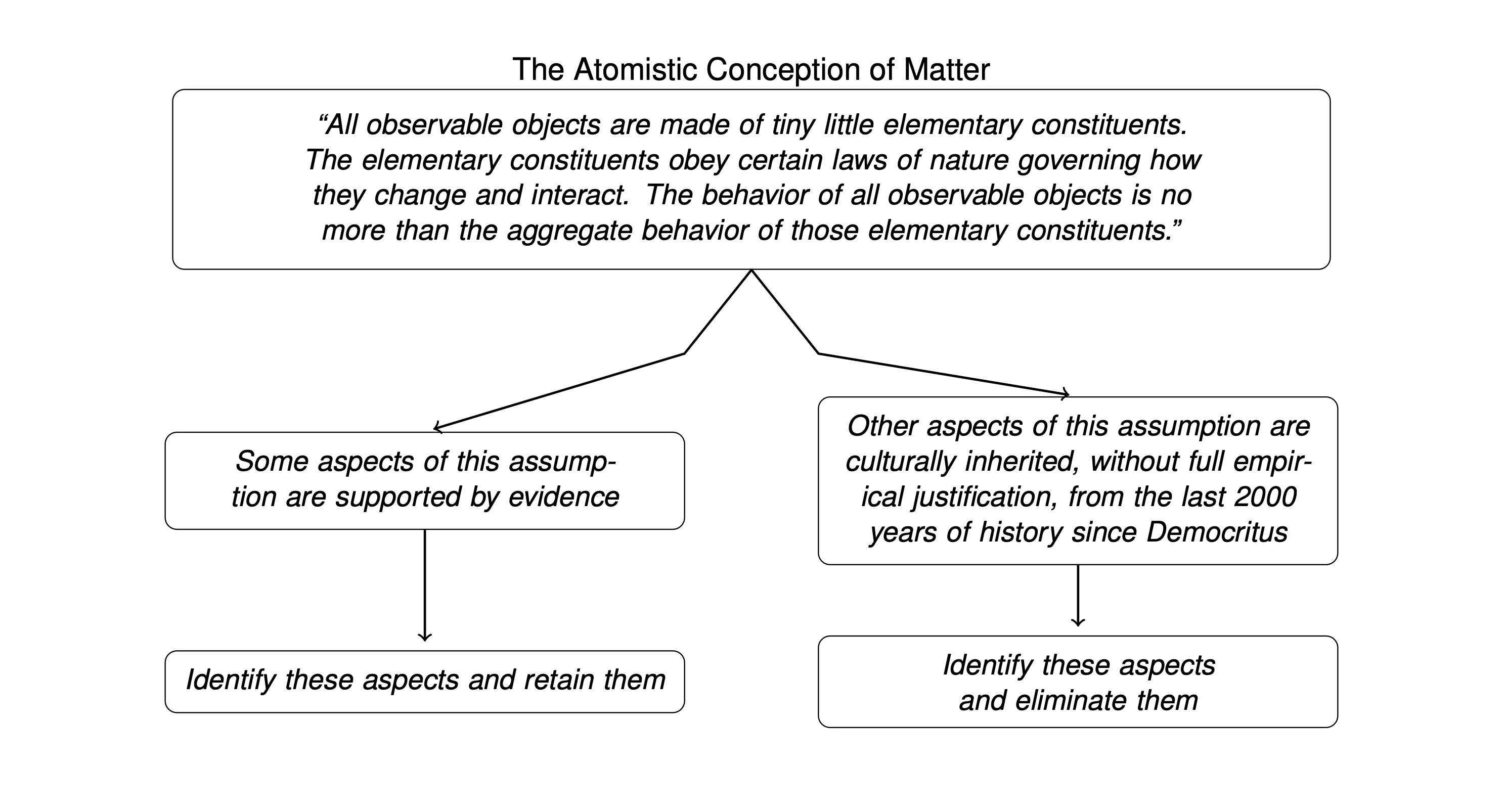} 
\caption{Revisiting the atomistic conception of matter}
\label{fig:atomistic-conception-of-matter}
\end{figure}

The aspects that need retention are evident without much difficulty in searching. It is the fact that there are universal, non-identical, classifiable experimental signatures in experiments that we call the `fundamental particles', that does indeed support the assumption that all matter, when probed at high enough energies, reveals experimental signatures that can be classified from among the same set of universal fundamental particles. And the properties of these elementary constituents and the dynamical equations governing their behavior, subject to epistemically constrained contexts, do place constraints on macroscopic observable effects, as evidenced in the role of the Pauli exclusion principle in understanding the behavior of neutron stars.

However, for the explanation of aggregate phenomena in condensed matter physics, in addition to the properties of the fundamental constituents and of the Schrodinger equation governing the dynamics of the state of fundamental constituents, epistemically constrained contexts are still nevertheless unavoidable in determining the Hamiltonian, the tensor product structure, in counting the time intervals, and in measuring spatial arrangements of the experimental outcomes \cite{Drossel2017TenReasons}.

Because of this reverse-supervenient role of the\textbf{ }epistemically constrained situations in every demonstrable quantum experiment, the rest of the metaphysical assumptions associated with the atomistic conception of matter, in particular the persistent hope that everything could eventually be reduced solely to the behavior of elementary constituents, seem unjustified and appear to be the manifestation of an instance of epistemic inertia rather than a need necessitated by experience.

The template supplied by the atomistic conception of matter originated along with classical physics, but the template is not necessarily classical mechanical. The same template can be retained with quantum mechanical state space and rules instead of classical mechanical state space and rules. The conceptual schema in which we implicitly embed quantum mechanics---in its orthodox version \cite{vonNeumann1932}\cite{Zurek2003Decoherence}\cite{Schlosshauer2004}---is arguably still embedded in the atomistic conception of nature because of the attempts to `explain' the Epistemically constrained situations \cite{Zurek2003Decoherence} from elementary entities.

In view of the role of epistemic constraints, the main point of departure from the atomistic conception of matter is that it is not simply systems, states, interactions, phase spaces, and transformations that are essential for the description of nature, but additionally, the `operational actions' of literal or counterfactual observers, and the associated semantics\footnote{the actions of actual or counterfactual observers and the meaning of those actions in a given context}, are also essential. These epistemically constrained contexts are functionally defined in terms of what an actual or counterfactual observer does in a lab, and these functionally defined maneuvers of an observer determine what state was prepared, what observable is measured, and so on. As we have already stressed, quantum states do not come `inbuilt' with a system in the way one may have envisioned classical coordinates to come `inbuilt' with every Newtonian point particle. It is the Epistemically constrained situations that provides the specification of the states and the Hamiltonians. There are no meaningful temporally persisting elementary entities that are the parts and that exist independently of the context, whether they are particles, fields, or strings. 

The campaigners for the stance that suggests taking the atomistic conception of matter, implicitly, in combination with quantum mechanics, often insist that any lingering preference for epistemic constraints and language is an indication of cultural inertia from the pre-quantum era \cite{Wallace2012}. These include allegations that insisting on definite outcomes, a single world, intersubjective agreeability, and ordinary-language describability is a relic from the pre-quantum worldview. However, on the contrary, it appears that it is this camp that is in fact insisting on a pre-quantum natural philosophy (namely the atomistic conception of matter), even when it is no longer fully justified by experience.

Definiteness of outcomes, intersubjective agreeability, and a single-world ontology are not relics of a pre-quantum era. They are instead irreducible aspects of language and communication that are fixed structural features of experimental practice that have remained in place at least as long as experimental practices have existed. What would be a stance with minimal cultural inertia from the pre-quantum era, it appears, is a stance that takes quantum mechanics to be universal, with caution about what universal means, and revises the atomistic conception of matter, to give a place to the irreducible aspects of language and communication as features in nature.

\section{Reasoning about Historical Events}

A likely objection from a cosmologist: ``But at the beginning of the universe, what existed was only what you are calling potentialities---quantum fields, and nothing else.'' 

My response to that is this: ``You are suggesting a model based on our CMB evidence and light-matter abundance \cite{tong2019cosmology}. But note that, for that evidence and the extrapolation, we nevertheless needed actualities, i.e., epistemic constraints; after all, that is how we can ever hope to make observations. Asserting that only potentialities existed back then is an extrapolation based on assuming a monist mode of existence, an inheritance of the atomistic conception of matter, which is an implicit assumption of our current fundamental physics. If we take the said dual mode of existence---of actualities and potentialities---our extrapolation will also be adjusted in light of the same set of CMB evidence and light-matter abundance evidence. The evidence stays the same, but the story we tell to make sense of that evidence will differ.''

A model is constructed out of evidence and implicit metaphysical assumptions; the failure of the latter, and thus the model, does not contradict the evidence.

All this does not mean we need to assume the literal existence of observers in the primordial universe. We only need to assume counterfactual, fictitious observers\footnote{Assuming that observers could have existed and functioned in the distant past is an assumption justified by virtue of there being no known law of nature prohibiting such an assumption. This is again similar to how we can imagine clocks as ``ticking'' in the distant past of the universe, as long as the functioning of a clock is not physically prohibited.} until the conditions back in the history of the universe when the existence and functioning of an observer as a physical system were physically possible. This kind of reasoning is not very different from counterfactually assuming fictitious clocks when talking about a duration in the history of the universe when no actual clocks were around. We just reason as to what a clock might have read, had it been positioned in that situation in the distant past.

The implicit metaphysical model we traditionally have in physics is supplied by Conway's Game of Life: the initial conditions of the universe---a set of primitive stuff with a set of primitive laws for that stuff---ensure, either probabilistically or deterministically, the universe we have, including life, complexity, and language. Had the initial conditions been different, we might have ended up with a different universe. But this model appears to make category error. Unlike in Conway's Game of Life, there are no counterfactual other universes; as far as we can tell, there is just one, and it has life, language, and other complex phenomena. Therefore, instead of the reductionist attitude that says ``the fundamental laws of nature should not care about life, language, or other complex phenomena, because this universe has those phenomena due to one choice of initial conditions,'' we must instead perhaps say, ``we know that life, language, and complexity exist as a matter of fact in this universe, so we cannot pretend otherwise if we are to completely describe the universe.'' 

This is not to suggest that the universe is strictly deterministic and that life was predestined in a dynamical sense given the initial conditions of the universe---it is not, as we know well (due to the indeterministic laws of nature)---but that there are higher-level, often probabilistic but fundamental laws of nature, such as the laws of evolution by natural selection, that start applying only when the relevant systems where those laws can meaningfully instantiate have already emerged. In the case of evolution, the relevant systems are the replicators  \cite{Kauffman2000}.

Before concluding this section, we should address a reductionist critique. 

If an object \(A\) is made of constituents \(a_1\), \(a_2\), \(a_3\), \(a_4\), \(\ldots\), then a reductionist concern arise: 1) if \(A\) is made of \(a_1\), \(a_2\), \(a_3\), \(a_4\), \(\ldots\), then any behavior that \(A\) exhibits can be identically re-expressed in terms of the aggregate of the detailed behavior of \(a_1\), \(a_2\), \(a_3\), \(a_4\), \(\ldots\), with the former only being a coarse-graining effect. Here, by identically re-expressed, we refer to the re-expression of observed macroscopic behavior purely in terms of the properties and behavior of the aggregate of constituents interacting with each other; and 2) if one demands a novel effect at the level of \(A\) that cannot be re-expressed by the method of 1), then we must supply a boundary or cut-off in the number of constituents and the complexity of the interactions between the constituents at which the effect happens. In the absence of such a cut-off, any irreducible property is ill defined.

However, the critique in 2) does not follow because the statement 1) is not a logical necessity; it is a methodological habit of thought that has nevertheless been successful, but can have limitations. It is consistent to have higher-level phenomena that naturally supervene on a lower-level substrate but are not logically necessitated by it \cite{Chalmers2006StrongWeakEmergence}. Moreover, the demand in 2) hold only when the parts \(a_1\), \(a_2\), \(a_3\), \(a_4\), \(\ldots\), and \(A\) have the same monist mode of existence. But that is not the case for situations involving a quantum mechanical system and the Epistemically constrained situations in which it is materialized. Thus, we can say that quantum mechanical systems are reverse-supervenient on the Epistemically constrained situations---there is no instance in which we observe a quantum mechanical system, the potentialities, without the epistemically constrained contexts, the actualities. No difference in lower-level properties occurs without some difference in higher-level properties. So it is not just that epistemically constrained contexts are supervenient on the familiar particles of the standard model, but also that the familiar models of the standard model are reverse-supervenient on the Epistemically constrained situations. 

\section{The Method of Integrating Strong-Emergent Features with Weak-Emergent Modeling}

The program we have been pursuing in this work can be seen as an instance of a broader method in which a strongly emergent feature is combined with a weakly emergent mode of model building, with the intent of seeking new physical conclusions.

In the usual mode of weak-emergence modeling, which goes all the way back to Descartes, we break a phenomenon into simpler analyzable parts, and from the understanding of how the parts behave and affect each other, we seek an understanding of the phenomenon. These parts are characterized as \textit{res extensa} of some form---particles, fields, etc.---entities that persist temporally, are individuated, and possess a state without reference to an external context.

Much of this method shall remain, given its immense usefulness.

However, informed by the practice of quantum theory, we do need a minor refinement, to incorporate into this picture the irreducible external context---the epistemic constraints. The method that this example suggests to us is one in which we have a top-down determination \cite{Ellis2012TopDown} to specify the state and evolution of the parts constituting a whole, followed by the usual bottom-up emergence once that specification has been made. We can call this the \textsc{The method of top-down determination followed by bottom-up emergence}. New conclusions can arise because new principles and generalities in nature are underexplored in this top-down determination process. In the case of quantum foundations, it is at this interface that we have the two postulates that we will utilize in the next section.

The original method of breaking a phenomenon into analyzable parts and understanding its behavior from knowledge of the parts and their interactions was conceived in the context of mechanics, but later proved useful in other sciences, such as in how we came to understand biology from the behavior of cells. Likewise, the method for integrating strongly emergent features with weak-emergence modeling, though informed by quantum processes, may find applications in other sciences.

This method can therefore likely be explored in other instances in science where other kinds of strong emergence are suspected. An indication of strong emergence is the presence of a recursive situation in which, to sensibly define\footnote{to operationally specify its relevant properties} the part of a whole, the whole in question itself is needed.

A few examples of where it is indicated and not indicated are as follows. There is no strong emergence in an ant colony, since, for the purpose of understanding an ant colony, we can define a couple of ants without reference to the ant colony as a whole. In contrast, we need epistemic constraints to characterize the changes in the pointer needles, and we need the very same epistemic constraints for any attempt to characterize the quantum system whose detection led to the changes in the pointer. Another distinct example where strong emergence might be indicated, as noted by Chalmers \cite{chalmers1996conscious}, is in consciousness. In this work, we have kept epistemic constraints and consciousness as distinct phenomena; however, whether or not there is any connection between the two is not fully demonstrated in this work.

In these two distinct example where strong emergence is suspected, a common critique might be, ``aren't you being circular by assuming what you want to explain?'' It must be noted that the question is no longer to ``explain'' epistemic constraints, or, in the case of consciousness, to explain how it emerges from non-conscious matter, but rather to seek what new conclusions can be made, or what constructive developments can be made, with the joint consideration of that which is irreducible and the usual means of reductionist, bottom-up reasoning, after appropriately incorporating the former into the latter. In the case of quantum foundations, this method leads us to the Bohrian Program. It can be speculated that, in the case of consciousness, this mode of reasoning can lead us to a criterion for certifying conscious machines, not by looking at their circuitry, nor by looking at their verbal interaction or subjective reports as in the Turing test, but by supposing, perhaps, that, for any pair of systems, one of which is conscious (the certifier) and the other (a test system) whose consciousness is being tested, it is always possible to find an interface such that the different components of the sensory and phenomenal experience of each can be transduced by the interface. An example of such an interface could be a brain--brain interface. In this way, there may exist a minimal set of experiences that, when experienced by one member of the pair, the test system, will cause the felt experience of a similar experience in the other system, the certifier. This can be used as a benchmark for consciousness certification. There is no circularity in all of this.

I will not speculate more on consciousness in this work, and shall now return to my main focus.

We now specifically illustrate this method of top-down determination followed by bottom-up emergence for many-component quantum systems by showing how epistemic constraints, on the one hand, and the usual means of studying the behavior of many-component quantum systems from knowledge of individual components, on the other hand, can be integrated to lead to new conclusions that cannot be arrived at from the latter alone.

Given a piece of matter, we seek to ascribe states to the parts of the matter, conditioned on the set of all possible operations that could be performed on the matter by an externally stationed observer. The set of all such possible operations itself is part of the analysis. It is at the interface of this top-down determination process that we have the two principles (1 and 2) and the two postulates that we introduced earlier. As for the set of all possible operations that can be performed from the outside on a piece of matter, there need not be an actual observer present; a counterfactual fictitious observer suffices. So we can theorize about a piece of crystalline material in deep space by picturing such a local observer in its vicinity, without there being an actual observer.

Now I apply this method.


\section{Two Observers in Relative Motion Optically Monitoring a Drifting Gas: An Application}


Consider a gas of atoms or molecules.

The gas is not confined to a container, and thus will expand freely in general. It can also, as a collective whole, drift in a biased direction due to an external potential. To simplify our consideration, we restrict to a specific kind of macroscopic phenomenology: a gas that predominantly drifts collectively and has negligible free expansion. This simplification, however, does not affect the validity of the conclusions we will draw and can later be relaxed if needed.

To this system, we now append two observers, \(A\) and \(B\). \(A\) intercepts the light from the drifting gas, and comoves with the drifting gas. \(A\)'s presence or possible intervention in the moving gas is assumed to cause only negligible effects on the dynamics of the moving gas. \(A\) need not be an actual observer confronting the drifting gas; it could be a fictional one. The gas is assumed to be subject to an external potential \(V(\vec{r})\), where \(\vec{r}\) is the position vector joining observer \(B\) and an arbitrary region in the bulk of the drifting gas, with \(B\) taken as the origin. This is in addition to the interparticle forces that may be acting between particles in the gas, or between a particle in the gas and the rest of the gaseous system\footnote{Note that despite our critique of the `particle' and `field' concepts, we continue to use the terms for pedagogical reasons. But a particle here is only a miniature model, in accordance with Assumption 8, for potentialities, under certain situations, and not a space-time event ontology inside the gas.}.

We suppose that \(A\) inserts, or counterfactually does so, a pair of probes at two distinct regions in the bulk of the gas. The probes can count particle number or perform localized measurement of any pair of local observables.

In typical many-body quantum systems with local interactions among constituents, the reduced state determined by the two local probes becomes increasingly mixed---less entangled---as the spatial separation between the probes increases. This makes the observation of a nearly pure bipartite entangled state overwhelmingly unlikely for large probe separations. Thus, we can assign a nearly pure quantum state ( which are also, in general, entangled) only to the detections by the pairs that are stationed in nearby locations. Performing this counterfactual operation for the set of all possible pairwise arrangements, we can write the total Hilbert space as a direct sum over the regions where we counterfactually could have inserted the pairs of probes, such that within each such region there is at least a pair of probes whose clicks can be described by a nearly pure quantum states.

We define a cluster as the collection of degrees of freedom or subsystems that is effectively described by a pure state, where any entanglement exists within the cluster but not between clusters. For instance, if we had 10 gas molecules that share an entangled state, then that constitutes a cluster, and any molecule besides these ten molecules is not part of the cluster. If there is a nearby group of six molecules in another entangled state, then that is another cluster.

At the one-particle level, since the regions are disjoint, the one-particle Hilbert space decomposes as a direct sum, since it has support in those regions:

\begin{equation}
\mathcal{H}_{1\text{p}}
  = \mathcal{H}_{R_1} \oplus \mathcal{H}_{R_2} \oplus \cdots \oplus \mathcal{H}_{R_k}.
\end{equation}

Here the subscript labels refer only to one possible choice of counterfactual probing. So the particular way of partitioning the volume into smaller volumes containing the pairs of probes does not matter. Furthermore, the number of such partitions can be as high as is operationally feasible.

For a variable number of particles, we can build the many-particle Fock space by symmetrizing, if bosons, or antisymmetrizing, if fermions, the unsymmetrized \(N\)-single-particle Hilbert space, and by letting the particle number not be fixed, in general,

\begin{equation}
\mathcal{F}(\mathcal{H}_{1\text{p}})
  = \bigoplus_{N=0}^{\infty}
    \mathrm{Sym/AntiSym}\!\left( \mathcal{H}_{1\text{p}}^{\otimes N} \right).
\end{equation}
({Here, the symmetrization and antisymmetrization account for both the bosonic and fermionic cases. The physical meaning of Eq. 6.2 is that the symmetrized or antisymmetrized subspace of the Hilbert space of \(N\) particles, formed from the single-particle Hilbert space, can be used to construct a Fock space with variable particle number by taking a direct sum over particle number. )

{The general protocol when one writes down a Fock space \(\mathcal{F}\) (of a Hilbert space) is as follows: first, take the \(N\)-particle tensor product of the Hilbert space in parentheses; then symmetrize or antisymmetrize it, depending on whether the particles are bosons or fermions; and finally, take the direct sum over particle number \(N\).}}

Now we can write the direct sum as a direct product

\begin{equation}
    \mathcal{F}\!\left( \bigoplus_i \mathcal{H}_{R_i} \right)= \mathcal{F}\!\left(  \mathcal{H}_{1p} \right)
  \simeq
  \bigotimes_i \mathcal{F}(\mathcal{H}_{R_i}),
\end{equation}

which means, we can equivalently use the direct-product factorization on the right-hand side to express the total Hilbert Space (the total Fock space can be written as a tensor product of regional Fock spaces), 

\begin{equation}
    \mathcal{H}_{\text{total}}
  \simeq
  \mathcal{F}(\mathcal{H}_{R_1})
  \otimes
  \mathcal{F}(\mathcal{H}_{R_2})
  \otimes
  \cdots
  \otimes
  \mathcal{F}(\mathcal{H}_{R_k})
\end{equation}

Now, if we identify \(\mathcal{F}(\mathcal{H}_{R_k})\) with a cluster \(C_k\) and write it as \(\mathcal{H}_{C_i}\), we can write

\begin{equation}
\mathcal{H}_{\text{total}}
  = \mathcal{H}_{C_1}
    \otimes \mathcal{H}_{C_2}
    \otimes \cdots
    \otimes \mathcal{H}_{C_k},
\end{equation}

where we account for the symmetrization or antisymmetrization of the physically allowed Hilbert space:

\begin{equation}
\mathcal{H}_{C_i}
  = \mathcal{F}(\mathcal{H}_{R_i})
  = \bigoplus_{N_i=0}^{\infty}
      \mathrm{Sym/AntiSym}\!\big(
        \mathcal{H}_{R_i}^{\otimes N_i}
      \big).
\end{equation}

Therefore,

\begin{equation}
    \mathcal{H}_{\text{total}}
  = \bigotimes_{i=1}^{k}
      \Bigg[
        \bigoplus_{N_i=0}^{\infty}
        \mathrm{Sym/AntiSym}\!\big(
          \mathcal{H}_{R_i}^{\otimes N_i}
        \big)
      \Bigg].
\end{equation}

A few remarks are in order about the usage of direct products and direct sums. If a system is composed of subsystems that are operationally distinct and are operationally accessible individually, then we use a direct product to compose the Hilbert spaces of the subsystems. Elementary examples include the direct product of the Hilbert spaces of two qubits. On the other hand, a direct sum composes the orthogonal sectors of the Hilbert space of a single system. The sectors are mutually exclusive. Elementary examples include the one-particle Hilbert space of a particle that could be in disjoint spatial regions, which is described by the direct sum of the Hilbert-space sectors corresponding to the two spatial modes.

And at time \(t\), given the fact that the accessible mode entanglement decays with distance, assign the pure state for each cluster:

\begin{equation}
|\Psi\rangle
  = |\psi_{C_1}\rangle \otimes |\psi_{C_2}\rangle
    \otimes \cdots \otimes |\psi_{C_k}\rangle .
\end{equation}

Now, with respect to the comoving observer \(A\), if the gas system as a whole is alignment measurable, then it is an actuality. In that case, we can invoke Postulate 1. And by Postulate 2, the gas counts as an actuality with respect to observer \(B\) as well, and so the subsystem partitioning made by observer \(A\) extends to observer \(B\). If one observer partitions a system into subsystems in a certain way, that partitioning stays fixed for all observers. Moreover, since the gas system counts as an actuality with respect to observer \(A\), observer \(B\) can also apply the actuality case of Postulate 1 to the gas system. This is why a Poisson process results, as we shall soon see in Eq. (6.10). But before we get there, we need to set up the machinery better.

Now observer \(B\) assigns an appropriately spacetime-transformed state for the same clusters in the gas. In our case, this would involve applying the Galilean transformation laws to the spatial and temporal coordinates while multiplying by the appropriate phase factors to ensure that the Schrodinger equation remains form-invariant \cite{landau1977qm}. The exact form of the transformation is not important for the current discussion, except for the fact that the number of subsystems remains the same and the identification of the subsystems stays the same between Eq. (6.8) and Eq. (6.9) . Here, by `identification of sub-system,' we mean that a subsystem that is labeled k by observer A is unambiguously subsystem k with respect to B as well.

\begin{equation}
|\Psi\rangle
  = |\psi_{C_1}\rangle_B \otimes |\psi_{C_2}\rangle_B
    \otimes \cdots \otimes |\psi_{C_k}\rangle_B .
\end{equation}

Now we have to look at the system's Hamiltonian and what it means operationally.

Classically, by the Hamiltonian of a system, we mean the Legendre transform of the Lagrangian of a closed system, which in many cases gives the total energy of the said closed system. In quantum theory, the Hamiltonian is the generator of time evolution, and a closed quantum system is one whose evolution is unitary under some Hamiltonian. Just as in classical mechanics, we can identify the time-dependent and time-independent Hamiltonian contributions from knowledge of the experimental context, such as the arrangement of magnets. In both the quantum and classical mechanical cases, the role of epistemic constraints is relevant to the specification of the Hamiltonian, but in classical mechanics, that role is, in some sense, a trivial part of the classical mechanical formalism itself, unlike in quantum theory. That's why a classical mechanics-based Laplacian worldview needs no new accommodation for epistemic constraints: it can be equated to the aggregate behavior of many classical mechanical particles.

The time-dependent Hamiltonian is typically specified by changing epistemically constrained contexts, such as a rotating magnet creating a time-dependent inhomogeneous magnetic field. As for the time-independent contributions to the Hamiltonian, they can arise from interparticle interaction terms from the model for the microscopic interaction, as prescribed by Assumption 8 in Chapter 4, and also from time-invariant experimental arrangements.

The specification of the Hamiltonian also brings with it a closer inspection of `closed system'. A closed system in a quantum experiment is not something arbitrarily determined by drawing an arbitrary border around a chosen volume and considering everything inside as closed, but rather whether a system is closed is primarily determined post facto by whether the dynamics is unitary---if it stays unitary, it is closed. So, unlike in classical physics, where, for example, all Hamiltonian contributions to two changed spheres arise only from the kinetic term and the interaction sources that are present within the enclosure containing the two spheres, a closed quantum system need not be associated with a single enclosing volume of space and everything inside it. In quantum theory, since the Hamiltonian is with respect to a closed system defined as that undergoing unitary dynamics, the contributions to the Hamiltonian need not be only from a connected enclosure, as in classical mechanics. This is because, when we have an entangled system, for example, we are allowed to have Hamiltonian contributions from spatially and temporally disconnected regions, such as one subsystem of an entangled pair undergoing a Pauli \(Z\) gate in one lab and another subsystem undergoing a Pauli \(X\) gate in another lab.

Since the closedness of a quantum system is determined post facto by whether the dynamics is unitary, there arises the question: how can we know whether the Hamiltonian that we wrote down is complete, in the sense of considering all the contributions?

This decision takes the form of the following iterative process: we write down a Hamiltonian based on what our experimental arrangement suggests, and we evolve unitarily given the reasoned guess of a Hamiltonian. If the probabilities and expectation values match the experiment, then our Hamiltonian was right. If not, we missed some contributions or subsystems; we go look for them, fix the Hamiltonian or Hilbert space or both, and then look again for evolved unitary dynamics. This process eventually converges on the right Hamiltonian when it is repeated iteratively. If a system does appear to evolve unitarily, consistently under repeated experiments, for a given guess at the Hamiltonian based on our knowledge of the experimental arrangement, then it is likely that we have identified a closed system or an effectively closed system.

With the observations about how the Hamiltonian\footnote{and thus the unitaries} is fixed and how the state is assigned, for the gas we considered, at any given time when an intervention is performed by inserting the probes, we found that the operationally meaningful unitaries and state assignments are only predominantly locally instantiable. Therefore, closed quantum systems are defined locally, for each cluster \(|\psi_{C_k}\rangle_B\), whose evolution is generated by a Hamiltonian \(\hat{H}_k\) fixed by the procedure we specified. We can denote the Hamiltonian for a cluster by \(\hat{H}_k\). The exact form of the Hamiltonian is not relevant for the present analysis; we only need to note that, for each cluster, \(\hat{H}_k\) is operationally meaningful locally, and should include a kinetic term, an interaction term for the interaction between the cluster and the external electromagnetic field \(V_{\mathrm{e.m.}}\), an interaction term describing the interaction between the clusters (we can assume nearest neighbour interaction), and an interaction term describing the interaction between the cluster and the external field \(V(\vec{r}_k)\) to which the gas as a whole might be subject (for instance, the gas as a whole might be under a gravitational influence). We expect the interaction term describing the interaction between the clusters to generate entanglement between neighbouring clusters, while \(V_{\mathrm{e.m.}}\) generates correlations with the electromagnetic field, which \(B\) intercepts to perform an alignment measurement of the drifting gas. We expect \(V(\vec{r}_k)\) to ``propel'' the cluster in accordance with the external potential acting on the entire gaseous system. This will become relevant when we study the mechanics of the whole system, which we undertake in Section 7.9.

Note that we have written down the Hamiltonian locally, for each cluster, and not for the entire gaseous system, because of Assumption 2, which emphasizes the role of epistemic constraints not just in the assignment of wavefunctions, but also in the assignment of unitaries, and thus Hamiltonians.


Now, given the experimental context we have considered, the drifting gas qualifies as an actuality. Therefore, the postulate of the probability of instantiability of the Born rule for the actuality case can be used by observer \(A\) for the drifting gas. And by the Postulate of objectivity of actuality vs. potentiality, the gas qualifies as an actuality also for observer \(B\), and \(B\) can invoke the same postulate for the drifting gas, i.e., for the actuality case. We expect that, in any given infinitesimal interval of time, each cluster in state \(|\psi_{C_k}\rangle_B\) undergoes a state reduction according to the Born rule \(Q\), as long as the probability of applicability of the Born state update \(P\) for the cluster is greater than 0. Defining the probability \(P\) in terms of a rate \(\lambda\) gives \(P=\lambda \Delta t\) for the infinitesimal time interval \(\Delta t\). So, for each infinitesimal interval, we can expect a Poisson process that yields

\begin{equation}
\Delta N(t) =
\begin{cases}
1, & \text{with probability } \lambda\Delta t,\\
0, & \text{with probability } 1 - \lambda\Delta t.
\end{cases}
\end{equation}

\(\Delta N(t)\) is the Poisson stochastic noise.

\(1-\lambda \Delta t\) is the probability that the cluster undergoes a unitary \(e^{-\frac{i\hat{H}_k\Delta t}{\hbar}}\).

At a given time \(t\), what characterizes the situation is not simply the state \(\bigotimes_k \ket{\psi _{C_k}}_B\), but what we can call the `clustered tensor product structure'. To understand that construction, we must look at tensor product structures. In the literature, a tensor product structure, TPS for short \cite{lloyd2004observable}, for a given Hilbert space \(\mathcal H\) is a collection of Hilbert spaces \(\{\mathcal H_k\}\) whose tensor product is identified with the original Hilbert space,
\[
\mathcal H \cong \bigotimes_k \mathcal H_k,
\]
(Here $\cong$ means isomorphism between the space on left hand side and right hand side)
and which specifies what states in \(\mathcal H\) are product states via a unitary identification
\[
\hat{U} : \bigotimes_k \mathcal H_k \to \mathcal H.
\]
We need this unitary mapping because we need to specify how vectors in \(\bigotimes_k \mathcal H_k\) correspond to states in \(\mathcal H\). This unitary identification is not a unitary evolution. That is, it is a unitary map, which is more general, of which unitary evolution and the unitary identification we saw are examples.

A tensor product structure by itself does not restrict the entanglement structure: states may still be entangled across the factors of the tensor product. That is, while a TPS does not constrain which states are entangled, it determines what counts as product and entangled states. For the given TPS, \(\bigotimes_k \mathcal H_k\), if a state can be written as \(\hat{U}(\bigotimes_k \ket{\Psi_k})\), then it is separable; otherwise, it is entangled with respect to that TPS. By the unitary mapping, the given TPS determines which states in the Hilbert space \(\mathcal H\) are product states and which are entangled. A different TPS may give a different determination of what states are entangled and what states are product states in the same Hilbert space \(\mathcal H\). Entanglement is relative to subsystem partitioning. But subsystem partitioning is not arbitrary; it is determined by operational access to subsystems \cite{lloyd2004observable}.

A second related way to characterize tensor product structure is in terms of what counts as the `subsystems' in a total system. This too is not arbitrary and is fixed by the operational access available in a given experimental arrangement.

A third related way to characterize a TPS is in terms of local operators \cite{Balachandran2013AlgebraicEntanglementEntropy}. There are individual subsystem local operators for the given TPS, which are of the form

\begin{equation}
  A_1\otimes  \mathbb{I}_2\otimes\mathbb{I}_3\otimes \mathbb{I}_4\otimes\mathbb{I}_5\ldots
\end{equation}
\begin{equation*}
   \mathbb{I}_1\otimes A_2\otimes \mathbb{I}_3\otimes \mathbb{I}_4\otimes\mathbb{I}_5\ldots
\end{equation*}
\begin{equation*}
   \mathbb{I}_1\otimes \mathbb{I}_2 \otimes A_3\otimes \mathbb{I}_4\otimes\mathbb{I}_5\ldots
\end{equation*}
\begin{equation*}
\vdots
\end{equation*}

\begin{equation*}
   \mathbb{I}_1\otimes \mathbb{I}_2 \otimes \mathbb{I}_3\otimes\mathbb{I}_4\ldots A_{k=N}
\end{equation*}

(Here, $\mathbb{I}_k$ is the identity operator acting on the $k^{th}$ sub-system)

Moreover, there is a commuting set of subalgebras \(\mathcal{L}_k\) \cite{Balachandran2013AlgebraicEntanglementEntropy}, such that

\begin{equation}
    [{A_k}, {A_l}]= 0 \quad \forall k\neq l,
\end{equation}

where \(A_k \in \mathcal{L}_k\) and \(A_l \in \mathcal{L}_l\).

The subalgebra is

\begin{equation}
\mathcal{L}_k=  \mathbb{I}_1\otimes \mathbb{I}_2 \otimes \ldots \mathcal{B}(\mathcal{H}_k) \otimes \mathbb{I}_{k+1} \otimes\mathbb{I}_{k+2}\ldots,
\end{equation}

with \(\mathcal{B}(\mathcal{H}_k)\) the subsystem subalgebra and \(\mathcal{L}_k\) its representation on the full space. \(\mathcal{B}(\mathcal{H}_k)\) is the space of bounded operators acting on the Hilbert space \(\mathcal{H}_k\), and it is sometimes denoted as the subalgebra \(\mathcal{A}_k\), as in \cite{lloyd2004observable}.

We can also have more general local operators applicable across subsystems, of the form

\begin{equation}
  A_1\otimes  A_2\otimes A_3\otimes A_4\otimes A_5\ldots
\end{equation}

Even more generally, after defining the set of commuting subalgebras of the form

\begin{equation}
    \mathcal{L}_k=  \{  \mathbb{I}_1\otimes \mathbb{I}_2 \otimes \ldots \mathcal{B}(\mathcal{H}_k) \otimes \mathbb{I}_{k+1} \otimes\mathbb{I}_{k+2}\ldots \}
\end{equation}

or, more precisely, as

\begin{equation*}
    \mathcal{L}_k=  \{  \mathbb{I}_1\otimes \mathbb{I}_2 \otimes \ldots {A}_k \otimes \mathbb{I}_{k+1} \otimes\mathbb{I}_{k+2}\ldots : {A}_k \in \mathcal{B}(\mathcal{H}_k) \},
\end{equation*}

we can, from the algebra \(\text{\textit{Alg}}(\mathcal{L}_1, \mathcal{L}_2, \mathcal{L}_3, \ldots) = \bigotimes_k \mathcal{B}(\mathcal{H}_k)\), which is closed under addition, scalar multiplication, and multiplication, get a class of operators in (6.11) and (6.14), but also operators that are in general entangling for the given TPS, such as

\begin{equation}
    \sum_i   A_1^{(i)}\otimes  A_2^{(i)}\otimes A_3^{(i)}\otimes A_4^{(i)}\otimes A_5^{(i)}\ldots
\end{equation}

or operators that are nonlocal only among a few subsystems, such as

\begin{equation}
   \Big( \sum_i   A_1^{(i)}\otimes  A_2^{(i)}\Big) \otimes A_3 \otimes A_4 \otimes A_5 \ldots,
\end{equation}

which are, in general, entangling only for subsystems 1 and 2.

At this juncture, I need to provide a criterion for what counts as changing the tensor product structure.

 I define the changing tensor product structure by the time-dependent identification \[
\hat{U_t} : \bigotimes_k \mathcal H_k (t) \to \mathcal H.
\]
That is, when the number of sub-systems, which is defined as the cardinality of the collection of distinguishable potentialities, each of which can be assigned a pure state, and the amount of accessible entanglement across all such sub-systems, evolve with time.

We will inherit all the above developments for the clustered tensor product structure described below.

The most general tensor product decomposition of a Hilbert space across the individual particles may look like
\[
\mathcal H = \bigotimes_{i=1}^N \mathcal H_i.
\]
Now we have a partition of the subsystem labels into clusters \(C_\alpha\), which is operationally decided by inserting pairs of counterfactual probes at any given time \(t\), as discussed earlier, such that the state can be written as
\[
|\psi\rangle
= \bigotimes_{\alpha} |\psi_{C_\alpha}\rangle,
\]
where each cluster state
\[
|\psi_{C_\alpha}\rangle \in \bigotimes_{i \in C_\alpha} \mathcal H_i
\]
may contain entanglement internally, but there is no entanglement between distinct clusters.

Given a fine-grained tensor product structure
\[
\mathcal H = \bigotimes_{i=1}^N \mathcal H_i,
\]
one may coarse-grain the decomposition by grouping indices into clusters \(C_\alpha\), defining cluster Hilbert spaces
\[
\mathcal H_{C_\alpha}
=
\bigotimes_{i \in C_\alpha} \mathcal H_i.
\]
The Hilbert space then admits the coarser decomposition
\[
\mathcal H = \bigotimes_\alpha \mathcal H_{C_\alpha}.
\]

We can define a cluster tensor product ($C\text{-}TPS$) structure by the identification \[
\hat{U} : \bigotimes_\alpha \mathcal H_{C_\alpha} \to \mathcal H.
\]
In a given time window \(\Delta t\), depending on which of the two possibilities specified by Eq.(6.10) happens, for each cluster, we have a different assignment of local operators, since one possibility involves unitary evolution of a cluster and the other possibility involves a state reduction of the entangled state describing a cluster.

It's easier to proceed from now on in an illustrative manner. But the general physical idea will be conveyed nevertheless.

In time window \(t+\Delta t\), let there be clusters for which the unitary possibility is realized, say for clusters labeled 1 to 5, with 1, 2, and 3 mutual neighbours, but 4 and 5 are not neighbours to 1, 2, and 3 or to each other. For these, the operators acting during \(t+\Delta t\) on the respective cluster states have the form given in \textbf{ }Eq.(6.17):

\begin{equation}
    \sum_i   ( U_{C_1}^{(i)}\otimes  U_{C_2}^{(i)}\otimes U_{C_3}^{(i)}) \otimes U_{C_4}\otimes U_{{C_5}} .
\end{equation}

Whereas for those clusters for which the state-reduction possibility is realized during \(t+\Delta t\), say for cluster 6, we will have in general measurement operators acting on the parts within a cluster, with the form given by Eq. (6.14):

\begin{equation}
  M_1^{\alpha}\otimes  M_2^{\beta} \otimes M_3^{\gamma} \otimes M_4^{\kappa} \otimes M_5^{\epsilon} \ldots \otimes M_k^{\zeta},
\end{equation}

where \(\{1, 2, 3, 4 \ldots\}\in C_6\) label the subsystems within cluster 6, and the superscripts represented by Greek letters denote the measurement outcomes that were realized for each of those subsystems.

These ideas can be generalized to any number of clusters with arbitrary mutual arrangement.

In the next short interval of time \(t+\Delta t+\Delta t\),  focusing on a cluster labeled \(k\), there is a probability \(1-\lambda \Delta t\) that the cluster undergoes a unitary \(e^{-\frac{i\hat{H}_k t}{\hbar}}\), and a probability \(\lambda \Delta t\) that the cluster undergoes a state reduction in the position basis according to Postulate 1. The probabilities \(\lambda \Delta t\) are identifiable with the probabilities for applicability of the Born rule that occur in the statement of Postulate 1.

With probability \(1-\lambda \Delta t\), we have the implementation of a unitary operation that can generate entanglement between cluster \(k\) and its neighbours because of the presence of interaction terms in the Hamiltonian \(H_k\). This effect amounts to an increment in the cluster size. And with probability \(\lambda \Delta t\), we have a Born-update operation that, in general, would reduce the entanglement within the cluster, and this effect amounts to a decrement in the cluster size. These two effects compete with each other, in the sense that the relative values of the probabilities \(\lambda \Delta t\) and \(1-\lambda \Delta t\) determine whether the merging of clusters dominates or the fragmentation of clusters dominates, in the entire system, as time progresses. Because the number of clusters at any given time, and the size of the clusters (characterizing how much entanglement is accessible within a cluster) change with time due to this process, the cluster tensor product structure $\bigotimes_\alpha \mathcal H_{C_\alpha} (t)$ changes with time.

Therefore, we expect the mathematical object characterizing the system to be
\[
\Xi(t) \coloneqq \{\bigotimes_{k} \ket{\psi_{C_k}}(t), C\text{-}TPS(t)\}.
\]
Note that the tensor product structure itself evolves at the coarse-grained level of clusters because what counts as the ``system'' (more precisely the sub-system) itself changes with time. The object $\Xi(t)$ encodes the evolution of states of each individual cluster, and of the tensor product structure. The states evolve, and the state space itself evolves.

There is one interesting observation that connects these results to the general theory of dynamical systems. In contrast to the Newtonian picture, where we assume physical systems come with pre-stated state spaces \cite{Kauffman2023ThirdTransition} and what evolves in time is simply the state of the totality of all systems, in the adjacent possible picture of Stuart Kauffman, originally motivated by the evolving biosphere, the state space itself evolves, in addition to the evolution of the state of the systems.

Note that here by Newtonian Picture, we do not necessarily mean Newtonian mechanics. Newtonian picture is a template for dynamical systems where we assume that systems exist with states (this could be classical, quantum, etc), and that the state space of all systems is pre-statable \cite{Kauffman2023ThirdTransition}.

Despite the replacement of Newtonian mechanics by quantum mechanics, we have largely retained this Newtonian picture of pre-stated state spaces, but for quantum systems. The developments in this thesis, and particularly the result in this section, suggest that this Newtonian picture, which we have largely retained in quantum mechanics, is replaced by the adjacent possible picture.

\chapter{ Origin of Mechanics and Some Applications in Statistical Physics}

Having shown that decoherence achieves a goal different from what is sometimes assumed \cite{Zurek2003Decoherence}, the goal in this chapter is to study the origin of classical mechanics by jointly considering quantum theory and epistemic constraints. 

The other goal in this chapter is to study the statistical physics of the heterogeneous systems consisting of potentialities and actualities. There are two kinds of epistemic: first, epistemic in the sense of ignorance about an underlying state, and second, epistemic in the sense of knowledge representation. The epistemic constraints apply to the latter kind, whereas probability theory applies to the former kind.
We shall come across these two kinds of epistemic playing a role in the formalism for the statistical physics of heterogeneous systems consisting of potentialities and actualities, unlike in ordinary statistical mechanics (of homogeneous systems of potentialities), where only one kind of epistemic plays a role: the ignorance kind. 

This chapter heuristically investigates these topics. These developments presently provide a programmatic direction and preliminary constructions rather than finalized derivations. The purpose is to motivate and outline the direction of research, but fully rigorous mathematical work is planned for future work.

\section{Statistical Mechanics and the Atomistic Conception of Matter}

Historically, it is well known that kinetic theory and later statistical mechanics were based on the atomistic conception of matter \cite{simonyi2012cultural}, and their subsequent empirical success is seen as justification of that conception.

As we already discussed, the atomistic conception of matter, the assumption that all matter is composed of tiny universal elementary constituents, originated with Democritus and Leucippus in ancient Greece, and was later made popular in Europe by Pierre Gassendi \cite{simonyi2012cultural} during the Enlightenment, which then influenced the early kinetic theories of matter.

From kinetic theory, statistical mechanics matured in the hands of James Clerk Maxwell and Ludwig Boltzmann \cite{simonyi2012cultural}. In particular, statistical mechanics developed at a time when the microscopic constituents were assumed to be subject to the same dynamical laws, namely Hamiltonian dynamics, and the same state variables, namely canonical position and momenta, as macroscopic objects. For \(N\) particles, the elementary constituents, the canonical position and momenta correspond to a point in \(6N\)-dimensional phase space, specified by a set of \(6N\) coordinates. Each point that is consistent with fixed macroscopic thermodynamic conditions is called a microstate. One then finds the probability distribution over the allowed microstates---this, for instance, could be obtained from the maximum entropy principle subject to normalization and expectation-value constraints---before interpreting and identifying the expectation values computed with respect to those probability distributions with the macroscopic thermodynamic quantities.

In this mathematical construction, developed through the works of Boltzmann, Maxwell, Gibbs, and later Jaynes \cite{MaxCal}, one can already see the mathematization of the once-philosophical idea of the atomistic conception of matter. This construction, which came to be called statistical mechanics, aimed to provide a bottom-up explanation for the phenomenological laws of thermodynamics.

When quantum mechanics was discovered, many of the insights from classical statistical mechanics provided guidance \cite{landau1980statistical} for constructing quantum statistical mechanics---there was no need to start from scratch. The first contribution in this direction arguably came from Satyendranath Bose, who derived the Planck formula from photon statistics. The next contribution came from Fermi and Dirac. In both cases, the single-particle eigenvalue problem was considered, and identical-particle occupation of the resulting eigenstates was assumed in order to find the relevant microstates and probability distributions over those microstates. This generalization also yields the familiar Boltzmann distribution in the limit where the particles become effectively distinguishable. In that limit, we have the eigenvalue problem for a single particle, and the resulting set of eigenstates is independently occupied by the \(N\) particles---each of which is subject to the same single-particle Hamiltonian---according to the Boltzmann distribution. These results were later formalized in the language of density matrices by von Neumann and Lev Landau \cite{landau1980statistical}.

Quantum statistical mechanics is a successful theoretical framework used to describe and predict condensed-matter phenomena. However, did the atomistic conception of matter that was foundational to classical statistical mechanics carry over to quantum statistical mechanics? That is, did the advent of quantum statistical mechanics allow for the retention of the ontology wherein we can suppose the atomistic conception of matter? As a reminder, to properly assess the situation, let us state again the atomistic conception of matter: ``matter is made of independently existing---that is, persisting in time as distinct entities---microscopic constituents, the microscopic constituents have a state, and the aggregate behavior of the microscopic constituents can be identified with the macroscopic behavior of matter.'' It becomes clear that this condition only partially holds when we look at how statistical mechanics is practiced. This was also already noticed by Drossel \cite{Drossel2017TenReasons}: the determination of the Hamiltonians, tensor-product structures, and measurement outcomes always takes on a top-down form, wherein the Epistemically constrained situations of the apparatus plays a role. This is also seen in the way we model electrons in a metal; the lattice structure determining the Bloch potential, for example, functions as the Epistemically constrained situations \cite{Drossel2017TenReasons} for the free electrons in the metal. Notice that this issue does not arise in classical statistical mechanics, wherein one can fully accommodate the atomistic conception of matter in its formulation. This is because, in the classical mechanical picture, we can assume the microscopic constituents to possess definite position and momentum at all times, and any top-down influence on those microscopic constituents can be viewed as merely an effective description, which can always be decomposed into interactions between the microscopic constituents with definite position and momentum. That is not possible in the quantum mechanical situation due to the unavoidability of epistemic constraints in providing the preconditions for quantum state description.

This is because the state vector has meaning only with respect to the experimental arrangement. In particular, there is a fundamental need for the role of the experimental arrangement, specified in plain language, in determining the quantum state of a particle. Surely, if the nature of quantum states has such a physical meaning, as specified in Chapter 4, involving the need for language, then that physical meaning has to be the same for all quantum mechanical applications, even for the application of quantum mechanics to account for thermodynamics, not just for individual quantum state preparation of a single particle, as in the Stern--Gerlach experiment.

Now, there is the well-known question \cite{popescu2006entanglement} in the foundations of statistical mechanics that seeks to explain the existence of equilibrium ensembles and the second law of thermodynamics, more colloquially, to explain the arrow of time, from more elementary considerations. Classically, this meant explaining how those conditions emerge from Newton's laws governing the microscopic constituents, but quantum mechanically, the same questions have been reinterpreted as seeking how those thermodynamic conditions emerge from quantum mechanics. We further refine this question in this work. Instead of asking how the canonical ensemble and the second law emerge from unitary dynamics alone---as is usually done in the literature \cite{popescu2009quantum}---we ask, based on the model developed in the previous chapters, how those thermodynamic conditions emerge from the interplay of unitary dynamics and Born state updates together, which we saw arises when we consider unitary quantum mechanics and Bohr's epistemic constraints as both primitive.

Another motivation for such a reformulation of the problem of explaining the equilibration of thermodynamic systems and the second law of thermodynamics is the desire to rederive thermodynamics without a metaphysics that is underdetermined by empirical evidence. An intelligent extraterrestrial lifeform arguably may never have shared our Democritean atomistic conception of matter, possibly due to its differing history of ideas, and yet it is arguable that it may have discovered statistical physics, given that statistical physics is a well-tested physical theory. They may have likely arrived at statistical mechanics without the atomistic conception of matter as we know it.

They may not have this intuitive metaphysics that we possess: ``that the world is made of tiny elementary constituents that exist and interact with each other, and all of reality arises from such elementary constituents.'' This intuition involves an excess metaphysical assumption that is not fully determined by current evidence. This is mainly because the elementary constituents we observe as changes in a classical experimental context always entail a reverse supervenience, in which the changes to those elementary constituents are supervenient on the classical experimental arrangement. As seen in Chapter 4, for an intelligent extraterrestrial, an atom may simply be the recognizable spectral pattern, along with the corresponding transition rules and intensities for the allowed transitions. Those are what we can conclude with certitude, given the assumptions concerning the universal applicability of mathematics and the universal character of the language of experimentation.

A ``universal method'' of reconstructing statistical mechanics must therefore be neutral about that part of our local history of ideas which is underdetermined by empirical evidence. For such a universal method, it seems we must rely only on the supposed universal validity of mathematics and the supposed universal principles concerning the preconditions of empirical knowledge, namely the two principles we discussed earlier in the thesis.

By the assumption that all possible knowledge-acquiring systems will exhibit certain universal characteristics, we mean the following: any laboratory maneuver that a human performs for a specific purpose, such as the polarization measurement of a photon, will be unambiguously recognizable to a hypothetical extraterrestrial intelligence, and similarly, the laboratory maneuver performed by a hypothetical extraterrestrial intelligence must be unambiguously recognizable to a human physicist on Earth.

The laboratory manual of the human physicist and that of the hypothetical extraterrestrial physicist will use a universal language, not only in the equations, i.e., the universality of the laws of nature, but also in the methods of empirical access. From that premise, we can study what the description of nature would be like for any evolved intelligence, anywhere in the universe, using only the minimal and sufficient metaphysics suggested by our empirical evidence.

Arguably, the conceptual development of our models is determined not only by experimental facts, but also by the unique social and cultural history of our species on Earth. The metaphysics that we attach to empirical evidence must therefore be minimal and reasonably acceptable to any intelligence across the universe. Note that we are not supporting a strict instrumentalist stance either: we do allow for electrons and other quantum systems as potentialities, as operationally defined in Chapter 2 and inspired by the original ideas of Heisenberg, but not as particles or fields existing as things in spacetime. With that qualification, explaining semiconductors or electron microscopes is not an issue, while at the same time, this remains at odds with the strong ontological reductionism of atomism, which supposes localized, independently existing constituents.

Therefore, it can be argued that we should eliminate aspects of our models that are not strictly dictated by empirical evidence. Anything less would make our description incomplete, and anything more would make our description decorative and distract from the essence that carves nature at her joints. The goal in the present chapter is to study how statistical mechanics could be reformulated with such a minimal metaphysics: namely, how statistical mechanics can be reformulated in such a way that it is neutral about the unique history of our civilization on Earth, and resembles a formulation that even an extraterrestrial could have arrived at.
\section{What's Epistemic about the Ensembles in Quantum Statistical Mechanics?}
In ensemble theory \cite{landau1980statistical} in statistical mechanics, we have epistemic probabilities in addition to the intrinsic quantum probabilities due to the Born rule. Usually, when we talk about epistemic probabilities, we mean a lack of knowledge about something: in classical statistical mechanics, this can mean a lack of knowledge about the positions and momenta of particles. In quantum mechanics, a straightforward extension is to say that the lack of knowledge concerns the quantum state.

But here is where the problem arises: unlike classical states, quantum states, at least in Bohr's reading, correspond to experimental arrangements. So it would seem that the lack of knowledge is about the experimental arrangement or preparation. While that is certainly true for a single quantum particle's state preparation, which is why we have density matrices, would that interpretation extend even to statistical mechanics? For a harmonious interpretation of `epistemic' in individual quantum systems versus an assembly of quantum systems, it appears that the `epistemic' should be taken as concerning the lack of knowledge about the preparation or Epistemically constrained situations in either case. This is in contrast to the prevailing view \cite{popescu2009quantum} in quantum statistical mechanics, which is arguably more Everettian in its foundations, where there are just quantum states `out there,' in a realist sense for particles or fields, without any reference to experimental arrangement, and the lack of knowledge is about those states without reference to the epistemic constraints that apply to the experimental arrangements.

Even for improper mixtures, such as in subsystems of entangled systems, when expressing a reduced state of subsystems, the `lack of knowledge' is arguably the local lack of knowledge about whether the other party made a measurement or not, i.e., about their experimental arrangement. If there were communication between the parties, allowing access to all subsystems, we would know that we can observe the purity of the global state, and that an improper mixture would become a local lack-of-knowledge artifact. The lack of a local pure state in such situations only reflects the fact that the totality of experimental arrangements needed to specify a quantum pure state need not have parts that are localized within a lightlike or timelike separation of each other.

This renewed emphasis on the experimental arrangement, even in the case of improper mixtures, may seem like a problem only if we tacitly picture quantum particles as `possessing' a state, like a label attached to them, which, as we discussed earlier, is not the right picture. Locally, yes, it may look like we have a mixture, with no apparent preparation procedure about which that mixture represents a lack of knowledge, seemingly contradicting the main point of the Bohrian view that we have been defending, namely, that lack of knowledge in density matrices concerns preparation procedures that are subject to the epistemic constraints. However, upon closer inspection, there is no contradiction, because, as emphasized earlier, in certain cases such as entangled states, the complete information about the experimental arrangement can only be specified globally and not locally---and there is no classical analog for this. The mixedness in such cases comes from incomplete access. Giving access to only a subsystem of an entangled system, even if one knows the full preparation locally, is still `incomplete knowledge' in the sense that we do not have a full description of the system locally.

In contrast to the ensemble theory that we intend to construct for statistical mechanics, based on this renewed emphasis on what `lack of knowledge' means, one can note that there is an element of classical-like metaphysics present in the usual way the ensemble theory underlying statistical mechanics is constructed. This is because it inherits certain features that were already present in classical ensemble theory in the nineteenth century, by supposing that states `exist' as features of microscopic constituents and that we are simply unaware of such states. This is the tacit belief that `tiny particles exist with a state that we do not know.' The point of view that we are supposing is more along the following lines: `counterfactual experimental probing of operationally accessible subsystems can be considered for a bulk system; such counterfactual probing reveals the distinction between potentialities and actualities, and the actualities provide the epistemic constraints for the potentialities; lack of knowledge about such preparations gives the epistemic contributions to quantum statistical mechanics.'

Notice that there have been two different senses in which we have been using, and will continue to use, the word ``epistemic'' in this chapter: one concerns the epistemic in the sense of lack of knowledge, and the other concerns the epistemic in the sense of knowledge representation and communication. It is with respect to the latter that our previously introduced principles of epistemic constraints become relevant. And the statistical physics of heterogeneous systems of potentialities and actualities will involve both kinds of epistemics.

\section{Work and Heat for the Elementary A-P-A unit}
A natural place to start such a reconstruction is to reconsider the nature of heat and work. Heat and work are the central, phenomenologically accessible quantities that one connects to more elementary descriptions, with the aim of reproducing, from the latter, the phenomenologically validated laws of thermodynamics concerning the constraints on work and heat transfer.

In the usual construction \cite{campbell2026roadmap}, whether in classical or quantum statistical mechanics, heat and work are not defined for the constituent particles, and the dynamics is fully reversible at the level of the constituent particles. Meanwhile, the aggregate exhibits phenomena involving heat, work, and thermodynamic irreversibility \cite{popescu2009quantum}. The goal in the usual construction is twofold: first, to derive microscopic descriptions of an aggregate of elementary constituents that is posited as being equivalent to these phenomena, even though the elementary constituents do not exhibit such phenomena individually \cite{landau1980statistical}\cite{popescu2009quantum}. A classical mechanical version of this step is the supposition that `heat is no more than the molecular kinetic energies of many molecules.' And second, to show that the laws of thermodynamics follow for such an equivalent microscopic description of the aggregate \cite{popescu2009quantum}.

(In what follows, we have put ``elementary'' in quotes because we shall define what ``elementary'' means shortly.)

The aim of this section is to show that work and heat are defined even for the most ``elementary'' quantum systems, and that a kind of irreversibility associated with the irreversibility of completed quantum measurements already applies at the level of the most elementary constituents. This we may call `Bohrian irreversibility,' which we take to be an in-principle irreversibility that is said to have happened once the epistemically constrained situations are amenable to alignment position measurement with respect to an actual or a counterfactual observer, resulting in the extraction of information about the outcome.

However, what does not in general apply at the level of the most ``elementary'' quantum systems is thermal behavior in the sense of the Gibbs distribution and thermodynamic irreversibility. Here, one may recall that thermodynamic irreversibility is not the same as what we have called ``Bohrian irreversibility,'' with the former applicable to processes that do not proceed through a continuous sequence of equilibrium states and do generate thermodynamic entropy \cite{landau1980statistical}.

Therefore, the goal of our reconstruction will be to derive thermal behavior and thermodynamic irreversibility for a sufficiently complicated system, given the work and heat associated with quantum measurements on elementary quantum systems and the Bohrian irreversibility associated with elementary systems.

To see the unavoidable role of what we have called `work and heat' associated with quantum measurements, even for pure quantum systems, consider the following setup: 

Adopting the usual quantum mechanical definition of heat and work \cite{deutsch1991quantum}\cite{campbell2026roadmap}\cite{potts2019introduction}, which are defined in terms of the density matrix, when we take the differential of average energy, which is \(\mathrm{Tr}(\rho H)\),

\begin{equation}
d\,\mathrm{Tr}(\rho H)
=
\mathrm{Tr}(d\rho\, H)
+
\mathrm{Tr}(\rho\, dH)
\end{equation}

The heat transfer is identified as \begin{equation}
\delta Q \equiv \mathrm{Tr}(d\rho\, H),
\end{equation} 
and the worktransfer as \[
\delta W \equiv \mathrm{Tr}(\rho\, dH),
\]
and the internal energy with \begin{equation}
U = \mathrm{Tr}(\rho H),
\end{equation}
So Eq.(7.1) expresses the first law of thermodynamics. However, with this machinery alone, if we were to look at a system prepared in a pure quantum state and later updated in a projective measurement, after getting an eigenvalue corresponding to an eigenstate as the outcome,  if the measurement is non-destructive, we get \begin{equation}
d\rho = -\frac{i}{\hbar}[H,\rho]\,dt,
\end{equation}
which is the von Neumann equation. It's easy to see that for such a dynamics, the heat is zero. Because, writing 
\begin{equation}
\dot{ Q} = \mathrm{Tr}(\dot{\rho} \, H),
\end{equation} 
and then using the Von Neumann equation, 
\begin{equation}
\dot{Q}
=
\mathrm{Tr}(H\dot{\rho})
=
-\frac{i}{\hbar}\mathrm{Tr}\!\left(H[H,\rho]\right),
\end{equation}
we notice that this commutator 
\begin{equation}
\mathrm{Tr}\!\left(H[H,\rho]\right)
=
\mathrm{Tr}(HH\rho - H\rho H),
\end{equation}
vanishes, because of the cyclic property of trace
\begin{equation}
\mathrm{Tr}(H\rho H)
=
\mathrm{Tr}(HH\rho).
\end{equation}

And so, \begin{equation}
\dot{Q}=0.
\end{equation}

However, the elementary meaningful unit of a quantum system, as we have shown based on operational arguments in Chapter 3, is a system of the type A-P-A. And for such elementary systems, since actualities are defined in terms of alignment position measurability by passive optical means, they function as heat reservoirs. And in a single instance of an A-P-A process, we seem to have stochastic heat exchange with the reservoir at the start and the end of the process. This is the heat that we can measure with a suitable calorimeter, and we can call this the stochastic heat \(\tilde{Q}\), and is stochastic because it is the eigenvalue problem describing the A-P-A process whose outcomes are intrinsically probabilistic with probabilities determined by the Born rule and the probability of instantiability of Born rule. And the stochastic heat is distinct from the heat we see in Eq.(7.2). In the finite case the latter is given by \(Q= \langle Q \rangle
=
\int \mathrm{Tr}(H\,d\rho)\), for a fixed Hamiltonian . What we measure in experiments after suitable repetitions of identically prepared experiments is the probability distribution \(P(\tilde{Q})\) and thence deduce the average heat Q which can be found to match the predicted heat. That is, we will find: 

\begin{equation}
\langle Q \rangle
=
\int \tilde{Q}\,P(\tilde{Q})\,d\tilde{Q}
\end{equation}

In addition to heat being definable for individual systems, work too can be defined for individual systems when the Hamiltonian of the system changes; however, work can be zero for a single unit of A-P-A, since it's in general possible to have \(\delta W \equiv \mathrm{Tr}(\rho\, dH)\) be zero when the Hamiltonian is fixed. Nevertheless, at least heat is a meaningful quantity measurable by a calorimeter for the elementary quantum system defined as an A-P-A.

\section{Re-Interpreting Canonical Typicality and ETH as Thermalization of Potentialities} 

There already exist two well-developed frameworks aiming to address the exact same foundational question that we discussed in the previous sections: how do systems thermalize, and how does the second law of thermodynamics follow from a more elementary description in terms of dynamical laws? These are the frameworks of canonical typicality \cite{goldstein2006canonical} and the eigenstate thermalization hypothesis (ETH) \cite{ETH2016}. But we wish to show in this section that what the two frameworks achieve is different from what they hope to accomplish.

Broadly, the ETH is a statement about certain closed many-body quantum systems \cite{deutsch1991quantum}, whereas canonical typicality \cite{goldstein2006canonical} is based on a style of reasoning similar to that of the decoherence framework.

There is this inter-theoretic relationship between thermodynamics and quantum mechanics, on the one hand, and between classical mechanics and quantum mechanics, on the other hand: just as certain quantum states, such as coherent states, can behave in a classical-mechanical-like manner, while other, more general states look nothing like classical mechanical evolution, certain classes of quantum Hamiltonians lead to dynamics that are locally thermodynamic-like, while more general many-body Hamiltonians do not lead to local dynamics that are anything like thermodynamics \cite{ETH2016}.

This intra-theoretic relationship appears not to be merely coincidental. Just as we reinterpreted the decoherence framework as supplying a mechanism for the classical-physics limit of potentialities, but not as addressing the origin of classical mechanics, in this section we reinterpret decoherence-based canonical typicality and the ETH as concerning the effective thermalization and entropy increase of homogeneous systems consisting of potentialities. The thermalization and the second law applicable to heterogeneous systems consisting of potentialities and actualities are addressed by neither canonical typicality nor the ETH, and for that we develop a suitable statistical mechanics in Sections 7.5 and 7.6.

 But before we do so, we need to elucidate quantum statistical mechanics, and the know solution to the emergence of thermodynamic limit for potentialities.

In the canonical ensemble 
 density matrix of a quantum system in a Hamiltonian $H$ is given by \cite{landau1980statistical}

\begin{equation}
    \hat{\rho} = \frac{e^{-\beta \hat{H}}}{Z},
\end{equation}

where $\beta$ is defined in terms of the temperature\textbf{ }

\begin{equation}
    \beta = \frac{1}{k_B T}.
\end{equation}

The partition function for the system is 

\begin{equation}
    Z = \mathrm{Tr}\!\left(e^{-\beta \hat{H}}\right),
\end{equation}

for the given eigenvalue problem

\begin{equation}
    \hat{H}\,|n\rangle = E_n\,|n\rangle.
\end{equation}

The density matrix of the system can be written as

\begin{equation}
    \hat{\rho}
= \sum_n p_n\,|n\rangle\langle n|,
\end{equation}

Where the probabilities are given by the Boltzmann distribution
\begin{equation}
p_n = \frac{e^{-\beta E_n}}{Z}.
\end{equation}

Average macro observables can then be found by taking the averages
\begin{equation}
\langle \hat{A} \rangle
= \mathrm{Tr}(\hat{\rho}\,\hat{A}).
\end{equation}

In particular, the internal energy of the system is

\begin{equation}
U = \langle \hat{H} \rangle
= -\frac{\partial}{\partial \beta}\ln Z.
\end{equation}

We can also find the free energy of the system from the partition function

\begin{equation}
F = -k_B T \ln Z.
\end{equation}

Similarly, the entropy can also be found once we know the partition function

\begin{equation}
S = -k_B\,\mathrm{Tr}(\hat{\rho}\ln\hat{\rho})
= k_B\left(\ln Z + \beta U\right).
\end{equation}

Once we know the internal energy, we can find the heat capacity

\begin{equation}
C = \frac{\partial U}{\partial T}.
\end{equation}

The canonical density matrix commutes with the Hamiltonian
\begin{equation}
[\hat{\rho},\hat{H}] = 0.
\end{equation}

So far, this is simply statistical mechanics, and it works in explaining many condensed matter phenomena. However, the foundational question concerns how the Boltzmann distribution over energy eigenstates emerges from purely unitary dynamics (or other more elementary principles). 

Closed quantum systems undergoing unitary evolution do not thermalize: there is no convergence to the Boltzmann distribution over energy eigenstates under unitary evolution. Canonical typicality \cite{goldstein2006canonical} aims to address this foundational question by pointing out that subsystems of a larger system undergoing unitary evolution can indeed, under general conditions, have reduced-density-matrix descriptions that become stationary and thermal. This criterion will now be made more precise.


To see how the usual mechanism takes us from unitary dynamics to the thermal distribution, let us first look at why we cannot have a thermal distribution and a temperature associated with a single particle. Consider a particle in a general state

\begin{equation}
    |\psi\rangle = \sum_i c_i \, |e_i\rangle.
\end{equation}

This state clearly has zero entropy because it's a pure state. And there is no meaning to have a temperature associated with it either. However if we now measure an ensemble of preparations of this particle each in state $|\psi\rangle$, and measure the energy basis on each of the members of the ensemble, 
we get outcomes consistent with the Born rule

\begin{equation}
   P_i = |c_i|^2.
\end{equation}
And the post-measurement density matrix for the ensemble will be

\begin{equation}
  \rho_{\text{post}} = \sum_i P_i \, |e_i\rangle\langle e_i|.
\end{equation}
To this state we can associate the Gibbs entropy
\begin{equation}
    S_{\text{Gibbs}} = -k_B \sum_i P_i \ln P_i.
\end{equation}
Or equivalently, the von-Neumann entropy
\begin{equation}
    S_{\text{vN}}(\rho_{\text{post}}) = -k_B \operatorname{Tr}\!\left( \rho_{\text{post}} \ln \rho_{\text{post}} \right).
\end{equation}

However, we still cannot associate a temperature with this entropy because, the probability distributions are in general not of the Thermal form; that is $P_i \propto e^{-\beta e_i}
$ is not the case, and the density matrix is not of the form $\rho_{\text{post}} \approx \frac{1}{Z} e^{-\beta H}
$ for some Hamiltonian $H$.

In the usual method \cite{goldstein2006canonical}\cite{ETH2016}, to meet the criterion for the successful derivation of the thermodynamic limit, for potentialities in our reinterpretation, we assume two systems, \(A\) and \(B\):
\begin{equation}
    S = A \cup B.
\end{equation}

We take the joint system to be in a pure state,
\begin{equation}
    |\Psi\rangle_{AB}.
\end{equation}
Now, for system \(A\) alone, all measurement outcomes can be predicted accurately from the reduced state of subsystem \(A\):
\begin{equation}
    \rho_A = \operatorname{Tr}_B \, |\Psi\rangle\langle\Psi|.
\end{equation}
It is then possible to associate the entanglement entropy with subsystem \(A\), which will become relevant when defining the temperature:
\begin{equation}
    S_A = -k_B \operatorname{Tr}\!\left( \rho_A \ln \rho_A \right).
\end{equation}
When system \(B\) is large compared to \(A\), that is, when
\begin{equation}
    \dim \mathcal{H}_B \gg \dim \mathcal{H}_A,
\end{equation}
it has previously been shown by Popescu and others \cite{popescu2006entanglement}\cite{popescu2009quantum} that the reduced state becomes canonical:
\begin{equation}
    \rho_A \approx \operatorname{Tr}_B \left( \rho_{\text{microcanonical}} \right)
= \frac{1}{Z} e^{-\beta H_A}.
\end{equation}

The entanglement entropy of \(A\) can then be taken to describe the thermodynamic entropy:
\begin{equation}
    S_A \approx S_{\text{th}}(E_A).
\end{equation}
Hence, the temperature can be found from the thermodynamic formula
\begin{equation}
    \frac{1}{T} = \frac{\partial S_A}{\partial E_A}.
\end{equation}

There are two closely related but independent results that we need. One is kinematical, and the other is dynamical.

Consider many-body quantum systems with non-integrable Hamiltonians \cite{ETH2016} (integrable systems have the same number of independent conserved quantities as the number of degrees of freedom; non-integrable Hamiltonians do not) evolving unitarily. For such systems, the eigenstates \(|E_n\rangle\) of the Hamiltonian are such that, for few-body local observables \(A\), the matrix elements in the energy eigenbasis take the form \cite{ETH2016}\cite{ETHreview} (this particular form of ETH is from the review paper by Mohsen Alishahiha and Mohammad Javad Vasli \cite{ETHreview}):
\begin{equation}
\langle E_m | A | E_n \rangle
=
A(\bar E)\,\delta_{mn}
+
e^{-S(\bar E)/2}\, f_A(\bar E,\omega)\, R_{mn},
\end{equation}
where \(\bar E = \tfrac{E_m + E_n}{2}\), \(\omega = E_m - E_n\), \(S(\bar E)\) is the thermodynamic entropy at energy \(\bar E\), \(f_A\) is a smooth function, and \(R_{mn}\) is a random variable with zero mean and unit variance.

In the thermodynamic limit \(N \to \infty\), the diagonal matrix elements become smooth functions of \(E_n\) and satisfy \cite{ETH2016}
\begin{equation}
\langle E_n | A | E_n \rangle
=
A(E_n)
\;\longrightarrow\;
\langle A \rangle_{\mathrm{microcanonical}}(E_n),
\end{equation}
while the off-diagonal fluctuation term is exponentially suppressed in system size, since \(S(E) \sim \mathcal{O}(N)\) \cite{deutsch1991quantum}. This is the eigenstate thermalization hypothesis \cite{deutsch1991quantum}. It concerns the dynamics, since it is a hypothesis about the solution to the\textbf{ }Schrodinger equation for many-body quantum systems. Eq.(7.37) is not an analytic solution; rather, it is the ansatz that can be plugged into the matrix elements that appear when computing the expectation values of the operator \(A\).

For the same problem of explaining thermalization under unitary evolution, canonical typicality \cite{goldstein2006canonical} provides a kinematical constraint. Namely, for a bipartite system \(\mathcal{H} = \mathcal{H}_A \otimes \mathcal{H}_B\), if subsystem \(B\) is much larger than subsystem \(A\), then, ``for almost all pure'' \cite{goldstein2006canonical} states in a narrow energy shell, the reduced density matrix of \(A\),
\begin{equation}
\rho_A = \mathrm{Tr}_B \, |\psi\rangle\langle\psi|,
\end{equation}
is approximately given by the canonical Gibbs state
\begin{equation}
\rho_A \approx \frac{e^{-\beta H_A}}{Z}.
\end{equation}

Thus, ETH addresses the dynamical mechanism of thermalization for non-integrable Hamiltonians, whereas canonical typicality addresses the kinematics of typical states in high-dimensional Hilbert spaces for more general systems.

In canonical typicality, when focusing on a subsystem of a larger system that is typical and in an energy shell, as defined by Popescu and others \cite{popescu2009quantum}, the subsystem becomes canonical. In ETH, on the other hand, for the subsystem of a larger non-integrable system, the expectation values of a local observable with respect to the full system's energy eigenstates become microcanonical in the large-\(N\) limit.

Consider, for example, a quantum system of 100 particles, described by a non-integrable Hamiltonian. If we focus on a few particles, or perhaps even one particle, we may observe predictions consistent with ETH; but in this case, we only have a 100-particle quantum system obeying the Schrodinger equation.

So, to match thermodynamics more closely, we can imagine something like the following: suppose we have a very large number \(N\) of particles described by some non-integrable Hamiltonian. If we have a subsystem consisting of, say, a fraction of those \(N\) particles, with \(M\) particles such that \(M \ll N\), but where \(M\) still contains nearly a thousand or a hundred thousand particles, then making coarse-grained energy measurements of these \(M\) particles can be expected to match the ETH prediction and also match what we operationally mean by the quantum thermodynamics of \(M\) particles \cite{campbell2026roadmap}.

However, note that both ETH and canonical typicality address physical situations that can be described by Class 2 dynamics, as discussed in Chapter 3. Both assume a global pure state and a subsystem that evolves non-unitarily. Therefore, in view of the discussions of Chapter 5 concerning the considerations of ``complex'' quantum systems, both ETH and canonical typicality are concerned with homogeneous complex systems of potentialities, in the sense of Chapter 5.

This naturally raises a question: What is the statistical mechanics of heterogeneous complex systems consisting of potentialities and actualities? Such heterogeneous consideration is needed for systems that qualify as actualities, as we have shown in Chapter 5. This is the statistical mechanics whose heuristic construction we have been pursuing in this chapter. We can call this \emph{heterogeneous quantum statistical mechanics}. And it is this statistical mechanics that departs from the atomistic conception of matter. It no longer supposes that there exist elementary constituents in some state without reference to an epistemically constrained situation, ECM, and ECU. The statistical mechanics of potentialities, which is the object of study of ETH and canonical typicality---which we can call \emph{homogeneous quantum statistical mechanics}---is also not in concordance with the atomistic conception of matter, since meaningful consideration of homogeneous complex systems of potentialities, including subsystem partitioning, specification of Hamiltonians, and determination of states, involves epistemically constrained situations too. Both heterogeneous and homogeneous quantum statistical mechanics are unlike classical statistical mechanics in this regard: the latter supposes the atomistic conception of matter, in that it supposes that there exist elementary entities that exist in some state without reference to epistemically constrained situations. The states in classical statistical mechanics are like time-dependent labels of those elementary entities, which exist as temporally persisting and individuated properties of temporally existing and individuated elementary entities.

To this end, we must first clarify what the members of the ensemble are.

Before we move on with that goal, we must say a word of caution to avoid a possibly misleading conclusion that may have been drawn from the terminology ``heterogeneous statistical mechanics'' and ``homogeneous statistical mechanics''. Homogeneous and heterogeneous here do not refer to different particle species of different chemical elements, each with a chemical potential of its own. Rather, homogeneous and heterogeneous here refer to the involvement of potentialities alone, or to the involvement of both potentialities and actualities, in the time evolution of the elementary constituents of the statistical mechanical system. Irrespective of the chemical species involved, potentialities and actualities are two distinct operational categories that we discussed in Chapter 2.


\section{Defining the Members of the Ensemble: From Microstates to Micro-processes}

Microstates are the members of the ensemble in the usual formulation of statistical mechanics \cite{landau1980statistical}. Standard statistical mechanics identifies microstates as instantaneous system states, such as phase-space points or quantum Hilbert-space vectors. Macrostates result from coarse-graining these states, that is, from looking at ensemble averages over many such possible microstates consistent with a given macrostate. The epistemic quality of being `unknown' applies to the microstates, and lack-of-knowledge probability distributions are defined over the microstates.

In contrast, the present framework takes what one may call completed experimental ``micro-processes'' to possess the epistemic quality of being ``unknown'', thereby constituting members of an ensemble to which we can assign a lack-of-knowledge probability distribution. Each process consists of a preparation, a unitary evolution, and a measurement outcome of a specific quantum problem, which we define later in the section---for now, take it to mean the complete specification of the experimental setup specifying the state and any intermediate operation. Micro-processes are thus defined as these elementary {`}`quantum problems.'' We have termed them ``micro-processes'', as opposed to simply processes, since we are referring to the quantum problems that can be instantiated within a given macroscopic system by considering counterfactual probes inside the said system, in a manner similar to the method we studied in the previous chapter when writing down the tensor-product structure of the gaseous system by considering pairwise counterfactual probes inside the gas.

Macrostates of the system at any two points in time, \(t_1\) and \(t_2\), counted by a clock positioned close to the macroscopic system, are found by collecting all such micro-processes, which are completed quantum problems between time \(t_1\) and \(t_2\), and taking the ensemble average over all those processes. That is, only those actualities at time \(t_2\) are to be counted that complete the quantum problem specified by the actualities at time \(t_1\). The potentialities at time \(t_1\), the potentialities at time \(t_2\), and the actualities at time \(t_2\) that do not complete the quantum problems specified by the actualities at time \(t_1\) are not to be counted.\\

\begin{table}
     \centering

     \begin{tabular}{ccc}
     & \(t_1\) & \(t_2\)\\
     \hline 
          Quantum Problem 1 & \(A_1\) & \(A'_1\)\\
          Quantum Problem 2 & \(A_2\) & \(A'_2\)\\
          Quantum Problem 3 & \(A_3\) & \(A'_3\)\\
          Quantum Problem 4 & \(A_4\) & \(A'_4\)\\
          \(\vdots\) & \(\vdots\) & \(\vdots\)\\
          Quantum Problem \(n\) & \(A_n\) & \(A'_n\)\\
     \end{tabular}
     \caption{Ensemble of quantum problems specifying micro-processes between two times}
     \label{tab:placeholder}
\end{table}

A given quantum problem starting in \(A_m\) and ending in \(A'_m\) in the collection may be of the \((A_m \!-\!P\!-\!A)\!-\!P\!-\!(A\!-\!P\!-\!A)\!-\!P\!-\!(A\!-\!P\!-\!A)\!-\!P \ldots -\!A'_m\) kind, or of the \((A_m \!-\!P\!-\!A'_m)\) kind. The lack of knowledge, and the associated lack-of-knowledge probabilities, will enter in three respects. First, we may in general lack knowledge of how many such quantum problems, and thus micro-processes, there are within the macroscopic system that can be probed by counterfactual probes. Second, we may lack knowledge of the exact form of the quantum description,
\[
    \psi(t_1) \rightarrow \hat{M}_k \psi(t_2),
\]
that describes each of the micro-processes, including whether it was obtained by one uninterrupted unitary evolution or was interlaced with weak measurements. Third, assuming that the measurement operators \(M_k\) commute with the Hamiltonian of the micro-process, we may lack knowledge of the energy eigenvalue obtained at \(t_2\) in a micro-process. The commutation here is assumed so that the measurement at time \(t_2\) can be interpreted as energy measurements.

To study the connection to macrostates, let us look at a simpler case where we assume that each of the micro-processes is of the form \((A_m \!-\!P\!-\!A'_m)\).

In the language of Chapter 2, we may have a set of distinct micro-processes that may be labeled \(A_1 \!-\! P_1 \!-\! A'_1,\; A_2 \!-\! P_2 \!-\! A'_2,\; A_3 \!-\! P_3 \!-\! A'_3,\; A_4 \!-\! P_4 \!-\! A'_4, \ldots\). Macrostates correspond to coarse-grained measurements of \(A_1, A_2, A_3, \ldots\), which give a coarse-grained observable \(\mathcal{M}\) at the preparation stage, and coarse-grained measurements of \(A'_1, A'_2, A'_3, \ldots\), which give a coarse-grained observable \(\mathcal{M}'\). The observables \(\mathcal{M}\) and \(\mathcal{M}'\) together form the macrostates.

For example, the preparation of a photon assembly in a classical electromagnetic field, followed by evolution and eventual detection, constitutes such a macrostate.

This approach shifts the statistical description from distributions over states to distributions over completed elementary micro-processes, which brings into relevance not just the states of the elementary constituents, but also the semantics of the context that specifies the states and the updating of the states. 

\section{What is a `Quantum System'?}

This section defines what a ``system'' is and what a ``quantum system'' is. First, we need a few terms. A type is an abstract concept unambiguously recognizable by a community of communicators, and a token is an instance of a type. A system can be defined as a token instantiating a type.

A quantum system is identical to a quantum problem, where a \textsc{quantum problem} is a specific instance of a process of any one of the four types, Classes 1 to 4, such that many instances of that process can be pooled to collect statistics for the ``same experiment.'' 

A simple example is the A--P--A type. An experiment that prepares an electron in a spin-up state along \(S_z\), and then measures \(S_x\) one unit of time later, is a quantum problem. Any instance of the same quantum problem, whether performed today or at any other time and place in the history of the universe, belongs to the same kind of quantum problem: all such instances can be pooled together to collect statistics for the same experiment.

In accordance with the definition introduced in Chapter 3, a closed quantum system corresponds to a single A--P--A type process, and an open system corresponds to Class 3 or Class 4.

To elucidate the notion of a quantum problem further, we need additional exposition.

Any two quantum preparations can share the same quantum state, making them fully operationally equivalent. In quantum mechanics, distinct laboratories can prepare the same state $\ket{\Psi}$, and the states in the two labs so prepared are fully operationally equivalent. So, in quantum theory, a spatio-temporal description is needed to prescribe state preprations, but distinct spatio-temporal locations do not necessarily distinguish quantum states. So a quantum state of a photon $\ket{\Psi}$ prepared in Vienna is the same as the quantum state $\ket{\Psi}$ of a photon prepared in Tokyo or even the same as the state of a photon $\ket{\Psi}$ that hypothetically may have been prepared in Bern in 1925. Which means that, if we can pool the data from the individual experimental trials conducted at each of these spatio-temporal locations on the state $\ket{\Psi}$, we can instantiate the probability distribution associated with that state. The quantum state functions as a type, associated with repeatable operational procedures (that in a way transcends spacetime), and not a spatiotemporal `token'. The token is produced by the location of the experimental apparatus. The deeper reason is that, in quantum mechanics, states are assigned by the preparation procedure that can be imitated at distinct spatio-temporal locations.

This operational equivalence suggests an answer to another question that one might have raised earlier when we introduced the conceptual construct of an observer: After all, clocks measure time, and magnetometers measure magnetic fields. Having introduced the construct of an observer as an analog to a clock and to magnetometers, what do observers that we have functionally defined help measure/quantify? 

It is in discerning meaning. Note that, in the program that we have been developing, meaning is not something peripheral to the physical process that we add on top of the physical process, but is also a feature of the physical process itself. Let us elaborate.

To set the stage, let \(\mathcal{L}\) be a formal sentence about what happens in Hilbert space corresponding to a sentence \(L\) concerning the preparation described in ordinary language, along with the numerical parameters that specify the preparation, the setting of the timer in a clock before an intervention is made to reveal the actual outcome, and the numerical eigenvalues that are realized when the potentiality intersects with an ECM, as defined in Chapter 2. We can construct sentences with meaning, of the form \(\mathcal{L} \; \blacksquare \; L\)\footnote{This is a notation that I have introduced. The black square is not a typo.}, that we can call ``quantum problems,'' where the black square represents the interpretation linking \(L\) and \(\mathcal{L}\). The \(\mathcal{L} \; \blacksquare \; L\) sentences can correspond to any one of the four classes of quantum dynamics.

Let sentences \(A\) and \(B\) be individual quantum problems. Then what an observer ``measures'' is the function
\begin{equation}
    f(A, B) \in \{0, 1\},
\end{equation}
which assigns a value of \(0\) if the meaning does not match and \(1\) if it does. The matching of meaning is determined by whether the numerical values from quantum problems \(A\) and \(B\) can be combined for the statistical analysis needed to determine the measured expectation values. That is, whether \(A\) and \(B\) correspond, in \(\mathcal{L}\), to the same state preparation \(\ket{\Psi}\), the same unitary application for the same local time interval \(\hat{U}(\Delta t)\ket{\Psi}\), and the interrogation of the same observable \(\hat{O}\) at the end of \(\Delta t\).

The function \(f(A, B)\) concerns the elementary quantum phenomenon for which a pure-state assignment and updating can be verified. We use the Latin alphabet for such statements.

These could be interpreted as the micro-processes. Because we can achieve macroscopic operational equivalence between experimental propositions \(\chi\) and \(\Upsilon\), each consisting of preparation, evolution, and measurement, such as measuring the same classical electromagnetic fields at two points along their paths of propagation in two different experimental setups. In such a case, we can write the experimental determination by the function
\begin{equation}
    F(\chi,\Upsilon) \in \{0, 1\}.
\end{equation}
We have \(1\)\textbf{ }when the experimental propositions are operationally equivalent up to the uncertainty of the experiment, \(0\) otherwise. These can be reinterpreted as the macrostates. We use the Greek alphabet for such statements. These statements also have the form \(\mathbf{\Pi} \; \blacksquare \; \zeta\), where \(\mathbf{\Pi}\), using bold Greek letters, concerns statements about what happens in the respective classical state space, and \(\zeta\) concerns the experimental maneuvers carried out by the observer in the lab corresponding to the statement \(\mathbf{\Pi}\). These are the definitions of micro-processes and macrostates that we have used.

These are macrostates because each of the macrostate propositions \(\chi\) and \(\Upsilon\) can be decomposed into many smaller elementary quantum propositions, \(A_1, A_2, A_3, \ldots\) and \(B_1, B_2, B_3, \ldots\), respectively. Here, each of the \(A_i\) and \(B_i\) is simply one of the micro-process propositions we defined earlier. An example of this micro-process--macrostate relationship is when a proper mixture of pure quantum states is prepared, evolved, and later intervened upon by an ECM. Let \(\rho\) be the density matrix representing the proper mixture. The preparation, evolution, and measurement at an ECM can be represented in \(\mathbf{\Pi}\) by
\begin{equation}
    \rho \rightarrow \hat{U}(\Delta t) \rho  \hat{U}^\dagger(\Delta t) \rightarrow \hat{M}_k \Bigg[\hat{U}(\Delta t)\rho \hat{U}^\dagger(\Delta t)\Bigg] \hat{M}_k^\dagger
\end{equation}
upon obtaining result \(k\), where we have ignored normalization. Now, a given density matrix can correspond to many possible ensembles \(\{p_i, \ket{\Psi_i}\}\). Consequently, we can have \(F(\chi,\Upsilon)=1\) even when \(f(A_i, B_j)=0\), for all \(i\) and \(j\). Note that, unlike in the usual definition of microstates and macrostates in statistical mechanics, micro-processes and macrostates are defined in terms of completed quantum problems. 

To distinguish this definition of micro-processes and macrostates from the usual definition, we can label it by the name \emph{semantic definition of micro-processes--macrostates}, as it alludes to the meaning of `what was done' and `what was found' by a counterfactual fictitious observer.

\section{Statistical Mechanics without the Atomistic Conception of Matter?}

In this section, we heuristically describe and propose a conjecture about how thermalization may arise in heterogeneous systems comprising potentialities and actualities. An advantage of the following construction is that it appears to provide a unified framework for non-equilibrium and equilibrium processes, whereas the current ensemble understanding of statistical mechanics is primarily an equilibrium theory. A fully worked-out mathematical theory is left for future work.

It is well known that, once an equilibrium state is assumed, we can use the maximum entropy principle to obtain the Gibbs distribution. In this work, we need to show equilibration. That is, we need to explain the success of quantum statistical mechanics---why it works---and thus of thermodynamics, starting from quantum-mechanical principles and epistemic constraints concerning language and communication, which we have taken to be primitive.

In the previous section, we defined the members of the ensemble as `micro-processes' instead of microstates. However, an astute reader who has read Chapter 6 may have noticed that we could have retained the microstate definition, albeit by choosing a different notion of what microstates mean. Recall that, in Chapter 6, we showed that the state of the system at any given time is specified both by the tensor-product structure at that time and by the locally assignable pure states of the subsystems in that tensor-product structure:
\[
\Xi(t+\Delta t) \coloneqq \left\{\bigotimes_k \ket{\psi_k}_B(t+\Delta t), C\text{-}TPS(t+\Delta t)\right\}.
\]

Now, these microstates could very well be thought of as the members of the ensemble. However, there appears to be no easy way to go from such microstates to observable macrostates.

The micro-processes that we defined in the previous section appear to yield an equivalent description and seem to have the advantage of facilitating the use of an existing mathematical framework.\\

By considering the \(n\) micro-processes, each denoted by the stochastic trajectory \(\Gamma_k\), described by a stochastic trajectory in Hilbert space, as the members of the ensemble, we can construct a family between the times \(t_1\) and \(t_2\):
\begin{equation}
    \Gamma = \{\Gamma_1, \Gamma_2, \Gamma_3, \ldots, \Gamma_n\}_{t_1< t<t_2}.
\end{equation}
We can now use the Maximum Caliber principle \cite{MaxCal}, a generalization of the familiar maximum entropy principle in equilibrium statistical mechanics. In the Maximum Caliber principle, we have an object called the path entropy, which can be defined in this case as \cite{MaxCal}

\begin{equation}
    \mathcal{C} = -\sum_{\Gamma_1, \Gamma_2, \Gamma_3, \ldots, \Gamma_n} P[\Gamma_1, \Gamma_2, \Gamma_3, \ldots, \Gamma_n]\log  P[\Gamma_1, \Gamma_2, \Gamma_3, \ldots, \Gamma_n].
\end{equation}

Note that MaxCal is, in this case, not the same as quantum path integrals, since in path integrals we have many possible paths that can be coherently, and this is important, taken by a single system, with a probability weight for each path. Here, instead, we have many possible stochastic paths for each system, i.e., for each micro-process in an assembly of systems, \(\Gamma = \{\Gamma_1, \Gamma_2, \Gamma_3, \ldots, \Gamma_n\}_{t_1< t<t_2}\), with probability assigned to each family of stochastic realizations of the systems in the assembly.

The procedure is then to maximize the path entropy subject to constraints. One is, obviously, the normalization constraint:
\begin{equation}
    \sum_{\Gamma_1, \Gamma_2, \Gamma_3, \ldots, \Gamma_n} P[\Gamma_1, \Gamma_2, \Gamma_3, \ldots, \Gamma_n] =1.
\end{equation}
The probability distribution obtained by maximizing the path entropy subject to this constraint can be used to describe non-equilibrium processes, since the description developed so far accounts for transport phenomena.

Another constraint is that the average energy of a closed system must stay fixed on average. For this, associated with a single micro-process \(\Gamma_1\), we can define a change in measured heat \(\Delta \tilde{Q}_1\) between the times \(t_1\) and \(t_2\), where \(\tilde{Q}\) is the stochastic heat we saw in Section 7.3. So, for a family of many micro-processes, \(\Gamma = \{\Gamma_1, \Gamma_2, \Gamma_3, \ldots, \Gamma_n\}_{t_1< t<t_2}\), we can associate the total heat
\[
\sum_{i\in\Gamma} \Delta \tilde{Q}_i \coloneqq Q_T(\Gamma)
=
Q_T(\Gamma_1, \Gamma_2, \Gamma_3, \ldots, \Gamma_n).
\]

So the energy constraint can be written as
\begin{equation}
    \sum_{\Gamma_1, \Gamma_2, \Gamma_3, \ldots, \Gamma_n} Q_T(\Gamma_1, \Gamma_2, \Gamma_3, \ldots, \Gamma_n) P[\Gamma_1, \Gamma_2, \Gamma_3, \ldots, \Gamma_n]
    =
    \frac{1}{t_2-t_1}\int_{t_1}^{t_2} E(t)\,dt
    =
    \bar{E}.
\end{equation}
Here, \(\bar{E}\) is the time-averaged energy of the closed system between the two times at which we measure. 

In the present framework, where the members of the ensemble are not instantaneous microstates but completed micro-processes, the analogue of the usual average-energy constraint that we use in MaxEnt \cite{MaxCal} is imposed on the stochastic heat functional associated with a family of micro-processes. Since the Hamiltonian is fixed, there is no work contribution, and the energy-valued trajectory functional is represented by just the heat \(Q_T(\Gamma)\).

For linear constraints, the probability distribution would have the form \cite{MaxCal}
\begin{equation}
     P[\Gamma]  \propto e^{-\sum_i \Lambda_i A_i[\Gamma]}.
\end{equation}

At this point, the route to recovering the Gibbs distribution seems open if we make a conjecture, which we leave unproven in this work.

The conjecture is this: when the measurable gradients of temperature, chemical potential, pressure, or other generalized potentials are zero, the MaxCal probability distribution, when marginalized over all possible realizations for the family of stochastic trajectories of the \(n\) micro-processes, such that the realizations end in a microstate of actualities at time \(t_2\), will result in a probability distribution over the microstates at the final time \(t_2\). This probability distribution is obtained from the maximum entropy principle over the energy eigenstates that correspond to the stochastic heat as the energy eigenvalues at the final time.

That is,
\begin{equation}
    \sum_{\Gamma: A'(t_2)}  P[\Gamma]
    \propto
    \sum_{\Gamma: A'(t_2)} e^{-\sum_i \Lambda_i A_i[\Gamma]}
    \propto
    e^{\frac{-E_i}{k_B T}}.
\end{equation}

That is, it is conjectured that the marginalization of the MaxCal probability distribution in Eq.(7.47) will give the usual thermal distribution for the final time \(t_2\).

The proof would require a separate mathematical treatment. Here, the aim is only to make a reasoned guess as to why, within the present construction, the Gibbs distribution is the natural equilibrium limit of the MaxCal distribution over a family of completed micro-processes.

\section{What Does It Operationally Mean for an Object to Follow \(m \ddot{\mathbf{r}}= -\nabla V\)?}

Imagine a classical object moving through space at a certain velocity relative to an observer, in a potential field \(V\). Operationally, the object's following Newton's laws is manifest to the observer only in terms of how the light from the moving body changes with time. If the object is at position \(\mathbf{r}_1\) relative to the observer, rather than at a different position \(\mathbf{r}_2\) relative to the observer, the light from the object will have different characteristics; for instance, the direction in which the light from the object is detectable changes. Similarly, other dynamical quantities of the moving object can be inferred from the frequency, amplitude, and phase of the light from the moving object. The relationship between the so-inferred dynamical quantities of the object will then be mathematically related by Newton's laws.

In that sense, that is all we can claim to mean operationally by ``the object follows Newton's laws.'' Now there are two objections to respond to. First, one may object that we can infer the dynamics of Newtonian objects by observing other signals, such as sound. That is indeed true. However, there are cases where the dynamics of the object can be inferred from the light from the object but not from any sound from the object, such as when the object is moving in a vacuum. At the same time, for ordinary matter, it appears that tracking the moving object using light from the object is a more versatile method and appears to work even when other signals from the object are nonexistent.

Second, one may object, from the kind of realist stance that mechanical philosophy suggests, that the body simply has a `position' and other dynamical quantities as inherent properties of the object, akin to time-dependent labels of the object, without reference to any optical means of measuring those labels. However, this realist stance is not supported by observations---we appear to have no case of an object having position, momentum, and acceleration without their being revealed to at least some observer through optical means. There appears to be no justified inherent essence of `position' and other mechanical quantities as inherent tokens of objects outside of this means of measuring them.

Note that this reinterpretation of position, momentum, and other dynamical quantities is not an anti-realist stance. It is simply anti-realist in the mechanical-philosophy sense, which assumes that point particles `exist' with definite position and momentum in space and time, without reference to light from the point particles. Such a stance is understandable given that Newtonian mechanical philosophy developed before our understanding of the thermodynamics and electrodynamics of moving objects took full shape.

This reinterpretation of mechanical quantities is realist in a ``post-mechanical sense''; that is, an object has position, momentum, and other dynamical quantities even before I see it, if a counterfactual observer in the local vicinity of the object, with respect to whom the object has those mechanical quantities inferred via optical means, can be consistently assumed.

Thus, the goal in deriving the origin of classical mechanics is not to obtain a density matrix with diagonal entries in some basis, which we classified as the classical limit of potentialities in Chapter 5, using only unitary quantum mechanics. Instead, the goal is to show that a moving object emits light in a way that matches a light-emitting Newtonian point particle. The assumptions will include quantum theory without any modification, but also epistemic constraints concerning language and communication that are true of any physical theory.

\section{A Simple Model for the Origin of Classical Mechanics:  The Joint Role of Quantum Theory and Epistemic Constraints}

This section explores the emergence of the classical dynamics of actualities, as opposed to the emergence of the classical dynamics of potentialities, which, as we saw in the previous chapter, is taken care of by decoherence and the correspondence principle. The emergence of the classical dynamics of actualities in the non-relativistic limit is the dynamics of Galileo and Newton. We intend to show heuristically that this emergence can be understood as starting from quantum theory and the natural principles governing epistemic constraints. Recall that the two principles, the one hypothesis, and the one postulate are the contents of ``epistemic constraints.'' 

The idea we have for understanding the origin of classical mechanics can be illustrated with reference to the same drifting gaseous system that we studied in Chapter 6. We have a drifting gas, and a co-moving observer \(A\) is imagined who makes an alignment position measurement of the gas by passively recording the thermal radiation from the gas. We imagine another observer, \(B\), who is in relative motion with respect to \(A\). We now want to show whether the gas emits light in the same way a light-emitting Newtonian body would emit light.

At any given time, we have an assembly of clusters, each in a pure state. In the next small time window \(\Delta t\), each cluster has a probability \(\lambda \Delta t\) of undergoing a Born update stochastically in the position basis, which will in general reduce the size of a cluster in one of many possible ways. In that same window of time, with probability \(1-\lambda \Delta t\), each cluster evolves unitarily under \(e^{-i \hat{H}_k \Delta t/\hbar}\), which can merge nearby clusters into a larger cluster, since the Hamiltonian can generate entanglement with neighboring clusters.

Since the clusters are distinguishable and can thus be labeled, the transition in time \(\Delta t\),
\[
    \Xi(t) \mapsto \Xi(t+\Delta t),
\]
will have a record of which clusters merged with which other clusters, and which ones split to yield ``daughter'' clusters.

To better imagine this, suppose we have three clusters at time \(t\), labeled \(1,2,3\), in the state
\[
    \bigotimes_{k=1}^{3} |\psi_k\rangle_B,
\]
where clusters \(1\) and \(2\) each contain one particle, and cluster \(3\) contains two particles. At time \(t\), with some probability, clusters \(1\) and \(2\) merge into one cluster, labeled \(1'\), while cluster \(3\) splits into two smaller clusters, each with one particle, which we label as \(2'\) and \(3'\). The final state can then be written as
\[
    |\psi_{1'}\rangle_B \otimes |\psi_{2'}\rangle_B \otimes |\psi_{3'}\rangle_B.
\]
There are, however, many possibilities for this merging and splitting of clusters, each realized with an appropriate probability. These probabilities are determined both by the Born-rule applicability probability computed for each cluster from the hypothesis concerning the probability of instantiability of the Born rule, and by the probabilities for the realization of different eigenvalues for the quantum-mechanical problem defined by the state of the cluster and the local Hamiltonian to which it is subject.

To make further progress, we need to make a few simplifying assumptions. First, instead of studying the Class 4 dynamics that characterize the system, we can use the reasoning we used in our consideration of the statistical mechanics of heterogeneous systems of potentialities and actualities. Namely, we can assume that we have many micro-processes between two times. In the context of this section, that would amount to many clusters corresponding to the many micro-processes. We can then carry out the following reasoning for one typical cluster and extrapolate the conclusion thus derived to \(N\) such typical clusters.

It is helpful to recall the It\^{o} rules once again (see Appendix A for a review of stochastic processes and It\^{o} calculus)\textbf{:}
\begin{align}
\langle dN \rangle &= \lambda\, dt, \qquad (dN)^2 = dN,\\
(dt)^2 &= 0, \qquad dt\, dN = 0,
\end{align}
and the chain rule for noise,
\begin{equation}
d(XY) = X\, dY + Y\, dX + dX\, dY.
\end{equation}

The Kraus operator for unitary evolution can be chosen to be proportional to the probability for unitary evolution to occur:
\begin{equation}
M_U = (1-\mu\, dt)^{1/2} U
\approx I - \frac{i}{\hbar}\hat{H}\,dt - \frac{\mu}{2}dt,
\end{equation}
and a similar choice can be made for the Kraus operators for measurements:
\begin{equation}
M_k = \sqrt{\mu\, dt}\, P_k,
\end{equation}
such that the normalization condition is satisfied,
\begin{equation}
\sum_{\alpha} M_\alpha^\dagger M_\alpha = I.
\end{equation}

In a small interval of time, the increment in the state will therefore be stochastic:
\begin{equation}
|\phi(t+\Delta t)\rangle
=
M_U |\phi(t)\rangle
+
\sum_k
(M_k - M_U)\, dN_k\, |\phi(t)\rangle .
\end{equation}
When the time interval is small, we can approximate the Kraus operators as
\begin{equation}
M_U = (1-\mu\, dt)^{1/2}U
\approx
I - \frac{i}{\hbar}\hat{H}\,dt - \frac{\mu}{2}dt,
\end{equation}
and
\begin{equation}
M_k = (\mu\, dt)^{1/2} P_k.
\end{equation}
Therefore, the term \(M_k - M_U\) can be written as
\begin{equation}
M_k - M_U
=
(\mu\, dt)^{1/2}P_k
-
I
+
\frac{i}{\hbar}\hat{H}\,dt
+
\frac{\mu}{2}dt .
\end{equation}
Using the It\^{o} rule \(dt\,dN_k=0\), we obtain
\begin{equation}
(M_k - M_U)dN_k
=
\left(
-I + (\mu\, dt)^{1/2}P_k
\right)dN_k .
\end{equation}
This leads to the stochastic Schr\"odinger equation for a Poisson process:
\begin{equation}
d|\phi\rangle
=
\left(
-\frac{i}{\hbar}\hat{H} - \frac{\mu}{2}I
\right)
|\phi\rangle\, dt
+
\sum_k
\left(
-I + (\mu\,dt)^{1/2}P_k
\right)
dN_k\, |\phi\rangle .
\end{equation}

In the diffusive limit, when the central limit theorem is applicable, we can expect the Poisson process to be well approximated by a Wiener process.

Now, this is just for one typical cluster. We have many such typical clusters undergoing a similar process.



Earlier, Bhattacharya, Jacobs, and Habib studied \cite{bhattacharya2003continuous}, in the context of optically
mediated continuous weak measurements, that Wiener quantum trajectory unravelings
can approximate classical Newtonian dynamics in the expectation values of position and
momentum under certain limiting conditions. When this limit is reached by a large number
of clusters, one can expect that the criterion we imposed for the derivation of classical mechanics
would be reached. That is, the characteristics of the intercepted light by $B$ would
resemble how a light-emitting Newtonian body would behave.



Let us make this picture a little more quantitative. There are two mathematical goals we have. In Bhattacharya and Jacobs's work, their stochastic Schrodinger equation, Eq.(53) in \cite{bhattacharya2003continuous}, was
\begin{equation}
d|\psi\rangle
=
\left[
-\frac{1}{\hbar}
\left(
iH(t)+\hbar k X^{2}
\right)dt
+
\sum_{i=1}^{N}
4\eta_i k X\, dr_i
\right]
|\psi\rangle ,
\end{equation}
with noise
\begin{equation}
dr_i
=
\langle X \rangle dt
+
\frac{dW_i}{\sqrt{8\eta_i k}} .
\end{equation}
The first goal is to derive Eq.(7.61) as a limit of Eq.(7.60). The second goal is to show that the several limiting conditions derived in their paper become naturally applicable to our system. If that happens, then we will have the result that, for a typical cluster, we have \cite{bhattacharya2003continuous} 

\[
\dot{x} = \frac{p}{m}
\]

\[
\dot{p} = \langle F(X)\rangle \approx F(x)
\]
for
\[
x \equiv \langle X \rangle_c,
\qquad
p \equiv \langle P \rangle_c .
\]

Operationally, what these equations mean in the context of the quantum trajectories studied here requires some nuance. In the unconditioned evolution of a quantum system, the quantities \(\langle X\rangle\) and \(\langle P\rangle\) do not make much sense for a single system; they are expectations that we can realize when we repeat the experiment many times and measure position and momentum. In quantum trajectories, we interpret \(\langle X\rangle_c\) and \(\langle P\rangle_c\) as the mean position of the center of the wavepacket and the rate of change of the center of the wavepacket, up to a factor of \(1/m\), respectively.

What this means for ``deriving the classical limit'' aligns with the criterion for the classical limit specified in the previous section. Namely, the following:

Forget quantum mechanics for now, and imagine a classical Newtonian particle of mass \(m\) that moves in a potential \(V\) while emitting light as it moves. Let this be Scenario 1. Now consider a quantum particle of mass \(m\) moving in the same potential, while the paper's three limits---low noise, narrow wavepacket, and faithful tracking---apply, and the Newtonian limit for expectation values holds. Let this be Scenario 2. For the observer, Scenario 1 and Scenario 2 can look interchangeable, insofar as the alignment-position record inferred from the emitted light traces the same trajectory in both cases.

However, the dynamics studied in their paper, in our terminology, is of the \((A\!-\!P\!-\!A)\!-\!P\!-\!(A\!-\!P\!-\!A)\!-\!P\!-\!(A\!-\!P\!-\!A)\!-\!P\ldots\) kind. The situation we have is of the \(\{A, P, A, \ldots\}\!-\!\{P, P, A, \ldots\}\!-\!\{A, A, A, \ldots\}\ldots\) kind, which we have approximated, as reasoned in Section 6.5, by a collection of many coexisting processes of the \((A\!-\!P\!-\!A)\!-\!P\!-\!(A\!-\!P\!-\!A)\!-\!P\!-\!(A\!-\!P\!-\!A)\!-\!P\ldots\) kind. For each realized trajectory in the collection, we assume that the limits derived in the paper \cite{bhattacharya2003continuous} become applicable. This is part of the second derivational goal.

Under this assumption, for the collection, assuming there are \(N\) such trajectories, we can write, using the arithmetic mean,
\[
\bar{x}_N(t)
=
\frac{1}{N}
\sum_{r=1}^{N}
x_c^{(r)}(t),
\qquad
\bar{p}_N(t)
=
\frac{1}{N}
\sum_{r=1}^{N}
p_c^{(r)}(t),
\]
where
\begin{equation}
\dot{x}_c^{(r)}
\approx
\frac{p_c^{(r)}}{m},
\qquad
\dot{p}_c^{(r)}
\approx
F\!\left(x_c^{(r)}\right).
\end{equation}

This could correspond to the position of the entire moving gas system, due to the light from all parts of the gas. Note that we have not used the center of mass here. We are focusing on the ensemble average over all the individual instances of \((A\!-\!P\!-\!A)\!-\!P\!-\!(A\!-\!P\!-\!A)\!-\!P\!-\!(A\!-\!P\!-\!A)\!-\!P\ldots\) of a typical cluster in the family of such processes of many typical clusters.

For \(\bar{x}_N(t)\) and \(\bar{p}_N(t)\) to follow Newton's laws, the momentum equation poses no problem, since
\[
\dot{\bar{x}}_N
=
\frac{1}{N}
\sum_{r=1}^{N}
\dot{x}_c^{(r)}
\approx
\frac{1}{m}
\frac{1}{N}
\sum_{r=1}^{N}
p_c^{(r)}
=
\frac{\bar{p}_N}{m}.
\]
However, for the second law, we have a hurdle. For the \(N\) members in the collection, we can have
\begin{equation}
\dot{\bar{p}}_N
=
\frac{1}{N}
\sum_{r=1}^{N}
\dot{p}_c^{(r)}
\approx
\frac{1}{N}
\sum_{r=1}^{N}
F\!\left(x_c^{(r)}\right).
\end{equation}
However,
\begin{equation}
\dot{\bar{p}}
\approx
\mathbb{E}\!\left[F(x_c)\right],
\qquad
N\to\infty .
\end{equation}
The right-hand side could be interpreted as the ensemble average of the force on each typical cluster.

But for recovering the Newtonian form, we need
\begin{equation}
\dot{\bar{p}}
\approx
F(\bar{x})
=
F\!\left(\mathbb{E}[x_c]\right).
\end{equation}
This is not true in general because
\[
\mathbb{E}[F(x)]
\neq
F(\mathbb{E}[x]).
\]

Despite the hurdle it raises, this is a good sign: it means that merely taking the ensemble average over the family and going to a master-equation description will not give Newtonian behavior, even if we use the paper's limits for the individual members in the family.

We additionally require some mechanism that imposes
\[
\mathbb{E}[F(x)]
\approx
F(\mathbb{E}[x]).
\]
Physically, this means that the clusters are somehow kept close to each other around some common average position, and that the localization of the typical clusters happens within that vicinity.\footnote{This is because, if that happens relative to the scale at which the nonlinearity of the force may be felt, then the linear approximation for the force may apply.} This is exactly the approximation we made earlier when we approximated the evolving localized states and the cluster tensor-product structure undergoing a \(\{A, P, A, \ldots\}\!-\!\{P, P, A, \ldots\}\!-\!\{A, A, A, \ldots\}\ldots\)-type process by a collection of many typical clusters undergoing an \((A\!-\!P\!-\!A)\!-\!P\!-\!(A\!-\!P\!-\!A)\!-\!P\!-\!(A\!-\!P\!-\!A)\!-\!P\ldots\)-type process that remains confined to a volume. This can be partly argued physically as follows: at any two times \(t_1\) and \(t_2>t_1\), as seen in Section 7.5, we can have actualities \(A_1\) and \(A_2\) that complete a quantum problem. The potentialities and actualities interlacing such a pair of actualities can be identified with a cluster's evolution. Between two times, there will be many such clusters, but one may identify typical clusters corresponding to a typical micro-process. However, there is still a need to explain why all the typical clusters are ``clumped'' together in the same system. Perhaps the clumping is due to the various collective intermolecular attractive forces. Once inside the volume that qualifies as an actuality, then, by the hypothesis concerning the probability of instantiability of the Born rule, the clusters must undergo localizations that keep them undergoing the \((A\!-\!P\!-\!A)\!-\!P\!-\!(A\!-\!P\!-\!A)\!-\!P\!-\!(A\!-\!P\!-\!A)\!-\!P\ldots\) type process. If this line of reasoning behind the approximation is right, which we have left only as a physically reasoned conjecture, and not yet as a proof, then we have the mechanism to impose
\[
\mathbb{E}[F(x)]
\approx
F(\mathbb{E}[x]),
\]
which ensures, for the momentum equation, that
\[
\dot{\bar{p}}
\approx
F(\bar{x})
=
F\!\left(\mathbb{E}[x_c]\right).
\]

Nevertheless, mathematically showing whether a \(\{A, P, A, \ldots\}\! - \!\{P, P, A, \ldots\}\!-\!\{A, A, A, \ldots\}\ldots\) type process can be approximated by a collection of many typical clusters undergoing an \((A\!-\!P\!-\!A)\!-\!P\!-\!(A\!-\!P\!-\!A)\!-\!P\!-\!(A\!-\!P\!-\!A)\!-\!P\ldots\) type process that remains confined to a volume is the third goal of the calculation. It appears that the percolation theory from mathematics \cite{PercolationTheory} could be useful for this task.



\chapter{Conclusion} 
\resetfootnote 


This dissertation develops a Bohr-inspired reframing of the quantum measurement problem by distinguishing epistemic constraints from classical mechanics and treating the former as primitive, in combination with quantum theory. It develops the conceptual vocabulary, analytical tools, and an organizational framework for studying the consequences of this reframing. The dissertation further argues that programs seeking to derive classicality and thermalization solely from unitary quantum theory, including decoherence-based approaches and canonical typicality, address a related but nevertheless distinct explanatory problem from what is identified here as the ``thermalization of heterogeneous systems with actualities and potentialities'' and the ``mechanics of heterogeneous systems with actualities and potentialities''; explaining these remains an open task. The framework introduced here---which treats both unitary quantum theory and epistemic constraints as primitive---is then applied to investigate these open explanatory questions.

Let us first condense the contributions we have made in this dissertation before offering a broader conclusion that gestures toward future work and exploration. The contributions in this dissertation involve many logically inter-dependent sequences of conceptual constructions, reinterpretations, and the mathematical modeling that results from them.

\begin{enumerate}
    \item Our starting point was reframing the quantum measurement problem by reconsidering our assumptions and what needs explanation. This led to the Bohrian Program, in contrast to the existing Schrodingerian measurement problem. This is done in Chapter 0. Chapter 1 provides further context.

    \item This reframing suggested a new strategy for making progress that is distinct from existing strategies, including interpretations, modifications of quantum mechanics, and the reconstruction program. This is done in Chapter 2.

    \item This strategy seeks to retain quantum theory and epistemic constraints as primitives, not just quantum theory as a primitive. To this end, we operationalize and formalize language and communication, namely, epistemic constraints, using the construct of an observer. This can also be thought of as a formalization of `experience' in describing physical phenomena: although `experience' itself is a relatively hazy concept not amenable to formalization, language, which is an aspect of experience, is amenable to formalization. This is done in Chapter 2.

    \item This construct of an observer led us to clarify three distinct usages of what is meant by `measurements': measurement as information copy, alignment measurements, and quantum state tomography, and to demonstrate that alignment measurements are physically more primary, in the sense that measurement by information copy cannot always replace alignment measurement, and in the sense that quantum state tomography requires alignment measurements. This is done in Chapter 2.

    \item The identification of alignment measurements led us to identify Heisenberg's potentiality and actuality as two distinct classes of operations that can be counterfactually performed by observers in reference to alignment position measurement. This is done in Chapter 2.

    \item Once we made these operational definitions, we were able to observe two empirical regularities concerning epistemic constraints, which we captured in Principle 0 (not reproduced here), Principle 1, Principle 2, Postulate 1 and Postulate 2. All these are done in Chapter 2: 
    
    \begin{itemize}
        \item Principle 1: The definiteness of the occurrence of an event specifiable by an experimental proposition \(E\), defined as the truth value \(T(E)\in \{0,1\}\), where \(T(E)=1\) if the event occurred and \(T(E)=0\) otherwise, is universally invariant across all observers. Here, an `event' is an individuated occurrence in a region of space that is alignment-position-measurable.
        
        \item Principle 2: All natural phenomena, even in regions of the universe without actual observers, will be found to abide by Principle 1, with counterfactual, fictitious observers, when a later evolved observer studies the past and those regions of the universe without observers.

        Furthermore, the following hypothesis and a postulate were posited as part of the framework: 

        \item Postulate 1:
\begin{itemize}
    \item 1) Removing classical ignorance, the probability of instantiability of the Born Rule has a Bernoulli form:
    \begin{align}
        P(\text{Born Rule instantiable situation present} \mid E)
        &= 1-Q(E), \\
        P(\text{Unitary evolution continues} \mid E)
        &= Q(E),
    \end{align}
    with
\begin{equation}
\begin{aligned}
&P(\text{Born Rule instantiable situation present} \mid E) \\
&\qquad
+
P(\text{Unitary evolution continues} \mid E)
= 1.
\end{aligned}
\end{equation}

    \item 2) For regions of space inside the box with potentiality,
    \begin{equation}
        \begin{aligned}
        &P\left(
        \text{Born Rule instantiable situation present}
        \mid
        \text{\(E\) specifies potentiality}
        \right) \\
        &\qquad = 1-Q(E)
        =
        0.
        \end{aligned}
    \end{equation}

    \item 3) For regions of space inside the box with actuality,
    \begin{equation}
    \small
        \begin{aligned}
        &P\left(
        \text{Born Rule instantiable situation present}
        \mid
        \text{\(E\) specifies actuality}
        \right) \\
        &\qquad = 1-Q(E) \\
        &\qquad =
        \sum_i \sum_l
        \left|
        \left\langle i^l \right|
        \hat{U}
        \left(
        \ket{d} \otimes \ket{j}
        \right)
        \right|^2
        \leq 1.
        \end{aligned}
    \end{equation}

    \item The label \(l\) denotes distinct classical information channels discernible in an actuality. The state \(\ket{d}\) is the initial state of the relevant degrees of freedom of the actuality with which the potentiality in state \(\ket{j}\) interacts. The states \(\ket{i^{l}}\) denote output states corresponding to the activation of channel \(l\).
\end{itemize}

        \item Postulate 2: What is an actuality with respect to one observer is an actuality with respect to all observers. Similarly, what is a potentiality with respect to one observer is a potentiality with respect to all observers. This means that any situation where epistemic constraints become relevant, such as in subsystem partitioning and in the application of unitary evolution versus state update, is invariant across all observers.
    \end{itemize}

    \item In light of these developments, to prepare the stage for the next sequence of developments in the second half of the thesis, on the physical side we then embarked on a few intermediate tasks in which we classified the known kinds of quantum dynamics, made clear some assumptions concerning what quantum theory means and what the state means, and finally clarified the different ways ``complex systems''\footnote{a term that we have used loosely, when compared to it's rigorous definition in complex system science \cite{Ellis2012TopDown} } with many quantum subsystems could be considered. This led us to reclassify the framework of decoherence as the study of the complex quantum system of potentialities. This raised the question of how to deal with complex systems involving both actualities and potentialities. This is done in Chapters 3 and 5.

    \item Similarly, on the philosophical side, we embarked on a few intermediate tasks to prepare the stage for the next sequence of developments in the second half of the thesis. We had to reassess the atomistic conception of matter that is implicitly assumed at the foundation of our conception of matter. Such a reassessment leads us to new methods for studying parts of a composite system. This is done in Chapter 6.

    \item With these developments, we then studied the behavior of systems that involved both potentialities and actualities. The model we found had evolving localized quantum states and a time-evolving tensor-product structure of subsystems, not just the state evolving with time. This is done in Chapter 6.

    \item This model then suggested studying the statistical mechanics of such systems and, finally, proposed a mechanism for the origin of Newton's mechanics. This is done in Chapter 7.
\end{enumerate}

Now the longer conclusion.

\section{Longer Conclusion}

In the chapter ``Physics of Semantics,'' Stuart Kauffman asks, ``Can semantics be part of a science of nature rather than merely human interpretation?'' \cite{kauffman2000physicsOfSemantics} The essence of the question can be reinterpreted as ``Can semantics be part of the fundamental laws of nature, rather than just a human construct imposed on an otherwise indifferent cosmos?''

Physics is based only on forces and states, with no fundamental role for functionally defined primitive entities. The arguments and results developed in the thesis show that states, forces, and functionally defined primitives together are essential for a complete description of the physical universe. This is not simply because of our epistemic limitations but, as we have argued, because of how nature is constituted.

Kauffman's argument for carving out a conceptual space for viewing function in a primitive sense is, mostly, motivated by biology \cite{Kauffman2023ThirdTransition}, where he examined goal-directedness in autonomous agents, including biological organisms. Function brings in semantics, because a stimulus means something, presents choices, and affects the course of action to maximize some value.

Kauffman's argument can also be motivated from within physics, which brought us to Bohr: determining what observable was measured is not fixed by the quantum states alone. To determine which physical quantity was measured, especially in quantum experiments, and whether measurement was performed at all, one needs a characteristic manipulation of the devices in the laboratory. Different maneuvers correspond to different measurements, and different maneuvers can determine whether a measurement is made at all. Only that, in the case of quantum experimentation, we do not need human observers. Machines, or even counterfactually present observers, as we have abstractly defined them, are sufficient.

From a naturalistic account of Bohr-inspired semantics that we have developed in this thesis, Kauffman-inspired semantics, which additionally requires purpose and value rather than just function, may follow without introducing anything new. But that was not the scope of this thesis.

To create a naturalistic conception of semantics, one that decouples it from any human reference, we need exactly this observation:  for every naturally occurring quantum experiment, even a billion years into the past, what was measured is fixed by what a \textit{counterfactual observer} would have written down, given the arrangement and the records available to them. This is similar to a naturalistic conception of time, decoupled from a human-made notion of a clock: the duration of a process, say, a chemical reaction that occurred millions of years ago, is determined counterfactually by what a nearby clock would have read.

Before we end, it is a good time to talk about the realist position of objects existing `out there.' Is it still a permissible position?

We believe we can still consider a `world out there,' but not in the sense specified by mechanical philosophy. The sense specified by mechanical philosophy posits that `systems possess states and properties' as though they are labels attached to objects. Such a kind of realism does not seem justified. But a more nuanced version of realism appears justified.

The meaning of physical quantities associated with any system can be considered ``real'' only in the counterfactual operational sense. The system does not have `positionness' or `massness,' for example. There is nothing more to the existence of those physical quantities than what can be specified by a counterfactual or actual observer based on counterfactual or actual operations to measure those physical quantities  \cite{tal2015measurement}. This implicitly assumes a certain principle of representability. Any distinction between things in nature can always be represented as a distinction in sense impressions or concepts in an observer, at least indirectly, with the aid of instruments. There is no distinction between things in nature that cannot, in principle, be mapped to distinctions in sense impressions or concepts in an observer.

On this view, however, ``nature'' can no longer be understood simply as ``everything out there'' independent of all observation, but rather as that portion of the totality of experience that is, in principle, shared among all actual or fictitious observers who can communicate with one another within some shared language.

In the discovery of the laws of motion \cite{NewtonPrincipia}, a key realization was that generalities extracted from terrestrial mechanical processes in our local environment should yield laws that apply equally to distant celestial bodies. Likewise, one may posit \cite{Wheeler1989} that generalities concerning the epistemic limitations of our access to quantum processes here in our labs can be formulated as laws that apply as well to physical situations in regions of the universe devoid of human observers---provided that those remote phenomena would, in principle, be describable and comprehensible by some local, hypothetical observer in ways that stay consistent with our observations today.

Systems such as the Moon \cite{Pais1982}, when no one is looking, or the fall of a tree in a forest without any human witness, satisfy this operational criterion for ``existing out there.'' By contrast, a single proton propagating through the interstellar medium following a supernova explosion does not, in general, satisfy this requirement, owing to the absence of any physically meaningful counterfactual measurement context for the proton in this situation.

\section{Three Criteria for Completion of the Bohrian Program}

The Bohrian Program, while open-ended in that it seeks to find new physical conclusions from the joint consideration of quantum theory and epistemic constraints, without being particular about addressing any specific problem, has three checkpoints to strike off from the to-do list in order to fully physically realize the situations where the conclusions made from this joint consideration become relevant.

\begin{enumerate}
    \item Address what follows from the joint consideration of quantum theory and epistemic constraints for physical phenomena that are presently operationally accessible, either by recasting known processes in a new light or by suggesting new processes that can be realized with presently known operational access.
    \item Specify the reproducible experimental protocol for creating macroscopic quantum superpositions of arbitrarily large objects.
    \item Address questions that may arise when such a possibility is realized---what happens to observers? What happens to the gravity of such superposed systems?---by pursuing the implications that follow for known physics, given the knowability (epistemic) constraints on physically allowed observers.
\end{enumerate}

This dissertation makes a partial contribution to 1) and 3), but not to 2). Future work will explore gravity and the second goal, 2), as well as fill some gaps present in the reasoning of the present work.

As for the discussions centered on the original measurement problem, namely, the Schrodingerian measurement problem, this thesis has made a contribution that can be considered a higher-level analysis. It has reinterpreted the assumptions and goals of the original measurement problem, removing the need either to solve it or to dissolve the original framing of the measurement problem. It recognizes certain non-trivial sources of unease in the original framing---such as the definiteness of experience and the intersubjective objectivity of facts---and finds new directions to metabolize the sources of that unease in combination with known physics.

Although one can view the tool that we introduced, the probability of instantiability of the Born rule, as having contributed in this direction if the Schrodingerian measurement problem is phrased as ``when to apply the Born rule and when to apply unitary transformation?'', instead of being phrased as ``how do definite outcomes arise?'' The latter is an ill-posed question in the higher-level reinterpretation of the measurement problem.

The emphasis is now on how to integrate, mathematically, what we commonly take to be qualitative and outside the domain of physics, namely, aspects of how we come to know the world, with what we come to know about in the world, namely, quantum phenomena at the most fundamental level. This is done by first positing that the means by which we acquire knowledge about the world are the best we can ever achieve, due to limitations on what kinds of observers are allowed to evolve or be engineered in the world. Hence, we seek what physical conclusions follow from the integration of these two seemingly disparate aspects of our experience.

\appendix

 \appendix
\chapter{ Appendix: Review of Stochastic Processes and Ito Rules}

This review is partly based on readings of \cite{Percival1998}, \cite{pfeiffer1978concepts}, \cite{wiseman2009quantum} and \cite{VanKampen2007}.
A stochastic process is a collection of random variables indexed by time

\[
\{X_t : t \in T\}.
\]
The index t can be discrete or continuous.

Each $X_t$ is a random variable on some probabilty space $(\Omega,\mathcal{F},\mathbb{P})$.

$\Omega$ is the sample space. $\mathcal{F}$ is the event space, which is a collection of subsets of $\Omega$; its elements are called events and satisfy: 

$$\Omega \in \mathcal{F}$$ and if $$ A \in \mathcal{F}$$ then $$ A^c \in \mathcal{F}$$ (closed under compliments).

Moreover, For $A_1, A_2, A_3, \ldots \in \mathcal{F}$, we have
\[
\bigcup_{n=1}^{\infty} A_n \in \mathcal{F}.
\]

$\mathcal{F}$ specifies which events we are allowed to assign probabilities to, and $\mathbb{P}$ is the probability measure that assigns probabilities to events:\[
\mathbb{P}:\mathcal{F}\to[0,1]
\]
such that:
\begin{enumerate}
    \item $\mathbb{P}(A)\ge 0$ for all $A\in\mathcal{F}$,
    \item $\mathbb{P}(\Omega)=1$,
    \item if $A_1,A_2,A_3,\ldots \in \mathcal{F}$ are pairwise disjoint, then
    \[
    \mathbb{P}\!\left(\bigcup_{n=1}^{\infty} A_n\right)
    = \sum_{n=1}^{\infty} \mathbb{P}(A_n).
    \]
\end{enumerate}

A process is characterized by all joint distribution 

\[
(X_{t_1},\ldots,X_{t_n})
\]

for any $n\in\mathbb{N}$ and any times $t_1<\cdots<t_n$. In practice, we don't specify this sequence every time; we simply say, for example, that it's ``Gaussian'', and that determines the sequence stochastically.

A stochastic process $\{X_t : t\in T\}$ is a collection of random variables, and a single outcome $\omega\in\Omega$ produces one sample path \[
t \mapsto X_t(\omega).
\] Randomness chooses $\omega$, and we observe the realized path over time: in the discrete time case, this is \[
\bigl(X_0(\omega),\,X_1(\omega),\,X_2(\omega),\,\ldots\bigr);
\]
and in the continuous time case it is a random function $t\mapsto X_t(\omega)$.

Note that each $$\omega\in\Omega$$ does not correspond to a single value  $X_t$ at one time $t$, instead, it determines the entire trajectory of the process 
\[
\omega \mapsto \bigl(X_t(\omega)\bigr)_{t\in T}.
\]

A special case is the white noise. At each discreet time interval, we pick a value from S, with the same probability distribution with mean 0 and finite variance–independent of past states of $X$ (where S is the state space of the process; the set of all possible values that $X_t$ can take for every t).

In more general cases, for an important class of processes called the `discrete time Markov process', we have 
\[
X_{t+1} = F\bigl(X_t,\xi_{t+1}\bigr),
\]

where $\xi_{t+1}$ is a random input (the noise) and $F$ is any measurable function (i.e for the input $\bigl(X_t,\xi_{t+1}\bigr)$ it produces a well-defined random variable $X_{t+1}$ as the output). Here, $\xi_{t+1}$ is sampled from the space of noise realizations with a distribution (which is called the {distribution for the noise} )over that space, which is in turn derived from the underlying probability distribution $\mathbb{P}$ of the process. More precisely  $\xi_{t+1}$ are independently and identically distributed, and does not depend on $X_0$.

If the process is a Gaussian random walk in one dimension, for example, the function $X_{t+1}=F\bigl(X_t,\xi_{t+1}\bigr)$ takes the simple form $X_{t+1}=X_t + \xi_{t+1}$ where the increment $ \xi_{t+1}$ can be taken to be a Gaussian random variable, for example, in which case the noise distribution is a Gaussian distribution. $$\xi_t \sim \mathcal{N}(0,\sigma^2)$$ with mean 0 and variance $\sigma^2$. $\xi_t $ are independently and identically distributed. The marginal distribution and the conditional distributions, on the other hand, are the distribution for the state variable $X_t$, depending respectively on whether the distribution is marginalized or not with respect to previous state values. The conditional distribution is \[
\mathcal{L}\bigl(X_{t+1}\mid X_t=x\bigr)=\mathcal{N}\!\left(x,\sigma^2\right).
\]
And the marginal distribution is the distribution of $X_t$. Since \[
X_t = \sum_{k=1}^{t} \xi_k,
\] with 

\(\mathbb{E}[\xi_k] = 0\) and\textbf{ } \(\operatorname{Var}(\xi_k) = \sigma^2
\) we have 
\[
\operatorname{Var}\!\left(\sum_{k=1}^{t} \xi_k\right)
= \sum_{k=1}^{t} \operatorname{Var}(\xi_k).
\]
so \[
\operatorname{Var}(X_t)
= \sum_{k=1}^{t} \sigma^2
= t\,\sigma^2.
\]
The marginal distribution is \[
X_t \sim \mathcal{N}\!\left(0,\,t\sigma^2\right).
\]

In continuous time, we can consider the Wiener process and the Poisson process as examples.

In wiener process $W_t$, such as the one that describes Brownian motion, the infinitesimal noise increments are \cite{VanKampen2007}

\begin{equation}
dW_t \sim \mathcal{N}(0,\,dt),
\end{equation}

Or more accurately, for each small time step $\Delta t>0$, 
\begin{equation}
\Delta W_t := W_{t+\Delta t} - W_t \sim \mathcal{N}(0,\,\Delta t).
\end{equation}

In continuous time, the Wiener stochastic process is described by the diffusion equation of the form 

\begin{equation}
dX_t = a(X_t,t)\,dt + b(X_t,t)\,dW_t .
\end{equation}

In contrast, a Poisson process $N_t$ describes random occurrance of discrete events, at a constant average rate $\lambda > 0$, in continuous time. An example is the counting of photons by a photodetector. $N_t$ counts the number of events that has occurred up to time t. For any $\Delta t>0$, 
\begin{equation}
\Delta N_t := N_{t+\Delta t} - N_t \sim \mathrm{Poisson}(\lambda\,\Delta t).
\end{equation}
That is $\Delta N_t $ is Poisson distributed with parameter $\lambda\,\Delta t$ , i.e: \[
P(\Delta N_t = k) = \frac{(\lambda \Delta t)^k e^{-\lambda \Delta t}}{k!}, \qquad k = 0,1,2,\ldots
\]

For small time 
\begin{align}
\mathbb{P}(\Delta N_t = 0)
&= e^{-\lambda \Delta t}
 \;\approx\; 1 - \lambda \Delta t, \\[6pt]
\mathbb{P}(\Delta N_t = 1)
&= \lambda \Delta t\, e^{-\lambda \Delta t}
 \;\approx\; \lambda \Delta t, \\[6pt]
\mathbb{P}(\Delta N_t \ge 2)
&= O\!\left((\Delta t)^2\right).
\end{align}

To first order in $\Delta t$,
\begin{equation}
\Delta N_t \in \{0,1\}.
\end{equation}

which is the Bernoulli distribution: \begin{equation}
\Delta N_t \sim \mathrm{Bernoulli}(\lambda\,\Delta t).
\end{equation}

$\lambda\,\Delta t$ is the probability of one jump in the small interval of time. \\

In continuous time the Poisson process is described by the Jump Stochastic differential equation of the form 

\begin{equation}
dX_t = a(X_t,t)\,dt + c(X_{t-},t)\,dN_t .
\end{equation}
here $X_{t-}$ denotes the prejump value.

To manipulate equations of the form (A.10) and (A.3) involving the stochastic differentials $dW_t$ and $dN_t$, we need Ito's stochastic calculus, which is different from the familiar calculus for deterministic processes.

Consider the following stochastic equation with a noise term $ \eta(t)$

\[\dot{x}(t) = F(x,t) + \eta(t)
\]
If $ \eta(t)$ were to be a function just like $ F(x,t)$, then ordinary rules of calculus would apply. But $\eta(t)$ is not a function, but a distribution. Ito's stochastic calculus, developed by Kiyosi Ito in the mid-twentieth century, was intended to handle the changes and rates of change in quantities that depend on a noise term.

Stochastic differentials do not obey the ordinary rules of deterministic calculus.  

 A Wiener process,
\begin{equation}
\Delta W_t := W_{t+\Delta t}-W_t
\end{equation}
has variance \(\Delta t\), so a typical increment scales as 
\begin{equation}
\Delta W_t = O\!\left(\sqrt{\Delta t}\right).
\end{equation}
This is because, of the property of Gaussian distribution. Consider a distribution $Z\sim \mathcal{N}(0,1)$ . We can then write $\sqrt{\Delta t}\,Z $ to 
\begin{equation}
\sqrt{\Delta t}\,Z \sim \mathcal{N}(0,\Delta t).
\end{equation}
Therefore, since
\begin{equation}
\Delta W_t \sim \mathcal{N}(0,\Delta t),
\end{equation}
we may write
\begin{equation}
\Delta W_t = \sqrt{\Delta t}\,Z,
\qquad Z\sim \mathcal{N}(0,1).
\end{equation}
Equivalently,
\begin{equation}
\Delta W_t
=
\sqrt{\Delta t}\times
\text{a standard normal random variable}.
\end{equation}
Thus, 
\begin{equation}
\Delta W_t = O\!\left(\sqrt{\Delta t}\right).
\end{equation}
Therefore,
\begin{equation}
(\Delta W_t)^2 = O(\Delta t),
\end{equation}
which is of the same order as an ordinary increment in calculus that we are familiar with. This is why quadratic terms in \(dW_t\) cannot be discarded. By contrast, in ordinary calculus one would neglect \((dt)^2\) and all higher-order terms as they become vanishingly small in the infinitesimal limit. In It\^o calculus, the stochastic increment \(dW_t\) is large enough, in the scaling sense defined above, \((dW_t)^2\) contributes at order \(dt\), which we do not neglect in ordinary calculus.

In the infinitesimal increment limit, the fundamental It\^o multiplication rules for Wiener noise are
\begin{equation}
dt\,dt = 0,\qquad dt\,dW_t = 0,\qquad dW_t\,dW_t = dt.
\end{equation}
We also need to look at the chain rule for functions of stochastic processes.

 If \(f(x,t)\) is a `sufficiently' smooth function of the stochastic process \(X_t\), then the ordinary chain rule is replaced by It\^o's formula. To see that we retain the second-order term in the two variable taylor expansion for $f$:
\begin{equation}
df(X_t,t)
=
\frac{\partial f}{\partial t}\,dt
+
\frac{\partial f}{\partial x}\,dX_t
+
\frac{1}{2}\frac{\partial^2 f}{\partial x^2}\,(dX_t)^2 .
\end{equation}
And use the It\^o rule \((dW_t)^2=dt\) in A.3, so we have 
\begin{equation}
(dX_t)^2
=
\left(a\,dt+b\,dW_t\right)^2
=
b^2(dW_t)^2
=
b^2\,dt,
\end{equation}
because \((dt)^2=0\) and \(dt\,dW_t=0\). Therefore, we are left with the It\^o's formula 
\begin{equation}
df(X_t,t)
=
\left(
\frac{\partial f}{\partial t}
+
a(X_t,t)\frac{\partial f}{\partial x}
+
\frac{1}{2}b^2(X_t,t)\frac{\partial^2 f}{\partial x^2}
\right)dt
+
b(X_t,t)\frac{\partial f}{\partial x}\,dW_t .
\end{equation}
The additional second-derivative term is the characteristic correction of It\^o calculus. It won't be there in the usual calculus.

Now we'll quickly look at a Poisson process. Poisson processes are different because the noise consists of jumps rather than continuous fluctuations like in Wiener. If \(N_t\) is a Poisson process with rate \(\lambda\), then over a small interval \(dt\),
\begin{equation}
dN_t =
\begin{cases}
1, & \text{with probability } \lambda dt + O(dt^2),\\
0, & \text{with probability } 1-\lambda dt + O(dt^2).
\end{cases}
\end{equation}
To first order in \(dt\),
\begin{equation}
dN_t^2 = dN_t .
\end{equation} because $0^2=0$ and $1^2=1$
The other basic It\^o multiplication rules for Poisson increments are provided without proof
\begin{equation}
dt\,dt=0,\qquad dt\,dN_t=0
\end{equation}
The mean increment is
\begin{equation}
\mathbb{E}[dN_t]=\lambda\,dt .
\end{equation}

Now consider a jump stochastic differential equation of the form
\begin{equation}
dX_t = a(X_t,t)\,dt + c(X_{t-},t)\,dN_t,
\end{equation}
where \(X_{t-}\) denotes the value of the process immediately before a possible jump at time \(t\). If no jump occurs, then \(dN_t=0\) and the process evolves according to the deterministic drift term. If a jump occurs, then \(dN_t=1\), and the state changes discontinuously according to
\begin{equation}
X_t = X_{t-} + c(X_{t-},t).
\end{equation}
For a smooth test function \(f(x,t)\), the jump version of It\^o's formula can be obtained from the conceptual picture that  \begin{equation}
df
=
(1-dN_t)
\left(\text{increment in } f \text{ if no jump occurs}\right)
+
dN_t
\left(\text{increment in } f \text{ if a jump occurs}\right).
\end{equation}
So that's,
\begin{equation}
df(X_t,t)
=
\frac{\partial f}{\partial t}(X_{t-},t)\,dt
+
\frac{\partial f}{\partial x}(X_{t-},t)\,a(X_{t-},t)\,dt
+
\left[
f\!\left(X_{t-}+c(X_{t-},t),t\right)
-
f(X_{t-},t)
\right]dN_t .
\end{equation}
Unlike the Wiener case, the jump contribution is not expanded only to second order. A jump can be finite, so the full finite difference
$
f(X_{t-}+c,t)-f(X_{t-},t)
$
must be retained. More on stochastic processes can be found in Kurt Jacob's excellent text, which is very Physicist-friendly!

\chapter{Appendix: An Experimentalist and a Philosopher of Language: Dialog}

This dialogue with two of my committee members, one of whom is an experimentalist (Professor Szydagis) and the other is a philosopher of language (Professor Armour-Garb), can be viewed as representative of some common objections to the central constructions in this dissertation, and my responses. Some of these responses can also be found in various other contexts within the dissertation. 

\vspace{0.5cm}

\textsc{Albany}

\vspace{0.5cm}

\textbf{Professor Mathew Szydagis:} You speak of human language invariance, but I am extremely skeptical of this. For example, some languages exist without tenses for their verbs, as we understand them in English and other languages with similar verb conjugation. How does one refer to the past, present, and future in such languages? Also, what about older languages like Latin, which would not have words for any of the subatomic particles as we know them today?

There may exist alien races that evolved with very different languages and have a correspondingly very different view of the universe. See, for instance, the movie \textit{Arrival}. I am skeptical about how you can generalize non-anthropomorphically to all possible ``observers,'' including both biological entities and AI, as another good example.

\textbf{V.I:} My intuition was to take the meaning of language abstractly, not necessarily as human language. Moreover, the language I mean concerns experimental descriptions of what was directly observed in a laboratory, not everyday communication. I mean things like, ``the Geiger counter clicked at 5:04:48 pm.''

And the invariances I argue for in language are precisely the core of the argument: no matter the culture, natural or artificial intelligence, terrestrial or alien intelligence, there are certain physically relevant invariant aspects of language and communication that describe experimental descriptions. What is invariant are the epistemic constraints that were identified, and the operational meaning of physical quantities.

The description of the experimental protocol to measure the spin of an electron should be unambiguous, and an alien, or an AI, or anyone from any culture witnessing what is being performed, would agree with the statement that ``the spin of the electron was measured,'' though the specific language, grammar, vocabulary, and the nature of embodiment of the cognition may differ.

What it means operationally to measure a specific physical quantity should be universally recognizable as belonging to at most an equivalence class of operations that correspond to the measurement of that physical quantity, irrespective of the embodied cognition that performs the operations or the language that is used to describe those operations.

There will also be invariances reflecting the epistemic constraints that I have posited to apply to any physically possible observer: experimental propositions will be found to satisfy semantic bivalence and universal intersubjective agreeability, regardless of the particular language in question or the particular nature of the observer's embodied cognition.

There will never be a situation, I argue, either among observers here on Earth or among all possible observers (artificial, terrestrial, or extraterrestrial) where an experiment was done on a system with a definite outcome occurring with respect to one observer, but no definite outcome occurring with respect to another observer for the same system. That is, there will never be a situation where a system undergoes a unitary transformation with respect to one observer while undergoing a quantum measurement with respect to another observer.

A similar impossibility argument can be extended to the other epistemically constrained situations I identified, namely subsystem partitioning.

I will now address the specific linguistic questions you raised.

There are languages that exist without tenses, but they can still distinguish between an event that occurred yesterday and an event that occurred today. The distinction is between how tenses occur grammatically, which varies across languages, and how they arise semantically, which is invariant across languages.

The same applies to older languages. It is true that they had no vocabulary for an alpha particle. But that is a matter of lacking vocabulary, not a difference in how a Latin speaker would account for an experiment involving an alpha particle. They would still agree on the classical records, such as, ``I heard a clicking sound from this device at this time.'' The device in question could be the Geiger counter for us. But for them to interpret what the click sound means, they would have to learn quantum mechanics.

And as for aliens, if they exist, they may have evolved differently, but the argument is that, just as mathematics would be the same for them, the operational meaning of physical quantities and the epistemic constraints would also be the same for them. They would still agree that an experiment measuring the spin of an electron and one measuring the polarization of a photon are different semantically and operationally, though the specific language, grammar, vocabulary, and nature of the embodiment of their cognition, in terms of which they express the same difference, may differ. And if we and they are witnessing the same experiment, we would still agree that the epistemic constraints hold for us mutually.

\textbf{Professor Mathew Szydagis:} You spend very little time, only a couple of pages here and there from what I have read so far---I have been jumping around and have not completed your text---on consciousness or intelligence. You say it is not needed, but you never try to define it. Impossible, I know, so I mean an operational definition for it, at least as you do with time, for example.

However, this has caused, as I am sure you know, a lot of concern in the past in quantum mechanics, and a lot of philosophical debate and argumentation. It seems, to me at least, like this is an elephant in the room that you pass over a little too quickly.

\textbf{V.I:} Yes, I agree. I have spent very little time discussing consciousness, and I think there were two reasons for that.

First, it seemed that an adequately sophisticated robot running a large language model in its processors, perhaps like Tesla's Optimus Robot, might very well be able to carry out at least simple quantum mechanical experiments on its own, in our absence. It seems that the consensus and evidence so far is that such robots are not yet conscious, though we can never be sure about this. That was one reason I defined ``observers'' as systems that use language, reason correctly, and navigate an environment, while remaining agnostic about consciousness.

The second reason was that a clear definition of consciousness is apparently lacking among cognitive scientists, even though it is something we as individuals can be most sure of experiencing, more than anything else. It seemed to be a problem of its own. 

To address whether language-using systems are also inevitably conscious, I think the key question is this: We can have phenomenal experience without language---toddlers and some animals probably have this phenomenon---but can we have autonomous language-using systems without phenomenal experience? If the answer to that question is yes, then I think consciousness is not needed.

\textbf{Professor Mathew Szydagis:} The universe got along OK with only fictitious observers for billions of years. :) That is something I always tell folks who claim to me that consciousness is required for the universe to exist at all, with wavefunctions collapsing into definite states all over the place. I believe this means you and I are very well agreed that ``counterfactual'' observers are all that are needed to discuss an objective reality, just like with counterfactual clocks---the main thrust of your thesis.

\textbf{V.I:} Yes.

\textbf{Professor Mathew Szydagis:} AI does lead us to a Wigner's-friend-style conundrum: if a non-conscious robot makes the measurement, is it not yet ``done'' until a human, or alien, etc., checks in? Can a cat, dog, or another ``lower'' animal be an experimenter in place of the robot?

\textbf{V.I:} As for the Wigner's-friend-style conundrum, for the experiment done under ordinary conditions, we can, I think, safely assume that the experiment was ``done'' long before a conscious experimenter looked at the result, as long as a counterfactual fictitious observer could be posited consistently to have already witnessed the result. 

This is already evident in our labs. Let us say we had an experiment designed to capture a cosmic ray in a bubble chamber, which is enclosed in a lab with no humans. Let us assume we have cameras attached to the setup. The cameras can capture particle tracks and the timestamps at which they formed. Let us say the computer time-stamped a very rare particle track of a rare particle species on June 30th at 7:43:52 AM. And let us assume a human experimenter came into the lab on August 30th after a long vacation. He can now very well assume the event happened on June 30th at 7:43:52 AM, without having to assume it happened on August 30th, and there is no contradiction in doing so. I think we should inform peers that the event occurred on June 30th at 7:43:52 AM, not on the day we entered the lab.

As for the question concerning whether simpler intelligent systems---dogs, cats, insects, plants, and so on---qualify as ``observers,'' I think a response can take an analogy from clocks: would we say a pendulum is a clock? Would we say a waterfall is a clock? Would we say a piece of rusting metal is a clock? I think yes, they do, but they are not ideal clocks because they do not have enough precision and do not have the ability to count arbitrarily long intervals of time. A pendulum may count the equivalent of 30 seconds, but not 30 years unless we deliberately engineer it to do so. We can call such clocks ``non-ideal clocks.'' Nevertheless, despite their limitations, in the immediate vicinity of such clocks, we can still say ``time passes,'' and we can talk about a spacetime metric, because we can always imagine placing a counterfactual clock of arbitrary precision close to that ``non-ideal clock.''

The point is that, however non-ideal a system may be at functioning as a clock, we can nevertheless talk about time in its vicinity with arbitrary precision, given the aforementioned counterfactual possibility of a high-precision fictitious clock positioned next to it.

I expect a similar argument to apply to the functionally defined observers. In fact, I can define the analog of a ``unit'' of a clock for observers: the completion of the simplest quantum-mechanical experiment, say involving a two-level system, like a photon, that started in a standard initial state \(\ket{0}\), evolved for 1 second under free evolution, and was probed at the end of 1 second for some fixed observable. Clearly, just as the unit of time is chosen arbitrarily, this unit is also chosen arbitrarily.

An insect, analogous to a non-ideal clock, can conceivably detect photon polarization using its compound eyes, thereby completing a simple quantum experiment. But just like a non-ideal clock, it has no sophistication to ``mark'' the completion of a simple quantum experiment. Nevertheless, we can consider the insect to be an observer if we can counterfactually position a sufficiently sophisticated counterfactual observer with language abilities to probe the insect and ``mark'' the completion of the simple quantum mechanical experiment.

So, simple intelligent systems, such as insects, cats, and plants, can be considered non-ideal experimenters to the extent that they consistently allow for the consideration of counterfactual fictitious observers who could ``mark'' the completion of the experiment on their behalf.

\textbf{Professor Bradley Armour-Garb}: I have a question about your alien thought experiment.
As I understand your thought experiment, you use it to argue that when we build epistemic
constraints into the foundations of physics, rather than trying to derive them from quantum
mechanics, we are not being anthropocentric; we are identifying genuinely universal features of
physical knowledge.

Your ``alien thought experiment'' appears in the dissertation as an argument for why epistemic
constraints must be treated as pan-theoretic, universal across all possible physics, rather
than anthropocentric.

You ask us to imagine discovering an alien civilization that has independently developed
physics. Your thought experiment proceeds as follows:

\begin{enumerate}
    \item Premise: These aliens have different sensory modalities, different evolutionary histories,
    different cognitive architectures, and presumably different languages than humans.

    \item Claim: Despite these differences, if the aliens have developed what we would recognize as
    physics---that is, systematic empirical knowledge of the natural world---then their physics must share certain structural features with ours.

    \item Key structural features:
    \begin{enumerate}
        \item Experimental propositions must have definite truth values, that is, semantic bivalence.

        \item Different alien observers must be able to agree on experimental outcomes, that is, intersubjective
        objectivity.

        \item Physical quantities must be operationally definable in ways that could, in principle, be
        communicated.

        \item The ``meaning'' of physical experiments must be expressible in some representational system,
        their analog of ``language.''
    \end{enumerate}

    \item Conclusion: These features---the epistemic constraints---are not accidents of human cognition
    or language, but are necessary conditions for \textit{any} possible physics. They are therefore
    candidates for being treated as fundamental physical principles rather than derived phenomena.
\end{enumerate}

Concern 1: Your thought experiment assumes that any beings who ``do physics'' must satisfy
epistemic constraints, including propositional representation and semantic bivalence. But this
seems to define ``doing physics'' in terms of those very constraints. That seems problematic to me.

Concern 2:
Consider an alternative: imagine beings who interact with quantum systems in sophisticated,
adaptive ways---predicting outcomes, manipulating systems, building technology---but who
process information non-propositionally, perhaps through something like distributed analog
computation without discrete symbolic states.
By your definition, they do not ``do physics'' because they lack language. But they do everything
physicists do except talk about it.

Questions:
\begin{enumerate}
    \item Does this not show that your ``epistemic constraints'' are really constraints on discourse
    about physics rather than on physical knowledge itself? And if so,

    \item Why should constraints on discourse be built into fundamental physics rather than
    treated as features of one particular linguistic way of engaging with the world?
\end{enumerate}

Reminder: Your argument has the following structure:

\begin{enumerate}
    \item Imagine aliens who do physics.

    \item They must have epistemic constraints, including language-like representation.

    \item Therefore, epistemic constraints are universal, not anthropocentric.
\end{enumerate}

My question challenges step 2 by proposing a symmetric counter-thought-experiment:

\begin{enumerate}
    \item Imagine aliens who discover atomism, or quantum mechanics.

    \item They do so \textit{without} language, in any sense we would recognize.

    \item Therefore, the link between physics and language is not necessary.
\end{enumerate}

If this counter-scenario is equally imaginable, the original thought experiment loses its force. It
becomes a battle of intuitions about what is ``really possible,'' with no clear winner.

\textbf{V.I}:   You are right to suspect, in your Concern 1, that my reasoning has the following pathological structure: ``All red-colored dots are red and have the property of being dots.'' But really, I think the reasoning has the structure of ``All apples are red,'' where the ``apple'' and the ``red'' are not identical to each other.

Let me explain. I do indeed assume that any being who ``does physics'' must satisfy the epistemic constraints, but that does not necessarily mean I have defined doing physics solely in terms of those very same constraints. Consider another principle with the same structure: ``All physical processes must satisfy the law of conservation of energy.'' This statement does not necessarily mean that we have defined physical processes in terms of the law of conservation of energy. The law of conservation of energy is one regularity that applies to any physical process, but there is more to a physical process than the law of conservation of energy, such as the law of conservation of momentum or the law of increase of total entropy. In a similar fashion, epistemic constraints are a regularity of any being ``doing physics,'' but there is more to doing physics, such as spacetime transformation laws among observers in different frames, and conservation laws relating the values of physical quantities before and after a process, than the epistemic constraints alone. \\

That is, ``doing physics'' is defined by epistemic constraints and other law-like constraints relating the observations of one or more observers, as opposed to being defined by epistemic constraints alone. It is also defined by the specific physical system that we may be considering in a particular experiment, such as a pendulum or an electron. This means that the meaning of ``doing physics'' is not identical to ``those that satisfy epistemic constraints.''

As for your second concern, regarding ``beings who interact with quantum systems in sophisticated, adaptive ways---predicting outcomes, manipulating systems, building technology,'' I think there is still an unavoidable role for language. If not, what do we even mean by ``predicting''? (A similar argument can be given for ``manipulating systems'' and ``building technology.'') After all, by predicting we mean that we have a symbolic system representing a phenomenon, and for certain values of the quantities represented in the symbolic system, we expect certain other quantities to realize a predeterminable value. All of this involves language.

Now I will come to your questions. My response to the first question is that this is where ``discourse about physics'' and ``physical knowledge itself'' begin to overlap. Can we think of an instance of physical knowledge that is not a discourse on physics? I think not, because any physical knowledge appears to be expressible as a declarative sentence. While that sentence may not yet be communicated to other observers, and thus may not yet enter discourse, it is in principle part of a discourse eventually. That is, there is no private meaning for declarative sentences corresponding to physical knowledge.

My response to your second question is that I am taking the constraints on discourse to be ``built into fundamental physics'' as a principle that I posit for physics-doing systems, but one that is supported by every known instance of intelligent engagement with the physical world. If there is a demonstrable counterexample, I am willing to revise my belief in such a principle, but it seems we lack a single counterexample. This is similar to how we posit any other principle in science: we cannot mathematically prove the principles we posit, but we nevertheless posit them based on regularities in nature, and they remain supported as long as we do not encounter counterevidence.

As for your counter-thought experiment, I think the issue lies in your second step. What do we mean by a theory? Can we even have a meaningful theory without language? I think we cannot, since a theory can only be expressed in some symbolic system. Now, if a theory cannot be sensible without language, I think it cannot be sensible to ``discover'' a theory, such as quantum mechanics, without language.

Moreover, the counter-thought experiment you suggested, concerning the possibility of doing physics without language, does not correspond to any known means of acquiring knowledge about the physical world.  This relates to my answer to your second question. That is, there is no known instance of an observer who acquires knowledge of physics through that means. Therefore, based on the existing evidence for how we acquire knowledge about the world—that is, how we do physics—we can postulate that this is the only way knowledge acquisition can happen in nature. However, this is not a deductive argument. It is similar to how we defend energy conservation. One can propose a hypothetical machine that violates energy conservation and runs perpetually, but the only way to defend against this is by saying that we lack a real instance of such a hypothetical machine.

\textbf{Professor Mathew Szydagis:} Why can't I measure the lonely supernova proton? A moving charge creates a B-field, so I can have a counterfactual B-field detector near it. What am I missing? You already mentioned a fictitious magnetometer earlier, in fact.

\textbf{V.I:} I think in such situations we cannot position the counterfactual experimenter because it would conflict with Principle 2: namely, such a placement near a system should not affect the future normal course of the system. Unlike in a classical case, for a supernova proton, positioning such a counterfactual experimenter would change the experimental context, in Bohr's sense, and may count as a detection already. It would also have obstructed the course the proton might otherwise have taken. More importantly, the undetected proton could exhibit quantum interference when it is observed later in its course, but if we imagine placing a counterfactual experimenter en route, we may never observe the interference we would otherwise have observed.

None of this applies to a space rock, for example. The presence of a fictitious observer co-moving with the rock, and alignment position measuring it, could be consistently assumed. Such a consideration would not significantly affect the otherwise normal course of the rock.

\textbf{Professor Mathew Szydagis:} Regarding the proton, that does remind me of the delayed quantum eraser on cosmological scales, which I do not recall if you mention in your thesis or not. I have a big-picture question, then: why would a counterfactual experimenter not always ``mess up'' a particle's future trajectory by causing a premature collapse of the wavefunction? My huge issue here also applies to the universe prior to the existence of any experimenters, minds, or languages, for billions of years.

``All this does not arise for a space rock, for example.'' This begs the question: at which scale do quantum effects matter? You do speak of a mesoscopic scale in your thesis already. Also, I am not sure what ``not significantly'' means. Are you referring to a ``weak'' quantum measurement here?

\textbf{V.I:} As suggested in Principle 2, the assignment of counterfactual experimenters is only allowed when it stays consistent with later observations. It cannot be assigned to a coherently prepared quantum system, for example. So a counterfactual experimenter does not always collapse the wavefunction prematurely, because in such cases we cannot assign counterfactual experimenters in the first place. I think this does relate to Wheeler's intuition, but I think the overlap is not exact.

In response to your concern about the ``cut-off,'' I have taken a stance similar to Bohr's. The ``cut-off'' is not determined by mass or particle number, but by the experimental context. Namely, there can be experimental contexts where even a big object like a rock is in a quantum state, but we can certify this only in the context of the experimental arrangement. Unlike the classical intuition, at least in Bohr's reading of quantum theory, which I adopt, systems do not come ``attached with a state''; rather, a state is assigned in the context of the experiment. 

To answer your second question, by ``not significantly,'' I mean non-invasive measurability---the ability to determine the state of a system without affecting the future course of evolution of the state of the system. This is an idealization even in classical physics, but in quantum systems, as we know, in addition to the noise from measurement-device random errors, which we can call measurement imprecision and which is true even in classical physics, the very act of measurement changes the state of the system, which can be called measurement back action.

So when I say ``not significantly,'' I am referring to the situations involving the determination of the state of the system under study, when measurement imprecision may still be there, but measurement back action is negligible when compared to classical measurement imprecision

\textbf{Professor Mathew Szydagis:} I am still hung up on Wigner's friend: is the wavefunction ``collapsed,'' Copenhagen-style, after the person in the room makes the measurement, and that is that? The friend, being a conscious being capable of language, cannot be in a state of superposition?

\textbf{V.I:} There are actually two possible scenarios to discuss. This can also be seen on Slide 28 in my upcoming Irvine talk. If the friend observes a definite outcome on his detector screen, then the experiment is also completed for Wigner. This is a consequence of Principle 1. In such cases, the friend, an observer capable of language, is not in any quantum superposition state.

However, there can arise circumstances, such as after sufficient shielding of the friend+detector system, when the friend can indeed become quantum-entangled with the detector. In such circumstances, however, the friend will stop functioning as an ``observer'' and will have no experience nor definite spacetime existence for the duration when he is in a quantum state. I have to impose this conclusion to ensure consistency with quantum mechanics and principle 1.

\textbf{Professor Bradley Armour-Garb}: You make two claims, and I am having trouble seeing how both claims can be true.

Claim 1 is that epistemic constraints are strongly emergent---irreducible to physics.

Claim 2 is that consciousness is not required because purely physical robots can be genuine
observers.

Questions:

\begin{enumerate}
    \item If a robot can be a full observer solely in virtue of its physical organization, then the
    epistemic constraints are fully realized by that physical organization. In what sense,
    then, are they irreducible?

    \item What blocks even an in-principle derivation of the constraints from complete physical
    knowledge of the robot?
\end{enumerate}

It seems like you face a trilemma: either (1) epistemic constraints require something nonphysical,
which would explain their irreducibility but which you deny; or (2) they are reducible to
the physics of systems that instantiate them, which you also deny; or (3) there is a third option I am
not seeing. What is that third option, and how does it escape both reducibility and nonphysicalism?

\textbf{V.I}: I think first I must clarify two priors, so that we can either agree on the definitions or agree to disagree on the definitions. First, it appears that by the word ``physicalism,'' you might be picturing the Newtonian ideal of rigid bodies, point particles, mechanical devices, or, more generally, ``stuff out there.'' But that is the definition of physicalism I have challenged. Physicalism, in this thesis, has always been the communicable and reproducible operations that observers can perform in the world. The older ideal of ``mechanical things out there'' can be accommodated within this more general reinterpretation of physicalism as a special case, because the meanings of physical quantities are nevertheless accessed by operations whose descriptions are communicable and reproducible.

So this also affects what ``reducible to physics'' means: a phenomenon is reducible to physics if the following is the case. The phenomenon can be described at a higher level by operationally accessing its properties and how they change. Meanwhile, there exists a lower-level phenomenon that can also be operationally accessed by accessing a different set of fine-grained properties and how they change. If we can now deduce the former phenomenon from the latter, then we have a reduction to physics. This works well in, say, explaining how a biological cell works: we can describe how the cell behaves, and we can also describe how the parts of the cell behave, and from knowledge of the latter we can infer the former to a good approximation. But epistemic constraints are not themselves reducible in that way, because in every such attempted reduction, we have epistemic constraints applicable both in the high-level description of the epistemic constraints themselves and in the lower-level description, say, of the brain's parts. I hope this addresses your two questions.

Second, the two claims you just highlighted in my argument do not necessarily refer to the same sort of strongly emergent feature. There are two supposed strongly emergent features being considered here, not one. Epistemic constraints concern language and communication, which are used to describe and communicate physical occurrences. Consciousness concerns phenomenal experience. I have deliberately left open the question of whether the two supposed strongly emergent features really refer to the same strongly emergent feature. As I pondered earlier, ``There can be phenomenal experience without self-directed language processing (animals do this all the time, and we do it too when picturing objects without attempting to articulate what they are); can there be self-directed language processing without phenomenal experience?'' If future research shows that autonomous language use always goes hand in hand with phenomenal experience, we would have to conclude that epistemic constraints and phenomenal experience refer to the same sort of strongly emergent feature. If that is the case, then Claim 1 holds and Claim 2 becomes false. If that is not the case, then Claim 1 and Claim 2 can both be true, provided we understand ``purely physical'' in the sense I have specified, because epistemic constraints are nevertheless a different kind of strongly emergent feature, not the same as phenomenal experience.

As for the trilemma, I think I must first clarify that I am not positing anything ``nonphysical''; I have simply defined ``physical'' more carefully. So I think the trilemma does not arise in the first place. If I must make a choice, I think I will choose your option 3 in your trilemma: redefine what physical even means, and thus what ``reducible to the physical'' means. With that redefinition, epistemic constraints can both be physical and non-reducible. 

Additionally, regarding physicalism, I agree that the way I have defined it may not align with how philosophers define physicalism. I think there is a long tradition stretching back to Descartes that has shaped philosophers’ definition of physicalism as res extensa. However, in recent times, there have been views—especially those influenced by Karen Barad—that have developed a position called onto-epistemology. I find this to be a more accurate description of the post-quantum picture of physical reality, in contrast to the picture given to us long ago by Descartes. I think this was better suited for Newtonian mechanics.

Epistemic constraints are physical because they are instantiated only in physically realized observers, such as humans, robots, or other possible observer-systems. But they are non-reducible because any attempted derivation of them from lower-level physics already presupposes the very epistemic conditions under which lower-level physical descriptions are formulated, communicated, and verified.

I think, in contrast to the redefinition merely being an escape from the issues you raised, that the redefinition arises from necessity, since it provides a more faithful account of how we actually do physics and how we reduce higher-level descriptions to lower-level ones.

\textbf{Professor Bradley Armour-Garb}: 

Your framework has a structural circularity that you acknowledge, but I wonder whether you fully resolve it. Here is the circularity, as I understand it. Please correct me if I misunderstand.

\begin{enumerate}
    \item Actualities are defined in terms of what a counterfactual observer could passively measure \textit{via} coincidence position measurement.

    \item Observers are defined as functionally specified systems with classical input/output channels involving actualities.

    \item The applicability of quantum mechanics---state assignment, unitary determination, and Born rule application---depends on the actuality/potentiality distinction.

    \item \textit{But determining whether something is an actuality} requires knowing whether a counterfactual observer could measure it, which requires knowing what observers are, which requires knowing what actualities are...\textit{ad infinitum}.
\end{enumerate}

You try to break this circle by treating observers as ``functionally defined primitives,'' like clocks. That is, you compare observers to clocks, suggesting that both can be taken as primitives. But it seems that there is a crucial disanalogy.

Clocks seem to admit of a purely physical characterization: just any periodic process. No semantic notions are required here. Your observers, by contrast, are defined by ``language-using capability'' and the production of ``meaningful classical information.'' These are semantic notions---they seem to involve things like reference, truth, and meaning.

As a result, it looks like you are faced with a dilemma: Either you can give a non-semantic characterization of observers---in which case the question arises as to what distinguishes them from thermostats and other systems that merely correlate with their environment---or you are committed to semantic properties being physically primitive, in which case you are claiming that meaning is as fundamental as mass or charge. This is a radical metaphysical commitment that requires substantial defense.

Which horn do you take? Or do you think that the dilemma itself does not really arise?

And is there not a worry that the measurement problem is precisely about how semantic content---determinate outcomes that mean something---arises from physical interactions? By taking semantic observers as primitive, are you not presupposing what needs to be explained?

\textbf{V.I}: I believe there is no circularity if we posit three kinds of primitives: actualities, potentialities, and observers, each of which is needed to define the others. Only when we demand one kind of primitive does the circularity, or problem of infinite regress, seem to arise. I think such a demand for one kind of primitive is not fully supported by experience, as I have argued throughout the thesis.

I think the disanalogy with clocks also does not arise: a clock is not simply ``any periodic system''; a clock is a periodic system along with a mechanism to count the increments in distinguishable periodic events. This counting mechanism does bring in semantic notions: the mechanism must be able to distinguish one event from another, represent each event as an elapsed instant, and increment how many such instants have elapsed. The clock does not need full-blown semantics involving propositions and language. But it does require a rudimentary mapping: a state transition in the periodic system \(\rightarrow\) an elapsed temporal interval marked as an elapsed count value recorded in some symbolic system, perhaps the Roman numeral system.

I think I choose the latter horn: to take semantic properties as physically primitive, albeit within the more carefully defined sense of ``physical'' that I described earlier. Additionally, I would not say their primitiveness is in the same sense as charge or mass, which we understand as temporally persisting entities. Rather, the primitiveness of semantic properties is of the same nature as the primitiveness of time: both are background preconditions for talking about phenomena and are defined in terms of their respective functionally defined units.

Finally, you are right: the measurement problem, as it is usually framed, does indeed seek to explain how semantic content---determinate outcomes that mean something---arises from physical interactions. But my target of analysis was not this original framing of the problem. I have rearranged where the problem itself lies, in the Bohrian Program. In this program, explaining semantic content is not the goal, but rather a feature to be accommodated alongside quantum theory. The target of analysis is what physical conclusions follow from such a joint consideration.

\textbf{Professor Mathew Szydagis\footnote{This comment is fictional}:} What counterevidence could I give you that might make you change your mind and accept, say, the Everettian view or the objective collapse theories?

\textbf{V.I:} Without getting into the Everettian view or objective collapse theories, I would change my mind to a more neutral stance, namely the ``shut up and calculate'' view, if the following counterevidence were presented to me:
\begin{enumerate}
    \item  Show me that there is an instance of a completed quantum measurement---where I define ``complete'' by the performance of an alignment measurement of a numerical quantity that results in knowledge of an eigenvalue---that can be done solely by an experimentally implementable unitary transformation that can also be reversed.
\item Show me an instance where one observer reported a definite alignment measurement value concerning a system's property \(P\), but for another observer, that same system's property \(P\), at the same time, is represented by a quantum superposition state with which they can demonstrably perform interference experiments. And these two observes must be able to meet later and corroborate their differences. After all, that's what we can do in the case of relativity, which is used as an inspiration for perspectivalism—two observers can later meet and verify their differences on the assignment of simultaneity, for example. In the absence of such corroboration to verify the difference, I think any claims for perspectivalism in quantum mechanics is physically vacuous even when philosophically appealing.

\item Show me that there can be an observer who can use language and cognize, who is fully engineered by a quantum state undergoing however complicated unitary transformations.

\vspace{0.5cm}
\end{enumerate}

\par

\vspace{2cm}

\begin{center}
    \large \textit{The End.}
\end{center}

\let\cleardoublepage\clearpage

\bibliographystyle{plain}

\bibliography{bibliography}

\end{document}